\documentclass[twocolumn]{aastex631}

\accepted{\today}

\submitjournal{ApJS}

\usepackage{graphicx}
\usepackage[utf8]{inputenc}
\usepackage{txfonts}
\usepackage{xcolor}
\usepackage{hyperref}
\usepackage{longtable}
\usepackage{comment}

\def\ha{H$\alpha$}
\def\hb{H$\beta$}
\def\mg{Mg{\sc ii}}
\def\rfe{R$_{\rm FeII}$}
\def\lagn{L$_{\rm 5100\AA}$}
\def\lagnuv{L$_{\rm 3000\AA}$}
\def\lagnir{L$_{\rm 6000\AA}$}
\def\feii{Fe{\sc ii}}
\def\mbh{M$_{\rm BH}$}
\def\lbol{L$_{\rm bol}$}
\def\ledd{$\lambda_{\rm Edd}$}
\def\msun{M$_{\odot}$}
\def\kms{km s$^{-1}$}

\hypersetup{linkcolor=blue,citecolor=blue,filecolor=blue,urlcolor=blue}

\shorttitle{CL-AGNs and Main Sequence of Quasars}
\shortauthors{Panda \& \'Sniegowska}

\begin{document}

\title{Changing-Look AGNs - I. Tracking the transition on the main sequence of quasars}

\author[0000-0002-5854-7426]{Swayamtrupta Panda}\email{spanda@lna.br}\thanks{CNPq fellow}
\affiliation{Laborat\'orio Nacional de Astrof\'isica - MCTI, R. dos Estados Unidos, 154 - Na\c{c}\~oes, Itajub\'a - MG, 37504-364, Brazil}

\author[0000-0003-2656-6726]{Marzena \'Sniegowska}
\affiliation{School of Physics and Astronomy, Tel Aviv University, Tel Aviv 69978, Israel}
\affiliation{Nicolaus Copernicus Astronomical Center, Polish Academy of Sciences, Bartycka 18, 00-716 Warsaw, Poland}

\begin{abstract}

This paper is the first in a series of preparing and analyzing spectral and other properties for a database of already discovered CL AGNs. Here, we focus on the spectral fitting and analysis of broad emission lines in a sample of 93 CL AGNs collected from the literature with existing SDSS/BOSS/eBOSS spectroscopy where the \hb{} emission line profile does not completely disappear in any epochs. Additionally, we have gathered older/newer spectral epochs from all the available SDSS data releases to make the database more complete. We use {\sc PyQSOFit} and perform a homogeneous spectral decomposition of all of our SDSS spectra and tabulate the AGN continuum and emission line properties per epoch per source, chronologically. This further allows us to categorize the sources in our sample as \textit{Turn-On} or \textit{Turn-Off} and subsequently check for repeated occurrences of such phases. We then estimate the black hole mass (\mbh{}) and the Eddington ratio (\ledd{}) per epoch per source where the required parameters are available and well-estimated. We realize the movement of the source in the \mbh{}-\ledd{} plane allowing us to check for systematic changes in the source's fundamental properties. We then track their transition along the optical plane of the Eigenvector 1 (EV1) schema and categorize sources that either stay within the same Population (A or B) or make an inter-population movement as a function of spectral epoch. We also test the Balmer decrement (\ha{}/\hb{}) of a subset of our sample of CL AGNs as a function of time and AGN luminosity.

\end{abstract}


\section{Introduction} \label{sec:intro}

Variability is among the prominent traits shown by active galaxies - one that ranges from a few seconds to a few tens of years \citep[see e.g.,][for review]{Ulrich_1997, Cackett_2021}. This feature is ubiquitous across the demographics of active galactic nuclei (AGN) and is observed and studied across a wide range of energies - in X-rays \citep[e.g.,][]{Alston_2020, Kammoun_2021}, optical \citep[e.g.,][]{Blandford_McKee_1982, Gaskell_1986}, and infrared \citep[e.g.,][]{Koshida_2014, Minezaki_2019}, to name a few. This crucial feature exhibited by these luminous sources has led to the estimation of the black hole masses for hundreds of AGNs using the technique of reverberation mapping \citep[e.g.,][]{Blandford_McKee_1982, Peterson_2004, Cackett_2021} supplemented with single/multi-epoch spectroscopy \citep[e.g.,][]{Kaspi_2000, Bentz_2013, Shen_etal_2019, Du_2019, 2023arXiv230501014S}.\\

Another aspect of AGNs is the broad range of spectral features they show across wavelengths. Following the AGN unification model \citep{Urry_Padovani_1995}, phenomenologically AGNs can be categorized based on whether the distant observer has (or has not) an unimpeded view of the central engine. If they do, then such sources are known as Type-1 AGNs, while the remaining are classified as Type-2. A primary feature of the Type-1 AGNs is the presence of both broad and narrow emission lines in their spectra, whereas Type-2 AGNs show only narrow lines \citep[see e.g.,][]{Weedman_1977, Antonucci_1993, Padovani_2017}. From the literature, we are aware that the intermediate AGN types (e.g., 1.5 or 1.8, \citealt{Osterbrock_1981}) are primarily based on the gradual weakening of the broad component of the H$\alpha$ and H$\beta$ emission lines in the spectra \citep[see also][]{Runco_etal_2016, Mickaelian_2021}. Such static classification based on the study of different AGNs and, more recently, with continuous follow-up of single sources, have blended. These changes to the broad features have been attributed to intrinsic changes in the accretion rate of the central supermassive black hole \citep{Elitzur_2014, lamassa2015, Guolo_2021, Mehdipour_2022}, apart from viewing-dependent obscuration - where passing dust clouds cover the line-of-sight to the emission region of the BEL, resulting in a decrease and reddening of flux \citep{Risaliti_etal_2009, Sheng_2017, Gaskell_2018, Wang_etal_2019, Yang2019} and, more recently, due to tidal disruption events \citep{Holoien_2016, Kool_2020, Short_2020, 2021A&A...656A..47P, 2023A&A...669A.140P}. However, the observed changes in the \hb\ broad emission line (BEL) can often not be explained by obscuration alone \citep{macleod2016, Sheng_2017, Marin_2020}.\\

However, we already know objects, which change their spectra significantly at different epochs and, consequently, the AGN type \citep[see e.g.,][]{Hon_etal_2022}. Noticeable differences in the spectra behavior (i.e. changes of continuum shape or changes in broad component intensity) have been noticed in a number of the AGN over the years. Stochastic variations that are relatively small in continuum and emission line fluxes are common in AGN (e.g., \citealt{MacLeod_etal_2010}, and references therein). On the other hand, a significant variation in continuum flux and the disappearance or appearance of BEL are also possible. The latter would effectively change the AGN type from type-1 to type-2 and vice versa.\\

The term `changing-look' AGN has become quite common in optical/infrared spectroscopy although its origin comes from X-ray astronomy where large changes in X-ray luminosity are typically associated with varying absorption, e.g. \citet{Matt_etal_2003} and \citet{Rivers_etal_2015b, Rivers_etal_2015a}. Changing-look quasars are relatively rare, but a notable well-researched example from the last 40 years is NGC 4151 \citep[a nearby AGN, e.g.,][]{Osterbrock_1977, Antonucci_1983, Penston_1984, Lyutyj_etal_1984, Shapovalova_etal_2010}, which has been observed ‘flickering’, as it transitioned back and forth between classification types. More recently, many authors have made use of the term `changing-state' AGN to classify sources attributing their cause to dimming/brightening of the AGN continuum - leading to a reduced supply of ionizing photons available to excite the gas in the immediate vicinity around the black hole \citep{lamassa2015, graham2020, Sanchez-Saez_etal_2021, claudio_benny}.\\

In the last decade, the number of changing-look/changing-state AGNs discovered has seen a steady rise owing to large, repeating spectroscopic campaigns, such as the multiple data releases of the SDSS \citep{lamassa2015, Runnoe2016, macleod2016, ruan2016, Yang2018, graham2020, potts2021, Green_etal_2022}. Previous works have made significant efforts to classify the origin behind the `changing-look' phenomena \citep{lamassa2015, Stern_etal_2018, Noda_Done_2018, Sniegowska_etal_2020, Begelman_etal_2022}. Numerical simulations of quasar accretion disks in steady state can account for luminosity fluctuations of the order of one magnitude on multiyear timescales \citep{Jiang_etal_2019, Jiang_etal_2020}. Such significant variations are suggested to follow the viscous timescale that is predicted to be on the order of 100 to 1000 years \citep[][]{King_etal_2008, macleod2016, Jafari_2019}. An alternative explanation for abrupt luminosity changes may be found in different processes that operate on much shorter timescales, such as the thermal time of the inner accretion disk or its hot corona that is thought to be the source of the X-ray emission \citep[e.g.,][]{lamassa2015}. \citet{Noda_Done_2018} suggest that changing-look/changing-state AGNs can be placed in one of three groups, depending on which particular aspect of the process they exhibit: (1) a factor of two to four decrease in luminosity associated with disc evaporation/condensation; (2) large mass accretion rate change due to thermal front propagation; and (3) a variability amplitude of more than 10 indicative of both phenomena. However, we still lack a clear picture of the overall behavior of these systems. Assessing the physical processes associated with the various timescales of variability in these AGNs is vital to our understanding of how these accreting systems work \citep[see][for a recent review]{claudio_benny}. Additionally, some fast QPO-type phenomena in galactic sources can be extended to timescales seen in AGNs \citep{Belloni_2010LNP...794...53B, deMarco_2022hxga.book...58D}.\\

Another way to look at the evolution of an AGN, specifically Type-1s, is to track their transition along the optical plane of the Eigenvector 1 (EV1) schema. The EV1 schema dates back to the seminal work by \citet{Boroson_Green_1992}, wherein the authors performed a principal component analysis using observed spectral properties for a sample of 87 low-redshift sources to realize the, now well-known, main sequence of quasars. The EV1 schema, or the optical plane of the main sequence of quasars, is the plane between the FWHM of the broad \hb\ emission line and the strength of the optical \feii\ emission to \hb{}. This has been a key subject of study spanning close to three decades that has advanced our knowledge of the diversity of Type-1 AGNs both from observational and theoretical aspects \citep[see][and references therein]{Boroson_Green_1992, Sulentic_etal_2000, Shen_Ho_2014, Marziani_etal_2018, Panda_etal_2018, Panda_etal_2019b}. An interesting aspect of the EV1 is the connection of the two parameters to fundamental BH parameters that make up the optical plane, i.e., the FWHM(\hb{}) is closely connected to the BH mass through the virial relation \citep{Czerny_Nikolajuk_2010, Shen_2013}, while the other parameter - the \feii{} strength (or \rfe{}) is closely connected to the Eddington ratio, and hence to the accretion rate of the BH \citep{Shen_Ho_2014, Marziani_etal_2018, Panda_etal_2018, Panda_etal_2019a, Panda_etal_2019b, Panda_2021_PhD-Thesis, Martinez-Aldama_etal_2021}. Thus, in principle, the EV1 schema can be a valuable tool to investigate the evolution of the BH's mass and accretion state by analyzing their movement along the optical plane of the main sequence of quasars.\\

This paper is the first in a series of preparing and analyzing spectral and other properties for a database of already discovered CL AGNs. Here, we focus on the spectral fitting and analysis of broad emission lines in a sizable sample of CL AGNs available in the literature. The paper is organized as follows:
The methodology to prepare a sizable sample of CL AGNs is reported in Section \ref{sec:sample}. In Section \ref{sec:analysis}, we present the spectral analysis of archival SDSS spectroscopic data using a publicly available automatic spectral fitting package {\sc PyQSOFit}\footnote{\href{https://github.com/legolason/PyQSOFit}{https://github.com/legolason/PyQSOFit}}. In Section \ref{sec:results} we estimate the BH masses and Eddington ratios for each epoch per source in our sample and study their transition along the quasar main sequence. We also study the Balmer decrement as a function of time and AGN luminosity for the sources in our sample. We discuss further possibilities and issues relevant to this study in Section \ref{sec:discussions} and summarize our findings through this study in Section \ref{sec:conclusion}. In this work we use $\Lambda$CDM cosmology ($\Omega_{\Lambda}$ = 0.7, $\Omega_{\mathrm M}$ = 0.3, $H_{0}$ = 70 km s$^{-1}$ Mpc$^{-1}$).

\section{Sample}
\label{sec:sample}

We search through the literature to collect and prepare a database of CL AGNs which have (existing) multiple spectroscopic observations using the Sloan Digital Sky Survey \citep[SDSS,][]{SDSS_2000AJ....120.1579Y} and later using the Baryon Oscillation Spectroscopic Survey \citep[BOSS,][]{BOSS_2013AJ....145...10D} and extended-BOSS \citep[eBOSS,][]{eBOSS_2016AJ....151...44D}. The choice of SDSS allows us to gather sources that can have a relatively high sampling rate (repeat spectra) for a given source (a better case scenario is shown in Figure \ref{fig:spectra-compare-Wang2018}). To prepare our sample, we utilize the following criteria - (a) the presence of at least 2 publicly available spectra from SDSS, BOSS, and eBOSS for the sources), and (b) the primary goal of this work involves tracking the movement of the source on the EV1 diagram, we need the coverage of the optical \feii{} emission between 4434-4684\AA\ and the full \hb{} profile. Due to the second criterion, we filter out known CL AGN with higher redshifts, e.g., from \citet{Ross2020MNRAS.498.2339R}.

We use objects that are reported as CL AGN in the following works (listed in order of decreasing number of sources): 

\begin{figure*}[!htb]
    \centering
    \includegraphics[width=\textwidth]{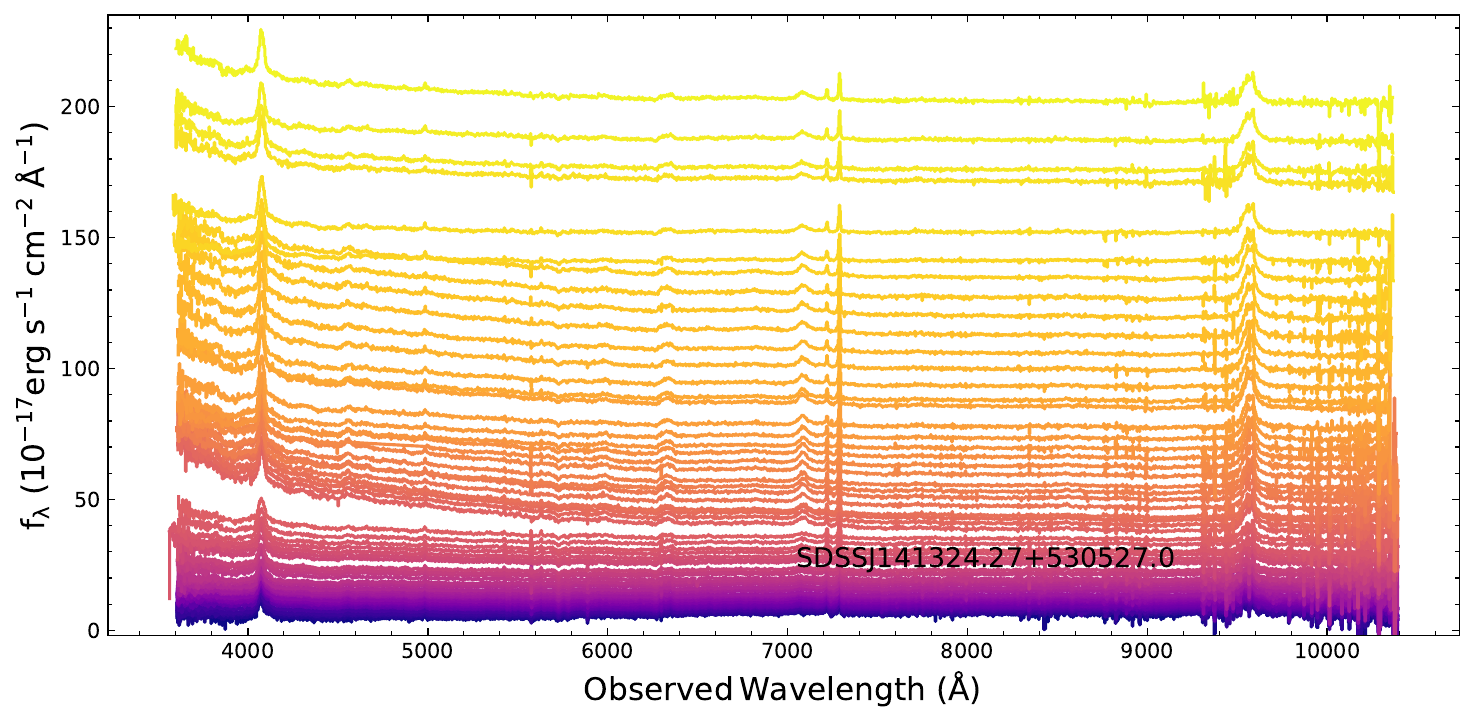}
    \caption{Binned spectra for the 72 available spectral epochs chronologically from bottom to top for the source SDSSJ141324.27+530527.0 from \citet{Wang2018ApJ...858...49W}. The spectra have been shifted by arbitrary factors for better visualization.}
    \label{fig:spectra-compare-Wang2018}
\end{figure*}

\begin{itemize}
      \item 58 sources from from \cite{Green_etal_2022} 
      \item 26 sources from \cite{graham2020}
      \item 7 sources from \cite{macleod2016}
      \item 5 sources from \cite{Yang2018}
      \item 4 sources from \cite{potts2021} 
       \item 2 sources from each work: \cite{ruan2016}, \cite{Runco_etal_2016} and \cite{Wang_etal_2019}
       \item one source from the following works: \cite{lamassa2015}, \cite{Runnoe2016}, \cite{Wang2018ApJ...858...49W}, and \cite{Yu_etal_2020}.
\end{itemize}

Thus, our parent sample contains 110 sources. For each candidate, we downloaded all available spectra from the SDSS DR16 website\footnote{The spectra for sources in our sample were retrieved using the following link: \href{https://skyserver.sdss.org/dr16/en/tools/explore/summary.aspx}{https://skyserver.sdss.org/dr16/en/tools/explore/summary.aspx}.}. We make a search using the sky coordinates listed in the respective works per source and then retrieve the spectra in FITS format using the option of ``All Spectra'' under the section \textit{Spec Summary}. This website lists all the existing spectra taken using the SDSS instruments (SDSS, BOSS, eBOSS) for a given source to date. Since we use the DR16 version, for some sources we have found more recent spectra than were analyzed in the previous works\footnote{Because some sources were included in DR16 \citep{SDSS-DR16_2020ApJS..249....3A} that were published after the original works.}. We then extract the basic information for each spectrum per source - e.g., MJD, sky coordinates, redshift (quoted by the SDSS pipeline), and the reference to the instrument used to collect the spectrum. We present those information in Table \ref{tab:table-observe}. In addition to this information, we report the original reference that first reported the discovery of a given source and the SDSS name for the source. For each source, we present the corresponding epochs of observations in chronological order.
    
From the Table, we can notice that spectra for different objects are sampled with a varied sampling rate. The most observed one is SDSSJ141324.27+530527.0 \citep{Wang2018ApJ...858...49W} with 72 epochs, which we plot in Figure \ref{fig:spectra-compare-Wang2018}. This source is included in the sample for the SDSS RM campaign, which is why the source is observed regularly (onward 56660 MJD) in comparison to most of the sources in our sample.

\begin{figure*}
\centering
\includegraphics[scale=0.375]{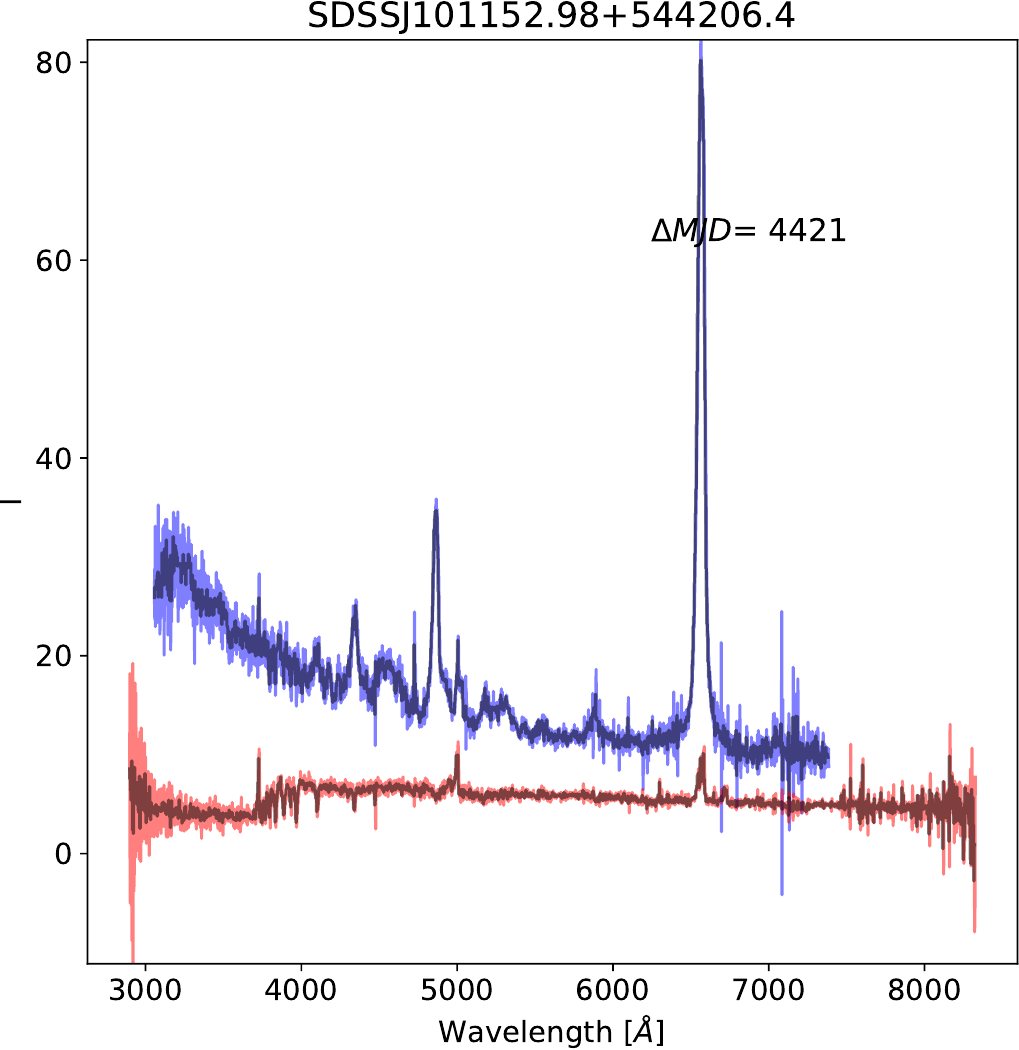}
\includegraphics[scale=0.375]{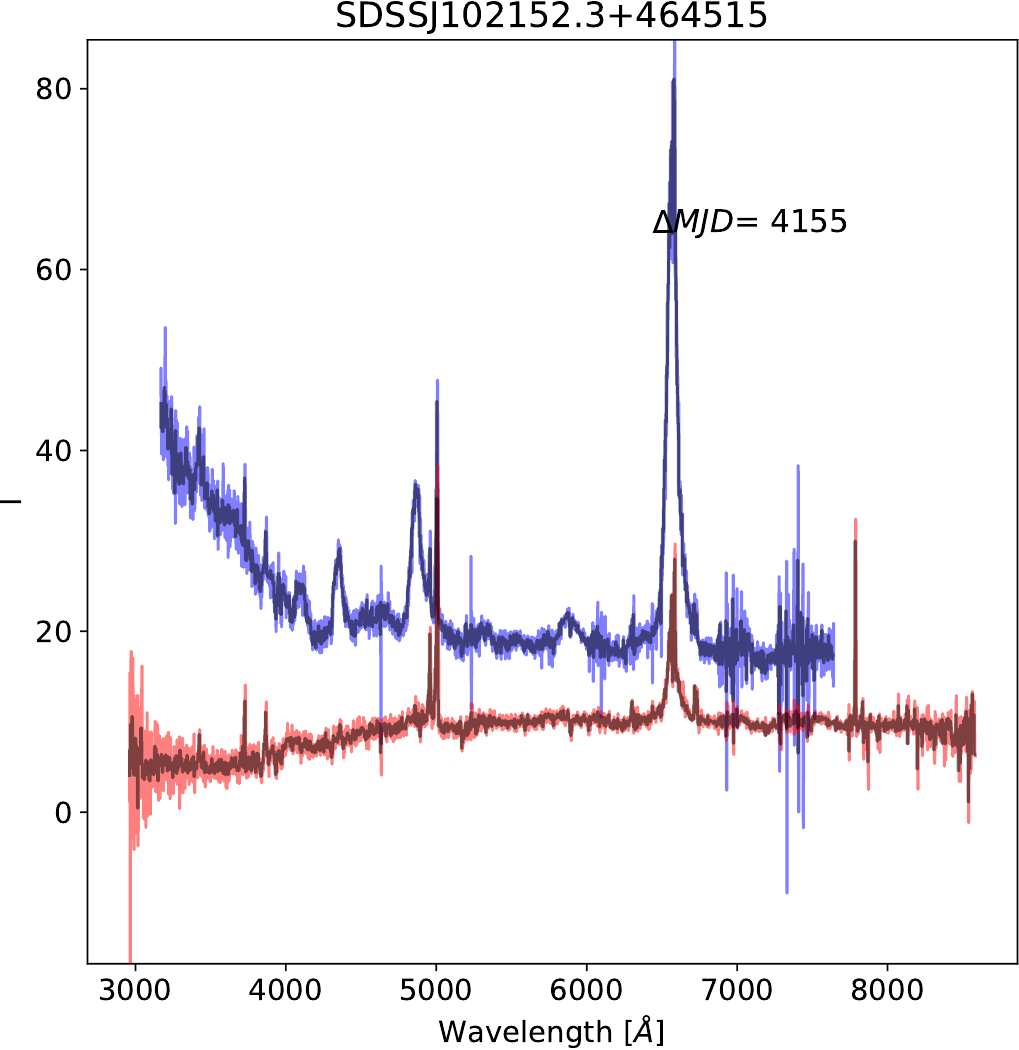}
\includegraphics[scale=0.375]{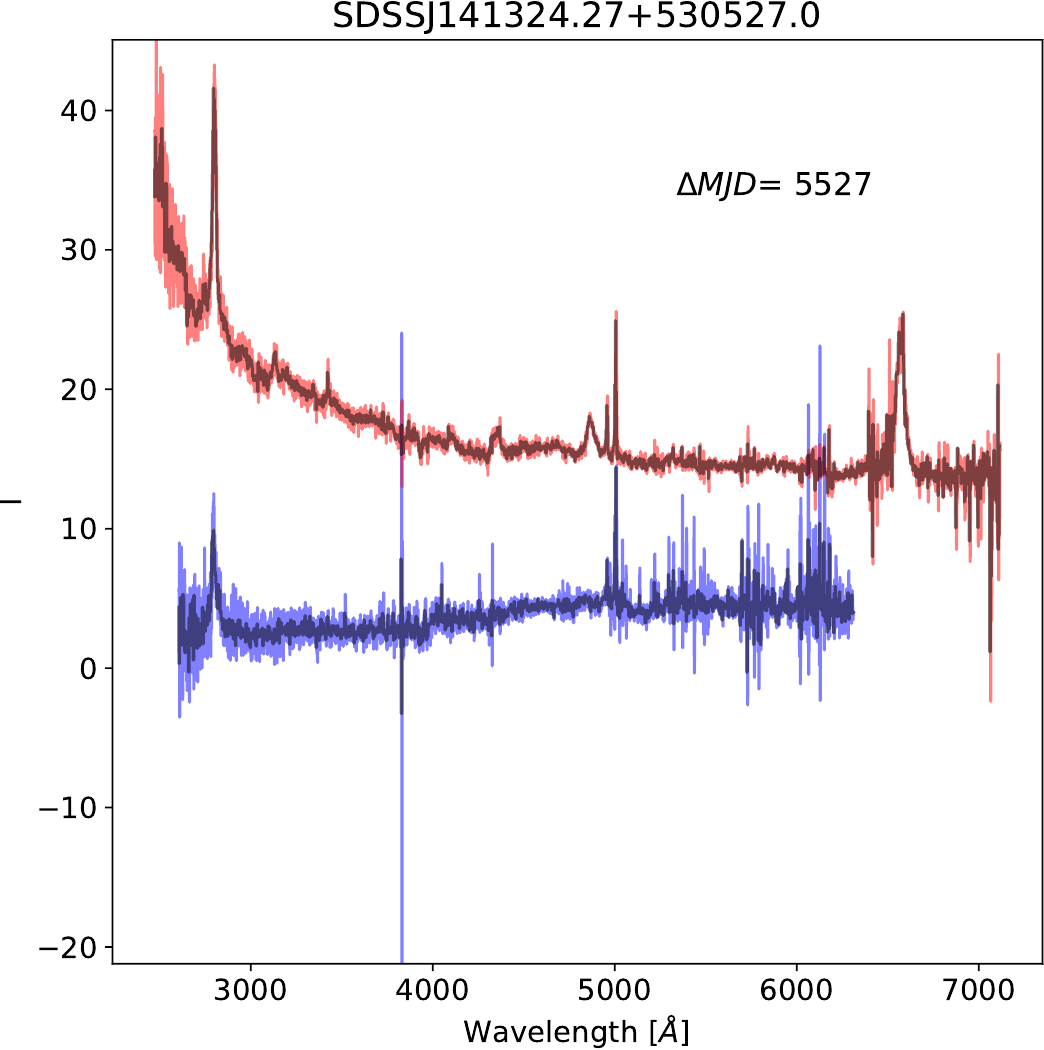}
\includegraphics[scale=0.375]{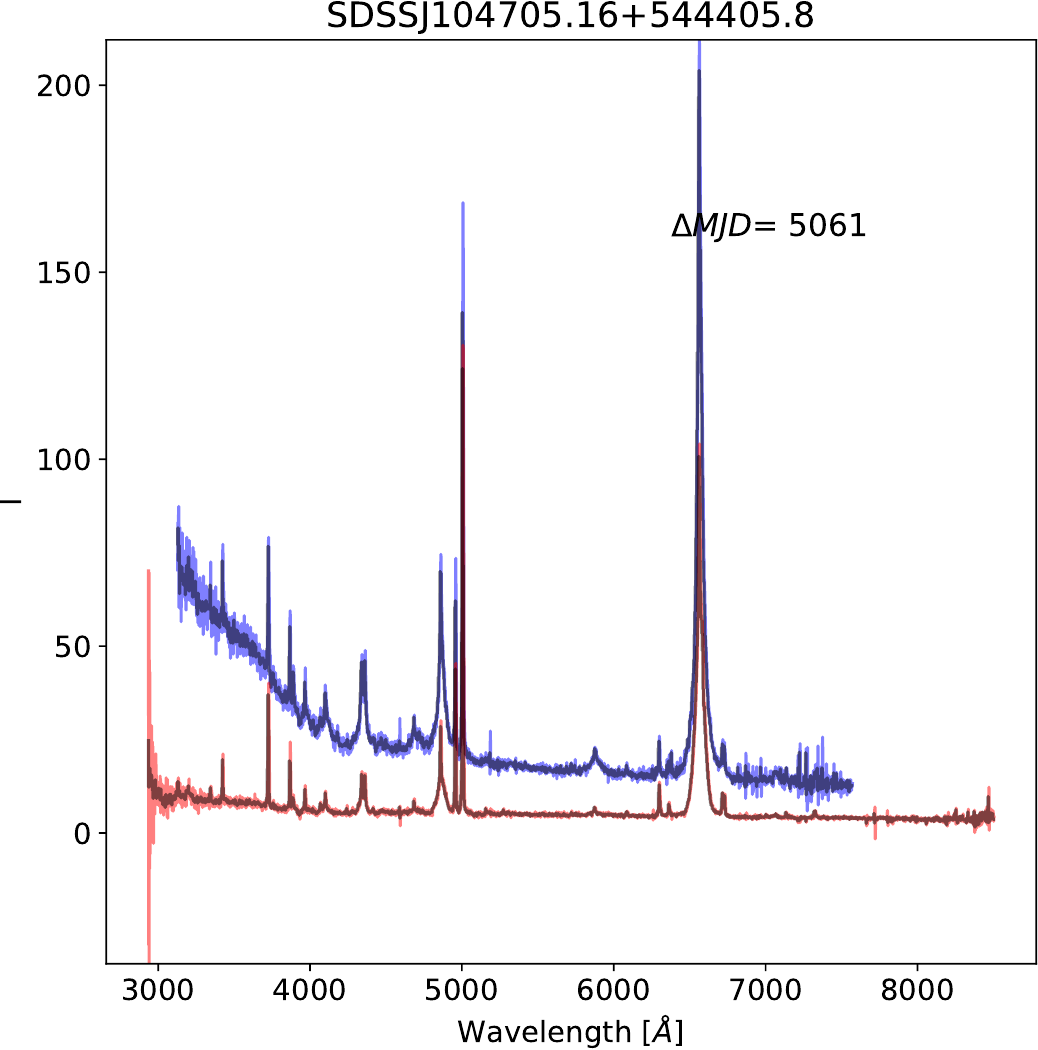}
\includegraphics[scale=0.375]{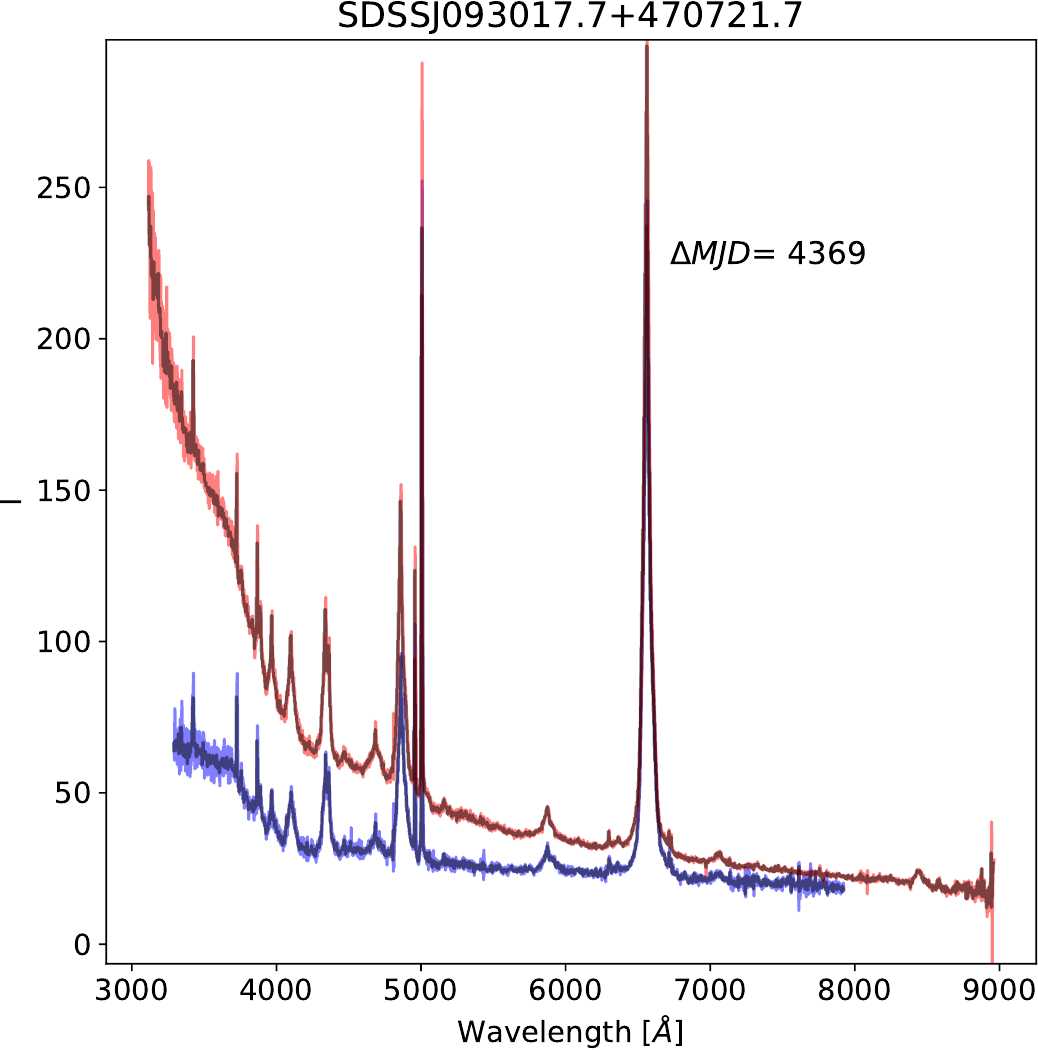} 
\includegraphics[scale=0.375]{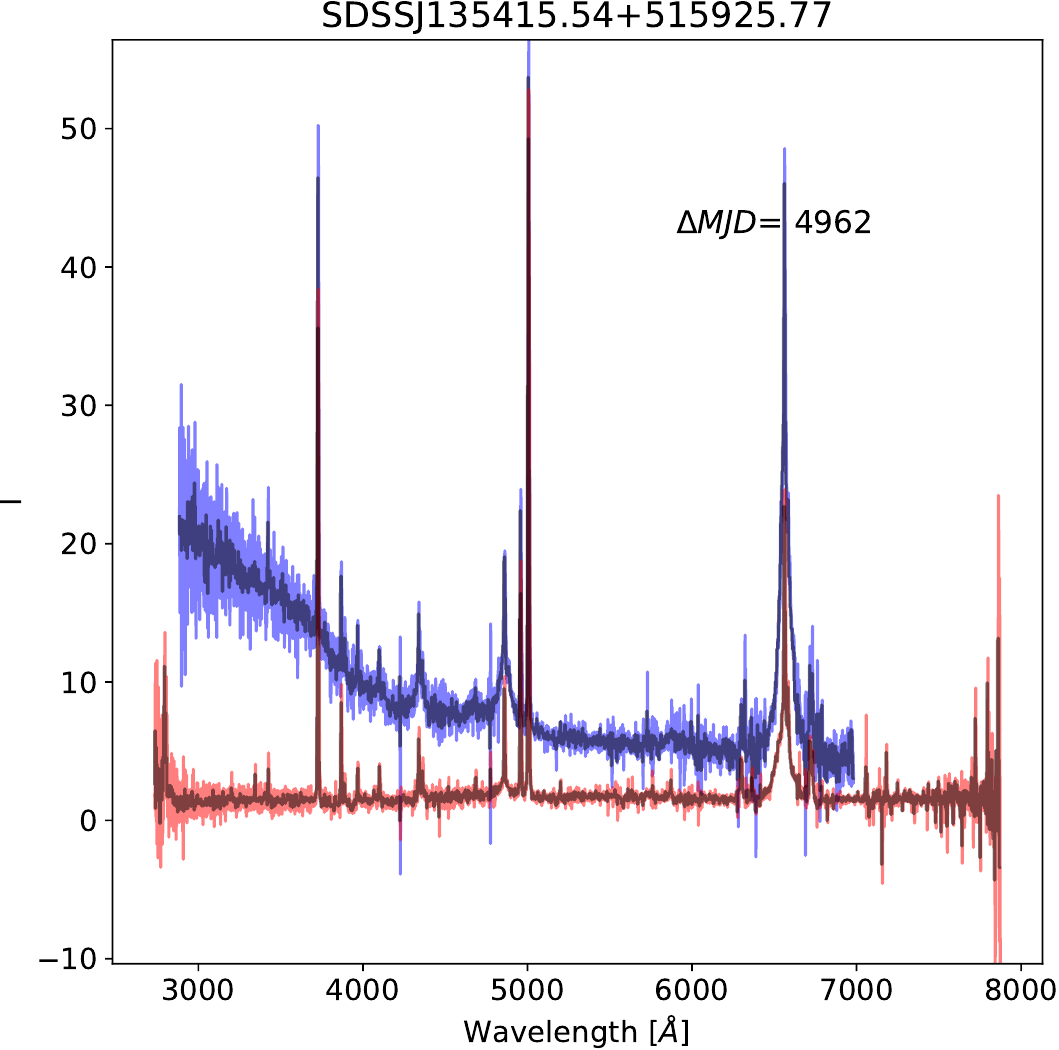}
\caption{The exemplary objects from our sample: SDSSJ101152.98+544206.4 from \citet{Runnoe2016}, SDSSJ102152.3+464515 from \citet{macleod2016}, SDSSJ141324.27+530527.0 from \citet{Wang2018ApJ...858...49W}, SDSSJ104705.16+544405.83 from \citet{Wang_etal_2019}, SDSSJ093017.7+470721.73 from \citet{graham2020}, and SDSSJ135415.54+515925.77 from \citet{Green_etal_2022}. The blue line represents the oldest spectrum, whereas the red one represents the latest one. Black lines present in both spectra are data binned with 3\AA.}
\label{fig:clagn_examples_one}
\end{figure*}

The spectra from the sample have different wavelength ranges for various objects, for example, due to the improvement of wavelength coverage between BOSS/eBOSS and SDSS (see plots of representative sources in Figure \ref{fig:clagn_examples_one}{\footnote{We discuss the implication of the continuum slope changes on the inferred monochromatic luminosities for our sources in Appendix \ref{sec:pl-slopes}}.}). Looking through Tab. \ref{tab:table-observe}, we can also notice that coordinates for the same source extracted from fits files may differ slightly. Thus, we utilize the redshift to help distinguish the source from any imposter.

\section{Analysis}
\label{sec:analysis}

\subsection{Spectral fitting with {\sc PyQSOFit}}

\begin{figure*}
    \centering
    \includegraphics[width=0.9\textwidth]{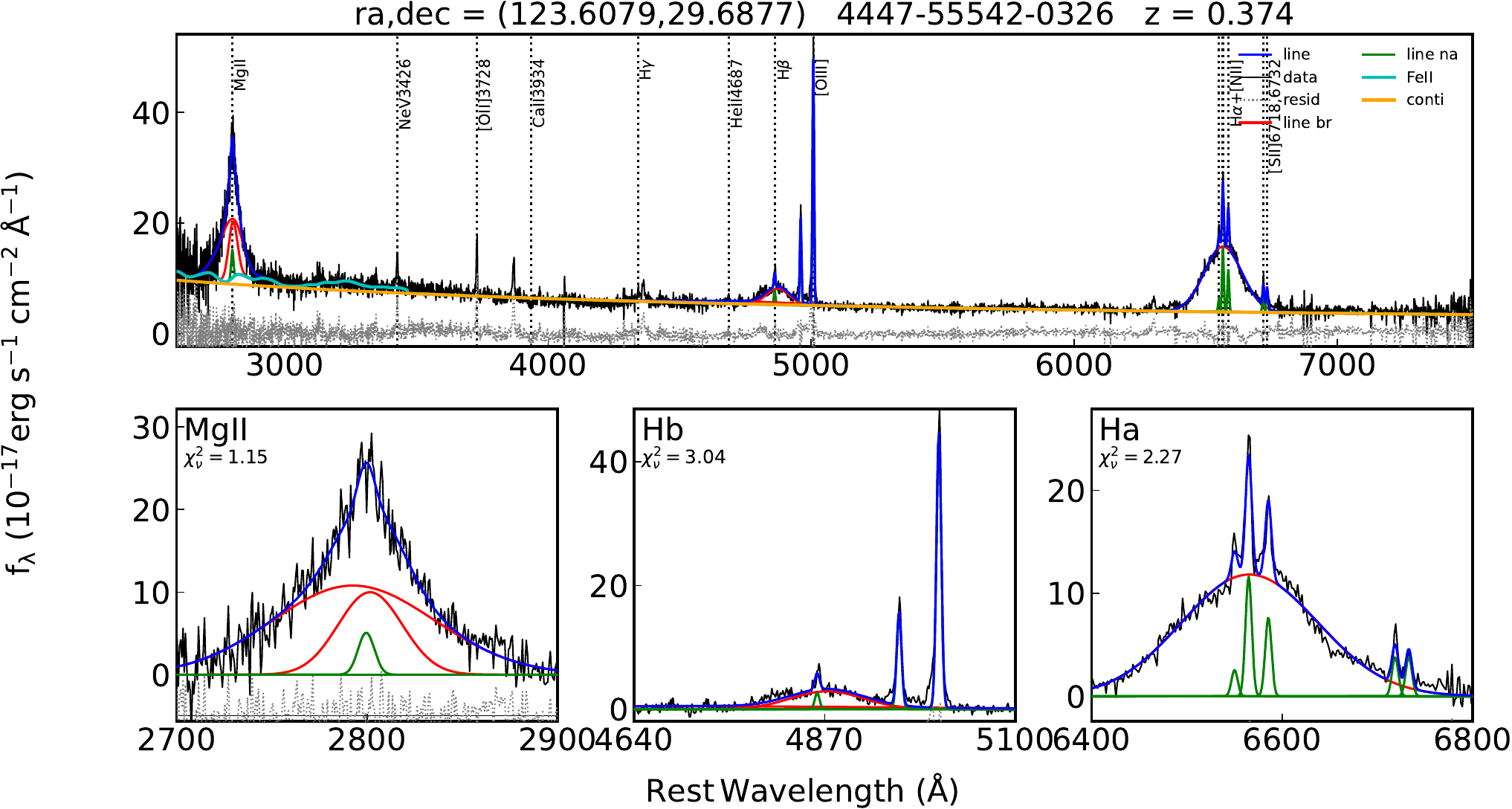}
    \caption{Exemplary fit using {\sc PyQSOFit} \citep{pyqsofit} for a quasar spectrum (SDSSJ081425.9+294116.3, \citealt{graham2020}) without significant host-galaxy contribution. Upper panel: we show in each panel the SDSS data (black), power-law continuum (yellow), \feii{} pseudo-continuum (light green), broad emission lines (red), narrow emission lines (dark green), and the total best-fit model (blue), which is the sum of continuum and emission lines. The sky coordinates (in degrees), the SDSS plate-MJD-fiber sequence and the redshift for the sources are quoted in the title of the figure. The gray horizontal bars on the top mark the continuum windows used for fitting. Lower panel: a zoomed version of individual line complexes is also shown along with the reduced $\chi^2$ obtained for each of the line complexes.}
    \label{fig:pyqsofit-example}
\end{figure*}

\begin{figure*}
    \centering
    \includegraphics[width=0.9\textwidth]{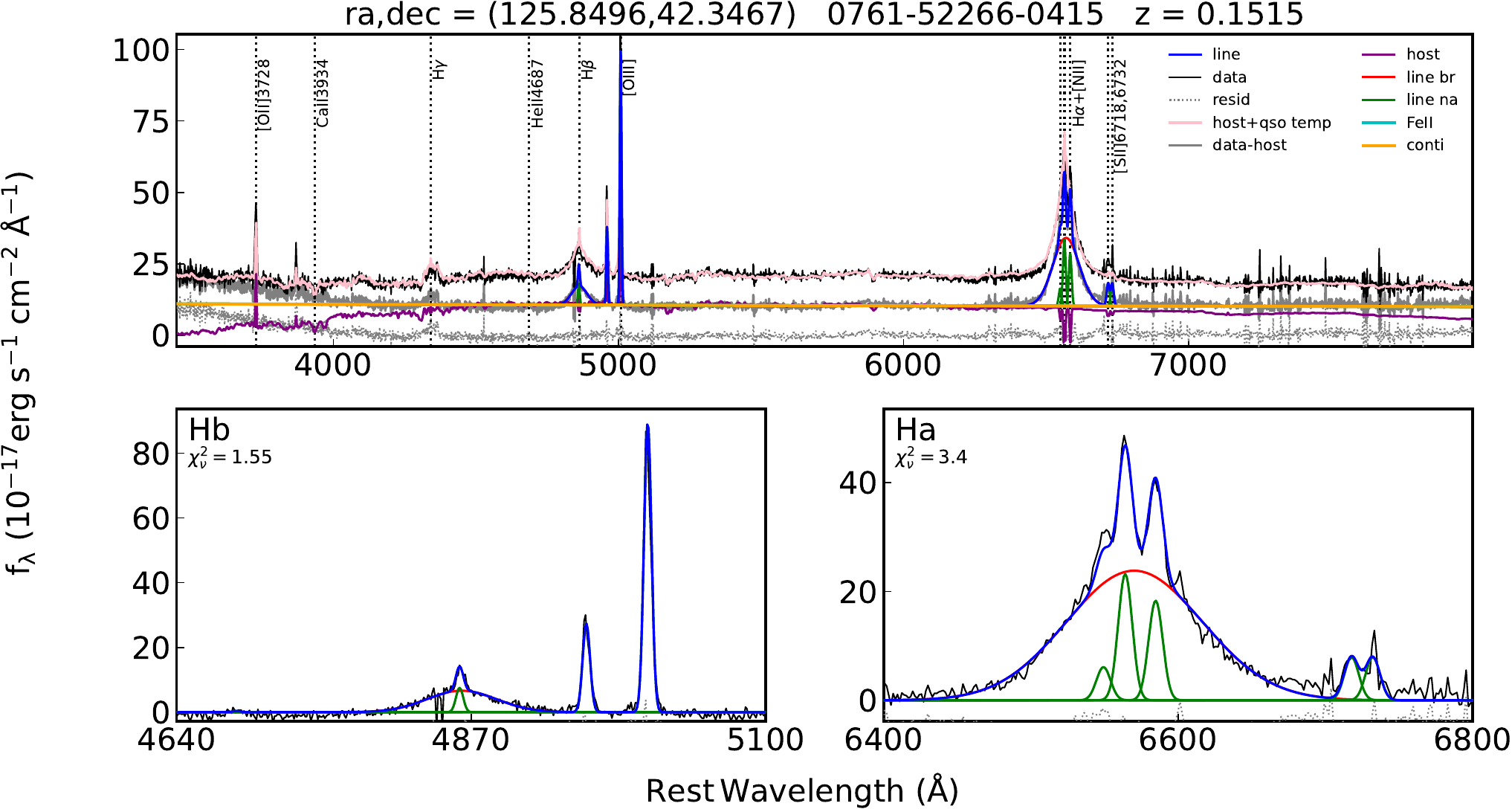}
    \caption{Exemplary fit using {\sc PyQSOFit} \citep{pyqsofit} for a quasar spectrum (SDSSJ082323.88+422048.2, \citealt{potts2021}) with significant host-galaxy contribution. In the upper panel, we show the host contribution (purple), the host+qso (pink) where qso = power-law continuum plus emission line. In gray, we show the residual between SDSS data and host contribution. The remaining components are identical to Figure \ref{fig:pyqsofit-example}.}
    \label{fig:pyqsofit-example-alt}
\end{figure*}

We employ {\sc PyQSOFit} to conduct spectral decomposition, as detailed in \citep{pyqsofit}, for all SDSS spectra in our study. Spectra are adjusted to the rest frame, and Galactic extinction corrections are applied using the extinction curve from \citet{Cardelli_etal_1989} and the dust map from \citet{Schlegel_etal_1998}. Subsequently, a host galaxy decomposition is performed utilizing galaxy eigenspectra from \citet{Yip_etal_2004a} and quasar eigenspectra from \citet{Yip_etal_2004b}, both implemented in the {\sc PyQSOFit} code. If more than half of the pixels in the resultant host galaxy fit exhibit negative values, the host galaxy eigenspectral fit is not implemented.\\

Following this, we fit power law, UV/Optical \feii{}, and Balmer continuum models. The optical \feii{} emission template, covering 3686-7484\AA, is adopted from \citet{Boroson_Green_1992}, while the UV \feii{} template spanning 1000-3500\AA is taken from \citet{Vestergaard_Wilkes_2001, Tsuzuki_etal_2006, Salviander_etal_2007}. {\sc PyQSOFit} fits these empirical \feii{} templates employing normalization, broadening, and wavelength shift. Subsequently, emission line fits are conducted using Gaussian profiles, as described in \citet{Shen_etal_2019, rakshitetal2020}. Depending on redshift and spectral coverage, the following emission lines are fitted: H$\alpha\lambda$6564.6 (broad and narrow), [NII]$\lambda$6549,6585, [SII]$\lambda$6718,6732, H$\beta\lambda$4863 (broad and narrow), [OIII]$\lambda$5007,4959, and Mg II$\lambda$2800 (broad and narrow). All fits are run using Monte Carlo simulation based on the actual observed spectral error array, yielding an error array for all decomposition fits. An illustrative example of spectral decomposition is provided in Figure \ref{fig:pyqsofit-example}\footnote{The fitted spectra from our sample are hosted on \href{https://zenodo.org/records/10729856?token=eyJhbGciOiJIUzUxMiJ9.eyJpZCI6ImEyMGJhZTBhLTYyZjgtNDI0Ni05ZWZmLWEzYmYzMjhhOTA5NiIsImRhdGEiOnt9LCJyYW5kb20iOiI1NGNlYmQ3ZmY3NGIyOTAyNTQ3YzhiMTUzZGQxYmU2NCJ9.fbQkKlrmCUYfZjkv8HkOH0-o4kdQFhS-MJRPn486f8C_ftTUwBc94m6Wgud9MgF0fPQzVDLIN9gN510kXXGAQA}{Zenodo} \citep{Panda_Sniegowska_zenodo}.}. While our methodology aligns with that used in \citet{Green_etal_2022}, we opted to re-analyze the sample from their paper along with other sources in our parent sample.\\

We highlight exemplary cases for (a) a quasar spectrum fitting without significant host-galaxy contribution (Figure \ref{fig:pyqsofit-example}), and (b) a quasar spectrum fitting with significant host-galaxy contribution (Figure \ref{fig:pyqsofit-example-alt}). The host galaxy fits used in {\sc PyQSOFit} are limited to rest-frame wavelengths between 3450 – 8000 \AA. Due to this limitation, to fit the \mg{} line complex, we follow the prescription of \citet{Green_etal_2022} but make a conditional execution of host decomposition in the same run, i.e., if z $<$ 0.25, then the host contribution is included. Otherwise, the host contribution is not accounted for. {We note here in passing that as shown in \citet{2023MNRAS.521L..11J}, the host fraction has a strong negative dependence on the redshift (and hence on the total luminosity) - the value decreases from 40\% $\rightarrow$ $\sim$20\% when the z increases from 0.2 to 0.8, suggesting that the host contribution becomes minimal at higher redshifts. Thus, our supposition to exclude the host contribution for z $<$ 0.25 does not affect significantly the inferred estimates that will be presented later in this work.}\\

\begin{figure}
    \centering
     \includegraphics[width=\columnwidth]{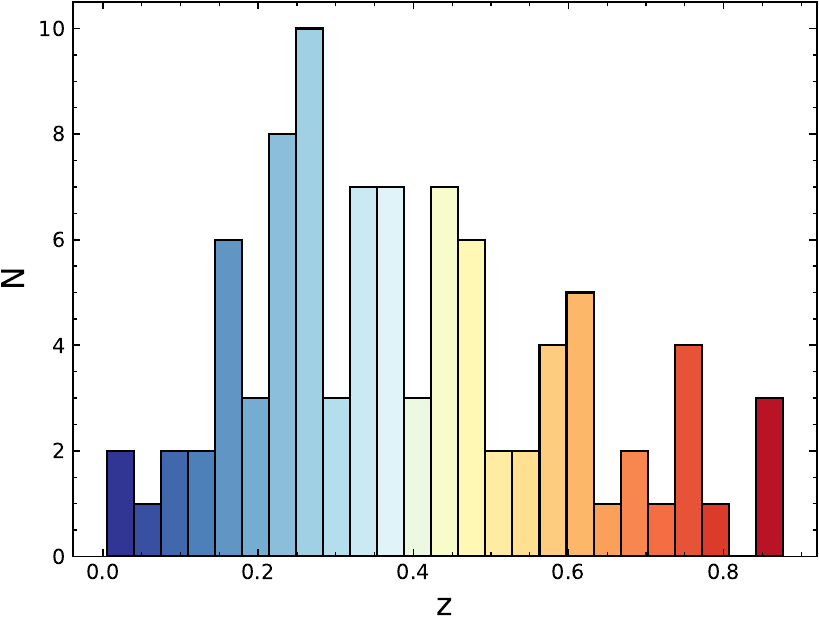}
    \caption{Redshift distribution for the 93 sources in the sample.}
\label{fig:Z-dist}
\end{figure}

After the fitting procedure, we perform manual checks to identify the spectra where the broad \hb{} emission line disappears and/or the spectrum is dominated by noise in this region and no reliable \hb{} measurements could be made. Our final working sample contains 93 AGNs:

\begin{itemize}
      \item 47 sources from from \cite{Green_etal_2022} 
      \item 25 sources from \cite{graham2020}
      \item 7 sources from \cite{macleod2016}
      \item 4 sources from \cite{potts2021} 
       \item 2 sources from each work: \cite{Runco_etal_2016}, \cite{Yang2018} and \cite{Wang_etal_2019}
       \item one source from the following works: \cite{lamassa2015}, \cite{Runnoe2016}, \cite{ruan2016}, and \cite{Wang2018ApJ...858...49W}.
\end{itemize}

Hereafter, we only consider these sources in further analysis. We present the redshift distribution for this sample in Figure \ref{fig:Z-dist}.\\

\begin{figure}[htb!]
    \centering
    \includegraphics[width=\columnwidth]{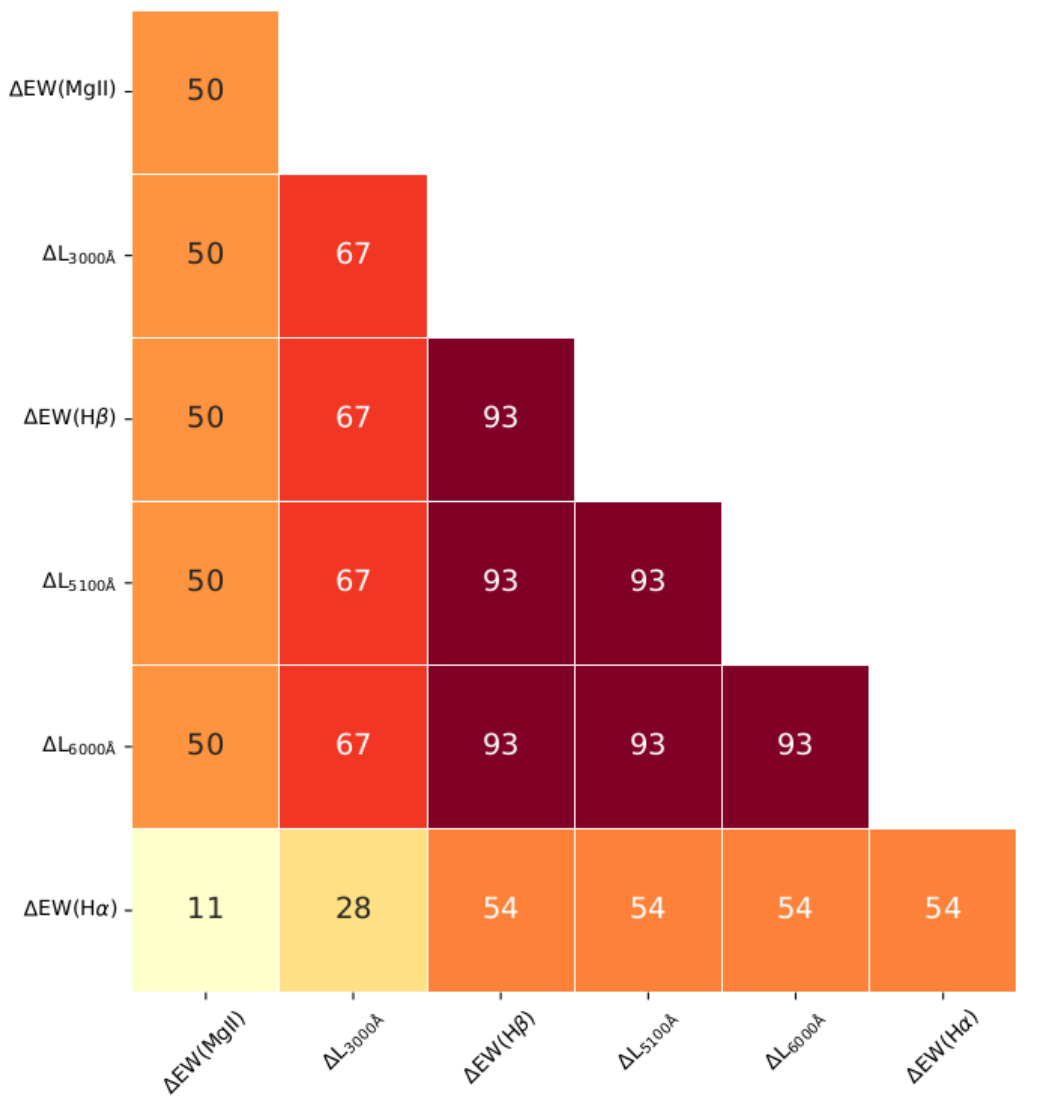}
    \caption{The heatmap between differences of fitted parameters between spectra from the first and the latest epoch for each source in our sample. The numbers in each square show how many sources we obtained this difference. The darkening of the cell represents a higher number of sources.}
    \label{fig:sample-description}
\end{figure}

In Table \ref{tab:table-pyqsofit-ew}, we report the continuum luminosities at 3000\AA{}, 5100\AA{}, and 6000\AA{}, and the corresponding equivalent widths (EWs) for the broad \mg{}, \hb{}, and \ha{} emission lines. These measurements are reported for the available epochs per source organized chronologically similar to Table \ref{tab:table-observe}. Figure \ref{fig:sample-description} shows the distribution of the number of sources with the particular measurements - \lagnuv{}, \lagn{}, \lagnir{}, EW(\mg{}), EW(\hb{}) and EW(\ha{}), present in the first and latest available spectra per source. Most of sources have \lagn{}, \lagnir{} or EW \hb{}. As per this figure, we have, for {93} sources where we recover - $\Delta$EW(\hb{}), $\Delta$\lagn{} and $\Delta$\lagnir{}, where the difference in the measured quantities is made using the first and the latest epoch of observation in each source from our sample. That is, we had non-zero measurements for the respective parameters in the first/last epoch using the {\sc PyQSOFit} analysis. More details on the correlation analyses on the EWs and continuum luminosities are reported in Appendix \ref{app:ews_lum}.

\section{Results}
\label{sec:results}

\subsection{Turn-on vs. Turn-off CL AGNs}
 
\begin{figure}
    \centering
     \includegraphics[width=\columnwidth]{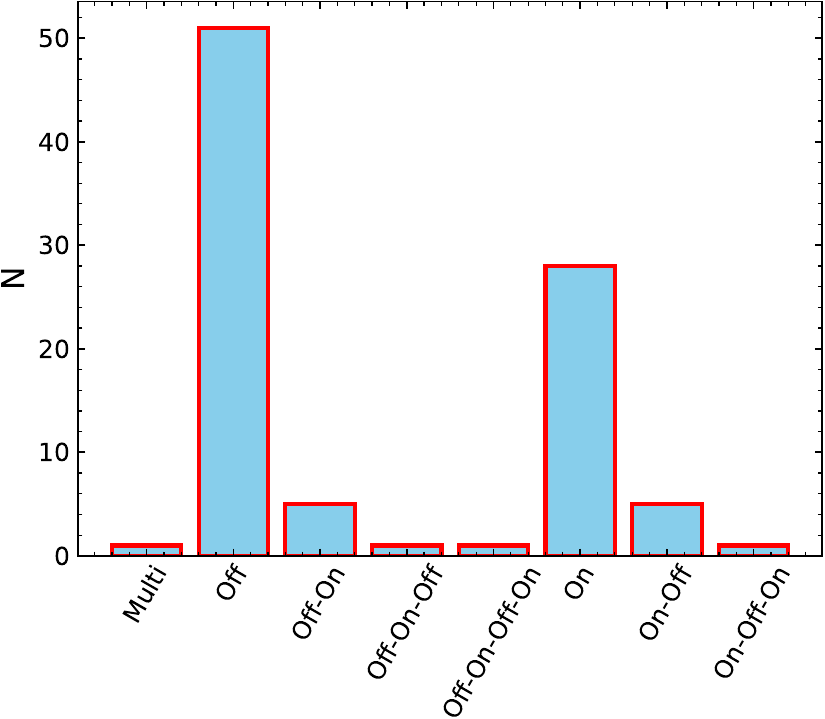}
    \caption{Distribution of state changes in our sample. Note that here the classification is based on the relative change in the \lagn{} obtained from the {\sc PyQSOFit} analysis of the epochs per spectra in chronological order.}
\label{fig:state-change-dist}
\end{figure}

In Table \ref{tab:table-pyqsofit}, we report the full-width at half maximum (FWHM) of the broad \hb{}, the ratio \rfe{}, the monochromatic luminosity at 5100\AA\ (\lagn{}), the bolometric luminosity (\lbol{}), the virial mass of the BH (\mbh{}), the corresponding Eddington ratio (\ledd{}), and, the relative change in the \lagn{} to the first epoch per source. {We describe how these parameters, especially \mbh{} and \ledd{} are estimated, in Section \ref{sec:BHmass}.} We denote this last parameter as $\Delta$\lagn{} and report the values for each epoch per source in the units of $10^{44}$ erg s$^{-1}$. We use this information to construct Figure \ref{fig:state-change-dist}, which shows the classification of the sources in our sample. The classification schema is as follows:

\begin{enumerate}

    \item A source is marked as "On" if the subsequent epochs after the first epoch keep increasing in their value of the \lagn{} which is confirmed by the increasing (positive) values for $\Delta$\lagn{},
    
    \item On the contrary, a source is marked as "Off" if the subsequent epochs after the first epoch keep decreasing in their value of the \lagn{} which is confirmed by the decreasing (negative) values for $\Delta$\lagn{},
    
    \item We then have sources that increase in their value of \lagn{} after the first epoch, but then reduce their \lagn{} below the value in the first epoch. This is confirmed with an initial positive value for the $\Delta$\lagn{} followed by a negative value. Such sources are classified under "On-Off",
    
    \item On the contrary, some sources first decrease in the \lagn{} value compared to the first epoch and later increase to attain a \lagn{} value higher than the first epoch. Such sources are classified under "Off-On",
    
    \item Then, some sources increase, then decrease, and again increase in their values of \lagn{}. Such sources are categorized as "On-Off-On",
    
    \item The complementary group of sources to the previous class is marked as "Off-On-Off",
    
    \item Sources, that have 2 cycles of the first decrease - followed by an increase in the \lagn{} values are clumped into "Off-On-Off-On", and
    
    \item Sources that show multiple (more than two cycles) epochs of rise-fall in their \lagn{} are put under the class "Multi".
    
\end{enumerate}

The above classification is made solely based on the spectroscopic epochs collected for each source in our sample. The distribution of sources is as follows - On (27), Off (52), On-Off (6), Off-On (5), On-Off-On (1), Off-On-Off-On (1), and Multi (1). Our results are consistent with what has been found in \citet{Shen_2021} - statistically, there are more detections of turning-off CL AGNs relative to the turning-on ones. This may be attributed to a selection effect with brighter AGNs being observed more frequently than their fainter counterparts.

\subsection{Black hole mass and Eddington ratios}
\label{sec:BHmass}

BH mass is estimated using the \citet{Vestergaard_Peterson_2006} relation:
\begin{equation}
    M_{\rm BH} = 10^{6.91}\left(\frac{\rm{FWHM(H\beta)}}{1000\; \rm{km\;s}^{-1}}\right)^2\left(\frac{L_{5100}}{10^{44}\; \rm{erg\;s}^{-1}}\right)^{0.5} M_{\odot}
    \label{eq1}
\end{equation}

\begin{figure}
    \centering
     \includegraphics[width=\columnwidth]{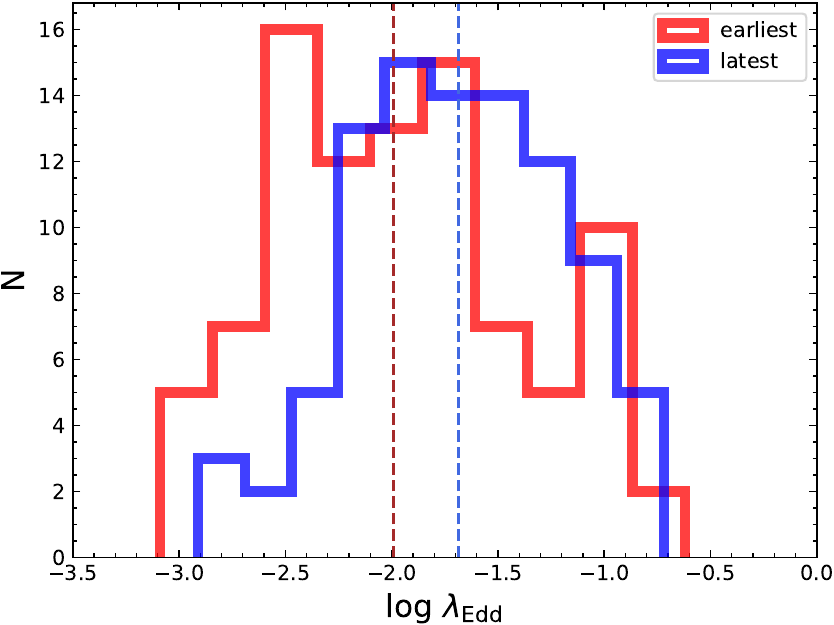}
    \caption{Distribution of Eddington ratios in our sample. We demonstrate the distributions for the earliest (in red) and the latest (in blue) epochs for the sources in our sample. The median values for the two distributions (red = -1.99, blue = -1.685) are marked with vertical dashed lines.}
\label{fig:eddr-compare}
\end{figure}

The bolometric luminosity (\lbol{}) is estimated using the standard \citet{Richards_etal_2006} prescription for sources with z$<$0.8, i.e., $L_{\rm bol}$ = 9.26$\times L_{5100}$. Subsequently, the Eddington ratio $\left(\lambda_{\rm Edd}=\frac{L_{\rm bol}}{L_{\rm Edd}}\right)$ is estimated, where $L_{\rm Edd}$ = 1.38$\times$10$^{38}\left(\frac{M_{\rm BH}}{M_{\odot}}\right)$. These estimates for the BH masses, the bolometric luminosity, and the Eddington ratios are reported in Table \ref{tab:table-pyqsofit}. We show the distribution of the Eddington ratios in our sample in Figure \ref{fig:eddr-compare}. Here, we demonstrate the distributions for the earliest (in red) and the latest (in blue) epochs for the sources in our sample. The median values for the distributions taken from the earliest and the latest epochs are -1.99 and -1.65, respectively. These values are consistent with previous studies that have considered samples of CL AGNs \citep[see e.g.,][]{lawrence_etal_2016, rumbaugh_etal_2018, graham2020, potts2021}. In a recent work by \citet{Zeltyn_etal_2024} where the authors make a compendium of newly discovered CL AGNs using proprietary SDSS V datasets, they too find a median Eddington ratio for their sample of 113 CL AGNs to be -1.522, while for their control sample of CL AGNs matched from the SDSS DR16 \citep{Wu_Shen_2022} the median Eddington ratio is -1.602, which are in close agreement to the median values for our sample. This furthers the hypothesis that CL AGNs if affected by intrinsic changes, are more frequent for sources with Eddington ratios around 10$^{-2}$. {Since all of the CL-AGNs considered in this work are a subset of the SDSS QSO sample, we find that the fraction of CL-AGNs to the total SDSS sample is quite minimal. However, the number of known CL-AGNs keeps growing \citep{claudio_benny, Zeltyn_etal_2024, 2024ApJS..270...26G} and helps support our findings from this work. {From the theoretical aspect, \citet{Noda_Done_2018} presented a scenario of the advection-dominated accretion flow (ADAF) \citep{ichimaru77, 1994narayan} appearance in the inner part of the accretion disk as a possibility of a decrease in accretion rate and disappearance of the broad emission lines. The transition between the standard accretion disk flow \citep{ss73} and ADAF is expected at a critical accretion rate of around 1\% of the Eddington rate. The distributions of CL AGN peak around this value of Eddington ratio (see for example \citet{macleod2019, Zeltyn_etal_2024}; this work). Also,
\citet{2019ruan, jin21} show that using spectral indices ($\alpha_{\rm OX}$) and Eddington ratios holds the potential to investigate the scenario of accretion state transitions for CL AGN populations in the future.}}\\ 

\begin{figure*}[!htb]
    \centering

    \includegraphics[width=0.495\textwidth]{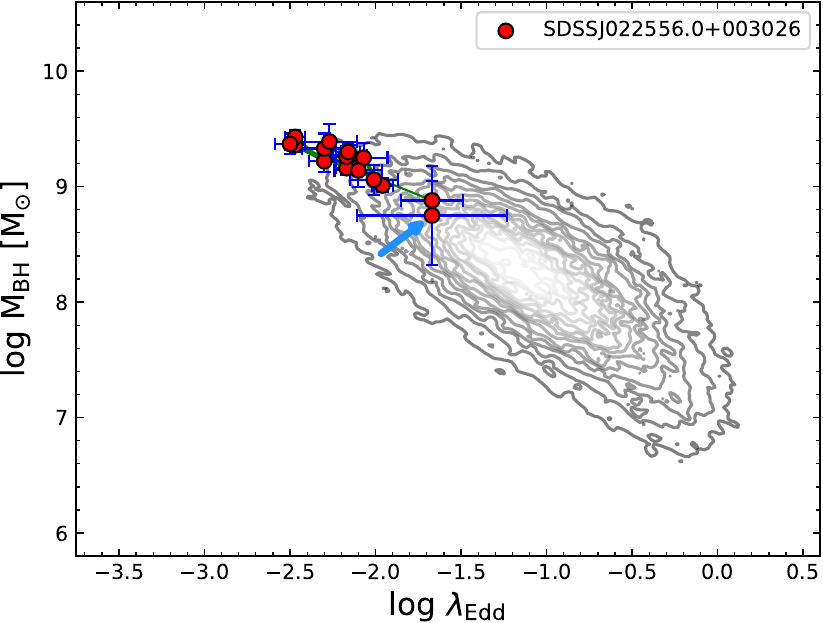}
    \includegraphics[width=0.495\textwidth]{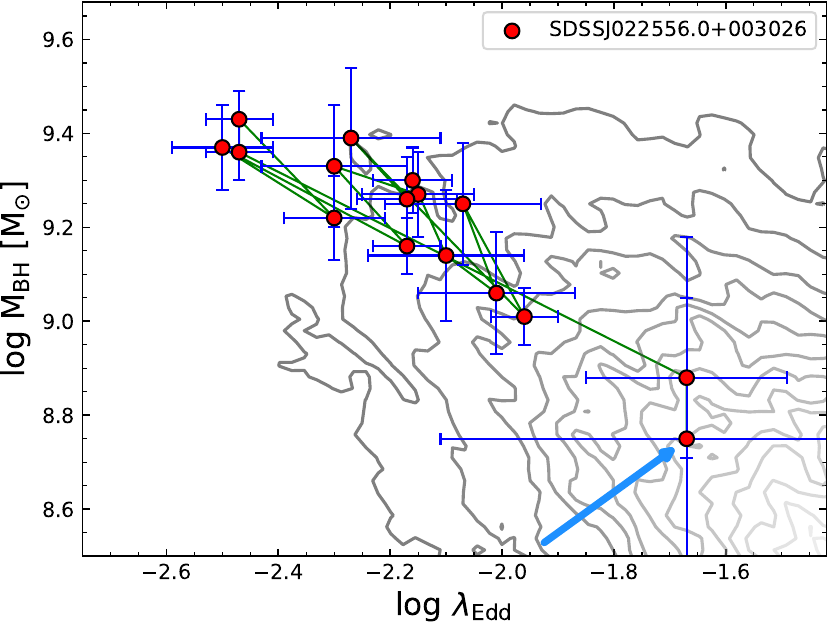}\\  
    \includegraphics[width=0.495\textwidth]{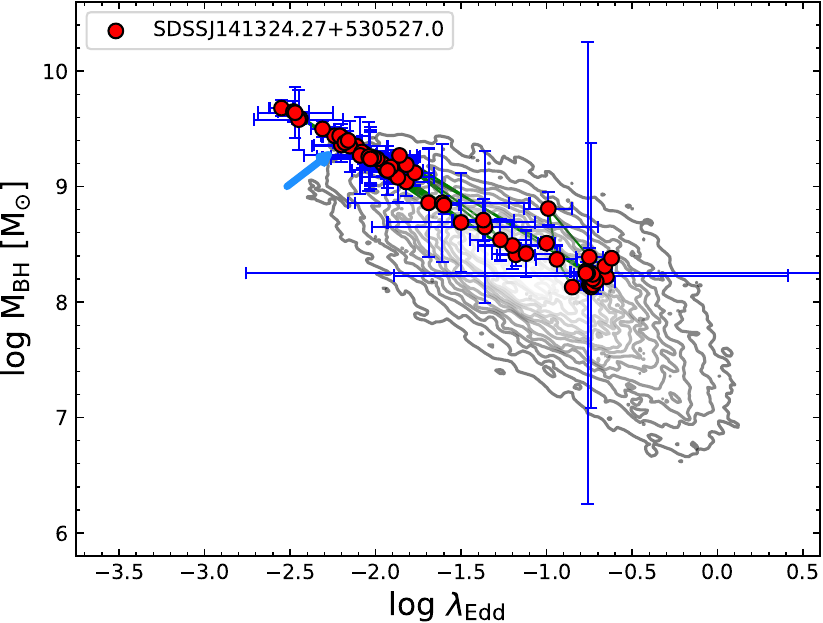}
    \includegraphics[width=0.495\textwidth]{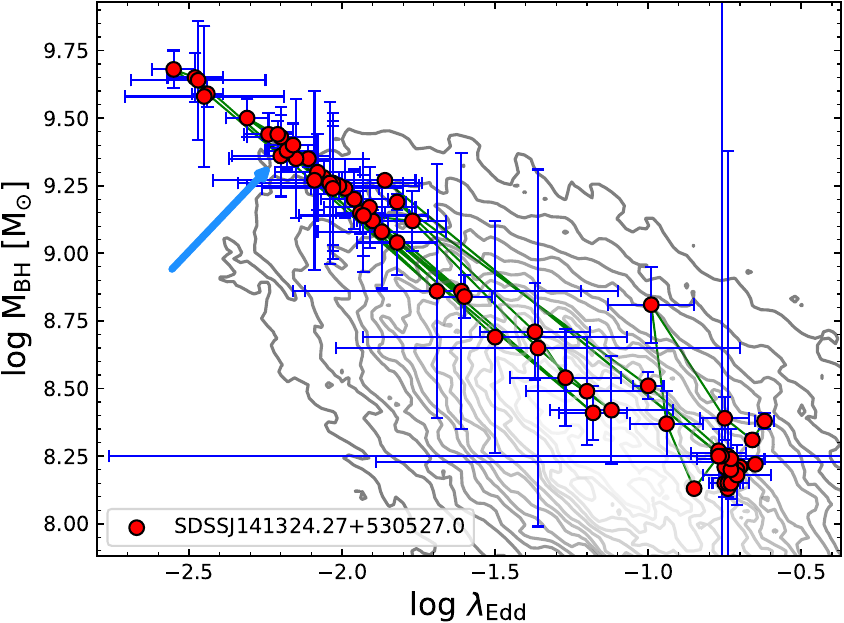}\\    
    
    \caption{Eddington ratio - black hole mass plane for the sources - (top:) SDSSJ022556.0+003026 \citep{macleod2016}, and (bottom:) SDSSJ141324.27+530527.0 \citep{Wang2018ApJ...858...49W}. Each data point from the sample (see Table \ref{tab:table-pyqsofit}) is shown using red dots with the respective error bars marked in blue. The arrow marks the location of the first data in the series (in terms of MJD) in these panels. The green line shows the trail followed by the series in each panel. The \citet{rakshitetal2020} full sample contains 526,265 spectroscopically observed SDSS quasars. Of these, there are 57,077 sources with \rfe{}$>$0, FWHM(\hb{})$>0$, and the errors associated with these two parameters non-zero. The extent of this subset is shown using the contours. The contours depict the iso-proportions of the density with a step size of 6.67\% (i.e., 15 levels). The right panels show a zoomed-in version of the plane for these two sources.}
    \label{fig:M-Edd}
\end{figure*}

\begin{figure*}[!htb]
    \centering

    \includegraphics[width=0.495\textwidth]{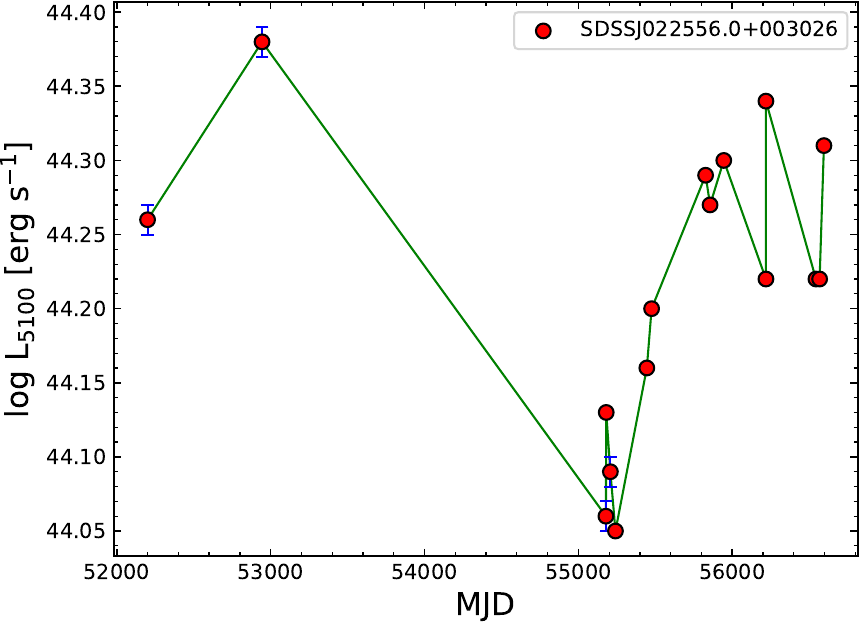}
    \includegraphics[width=0.495\textwidth]{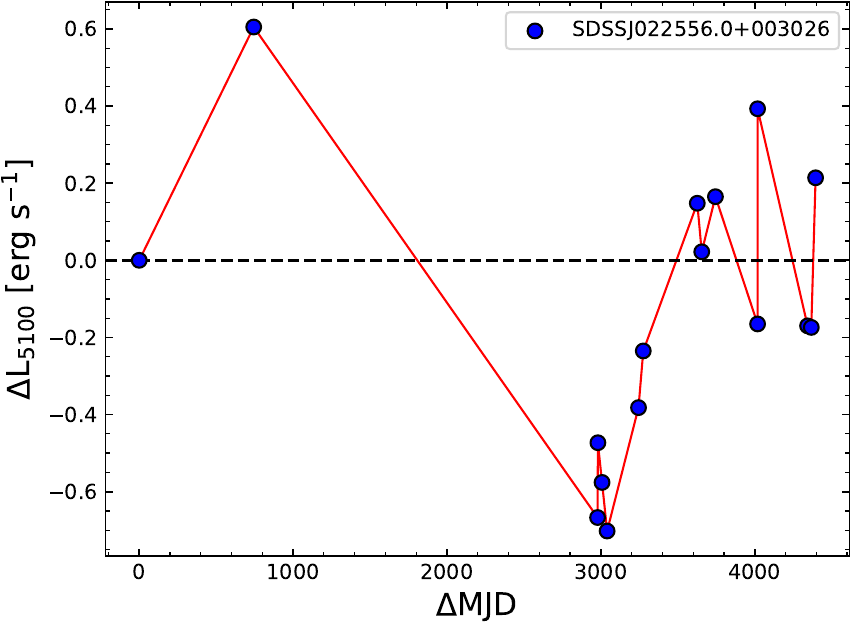}\\  
    \includegraphics[width=0.495\textwidth]{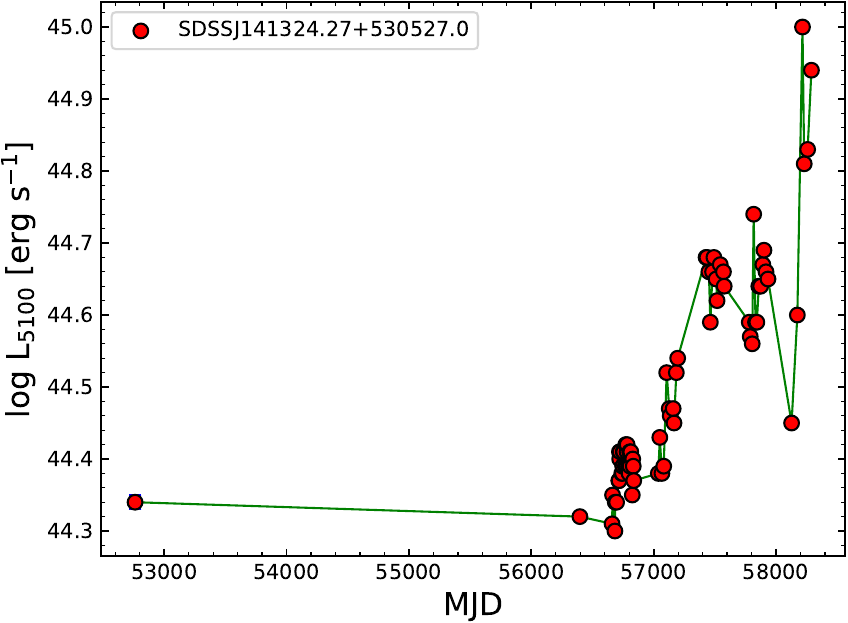}
    \includegraphics[width=0.495\textwidth]{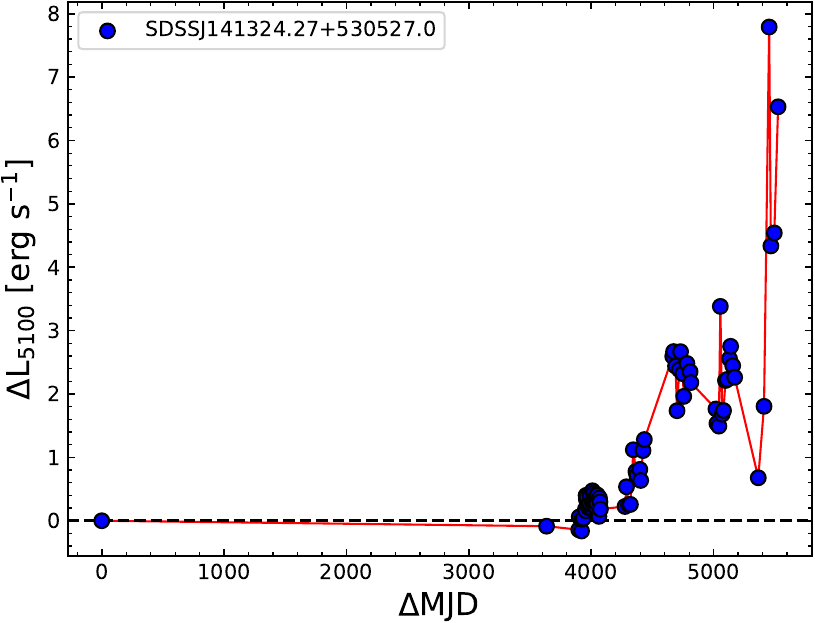}\\    
    
    \caption{AGN monochromatic luminosity at 5100\AA\ versus observed epoch (in MJD) for the two sources shown in Figure \ref{fig:M-Edd}. The right panels show the change in the \lagn{} versus the time difference (in days). The change is estimated to the first epoch for the respective source.}
    \label{fig:delL5100}
\end{figure*}

Since we have the compendium of the estimates for the BH masses and Eddington ratios for all available spectroscopic epochs per source, we can check the movement of the source in the \mbh{}-\ledd{} plane. We select the sources which have at least 3 epochs of SDSS spectroscopy. This criterion allows us to check for systematic changes in the source's fundamental properties. Based on this criterion, we have 32 sources. In Figure \ref{fig:M-Edd}, we show two examples from this sub-sample - SDSSJ022556.0+003026 \citep{macleod2016}, and SDSSJ141324.27+530527.0 \citep{Wang2018ApJ...858...49W}. We locate the transitions made by these sources on top of the SDSS quasar catalog with spectroscopic properties \citep[DR14 QSO catalog,][]{rakshitetal2020}. The \citet{rakshitetal2020} full sample contains 526,265 spectroscopically observed SDSS quasars from the SDSS data release 14 \citep{Paris_etal_2018}. Of these, there are 57,077 sources with \rfe{}$>$0, FWHM(\hb{})$>$0, and the errors associated with these two parameters are non-zero. These criteria allow us to constrain the sample with robust \mbh{} and \ledd{} estimates. The extent of this subset is shown using the contours in Figures \ref{fig:M-Edd} and \ref{fig:EV1}. We note that the fluctuations in the \mbh{} observed in the sources in our sample are primarily due to the changes in the FWHM of the broad \hb{} emission profile (see Section \ref{sec:ev1} for more details). From the theoretical point of view, one expects the changes in the BH mass to occur in Salpeter's timescale (or e-folding timescale, see \citealt{Shen_2013}). This characteristic timescale, $t = \frac{4.5\times 10^8 \eta}{\lambda_{\rm Edd}(1 - \eta)}$ [yr], is dependent on the radiative efficiency ($\eta$) and the Eddington ratio (\ledd{}), and ranges between 10$^{7}$ - 10$^{9}$ years (for an $\eta$ $\sim$ 0.1, and \ledd{} = [1, 10$^{-2}$]). This has been confirmed through studies that estimated the lifetime of AGN's "active phase" \citep[see e.g.,][]{Schawinski_etal_2015, Moravec_etal_2022}. Another useful way to estimate the \mbh{} would be to incorporate a proper {virial} factor \citep{Collin_etal_2006, Mejia-Restrepo_etal_2018, Panda_etal_2019b} and use the virial relation. But this process necessitates the information of the distance between the ionizing source and the BLR, i.e., the $\rm{R_{BLR}}$. One thus needs reverberation mapping monitoring for these sources to obtain more accurate \mbh{}. Until then, we suffice with the estimates for the \mbh{} obtained using the empirical relation as shown in Equation \ref{eq1}.\\

In the first case, i.e., SDSSJ022556.0+003026, we have 16 epochs of spectral information from which we estimate the \mbh{} and \ledd{} per epoch. The source starts (MJD = 52200) from a relatively high BH mass, log \mbh{} = 8.75$\pm$0.43 (in units of \msun{}), and high AGN luminosity, log \lagn{} = 44.26$\pm$0.01 (in units of erg s$^{-1}$). The corresponding Eddington ratio is log \ledd{} = -1.67$\pm$0.44. In the subsequent epochs, the \mbh{} varies within a 0.51 dex range, i.e., between log \mbh{} = 8.88$\pm$0.17 (second epoch, MJD = 52944, 744 days after the first epoch) and 9.43$\pm$0.06 (4$^{\rm th}$ epoch, MJD = 55181, 2981 days after the first epoch). The corresponding change in the Eddington ratio ranges between log \ledd{} = -1.67$\pm$0.18 (second epoch) and -2.50$\pm$0.09 (6$^{\rm th}$ epoch, MJD = 55241, 3041 days after the first epoch). This source is special as it is the only one that demonstrates multiple episodes of brightening and dimming phases (6 phases) and is categorized under the class ``Multi'' in Figure \ref{fig:state-change-dist}. We demonstrate the change in the luminosity for this source across the 16 epochs in the top panel of Figure \ref{fig:delL5100} where we can see the modulation of the source.\\

For the second source, i.e., SDSSJ141324.27+530527.0, we have 72 epochs of spectral information from which we estimate the \mbh{} and \ledd{} per epoch. The source starts (MJD = 52762) with a high BH mass, log \mbh{} = 9.36$\pm$0.15, and one of the lowest AGN luminosity for this source among all available epochs, log \lagn{} = 44.34$\pm$0.01. The corresponding Eddington ratio is log \ledd{} = -2.20$\pm$0.16, which is 0.72 dex smaller than the average for this source. The highest value for the \mbh{} for this source is estimated to be log \mbh{} = 9.68$\pm$0.07 (third epoch, MJD = 56660, 3898 days after the first epoch), while the lowest value for the \mbh{} is estimated to be log \mbh{} = 8.13$\pm$0.01 (56$^{\rm th}$ epoch, MJD = 57789, 5027 days after the first epoch). Correspondingly the highest value for the \ledd{} is estimated to be log \ledd{} = -0.62$\pm$0.03 (MJD = 58289, 5527 days after the first epoch, and is the last spectral epoch for this source), while the lowest value for the \ledd{} is estimated to be log \ledd{} = -2.55$\pm$0.07 which corresponds to the highest \mbh{} value reported above. Correspondingly, in Figure \ref{fig:spectra-compare-Wang2018}, we see a clear suppression in the broad \hb\ profile in the latter epochs relative to the first epoch. Overall, this source has brightened by nearly 6 times relative to its first epoch (see also the last column in Table \ref{tab:table-pyqsofit}). We highlight the overall change in the luminosity for this source across the 72 spectral epochs in the bottom panel of Figure \ref{fig:delL5100}. Contrary to the previous source, this does show modulation in the luminosity but with an overall rise in the AGN luminosity which supports the source to be at its highest accretion rate since the monitoring.\\

The remainder of the sources where we have at least 3 epochs of SDSS spectroscopy with \mbh{} and \ledd{} estimated per epoch are reported in Figures \ref{fig:M-Edd-others}, and \ref{fig:M-Edd-others2} {(see Appendix \ref{app:plots})}. Mostly, the various epochs for each source are well located within the contours of the subset of the DR14 QSO catalog, albeit a few. Generally speaking, the sources occupy the upper-left end of the contours, i.e., with high BH masses and low Eddington ratios. Occasionally, there are epochs for some sources where the source starts with (goes to) a relatively low BH mass/high Eddington ratio regime but moves (back) to the high BH mass/low Eddington ratio region. Individually, there are a few epochs for these sources where the estimates for the BH mass and Eddington ratios are well outside the contours (the last contour in Figures \ref{fig:M-Edd}, \ref{fig:M-Edd-others}, and \ref{fig:M-Edd-others2} is drawn at 6.67\% of the total number of sources in the sample) typically giving BH masses, \mbh{} $\gtrsim$ 10$^{9-10}$ \msun{}, and Eddington ratios, log \ledd{} $\lesssim$ -3. This agrees well with conclusions from previous works \citep[e.g.,][]{lawrence_etal_2016, rumbaugh_etal_2018}.

\subsection{Movement on the Eigenvector 1 optical plane}
\label{sec:ev1}

Another way to look at the evolution of an AGN, specifically Type-1s, is to track their transition along the optical plane of the Eigenvector 1 (EV1) schema. The EV1 schema dates back to the seminal work by \citet{Boroson_Green_1992}, wherein the authors performed a principal component analysis using observed spectral properties for a sample of 87 low-redshift sources to realize the, now well-known, main sequence of quasars. The EV1 schema, or the optical plane of the main sequence of quasars is the plane between the FWHM of broad \hb\ emission line and the strength of the optical \feii\ emission to \hb{}. This has been a key subject of study spanning close to three decades that has advanced our knowledge of the diversity of Type-1 AGNs both from observational and theoretical aspects \citep[see][and references therein]{Boroson_Green_1992, Sulentic_etal_2000, Shen_Ho_2014, Marziani_etal_2018, Panda_etal_2018, Panda_etal_2019b}. An interesting aspect of the EV1 is the connection of the two parameters to fundamental BH parameters that make up the optical plane, i.e., the FWHM(\hb{}) is closely connected to the BH mass through the virial relation \citep{Czerny_Nikolajuk_2010, Shen_2013}, while the other parameter - the \feii{} strength (or \rfe{}) is closely connected to the Eddington ratio, and hence to the accretion rate of the BH \citep{Shen_Ho_2014, Marziani_etal_2018, Panda_etal_2018, Panda_etal_2019a, Panda_etal_2019b, Martinez-Aldama_etal_2021}. Thus, in principle, the EV1 schema can be a valuable tool to investigate the evolution of the BH's mass and accretion state by analyzing their movement along the optical plane of the main sequence of quasars.\\

\begin{figure*}[!htb]
    \centering
    \includegraphics[width=0.495\textwidth]{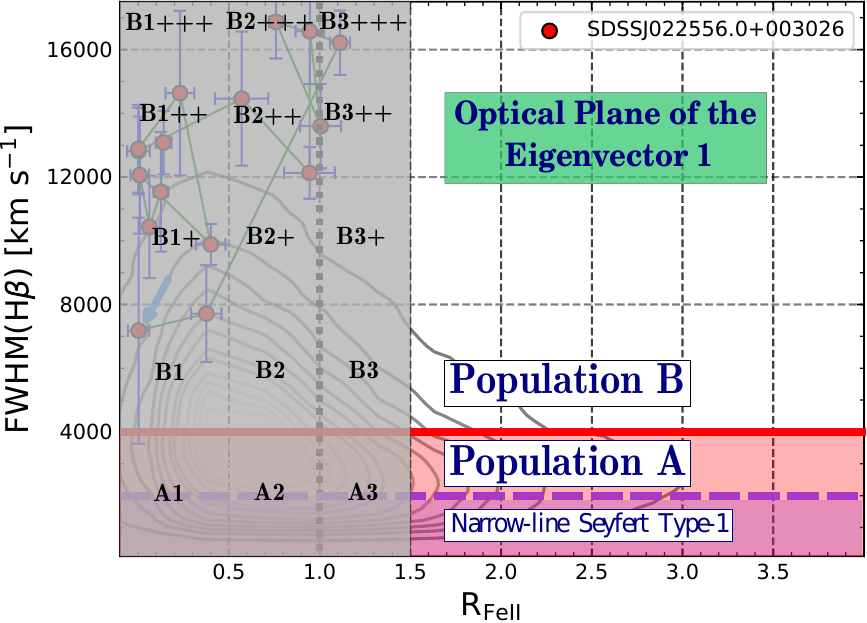}
    \includegraphics[width=0.495\textwidth]{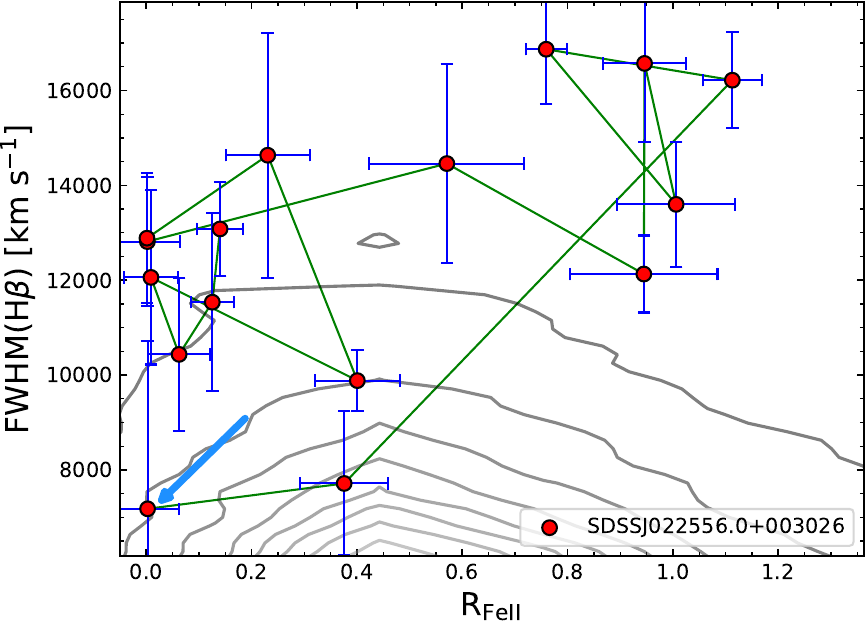}\\  
    \includegraphics[width=0.495\textwidth]{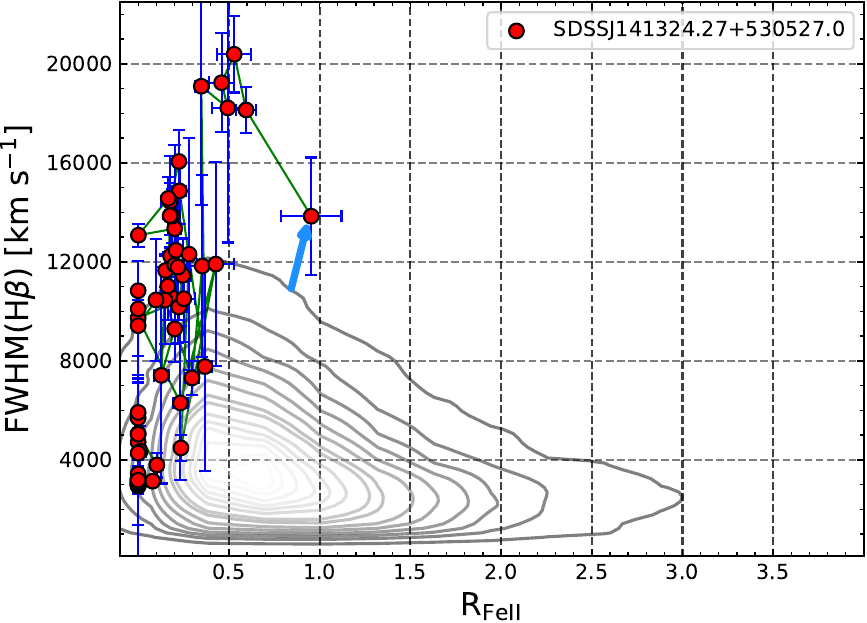}
    \includegraphics[width=0.495\textwidth]{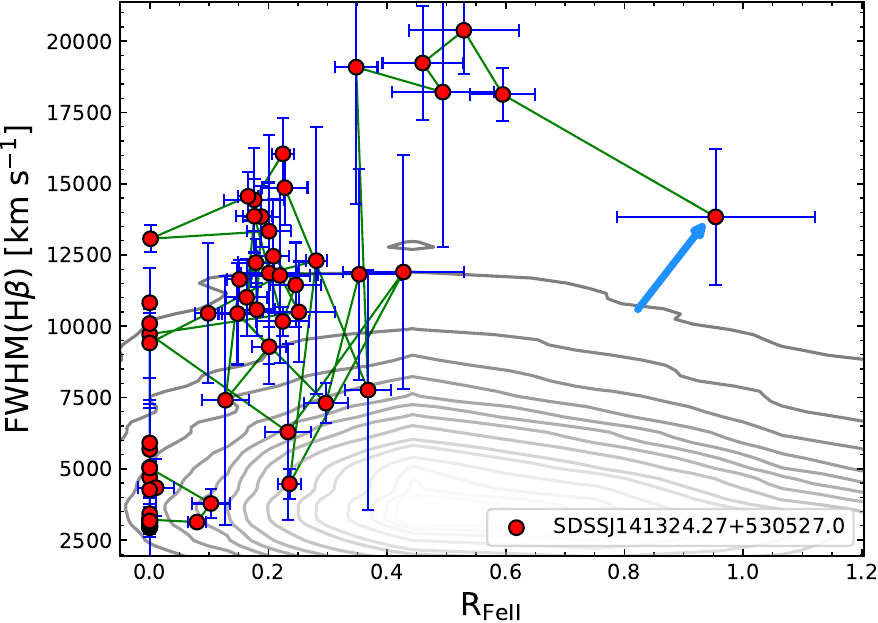}\\
    
    \caption{Optical plane of the main sequence of quasars for two exemplary sources with at least 3 spectroscopic epochs in our sample with \rfe{}$>$0 and FWHM(\hb{})$>$0 - (top:) SDSSJ022556.0+003026 \citep{macleod2016}, and (bottom:) SDSSJ141324.27+530527.0 \citep{Wang2018ApJ...858...49W}. {In the top-left panel, the red line denotes the threshold in FWHM(\hb{}) at 4000 \kms{} which separates the Population A and Population B sources. The ``classical'' Narrow-line Seyfert Type-1 (NLS1s) are located below the FWHM(\hb{}) $\leq$ 2000 \kms{} (dashed line). The vertical black dotted line marks the limit for \rfe{} = 1 separating the weak and strong \feii{} emitters (or xA sources). We also highlight the spectral types, i.e., A1, B1, B1+, etc. in this figure as described in Section \ref{sec:ev1}.} Each data point from the sample (see Table \ref{tab:table-pyqsofit}) is shown using red dots with the respective error bars marked in blue. The arrow marks the location of the first data in the series (in terms of MJD) in these panels. The green line shows the trail followed by the series in each panel. The \citet{rakshitetal2020} full sample contains 526,265 spectroscopically observed SDSS quasars. Of these, there are 57,077 sources with \rfe{}$>$0, FWHM(\hb{})$>0$, and the errors associated with these two parameters non-zero. The extent of this subset is shown using the contours. The contours depict the iso-proportions of the density with a step size of 6.67\% (i.e., 15 levels). The right panels show a zoomed-in version of the plane for these two sources.}
    \label{fig:EV1}
\end{figure*}

The onset of the 1990s and the emergence of larger surveys further led to the classification of Type-1 AGNs based on the ``broadness'' of their \hb{} emission line profile, i.e., Population A and Population B. The concept of the two quasar populations (A and B) was introduced as a simplified analog of the seven principal spectral types (OBAFGKM) identified in the HR Diagram for classification of stars \citep[e.g.,][]{Sulentic_etal_2000}. Population A sources can be understood as the class that includes local NLS1s as well as more massive high accretors which are mostly classified as radio-quiet \citep[][]{Marziani_Sulentic_2014} and have FWHM(\hb{}) $\lesssim$ 4000 \kms{}. On the other hand, Population B sources have broader \hb{} profiles, FWHM(\hb{}) $\gtrsim$ 4000 \kms{} {(see top-left panel in Figure \ref{fig:EV1})}. In addition, \citet[][and references therein]{Marziani_etal_2018} sub-categorizes the optical plane into spectral types. These spectral types have bin-width, $\Delta$\rfe{} = 0.5, and bin-height, $\Delta$FWHM(\hb{}) = 2,000 \kms{}. Hence, a source belonging to the bin - 0 $\leq$ FWHM(\hb{}) $\leq$ 4000, and, 0 $\leq$ \rfe{} $\leq$ 0.5, will be classified as type A1. A source with 0 $\leq$ FWHM(\hb{}) $\leq$ 4000, and, 0.5 $\leq$ \rfe{} $\leq$ 1.0, will be classified as type A2 and so on. Similarly, a source belonging to the bin - 4000 $\leq$ FWHM(\hb{}) $\leq$ 8000, and, 0 $\leq$ \rfe{} $\leq$ 0.5, will be classified as type B1; while a source belonging to the bin - 8000 $\leq$ FWHM(\hb{}) $\leq$ 12000, and, 0 $\leq$ \rfe{} $\leq$ 0.5, will be classified as type B1+; and, a source belonging to the bin - 12000 $\leq$ FWHM(\hb{}) $\leq$ 16000, and, 0 $\leq$ \rfe{} $\leq$ 0.5, will be classified as type B1++.\\

Typically, a transition of a source from Population A to Population B is rarely observed \citep[see e.g.,][]{Bon_etal_2018}. \citet{Sulentic_Marziani_2015} in their review posed an important question - \textit{``Are population A and B simply two extreme ends of the main sequence or do they represent two distinct quasar populations? Or are they tied via a smooth transition in the accretion mode?''} With our multi-epoch subset, we would like to test this scenario if any of our sources make a transition from Population A to Population B or vice versa. To gauge the transition of our sources in the optical plane of the EV1, similar to our approach from the previous section, we select sources that have at least 3 epochs of SDSS spectroscopy available and where we have non-zero FWHM(\hb{}) and \rfe{}. Altogether we have 32 such sources. In Figure \ref{fig:EV1}, we show the movement of two exemplary sources (these are the same sources shown in Figure \ref{fig:M-Edd}) - SDSSJ022556.0+003026 \citep{macleod2016}, and SDSSJ141324.27+530527.0 \citep{Wang2018ApJ...858...49W}.\\ 

For the first source (SDSSJ022556.0+003026 with 16 epochs of observations), the minimum value for the \rfe{} is 0.001$\pm$0.000 obtained at the 10$^{\rm th}$ epoch (MJD=55856, 3656 days after the first epoch). This corresponds to a FWHM(\hb{}) of 12886$\pm$1376 \kms{}. This puts the source in the sub-class B1++ on the EV1 diagram. The corresponding value for the \mbh{} for this epoch is estimated to be log \mbh{} = 8.75$\pm$0.06, and quite low Eddington ratio, log \ledd{} = -2.50$\pm$0.06. We note that the AGN luminosity in this case is moderate, i.e., log \lagn{} = 44.27$\pm$0.00. The maximum value for the \rfe{} is 1.11$\pm$0.06 obtained at the 3$^{\rm rd}$ epoch (MJD=55179, 2979 days after the first epoch). This corresponds to a FWHM(\hb{}) of 16218$\pm$1009 \kms{}. This puts the source in the sub-class B3+++. The corresponding value for the \mbh{} for this epoch is estimated to be log \mbh{} = 9.36$\pm$0.06, with a similar low Eddington ratio to the former case, i.e., log \ledd{} = -2.47$\pm$0.06. We note that the AGN luminosity in this case is moderate, i.e., log \lagn{} = 44.06$\pm$0.01. We however note that this particular spectrum has a relatively low S/N=4.6, especially around the \hb{} profile which may suggest that the deduced FWHM value is an upper limit. In any case, the \feii{} emission is quite prominent. The source, even with the significant variations in the FWHM(\hb{}) and \rfe{}, stays in Population B throughout all epochs and moves on the EV1 diagram in the following track: B1 $\rightarrow{}$ B1 $\rightarrow{}$ B3+++ $\rightarrow{}$ B2+++ $\rightarrow{}$ B3++ $\rightarrow{}$ B2+++ $\rightarrow{}$ B2++ $\rightarrow{}$ B2++ $\rightarrow{}$ B1++ $\rightarrow{}$ B1++ $\rightarrow{}$ B1++ $\rightarrow{}$ B1+ $\rightarrow{}$ B1++ $\rightarrow{}$ B1+ $\rightarrow{}$ B1+ $\rightarrow{}$ B1++.\\

For the second source (SDSSJ141324.27+530527.0 with 72 epochs of observations), the minimum value for the \rfe{} is 0 obtained 14 times throughout the monitoring. We refer the readers to the Table \ref{tab:table-pyqsofit} for the complete tabulation of the \rfe{} and FWHM(\hb{}), and the corresponding values for the \mbh{} and \ledd{} for this source. We note here in passing that the epochs with \rfe{}=0 correspond to a wide range in FWHM(\hb{}), i.e., 2947 - 10827 \kms{}. Hence, we are left with 58 observations with non-zero \rfe{} measurements where the \rfe{} ranges between 
{$\lesssim$0.01} - 0.95, and the corresponding range in FWHM(\hb{}) varies between 2935 - 20380 \kms{}. As we can see, there are epochs with \rfe{} values tending to zero, which suggests that the \feii{} emission for the source does go to zero in reality. Thus, for completeness, we show in the bottom panel of Figure \ref{fig:EV1} the full range, including the epochs where \rfe{} = 0. The source, unlike the previous one, starts off as a Population B source but eventually transitions to Population A, and moves on the EV1 diagram in the following track: B2++ $\rightarrow{}$ B2+++ $\rightarrow{}$ B2++++ $\rightarrow{}$ B1+++ $\rightarrow{}$ B1+++ $\rightarrow{}$ B1+++ $\rightarrow{}$ B1 $\rightarrow{}$ B1++ $\rightarrow{}$ B1+++ $\rightarrow{}$ B1++ $\rightarrow{}$ B1++ $\rightarrow{}$ B1++ $\rightarrow{}$ B1++ $\rightarrow{}$ B1++ $\rightarrow{}$ B1+ $\rightarrow{}$ B1++ $\rightarrow{}$ B1+ $\rightarrow{}$ B1++ $\rightarrow{}$ B1+ $\rightarrow{}$ B1+ $\rightarrow{}$ B1++ $\rightarrow{}$ B1+ $\rightarrow{}$ B1 $\rightarrow{}$ B1+ $\rightarrow{}$ B1+ $\rightarrow{}$ B1 $\rightarrow{}$ B1+ $\rightarrow{}$ B1+ $\rightarrow{}$ B1++ $\rightarrow{}$ B1 $\rightarrow{}$ B1+ $\rightarrow{}$ B1 $\rightarrow{}$ B1+ $\rightarrow{}$ B1+ $\rightarrow{}$ B1+ $\rightarrow{}$ B1+ $\rightarrow{}$ B1+ $\rightarrow{}$ B1 $\rightarrow{}$ B1 $\rightarrow{}$ B1 $\rightarrow{}$ B1 $\rightarrow{}$ B1 $\rightarrow{}$ B1+ $\rightarrow{}$ B1 $\rightarrow{}$ A1 $\rightarrow{}$ A1 $\rightarrow{}$ B1+ $\rightarrow{}$ A1 $\rightarrow{}$ A1 $\rightarrow{}$ A1 $\rightarrow{}$ A1 $\rightarrow{}$ A1 $\rightarrow{}$ A1 $\rightarrow{}$ A1 $\rightarrow{}$ A1 $\rightarrow{}$ A1 $\rightarrow{}$ A1 $\rightarrow{}$ A1 $\rightarrow{}$ A1 $\rightarrow{}$ A1 $\rightarrow{}$ A1 $\rightarrow{}$ A1 $\rightarrow{}$ A1 $\rightarrow{}$ A1 $\rightarrow{}$ A1 $\rightarrow{}$ A1 $\rightarrow{}$ A1 $\rightarrow{}$ A1 $\rightarrow{}$ B1 $\rightarrow{}$ A1 $\rightarrow{}$ A1 $\rightarrow{}$ A1. As we can see, in the beginning, the source modulated within Population B (primarily in the sub-classes B2++ and B2+++). The \rfe{} went done in the next few epochs and the source made modulation within B1, B1+, and B1++. In the last 28 epochs, the FWHM(\hb{}) further decreased and the source has moved along the B1 and A1 sub-classes, and eventually stays as a Population A source. This is a great example where the availability of such a large number of observations stretched across 5527 days ($\approx$15 years) allows us to peer through the strong, variable nature of the source using the EV1 diagram. The remaining sources from this sub-sample are shown in Figures \ref{fig:EV1-others}, and \ref{fig:EV1-others2} {(see Appendix \ref{app:plots})}.\\

From the sample of 32 sources, we find that the majority (26/32) modulate within the Population B class. Only one source (SDSSJ114408.9+424357.5 from \citealt{graham2020}) stays within Population A across epochs. Only one source (SDSSJ233602.9+001728 from \citealt{ruan2016}) does the transition from Population B to Population A. Similarly, we have only a single object that makes the following transitions: A $\rightarrow{}$ B $\rightarrow{}$ A (SDSSJ110349.2+312416.7 from \citealt{graham2020}), and B $\rightarrow{}$ A $\rightarrow{}$ B $\rightarrow{}$ A $\rightarrow{}$ B $\rightarrow{}$ A (SDSSJ141324.27+530527.0 from \citealt{Wang2018ApJ...858...49W}), while we have the remaining two sources (SDSSJ105058.42+241351.18 and SDSSJ230614.18-010024.45, both are from \citealt{Green_etal_2022}) which make a B $\rightarrow{}$ A $\rightarrow{}$ B transition. It is interesting to note that none of these sources that transition to Population A or stay in Population A class show an FWHM(\hb{}) $\lesssim$ 2000 \kms{} - that is, the classical threshold for the Narrow Line Seyfert Type-1 sources \citep{Osterbrock_Pogge_1985, Boroson_Green_1992, Mathur_2000, Collin_etal_2006, Zhou_etal_2006, Rakshit_etal_2017, Marziani_etal_2018, Chen_etal_2018, Sniegowska_etal_2023}.\\

\subsection{Balmer decrement}

\begin{figure*}[!htb]
    \centering
    \includegraphics[width=0.3\textwidth]{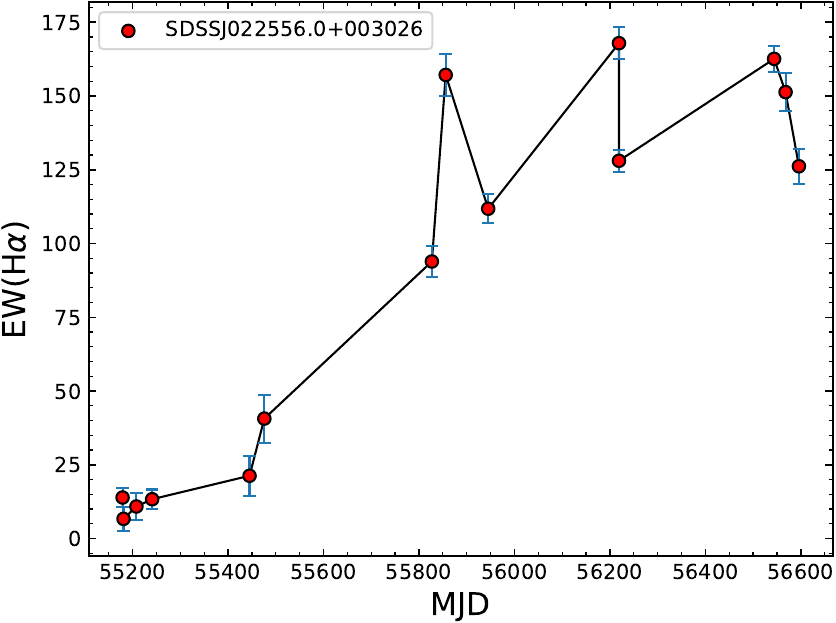}
    \includegraphics[width=0.3\textwidth]{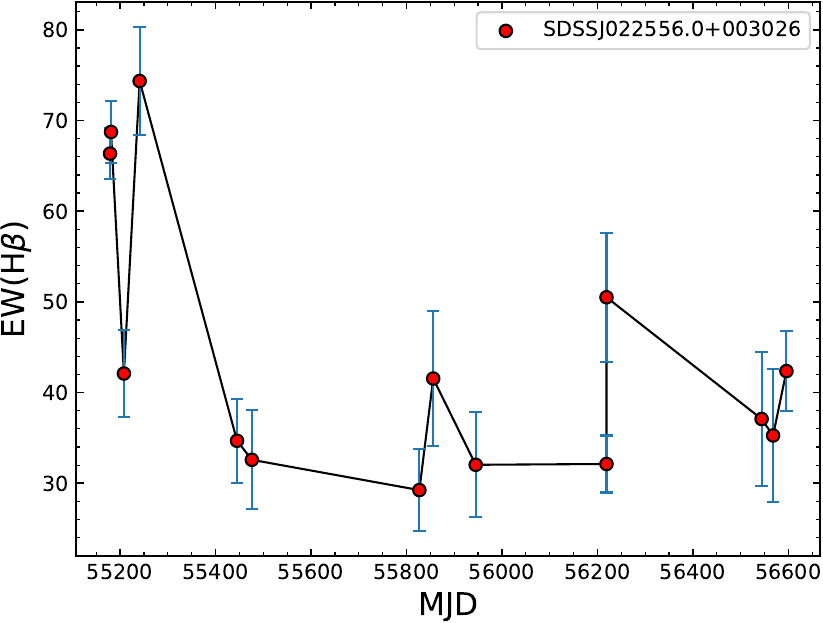}
    \includegraphics[width=0.3\textwidth]{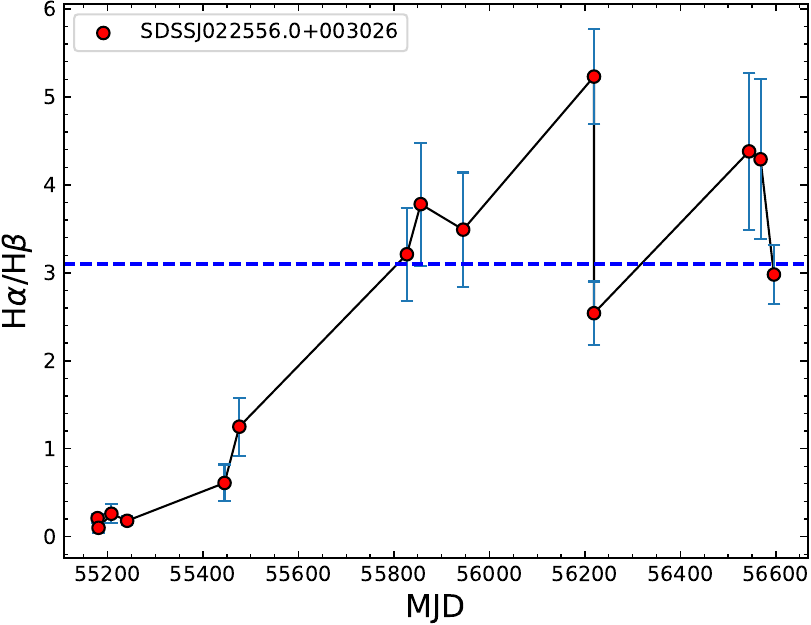}\\
    
    \includegraphics[width=0.3\textwidth]{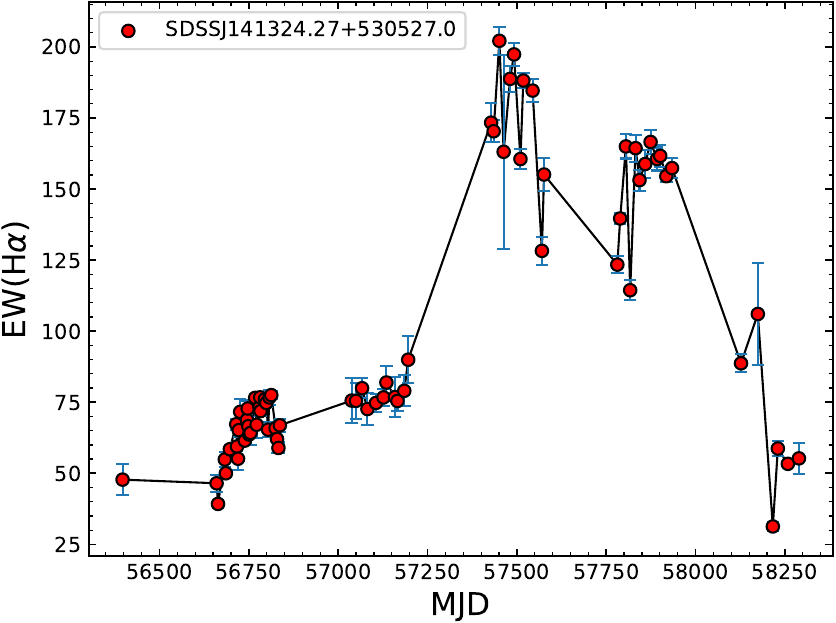}
    \includegraphics[width=0.3\textwidth]{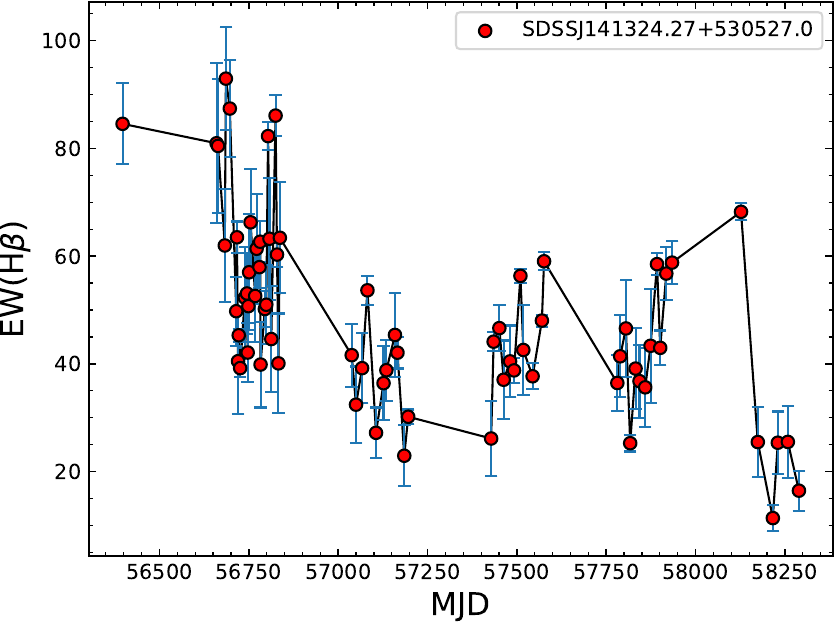}
    \includegraphics[width=0.3\textwidth]{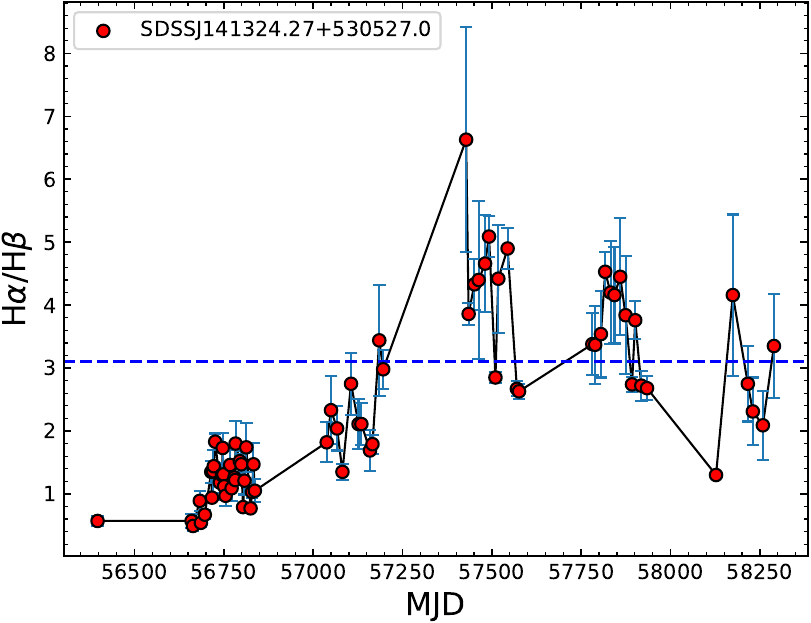}

    \caption{Trends of (left:) EW(\ha{}), (middle:) EW(\hb{}), and (right:) EW(\ha{})/EW(\hb{}), versus the observed data (in MJD) for (top:) SDSSJ022556.0+003026 \citep{macleod2016}; and (bottom:) SDSSJ141324.27+530527.0 \citep{Wang2018ApJ...858...49W}. The dashed blue line marks the fiducial value of 3.1 for the \ha{}/\hb{} ratio.}
    \label{fig:balmer_decre}
\end{figure*}

\begin{figure*}[!htb]
    \centering
    \includegraphics[width=\columnwidth]{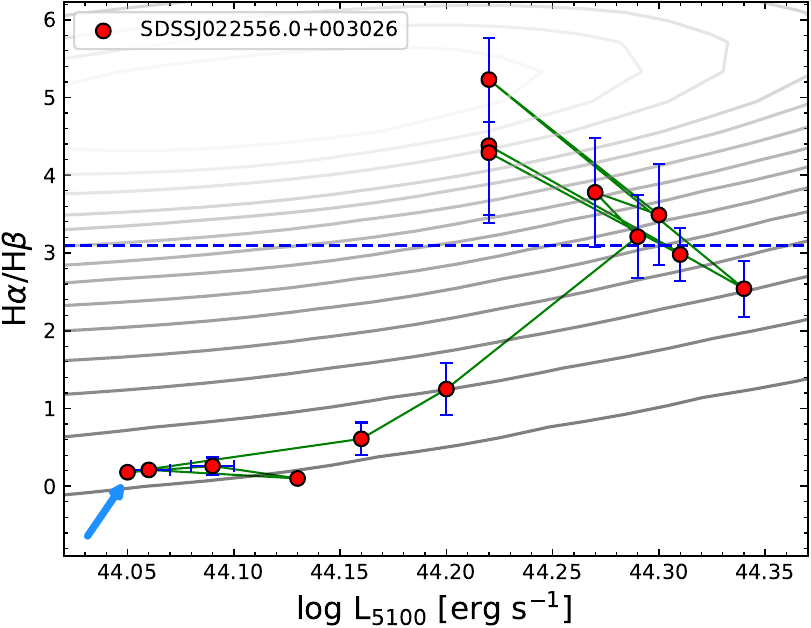}
    \includegraphics[width=\columnwidth]{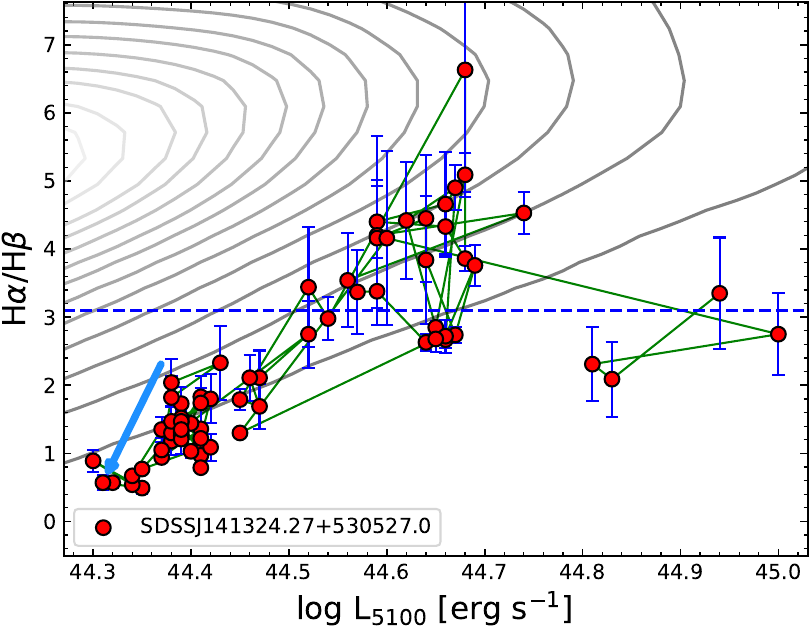}
    \caption{Trend between the \ha{}/\hb{} ratio and AGN luminosity at 5100\AA\ (\lagn{}) for the same two sources as shown in Figure \ref{fig:balmer_decre}. The location of the first epoch in each case is marked with an orange arrow. The contours in the background show the density distribution for the two parameters from the SDSS DR14 QSO catalog \citep{rakshitetal2020}.}
    \label{fig:balmer_decre-lagn}
\end{figure*}

The Balmer decrement (specifically, the ratio of \ha{} to \hb{}) is useful to probe the physics of the ionized medium from which these lines originate \citep{Kwan_Krolik_1979,Kwan_Krolik_1981, Mathews_etal_1980, Canfield_Puetter_1981, Dong_etal_2008, Jin_etal_2012, Ilic_etal_2012}.\\

The \ha{}/\hb{} ratio allows us to determine the extent of dust extinction for the low-density gas of the narrow-line region (NLR) in AGNs \citep{Osterbrock_Ferland_2006}. In this case, a value of 3.1 is generally adopted. While in the case of the BLR, the densities are much larger which in turn affects the \ha{}/\hb{} ratio, in addition to wavelength-dependent extinction by dust \citep{Osterbrock_1984, Goodrich_1995}. The observed broad-line \ha{}/\hb{} ratio is usually larger (steeper) than the Case B recombination value \citep[see e.g.,][]{Osterbrock_1977, Rafanelli_1985, Dong_etal_2005}, sometimes even as steep as ten or higher \citep{Osterbrock_1981, Crenshaw_Peterson_1988}. Considering a large, homogeneous sample of 446 low-redshift (z $\leq$ 0.35) blue type-1 AGNs, \citet{Dong_etal_2008} recovered an average value of 3.06, and suggest that the broad-line \ha{}/\hb{} ratio can be used as a dust extinction estimator, especially for radio-quiet AGNs.

\subsubsection{Temporal variations}

In our sample, we have 32 sources for which we have simultaneous measurements for \ha{} and \hb{}, and we have at least 3 epochs of observations for the source. To assess the change in the behavior of the \ha{}/\hb{} ratio as a function of time, in Figure \ref{fig:balmer_decre} we show the trends of EW(\ha{}), EW(\hb{}), and EW(\ha{})/EW(\hb{}), versus the observed date (in MJD) for the same exemplary sources - SDSSJ022556.0+003026 \citep{macleod2016}; and SDSSJ141324.27+530527.0 \citep{Wang2018ApJ...858...49W}. For the first source (SDSSJ022556.0+003026), we see that the {trends} between \ha{} and \hb{} are contrary to each other although the trend with \ha{}/\hb{} ratio follows mostly the trend shown by EW(\ha{}), as a function of time. There is an indication of saturation in the EW(\hb{}) closer to the latest epoch - we see a displacement towards lower values in the EW(\hb{}) after MJD=55445. Although, the EW(\ha{}) shows no such indication of saturation, rather shows an increase in the latter epochs. This may be attributed to the longer response time for \ha{} relative to \hb{} emission \citep[see e.g.,][]{2020ApJ...903...51W} although the trends in this particular source are complementary. \\

On the other hand, for SDSSJ141324.27+530527.0, the EW(\ha{}) as a function of time looks like a shifted version of the EW(\hb{}). As noted in Table 1 of \citet{2020ApJ...903...51W, Grier_etal_2017}, the rest-frame time lag for \hb{} (25.5$^{+10.9}_{-5.8}$ days) is about half the time lag estimated for \ha{} (56.6$^{+7.3}_{-15.1}$ days) which could be an added reason for such behavior. In the latest epochs, the two sources discussed above, modulate about 3.1. In our sample, there are 6 sources where all epochs have a \ha{}/\hb{} $<$ 3.1. On the contrary, there are 9 sources where all epochs have a \ha{}/\hb{} $>$ 3.1, even with the demonstrated modulations.

\subsubsection{Dependence on the AGN luminosity}

Monitoring campaigns focusing on single objects have found that the Balmer decrement is anti-correlated with the continuum flux \citep{Shapovalova_etal_2004, Shapovalova_etal_2010, Popovic_etal_2011}, although often show complicated behavior depending on the state of the AGN activity (e.g., NGC 4151, \citealt{Shapovalova_etal_2008}). Sources show the anti-correlation between the \ha{}/\hb{} ratio and AGN luminosity in the low state but the correlation flattens to $\sim$4.5 at higher values of continuum flux \citep{Popovic_etal_2011}. We tried to assess the behavior of our sources similarly. In Figure \ref{fig:balmer_decre-lagn}, we highlight two exemplary cases -  SDSSJ022556.0+003026 \citep{macleod2016}; and SDSSJ141324.27+530527.0 \citep{Wang2018ApJ...858...49W}. The behavior for both these sources and the remaining (see Figures \ref{fig:balmer_decre_others}, and \ref{fig:balmer_decre_others2} {in Appendix \ref{app:plots}}) is complicated with no obvious systematic increasing/decreasing trend. Yet, the \ha{}/\hb{} ratio seems to obtain the highest values close to the predicted value of $\sim$4.5 as per \citet{Popovic_etal_2011}. This highest value corresponds generally to the highest AGN luminosity (here, we consider the monochromatic luminosity at 5100\AA\ as an indicator of the AGN luminosity), although there are some systematic differences from case to case.

\section{Discussions}
\label{sec:discussions}

This paper is the first in a series of preparing and analyzing spectral and other properties for a database of already discovered CL AGNs. Here, we focused on the spectral fitting and analysis of broad emission lines in the optical wavelengths ($\sim$2500-9200\AA) in a sizable sample of CL AGNs assembled from existing literature. A complimentary work to assess the photometric light curves for these CL AGNs is ongoing and will be presented in a forthcoming work. This will allow us to study and constrain the overall transition timescales for the sources in our sample.\\

Keeping account of the changes in the spectral features in AGN spectra would allow us to characterize their response to the changes in the ionizing continuum emanating from the very centers of the SMBH. Dense and regular monitoring campaigns, e.g., AGN Watch \citep{Peterson_etal_2002, Peterson_2004}, LAMP \citep{Barth_etal_2015}, AGN STORM \citep{DeRosa_etal_2015, Pei_etal_2017, Horne_etal_2021}, are crucial to improving our understanding of the temporal and physical variations in the immediate surroundings of the SMBH. Such regular campaigns help to probe the possible processes - intrinsic changes to the accretion disk structure and accretion flows, and/or extrinsic effects, such as cloud passages leading to obscuration of the line emitting medium, or tidal disruption events that may lead to an episodic rise in the AGN luminosity and accretion state \citep[see e.g.,][]{lamassa2015, macleod2016, graham2020, Short_2020, 2023A&A...669A.140P}\footnote{Another interesting scenario could be the presence of accretion-modified stars (AMS, \citealt{Wang_etal_2021}) that could be fuelled by the infalling matter in the accretion disk. These AMS stars can end up taking in a notable fraction of the energy from the accretion flow that would have otherwise been radiated away.}. The advent of large spectroscopic surveys (e.g., SDSS) coupled with the realization of refined algorithms to discover newer CL AGNs that exhibit such vibrant changes has been on the rise over the past decade \citep{lamassa2015, macleod2016, Yang2018, graham2020, potts2021, Green_etal_2022, Lopez-Navas_etal_2022}. Although many authors have come up with newer algorithms to identify newer CL AGNs and populate the ever-growing list of such sources, we still lack a unified compendium of such sources. Our paper is an effort to address this issue. With a meticulous search through the literature, we have compiled a sample of 93 known CL AGNs and quantified their spectral properties by a homogeneous spectral fitting procedure using {\sc PyQSOFit}. Additionally, we have gathered older/newer spectral epochs from all the available SDSS data releases to make the database more complete. This then allowed us to track the evolution of these CL AGNs on the Eddington ratio-black hole mass plane and the optical plane of the Eigenvector 1 (Quasar Main Sequence) as a function of their spectroscopic epochs. Our study has allowed us to identify targets that, if monitored regularly, can shed light on their evolving nature and the overall demographics of AGNs as a whole.\\

Currently, the sample we have compiled contains spectroscopic information from SDSS only. There are some studies \citep[e.g.][]{graham2020, Green_etal_2022} where the authors have utilized other spectroscopic instruments at their disposal to procure newer spectra for some CL AGNs. We intend to incorporate spectra for the CL AGNs observed using other instruments for our sample and other sources studied in the prior works, although a significant hurdle in such complete compilation is the absence of public databases and proprietary data which makes it difficult to access them. {Another important aspect of the work is related to the use of the classical \feii{} template by \citet{Boroson_Green_1992}. More recently, improved empirical/semi-empirical \feii{} templates for AGNs in the optical region are available at our disposition \citep{2010ApJS..189...15K, 2022ApJS..258...38P}. In their work, \citet{2022ApJS..258...38P} show the comparison between the classical \citet{Boroson_Green_1992} \feii{} template extracted from I Zw 1 (the prototypical NLS1 galaxy) and their new \feii{} template based on the HST observations for another NLS1 galaxy, Mrk493. The two templates agree well overall with some salient transitions being stronger/weaker between the two templates, especially under the \hb{} profile. It is important to note that Mrk493 has a smaller black hole mass (log \mbh{} $\approx$ 7.13) than I Zw 1 (log \mbh{} $\approx$ 7.94) and the overall \rfe{} estimate for Mrk493 ($\approx$ 0.78) is lower than I Zw 1 ($\approx$ 1.78). These estimates are taken from our previous work where we compile the optical (and near-infrared) properties of well-known Type-1 Seyfert galaxies \citep{Martinez-Aldama_etal_2021}. Thus, apart from the higher resolution due to the HST for Mrk493, the two templates are relatively similar in performance, especially when applied to low/medium-resolution data such as presented in our current analysis. For completeness, we test the difference in the \rfe{} recovery using the new template from \citet{2022ApJS..258...38P} versus the default template \citet{Boroson_Green_1992} for one of the epochs for the source SDSSJ022556.0+003026 \citep[Plate-MJD-Fiber = 3615-55179-0641,][]{macleod2016} where the \rfe{} was estimated to be larger than 1. We find that the ratio with the new template, \rfe{} = 1.207$\pm$0.120, is comparable to what we estimated with the default template, i.e., 1.113$\pm$0.056. We thus, conclude that the change of template does not alter our findings in this work.}\\

A direct extension of this work would be to estimate the physical conditions of the BLR from where these emission lines are originating \citep[e.g.,][]{Negrete_etal_2013, Marziani_etal_2018, Panda_etal_2018, Panda_etal_2019b, panda_2021_cafe}. This requires the preparation of the broad-band spectral energy distribution (SED) per epoch for these CL AGNs. With the realization of such SEDs for a given source, one can perform photoionization models to better understand the effect of the changes in the ionizing continuum leading to variation in the emission lines that we observe. As we have seen in this paper, the changes in the accretion rates (and Eddington ratios) are closely linked to the changing location of the source on the Eigenvector 1 sequence. This result has been known for quite some time \citep{Sulentic_etal_2000, Shen_Ho_2014, Marziani_etal_2018, Panda_etal_2018, Panda_etal_2019b} and has important consequences in studies of quasars for cosmological purposes \citep[e.g.,][]{Czerny_etal_2019, Martinez-Aldama_etal_2019, Panda_etal_2019c, Lusso_etal_2020, Khadka_etal_2022}. In one such recent study \citep{Panda_etal_2022}, the authors reintroduced the Pronik-Chuvaev effect, i.e., the gradual flattening in the \hb\ emission with rising AGN continuum. This was shown using the long-term spectroscopic monitoring for a prototypical Population B source - NGC 5548, compiled from $\sim$17 individual observing campaigns including the AGN Watch and AGN STORM projects \citep[e.g.,][]{Peterson_etal_2002, DeRosa_etal_2015, Lu_etal_2016}. The original paper of \citet{Pronik_Chuvaev_1972} discovered this phenomenon for Mrk 6 and later other works followed suit (NGC4051: \citealt{Wang_etal_2005}, NGC4151: \citealt{Shapovalova_etal_2008}). {Another interesting case is of Mrk1018. It was first classified as a type 1.9 Seyfert galaxy and transitioned to a type 1 Seyfert galaxy a few years later (mid-2020) before returning to its initial classification as a type 1.9 Seyfert galaxy after $\sim$30 years using multi-wavelength observations \citep{2016A&A...593L...9H, 2023A&A...677A.116B, 2021MNRAS.506.4188L, 2023arXiv231207663V}.} A key element in these works was the long, dedicated, quasi-continuous, spectrophotometric monitoring of the sources {allowing us to gain a better understanding of the physical processes at play that drive such changes in these AGNs. Variability often occurs on thermal timescales, reflecting changes in the accretion disk's temperature \citep{Noda_Done_2018, Stern_etal_2018, Ross2020MNRAS.498.2339R}. Accretion rate changes, however, unfold over longer viscous timescales (decades/centuries). The heating/cooling front crossing timescale, bridges thermal and viscous timescales, and each is parameterized by fundamental black hole properties. Another important timescale in this regard is the disk evaporation timescale, which describes the removal of cold disk material from the disk \citep{2006ASPC..360..265C}. Dust cloud movement in the torus can cause extinction changes, leading to extreme broad line variability \citep{lamassa2015} and are invoked to explain short-term (days/months) emission line disappearance and a sudden change in the AGN continuum. Additionally, the inflow timescale, governing gas movement in the disk, impacts continuum luminosity fluctuations due to accretion rate shifts \citep{lamassa2015, claudio_benny}, likely on a few-year scale, consistent with observed variability timescales. Such studies are slowly growing in number which will enable us to decipher the nature of variability in these CL AGNs.} The study of CL AGN phenomena will benefit from the next generation of high cadence telescopes and surveys such as the 4-meter Multi-Object Spectroscopic Telescope (4MOST, \citealt{4MOST_2012}) and the Legacy Survey for Space and Time (LSST, \citealt{Ivezic_etal_2019}). These next-generation telescopes will facilitate a more complete and systematic discovery of CL AGN. With the successful deployment of JWST \citep{JWST_2006, JWST_2022}, we are entering a new era of large spectroscopic surveys that will enable us to sample sources at higher redshifts and monitoring them will be useful to populate the existing sample of CL AGNs complementing the SDSS surveys.\\

\section{Conclusions}

This paper is the first in a series of preparing and analyzing spectral and other properties for a database of already discovered CL AGNs. Here, we focus on the spectral fitting and analysis of broad emission lines in a sizable sample of CL AGNs available in the literature. We summarize our findings from this study below:

\label{sec:conclusion}

\begin{itemize}

    \item We compile a sample of 110 known CL AGNs with multi-epoch SDSS/BOSS/eBOSS spectroscopic observations from the literature. To prepare our sample, we check for (a) the presence of at least 2 publicly available spectra from SDSS, BOSS, and eBOSS for the sources, and (b) the coverage of the optical \feii{} emission between 4434-4684\AA\ and the full \hb{} profile, the latter allowing to achieve one of the primary goals of this work - tracking the movement of the source on the optical plane of the Eigenvector 1 sequence. Our working sample contains 93 AGNs where the \hb{} emission line profile does not completely disappear in any epochs.\\
    
    \item We use {\sc PyQSOFit} for spectral decomposition \citep{pyqsofit} of all of our SDSS spectra. After the fitting procedure, we perform manual checks to identify the spectra where the broad \hb{} emission line disappears and/or the spectrum is dominated by noise in this region and no reliable \hb{} measurements could be made. Our final working sample contains 93 AGNs. We tabulate the AGN continuum luminosities (at 3000\AA, 5100\AA and 6000\AA), the EWs of the broad emission lines (\mg{}, \hb{}, and \ha{}), the FWHM of \hb{}, the strength of the optical \feii{} emission (\rfe{}), per epoch per source, chronologically. This then allows us to estimate the black hole mass (\mbh{}) and the Eddington ratio (\ledd{}) per epoch per source where the required parameters are available and well-estimated. This further allows us to categorize the sources in our sample as Turn-On or Turn-Off and subsequently check for repeated occurrences of such phases. 27 sources are classified under pure Turn-On, while 52 sources are put under the category of pure Turn-Off CL AGNs. The remainder of the sources show multiple episodes of Turn-On/Turn-Off phases.\\

    \item We see strong and significant correlations between luminosities: \textit{r} = 0.93 (\textit{p-value} = 1.0$\times 10^{-29}$) in the case of $\Delta$\lagnuv\ vs $\Delta$\lagn\ and \textit{r} = 0.98 (\textit{p-value} = 1.24$\times 10^{-60}$) for $\Delta$\lagnir\ vs $\Delta$\lagn{}. The slope for $\Delta$\lagnuv\ vs $\Delta$\lagn\ is steeper (1.75) relative to the slope for $\Delta$\lagn\ vs $\Delta$\lagnir{} (0.89) which suggests a systematic increased change in the bluer wavelengths around 3000\AA\ compared to the redder wavelengths (around 5100\AA\ and 6000\AA), i.e., a stronger AGN contribution in the UV between the epochs. In the case of correlations between the $\Delta$EW of the emission line and the corresponding change in the continuum luminosity, we notice strong anti-correlation (\textit{r} = -0.79, \textit{p-value} = 9.32$\times 10^{-12}$) between $\Delta$\lagnuv\ and $\Delta$EW(\mg{}). The anti-correlation suggests the validity of the Baldwin effect - the anti-correlation between the continuum luminosities and the EW of the broad emission lines. For the other two cases - $\Delta$\lagnir\ and $\Delta$EW(\ha{}), and $\Delta$\lagn\ and $\Delta$EW(\hb{}), we do not find any significant correlation between continuum and emission line properties. And, for the correlations between the $\Delta$EWs, we find weak correlations overall between the lines EWs. There is a hint of similarity between \ha{} and \hb{} EWs - similarity of the emission may be attributed primarily to the recombination process, while \mg{} - a resonant line, being produced mainly via collisional excitation shows marked deviation from the trends shown by the Balmer lines.\\
    
    \item With the compilation of a compendium of the estimates for the BH masses and Eddington ratios for all available spectroscopic epochs per source, we can realize the movement of the source in the \mbh{}-\ledd{} plane allowing us to check for systematic changes in the source's fundamental properties. The median values for the Eddington ratio distributions taken from the earliest and the latest epochs for sources in our sample are -1.99 and -1.685, respectively. These values are consistent with previous studies that have considered samples of CL AGNs \citep[see e.g.,][]{graham2020, potts2021, Zeltyn_etal_2024}. Thereafter, we then track their transition along the optical plane of the Eigenvector 1 (EV1) schema and categorize sources that either stay within the same Population (A or B) or make an inter-population movement as a function of spectral epoch. Considering only sources with at least 3 epochs of spectroscopic observations (32), we have a clear majority (26 sources) making the transition within the Population B class, while only 1 source makes the transition within the Population A class, and, 5 sources making the transition between Population A and Population B classes.\\
    
    \item We test the Balmer decrement (\ha{}/\hb{}) of a subset of our sample of CL AGNs as a function of time and AGN luminosity. In our sample, we have 32 sources for which we have simultaneous measurements for \ha{} and \hb{}, and we have at least 3 epochs of observations for the source. Among them, 4 reflect a rising trend with time, while the remaining sources show a more complicated behavior. We find a similar, complicated behavior for the Balmer decrement as a function of AGN luminosity for these sources with no obvious increasing/decreasing trend. We find that, generally, the highest AGN luminosity corresponds to the maximum \ha{}/\hb{} value for the sources considered in our analysis.
    
\end{itemize}

\begin{acknowledgments}
{We thank the anonymous referee for their constructive suggestions that helped improve the manuscript.} We thank Bo\.zena Czerny for fruitful discussions and suggestions. We are thankful to Benny Trakhtenbrot for financial support. We acknowledge support from the European Research Council (ERC) under the European Union’s Horizon 2020 research and innovation program (grant agreement number 950533), and the Israel Science Foundation (grant number 1849/19). 
SP acknowledges the financial support of the Conselho Nacional de Desenvolvimento Científico e Tecnológico (CNPq) Fellowships {300936/2023-0} and {301628/2024-6}. MS acknowledges the financial support from the Polish Funding Agency National Science Centre, project 2021/41/N/ST9/02280 (PRELUDIUM 20). 

Funding for the Sloan Digital Sky Survey IV has been provided by the Alfred P. Sloan Foundation, the U.S. Department of Energy Office of Science, and the Participating Institutions. SDSS-IV acknowledges the support and resources from the Center for High-Performance Computing at the University of Utah. The SDSS website is \url{www.sdss4.org}. SDSS-IV is managed by the Astrophysical Research Consortium for the Participating Institutions of the SDSS Collaboration including the Brazilian Participation Group, the Carnegie Institution for Science, Carnegie Mellon University, Center for Astrophysics | Harvard \& Smithsonian, the Chilean Participation Group, the French Participation Group, Instituto de Astrof\'isica de Canarias, The Johns Hopkins University, Kavli Institute for the Physics and Mathematics of the Universe (IPMU) / University of Tokyo, the Korean Participation Group, Lawrence Berkeley National Laboratory, Leibniz Institut f\"ur Astrophysik Potsdam (AIP), Max-Planck-Institut f\"ur Astronomie (MPIA Heidelberg), Max-Planck-Institut f\"ur Astrophysik (MPA Garching), Max-Planck-Institut f\"ur Extraterrestrische Physik (MPE), National Astronomical Observatories of China, New Mexico State University, New York University, University of Notre Dame, Observat\'ario Nacional / MCTI, The Ohio State University, Pennsylvania State University, Shanghai Astronomical Observatory, United Kingdom Participation Group, Universidad Nacional Aut\'onoma de M\'exico, University of Arizona, University of Colorado Boulder, University of Oxford, University of Portsmouth, University of Utah, University of Virginia, University of Washington, University of Wisconsin, Vanderbilt University, and Yale University.
\end{acknowledgments}

\vspace{5mm}
\facilities{SDSS}

\software{{\sc MATPLOTLIB}  (\citealt{hunter07}); {\sc NUMPY} (\citealt{numpy}); {\sc SCIPY} (\citealt{scipy}); {\sc ASTROPY} (\citealt{astropy}); {\sc TOPCAT} (\citealt{topcat}); {\sc PYQSOFIT} (\citealt{pyqsofit})}

\bibliography{main}{}
\bibliographystyle{aasjournal}


\appendix

\section{Characterizing the changes in the continuum luminosity and the power-law spectral slope}
\label{sec:pl-slopes}

\begin{figure*}[b!]
\centering   
\includegraphics[width=0.75\columnwidth]{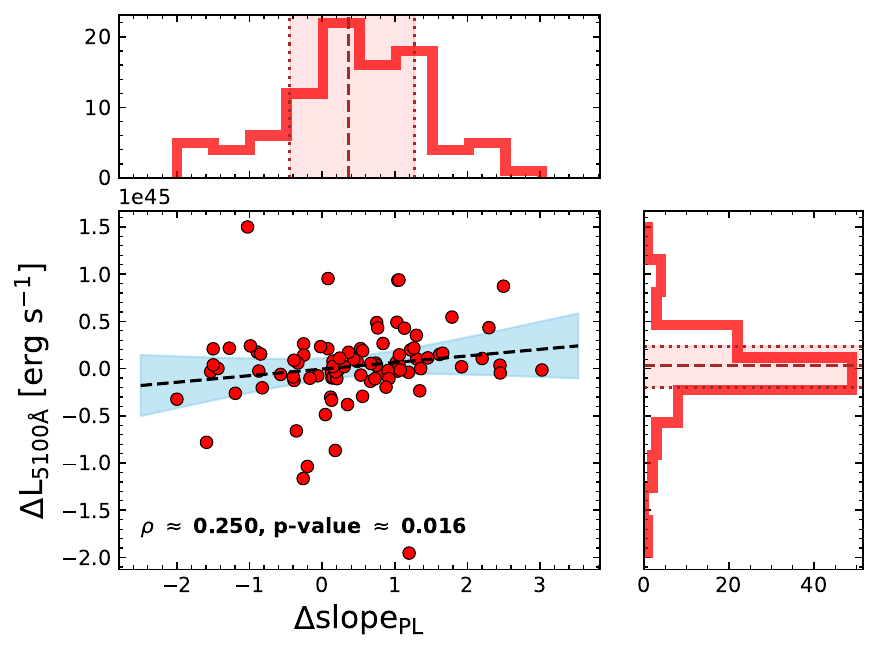}
\caption{{Joint distribution of the change in the power-law slope ($\Delta$slope = slope$_{\rm last-epoch}$ - slope$_{\rm first-epoch}$) estimated during the spectral fitting procedure, versus the change in the corresponding value for the AGN monochromatic luminosity at 5100\AA, i.e., $\Delta$\lagn{} = \lagn{}$_{\rm last-epoch}$ - \lagn{}$_{\rm first-epoch}$. For simplicity, we only consider the earliest (first) and the latest (last) spectral epochs for each of our sources in the sample. The linear best-fit relation (dashed line) along with 99\% confidence intervals (blue-shaded region) are shown in the main panel. The Spearman's correlation coefficient and significance of the fit are also reported. The marginal distributions in the form of histograms are shown for each variable with the respective medians (red dashed lines), 16$^{\rm th}$, and 84$^{\rm th}$ percentiles (red dotted lines) marked.}}
\end{figure*}

{We estimated the change in the power-law slope for each of the sources in our final sample, where $\Delta$slope = slope$_{\rm last-epoch}$ - slope$_{\rm first-epoch}$. Given that the change in slope affects directly the luminosity measurements from the spectrum, we also computed the change in the monochromatic luminosity at 5100\AA, where $\Delta$\lagn{} = \lagn{}$_{\rm last-epoch}$ - \lagn{}$_{\rm first-epoch}$. We show the correlation between the two parameters in the plot above. As per expectation, there is a correlation between the two parameters, when the power-law slope increases, it is related to a rise (or brightening) in the monochromatic luminosity. However, the trend is weak (Spearman's correlation coefficient, $\rho$ $\approx$ 0.25 and with a low significance, i.e., p-value $\approx$ 0.016). The median value for the 
$\Delta$slope = 0.36, with 16$^{\rm th}$, and 84$^{\rm th}$ percentiles as 0.45 and 1.27, respectively. Correspondingly, the median value for the $\Delta$\lagn{} = 3.45$\times 10^{43}$ erg s$^{-1}$, with 16$^{\rm th}$, and 84$^{\rm th}$ percentiles as -1.98$\times 10^{44}$ erg s$^{-1}$ and 2.35$\times 10^{44}$ erg s$^{-1}$, respectively. For a majority of sources (58/92), the change in slope is within $\pm$1, with only a few sources showing a drastic change in slope associated with a significant brightening/dimming in the continuum luminosity.}
\newpage

\section{Analysis of the continuum luminosities and the emission line EWs}
\label{app:ews_lum}

For 5/93 sources, we have measurements of $\Delta$EW(\mg{}) and $\Delta$EW(\ha{}), inclusive of the measurements in \hb{}, which means that all fitted parameters (continuum luminosities and EWs) are available for these sources.\\

In Table \ref{tab:coef_sp}, we present the Spearman's correlation coefficient (\textit{r}), to highlight the relations between continuum luminosities and EWs of emission lines (\mg{}, \hb{} and \ha{}) quantitatively. For $|r| >$ 0.5, we bold-face the values of \textit{r} and p-values.\\

\begin{table*}[!htb]
\caption{The Spearman's correlation coefficients and correlation significance for the broad emission line EWs and corresponding continuum luminosities for the sources in our sample} 
\label{tab:coef_sp}
\centering         
\begin{tabular}{l r|r r r r r r}
\hline\hline
&     & $\Delta$EW(MgII)  &  $\Delta$\lagnuv\ &  $\Delta$EW(\hb)   &  $\Delta$\lagn\ & $\Delta$\lagnir\ &  $\Delta$EW(\ha)  \\
\hline
$\Delta$EW(MgII)   & r& 1& {-0.7897} & 0.2274 & $\cdots$  & $\cdots$  & $\cdots$ \\
& P & 0 & {9.32e-12} &0.1121 & $\cdots$ & $\cdots$  & $\cdots$  \\ \hline 
$\Delta$\lagnuv\ & r &  {-0.7897} & 1&$\cdots$&{0.9310}&$\cdots$&$\cdots$ \\
& P &  {9.32e-12} &0 & $\cdots$& {1.0e-29} &$\cdots$&$\cdots$\\ \hline 
$\Delta$EW(\hb) &r &0.2274 & $\cdots$  & 1& -0.2906&$\cdots$&0.2839  \\
& P & 0.1121&$\cdots$& 0& 0.0049&$\cdots$&0.0393\\ \hline 
$\Delta$\lagn\ & r &$\cdots$ &{0.9310} &-0.2906 & 1& {0.9750 }&$\cdots$  \\
& P &  $\cdots$& {1.0e-29} &0.0049 & 0 & {1.24e-60}&$\cdots$\\ \hline 
$\Delta$\lagnir\ & r &$\cdots$ &  $\cdots$& $\cdots$ & {0.9750 } &1& -0.1072 \\
& P &  $\cdots$& $\cdots$ & $\cdots$ &  {1.24e-60}&0& 0.4361\\ \hline 
$\Delta$EW(\ha) & r &$\cdots$ & $\cdots$  &0.2839 & $\cdots$ &-0.1072& 1\\
& P &  $\cdots$&$\cdots$&0.0393&  $\cdots$&0.4361&0\\ \hline 
\hline
\end{tabular}
\end{table*}

\begin{figure*}[!htb]
    \centering  
    \includegraphics[scale=0.35]{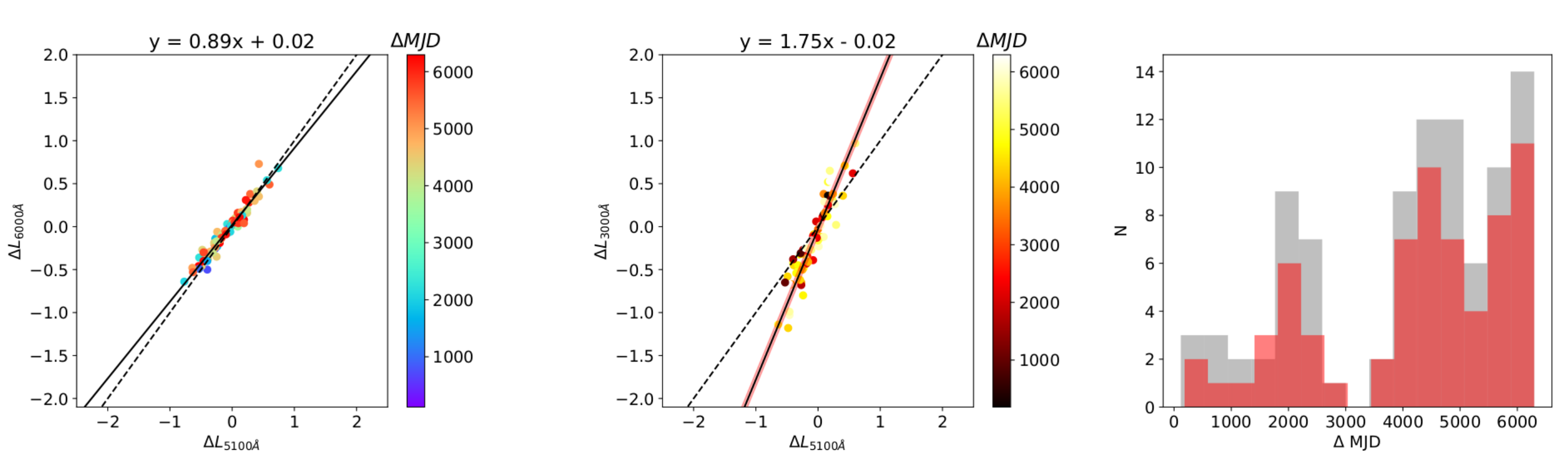}
       \includegraphics[scale=0.35]{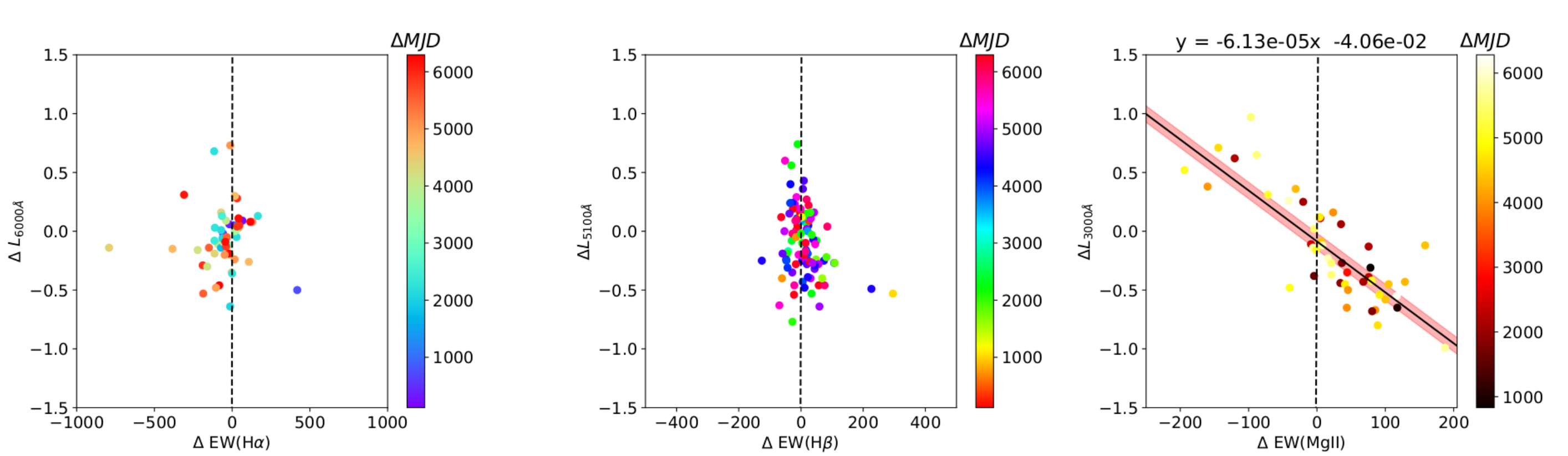}
     \includegraphics[scale=0.35]{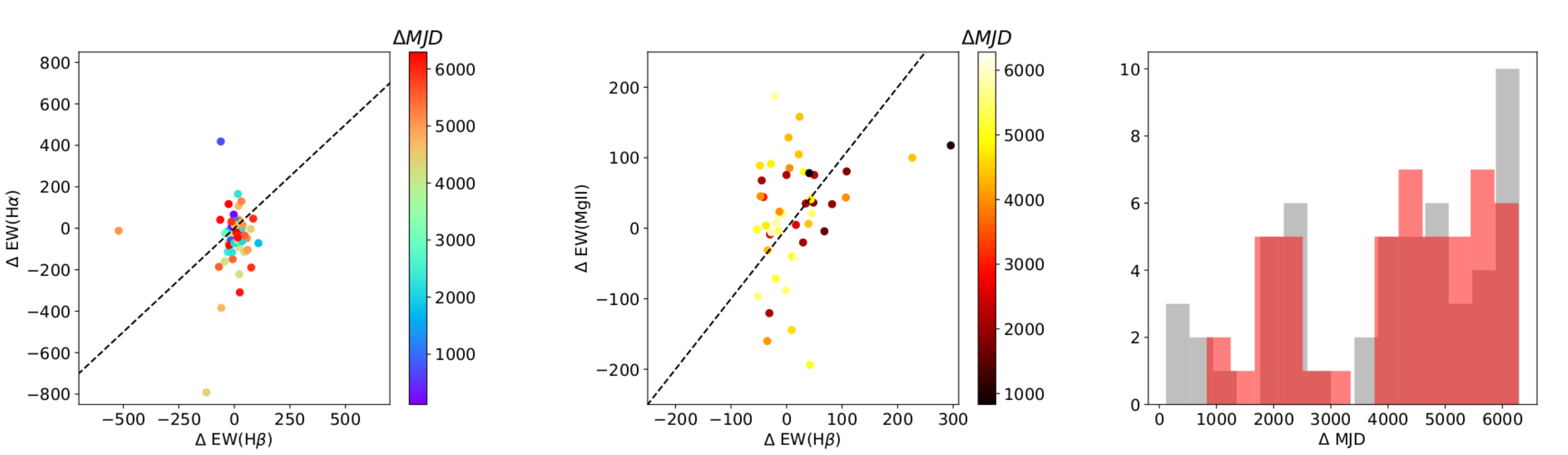}
    \caption{The dependence between fitted parameters between the latest and the earliest spectrum for our sample. For statistically significant correlation, we report Spearman's rank correlation coefficients and the correlation significance (p-value) and we show the best-fit line with 1$\sigma$ confidence intervals. For plots with no significant correlation, we overplot the zoomed version of each. We mark the 1-1 relation using a dashed black line in each correlation plot and color-code to the $\Delta$MJD corresponding to the difference between the latest and the earliest spectrum per source. From the top panel: the correlation between \lagnir\ and \lagn\ (left panel), \lagnuv\ and \lagn\ (middle panel), and the distribution of the $\Delta$MJD for the left plot (gray color) and the middle plot (red color). Middle panel: the correlation between the continuum and the corresponding emission line:  \lagnir\ and EW(\ha{}) (left panel),  \lagn\ and EW(\hb{}) (middle panel), and \lagnuv\ and EW(\mg{}) (right panel). Bottom panel: the correlation between EW(\ha{}) and EW(\hb{}) (left panel), EW(\mg{}) and EW(\hb{}) (middle panel), and the distribution of the $\Delta$MJD for the left plot (gray color) and the middle plot (red color).}
    \label{fig:color-code-corr}
\end{figure*}

We utilize these measurements and prepare correlation plots as shown in the panels of Figure \ref{fig:color-code-corr} and for correlations with $|r| >$ 0.5, we show the best fit. We use the orthogonal distance regression (ODR) method for line fitting \citep{Boggs_1989}. We see strong correlations between luminosities: \textit{r} = 0.93 in the case of $\Delta$\lagnuv\ vs $\Delta$\lagn\ and \textit{r} = 0.98 for $\Delta$\lagnir\ vs $\Delta$\lagn. In the former case, the best-fit correlation has a slope = 1.75 (see dashed line in each correlation plot), and the correlation becomes shallower in the latter case (slope = 0.89). This suggests a systematic increased change in the bluer wavelengths around 3000\AA\ compared to the redder wavelengths (around 5100\AA\ and 6000\AA), i.e., a stronger AGN contribution in the UV between the epochs \citep{Czerny_Elvis_1987, Padovani_2017}.\\

In the case of correlations between the $\Delta$EW of the emission line and the corresponding change in the continuum luminosity, we notice strong anti-correlation (\textit{r} = -0.79) between $\Delta$\lagnuv\ and $\Delta$EW(\mg{}). For the other two cases - $\Delta$\lagnir\ and $\Delta$EW(\ha{}), and $\Delta$\lagn\ and $\Delta$EW(\hb{}), we do not find any significant correlation between continuum and emission line properties. Removing the outliers for the change in continuum luminosity vs $\Delta$EWs for these latter cases does not lead to any significant change in our results, although, there is a smooth transition between the correlations going from the \ha{} (\textit{r} = -0.11, weakest), to \hb{} (\textit{r} = -0.29, weak) and \mg{} (\textit{r} = -0.79, strong and significant). The anti-correlation suggests the validity of the Baldwin effect - the anti-correlation between the continuum luminosities and the EW of the broad emission lines \citep{Wandel_1999, Wang_etal_2021, Martinez-Aldama_etal_2021}. \citet[][and references therein]{Shen_2021, Wang_etal_2021} have found from their statistical studies that \mg{} rarely changed when the CL AGN made a transition - on the other hand, the Balmer lines do change, which has been connected to the breathing effect in broad emission lines \citep{Guo_etal_2020, Wang_etal_2021}. \\

And finally, for the correlations between the $\Delta$EWs, we find weak correlations overall between the lines EWs. There is a hint of similarity between \ha{} and \hb{} EWs - similarity of the emission may be attributed primarily to recombination process, while for the \mg{} the primary channel of emission is via collisional excitation since this is a resonant emission line \citep{Czerny_etal_2019, Panda_etal_2019a, Sniegowska_etal_2020}.

\newpage

\section{Supplementary plots}
\label{app:plots}

\begin{figure*}[!htb]
    \centering
    \includegraphics[width=0.21\textwidth]{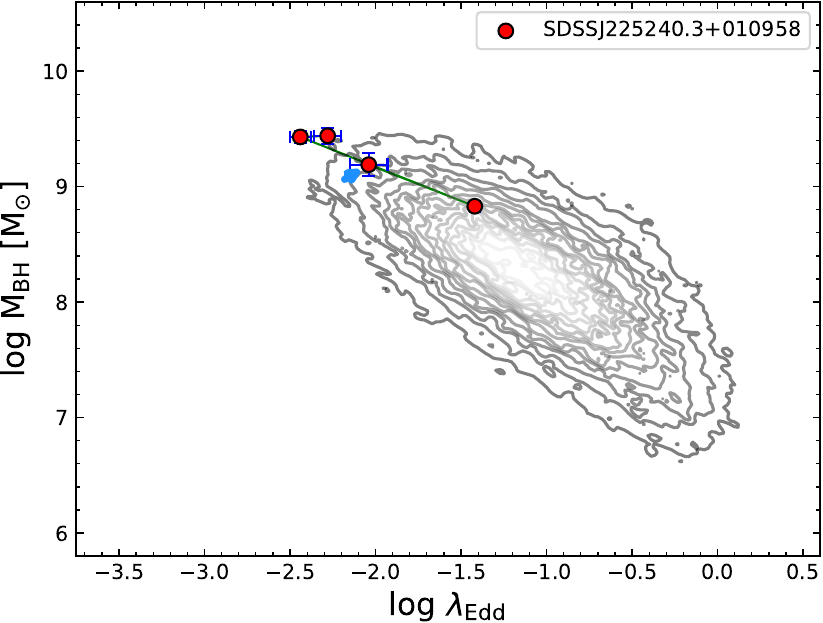}
    \includegraphics[width=0.21\textwidth]{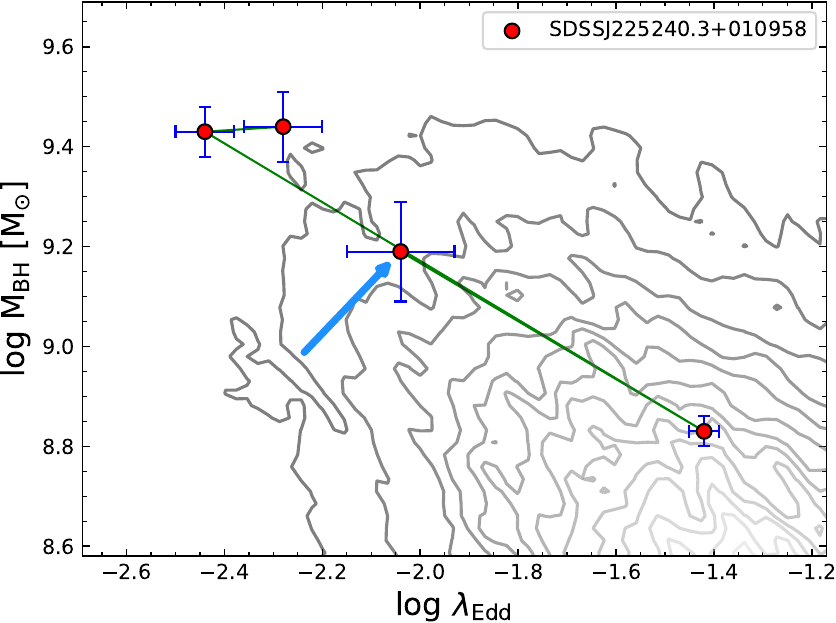}
    \includegraphics[width=0.21\textwidth]{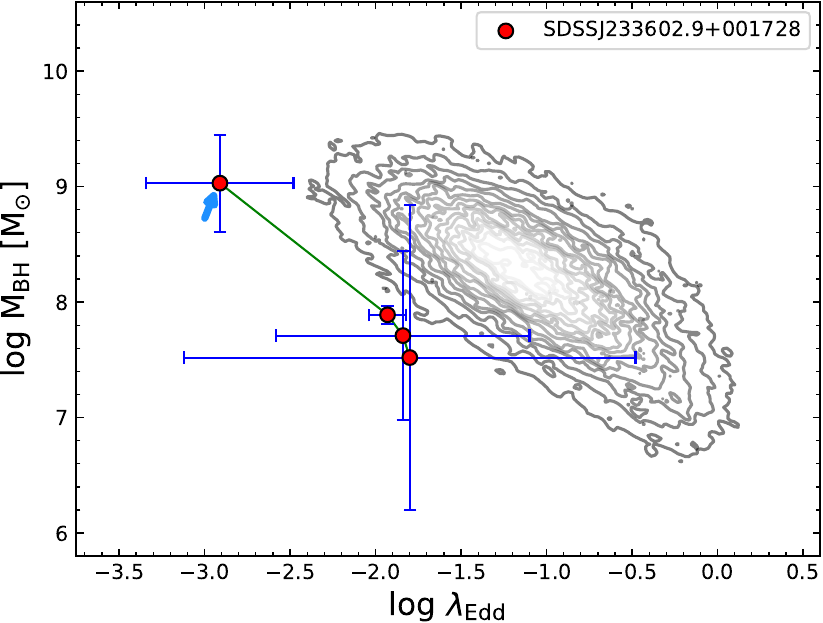}
    \includegraphics[width=0.21\textwidth]{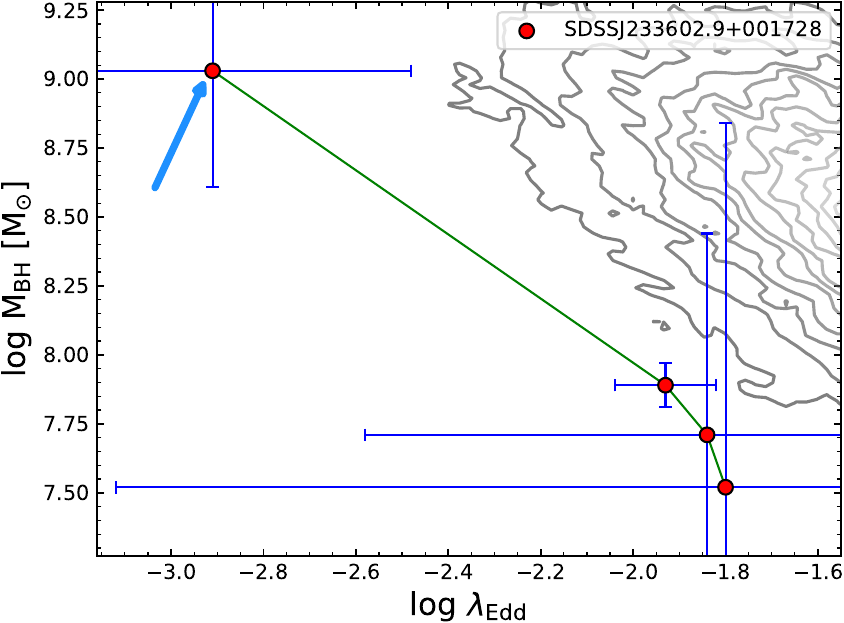}

    \includegraphics[width=0.21\textwidth]{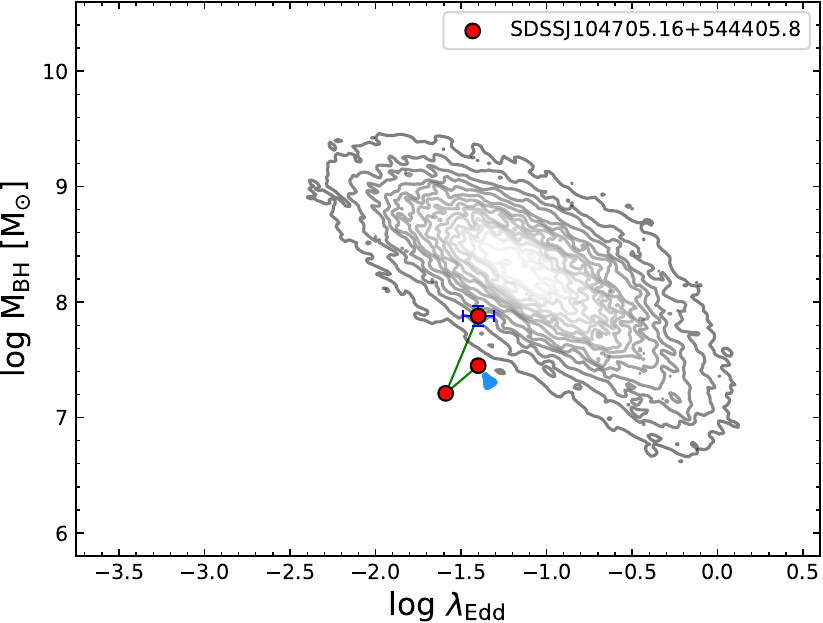}
    \includegraphics[width=0.21\textwidth]{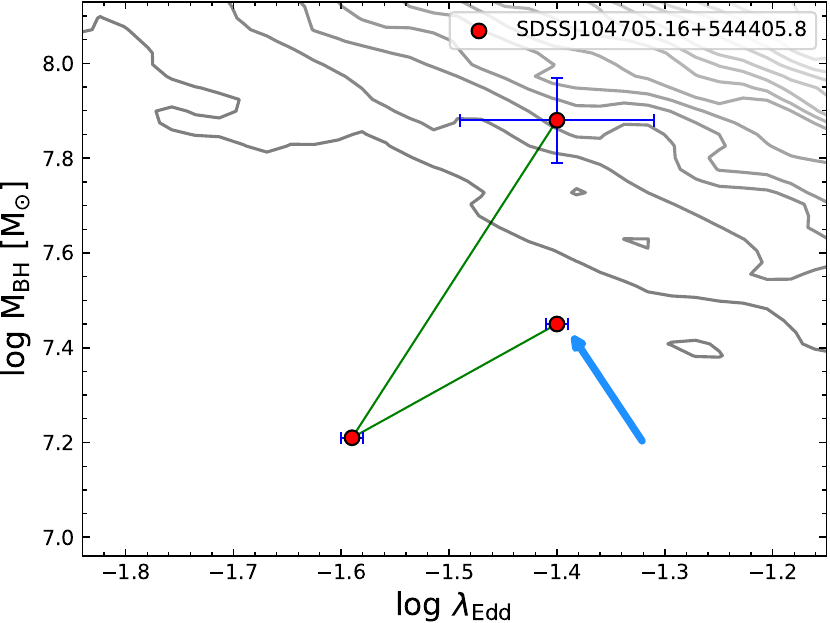}
    \includegraphics[width=0.21\textwidth]{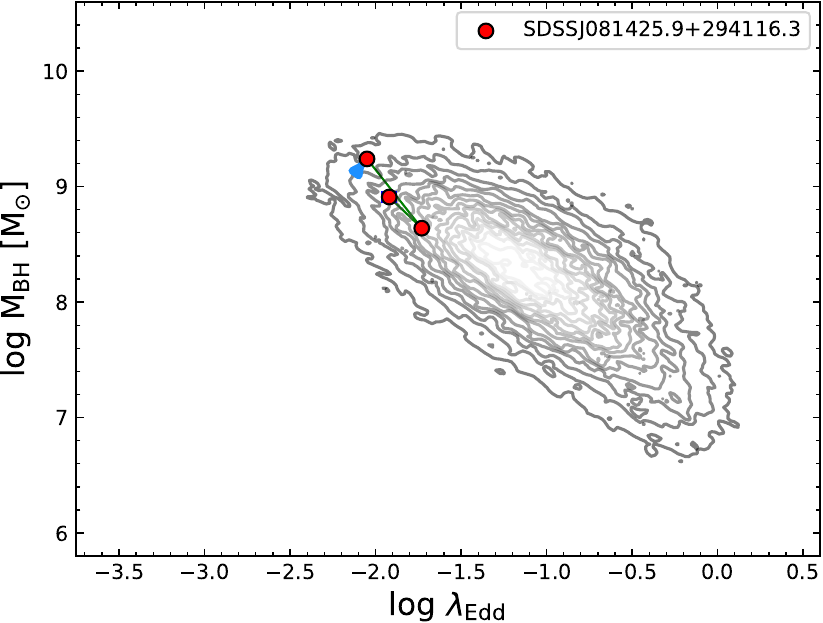}
    \includegraphics[width=0.21\textwidth]{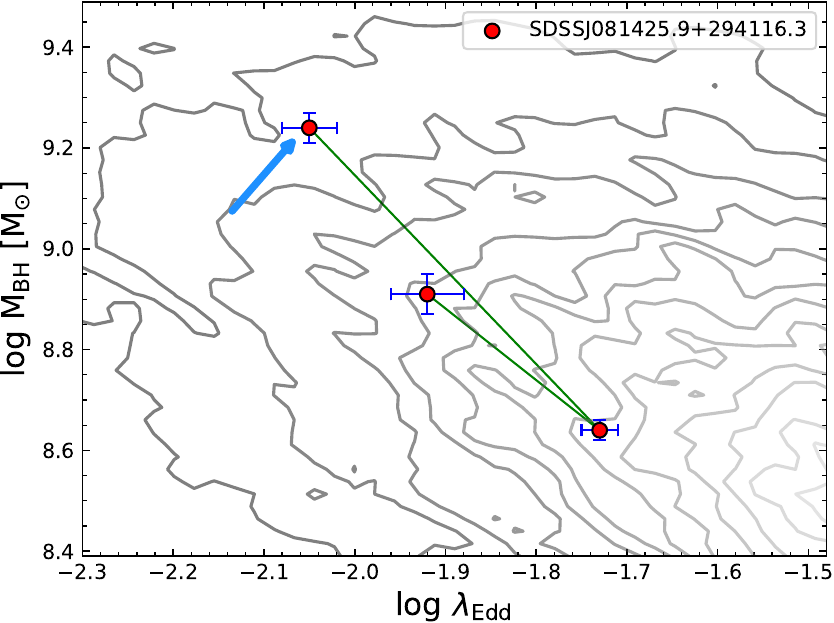}

    \includegraphics[width=0.21\textwidth]{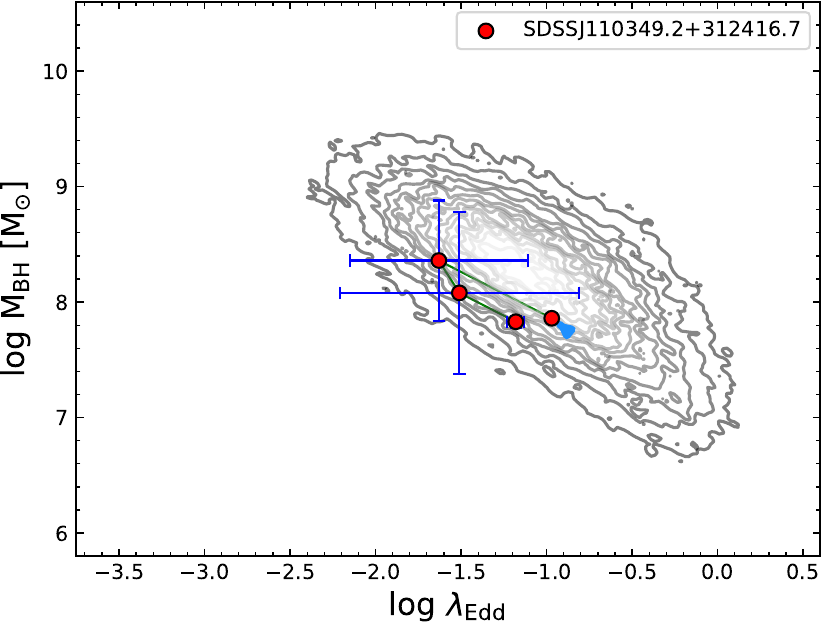}
    \includegraphics[width=0.21\textwidth]{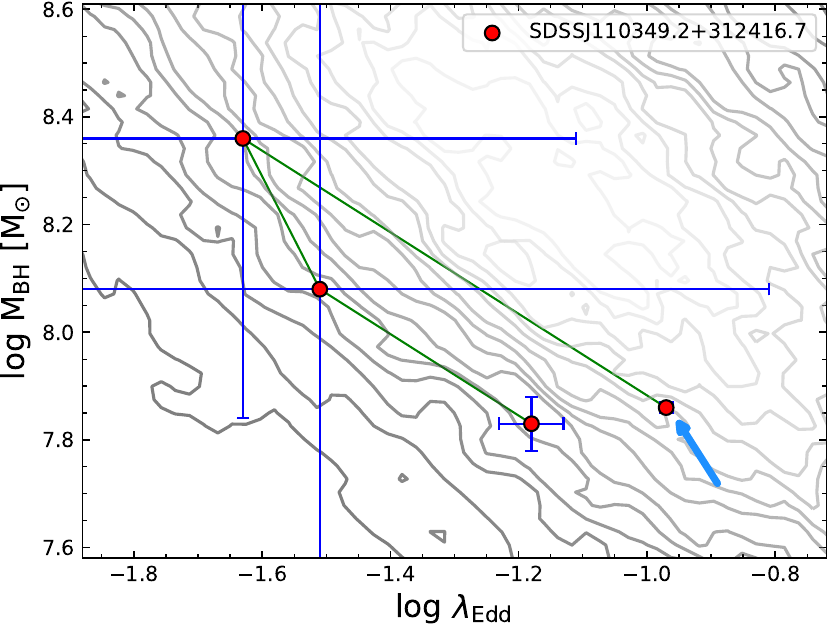}
    \includegraphics[width=0.21\textwidth]{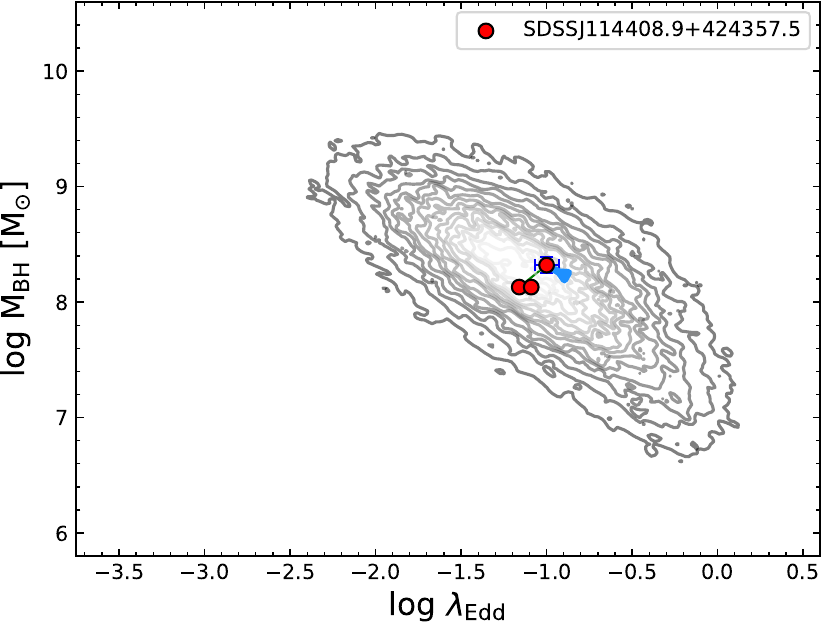}
    \includegraphics[width=0.21\textwidth]{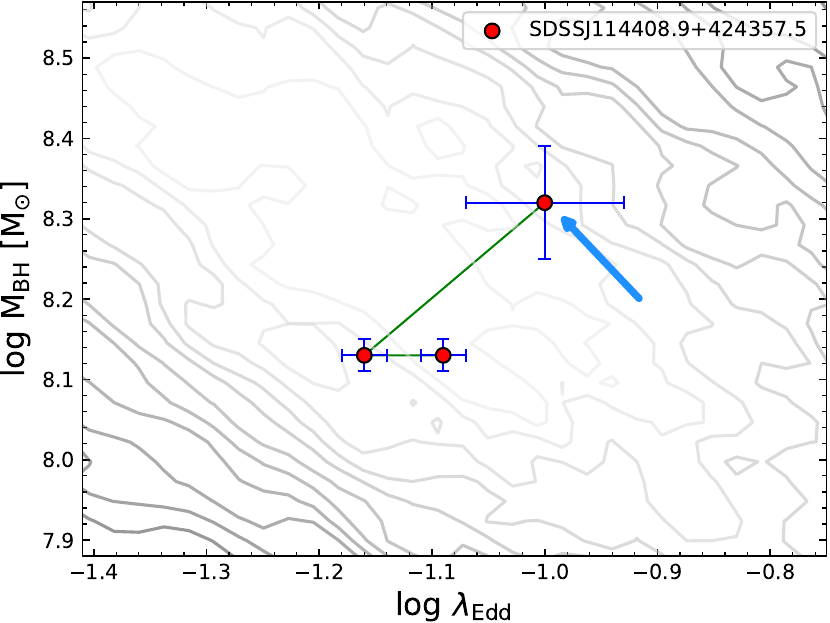}

    \includegraphics[width=0.21\textwidth]{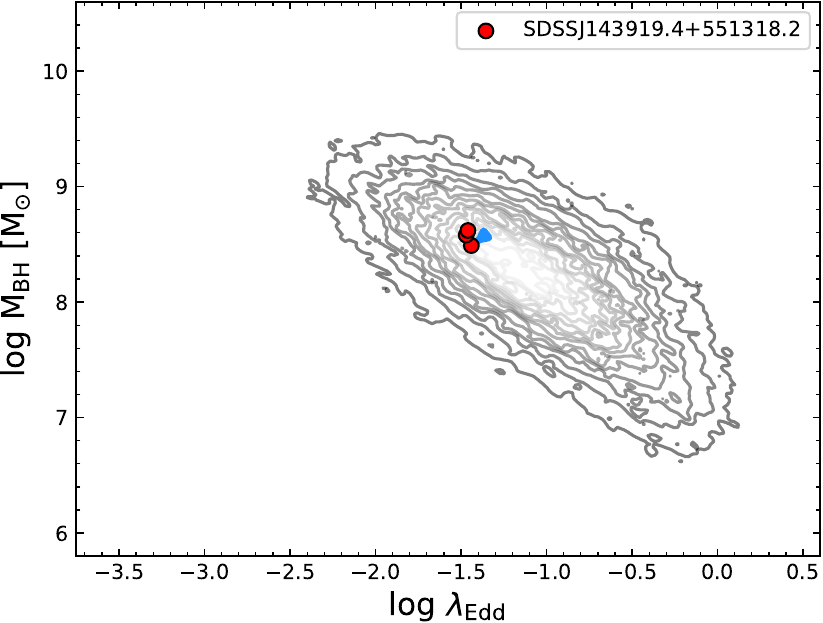}
    \includegraphics[width=0.21\textwidth]{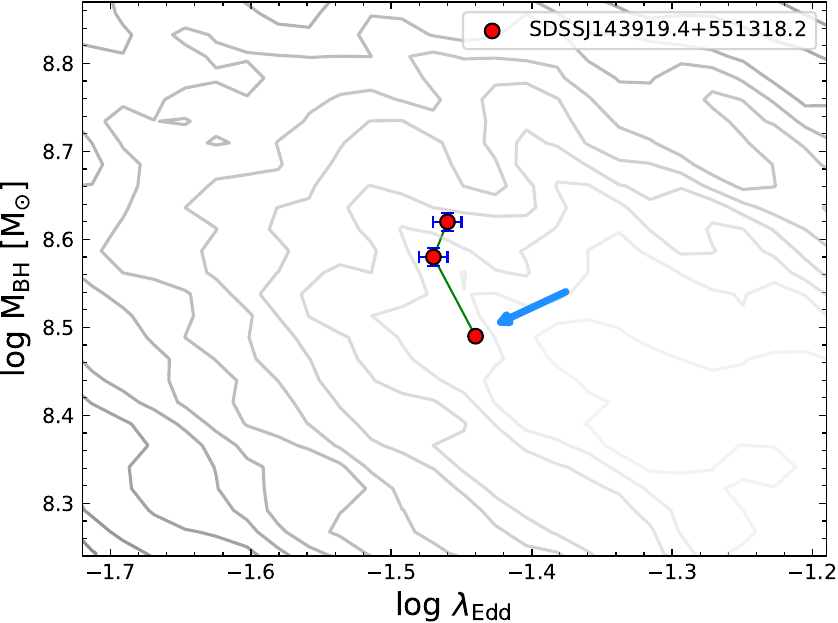}
    \includegraphics[width=0.21\textwidth]{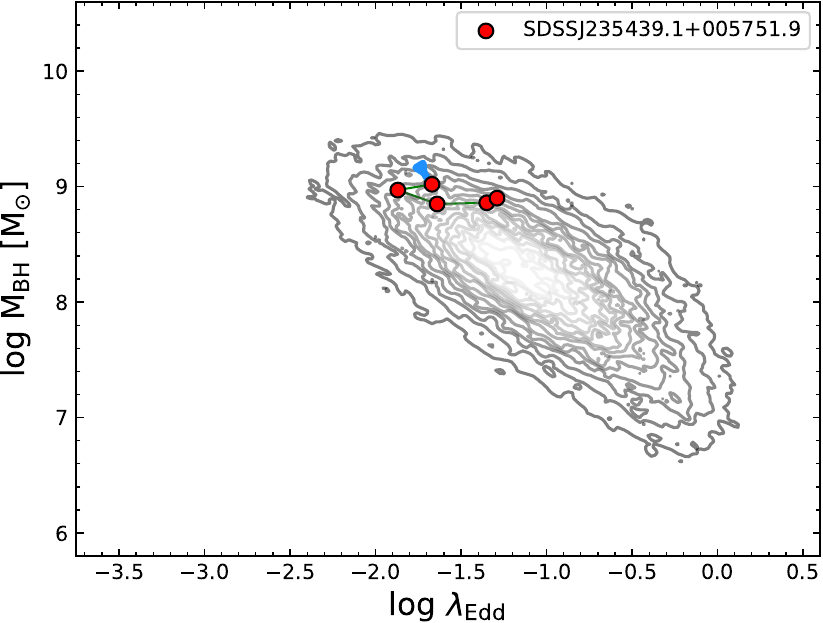}
    \includegraphics[width=0.21\textwidth]{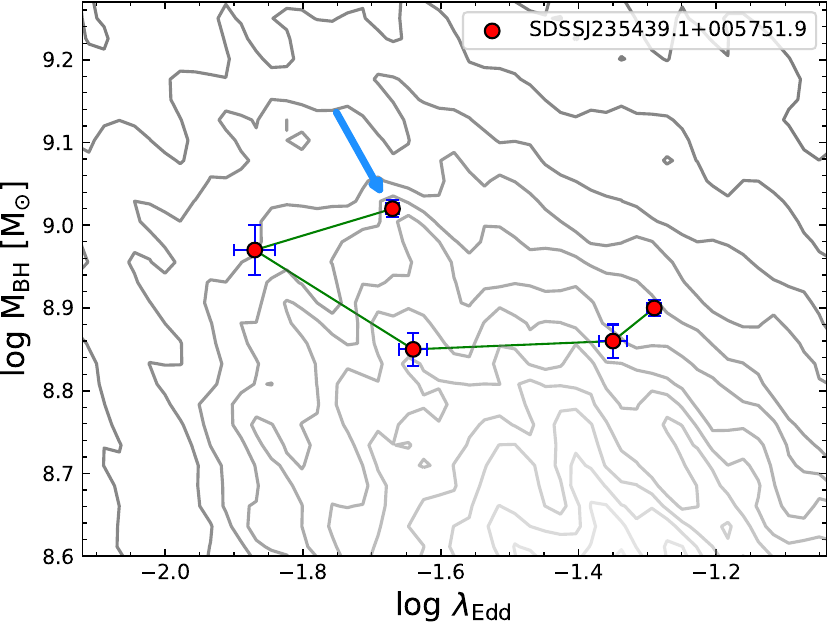}

    \includegraphics[width=0.21\textwidth]{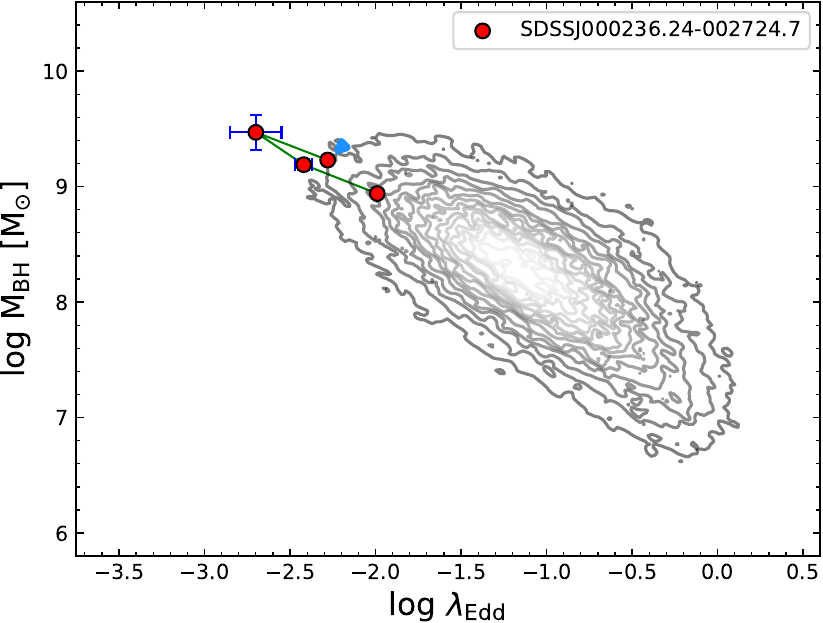}
    \includegraphics[width=0.21\textwidth]{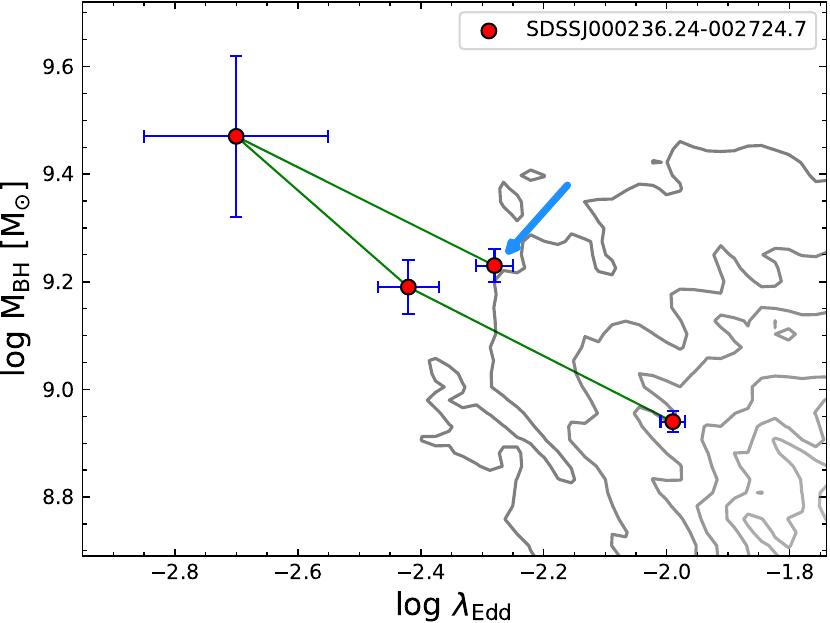}
    \includegraphics[width=0.21\textwidth]{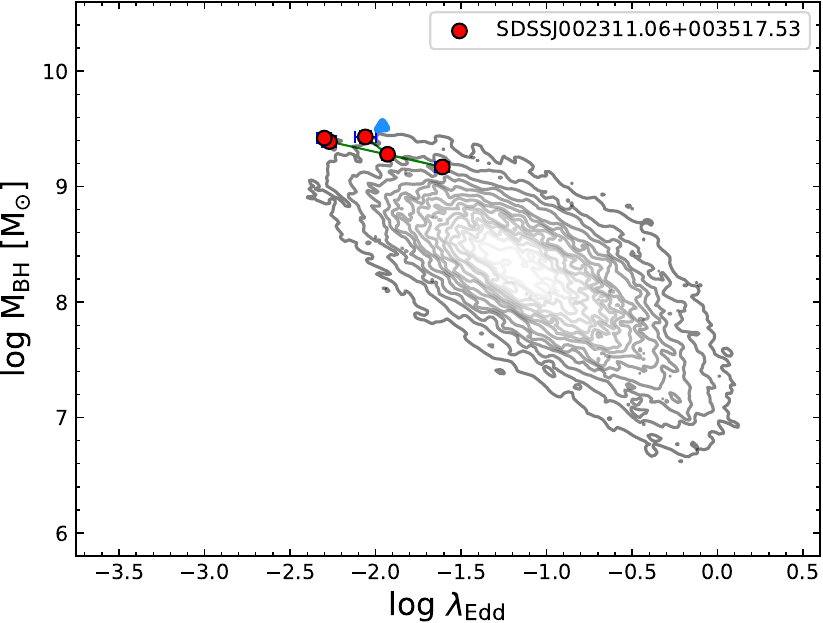}
    \includegraphics[width=0.21\textwidth]{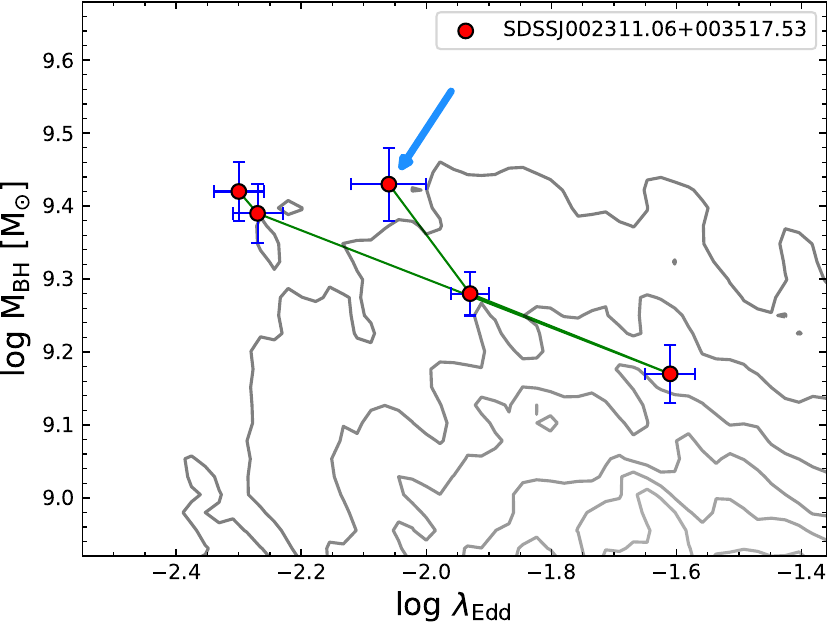}

    \includegraphics[width=0.21\textwidth]{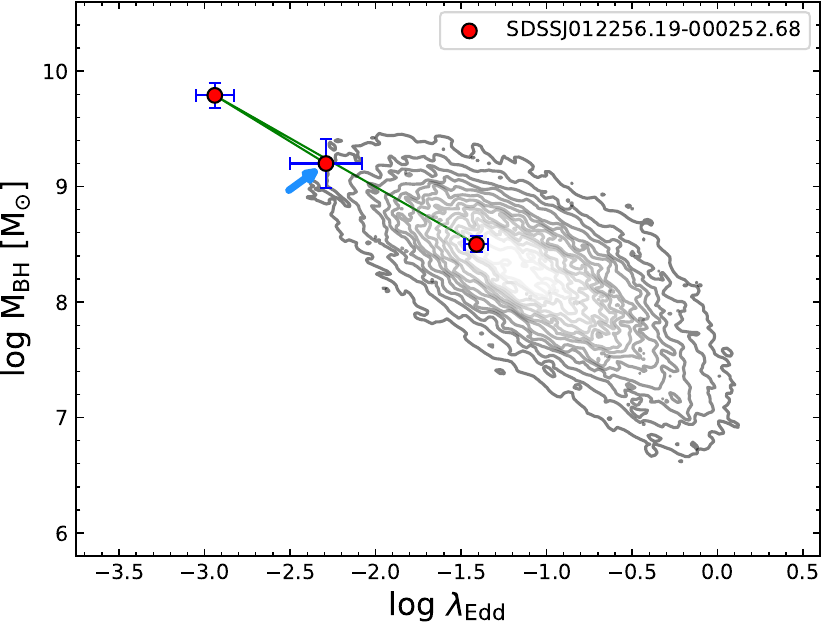}
    \includegraphics[width=0.21\textwidth]{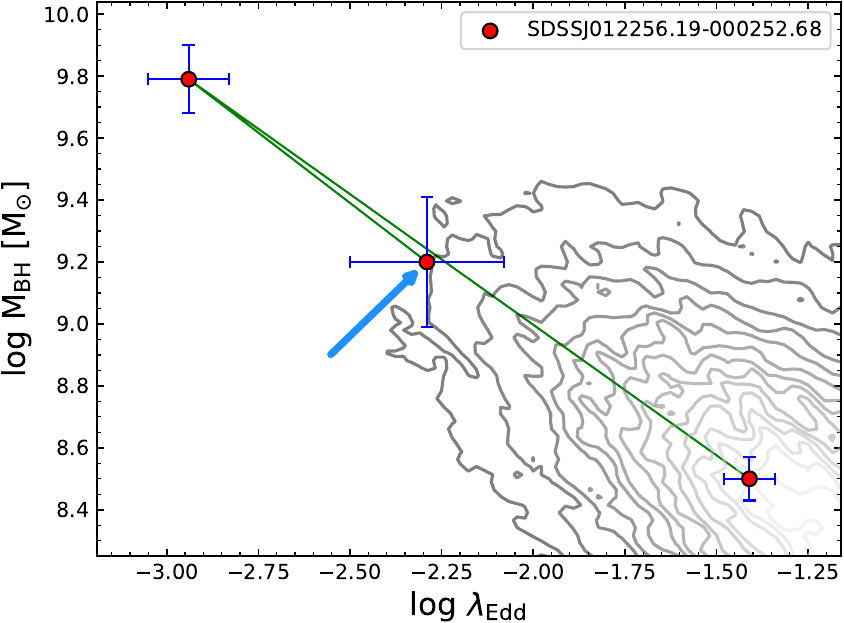}
    \includegraphics[width=0.21\textwidth]{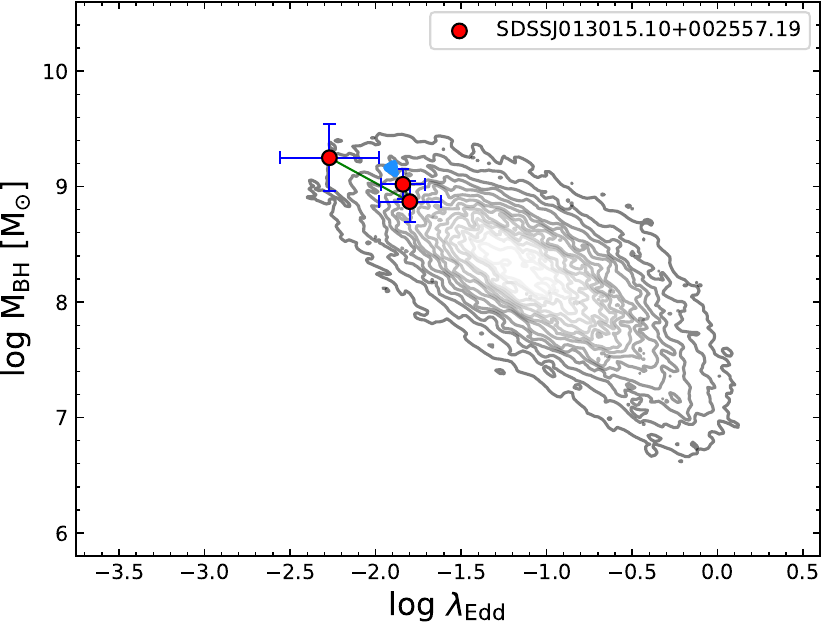}
    \includegraphics[width=0.21\textwidth]{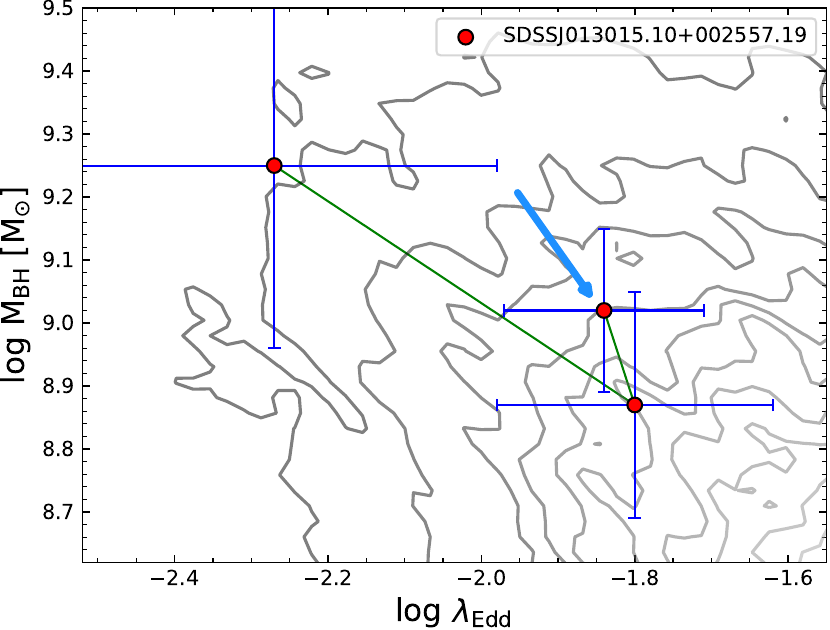}

    \includegraphics[width=0.21\textwidth]{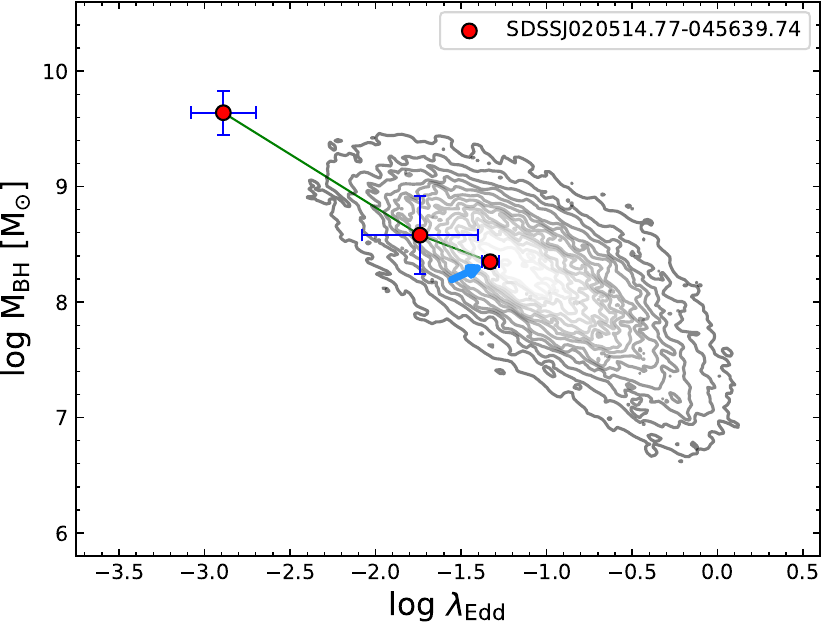}
    \includegraphics[width=0.21\textwidth]{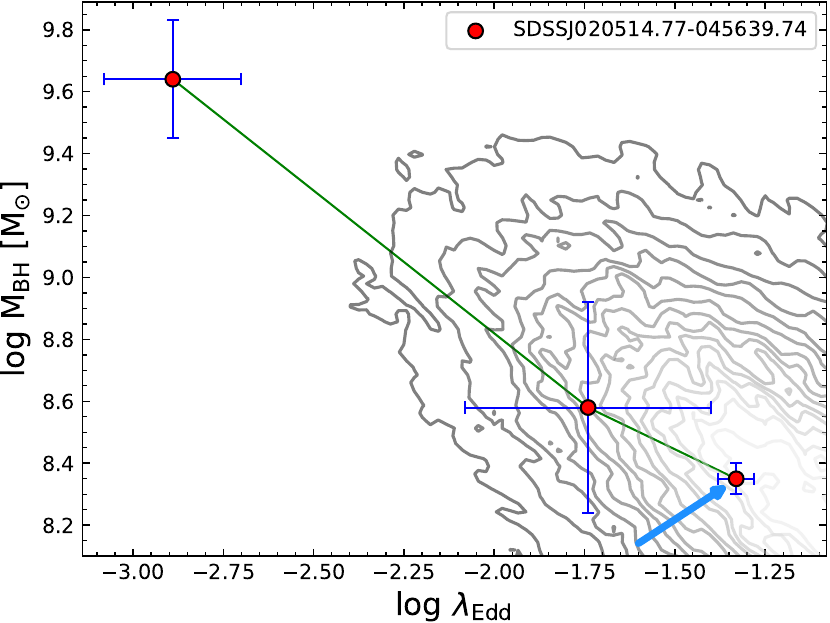}
    \includegraphics[width=0.21\textwidth]{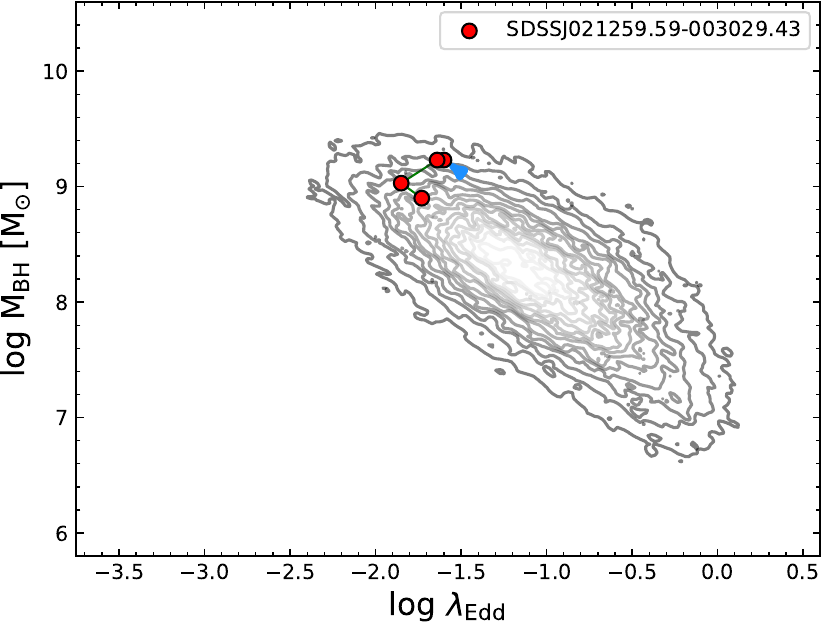}
    \includegraphics[width=0.21\textwidth]{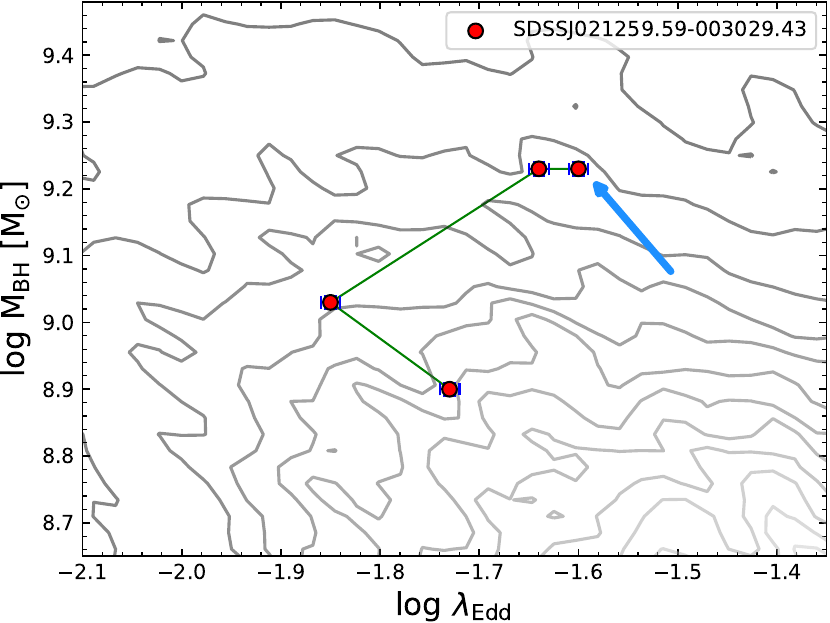}

    \caption{Panels are similar to Figures \ref{fig:M-Edd} for the remaining sources.}
    \label{fig:M-Edd-others}
\end{figure*}

\begin{figure*}[!htb]
    \centering
    \includegraphics[width=0.21\textwidth]{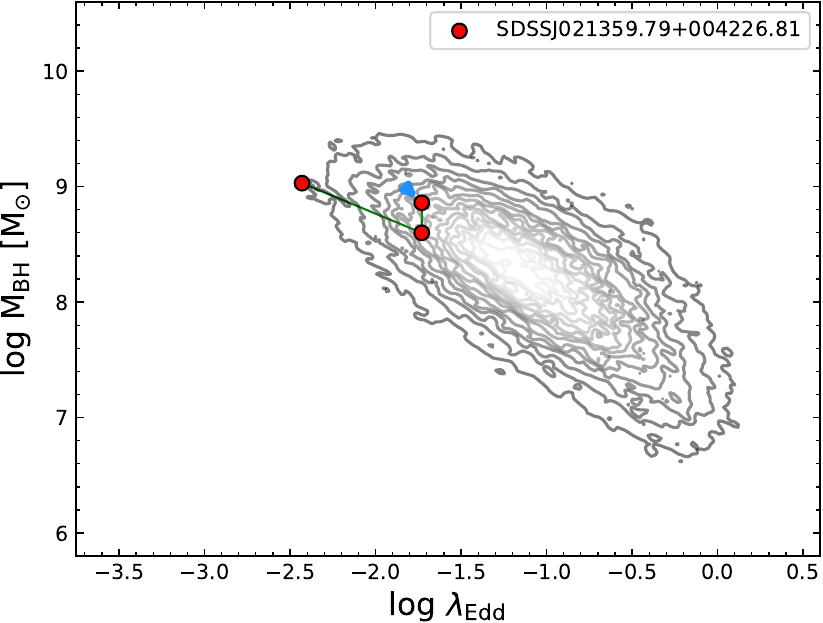}
    \includegraphics[width=0.21\textwidth]{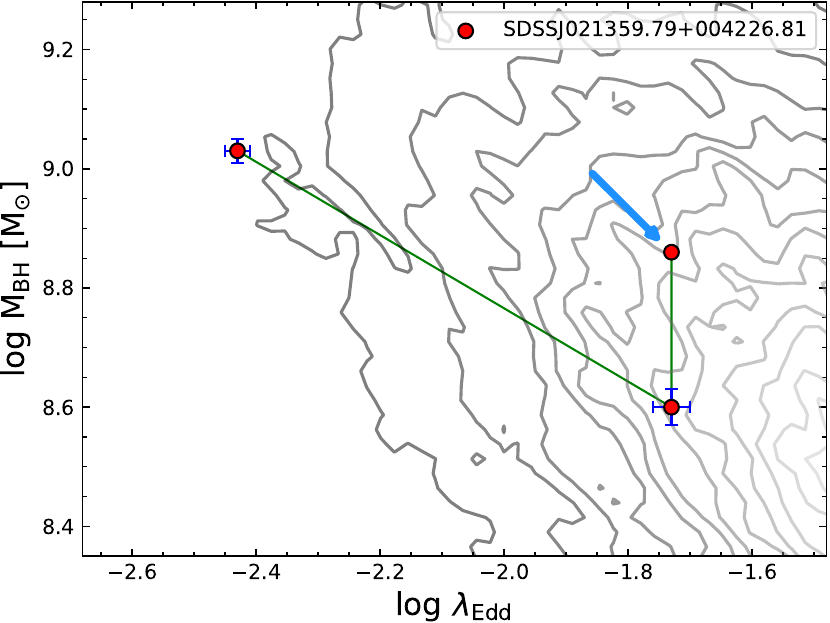}
    \includegraphics[width=0.21\textwidth]{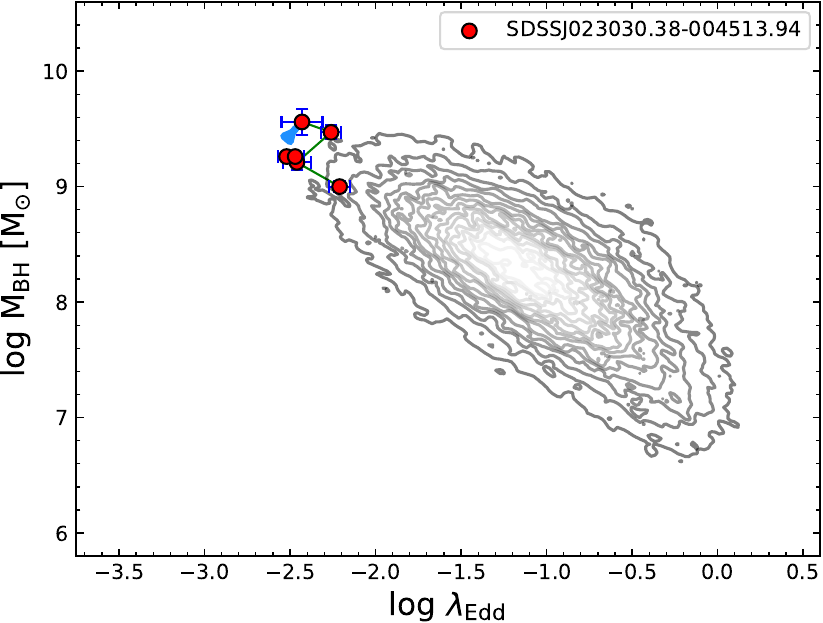}
    \includegraphics[width=0.21\textwidth]{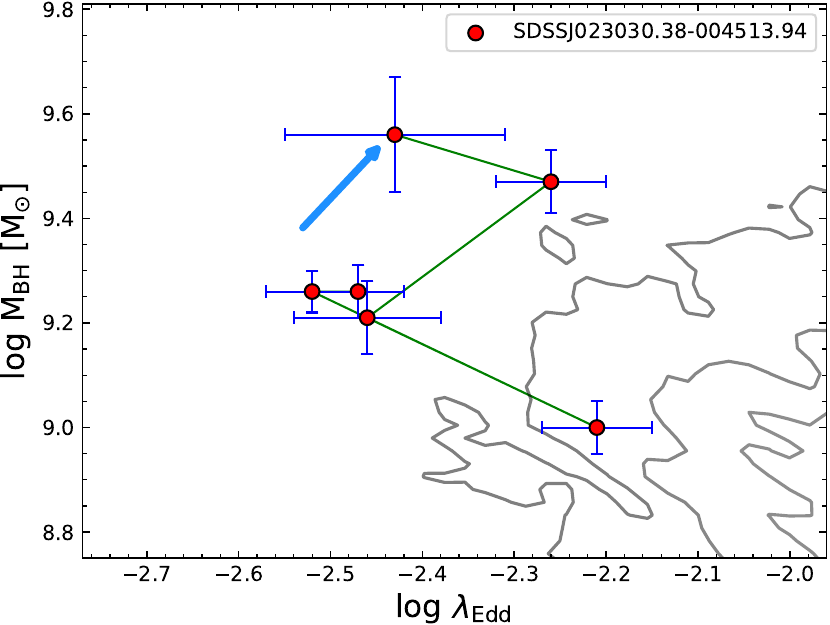}

    \includegraphics[width=0.21\textwidth]{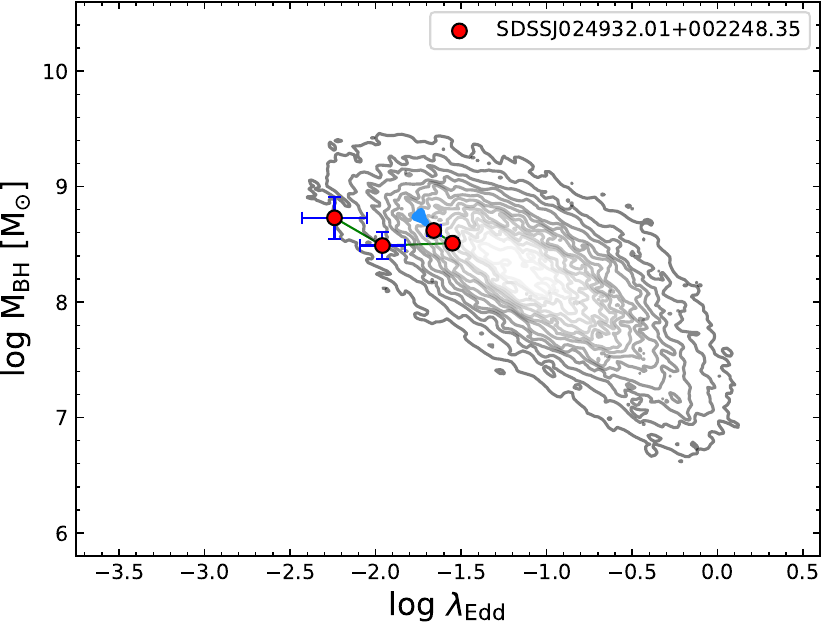}
    \includegraphics[width=0.21\textwidth]{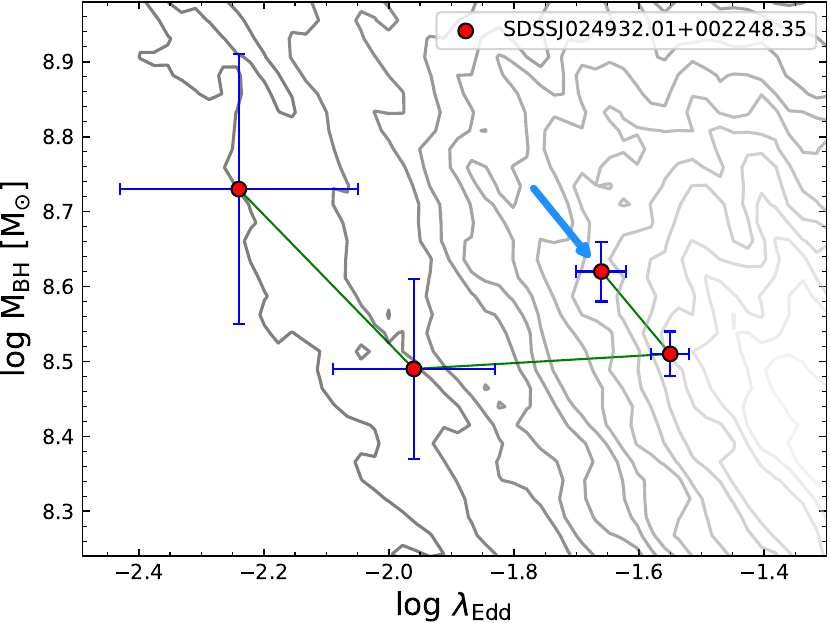}
    \includegraphics[width=0.21\textwidth]{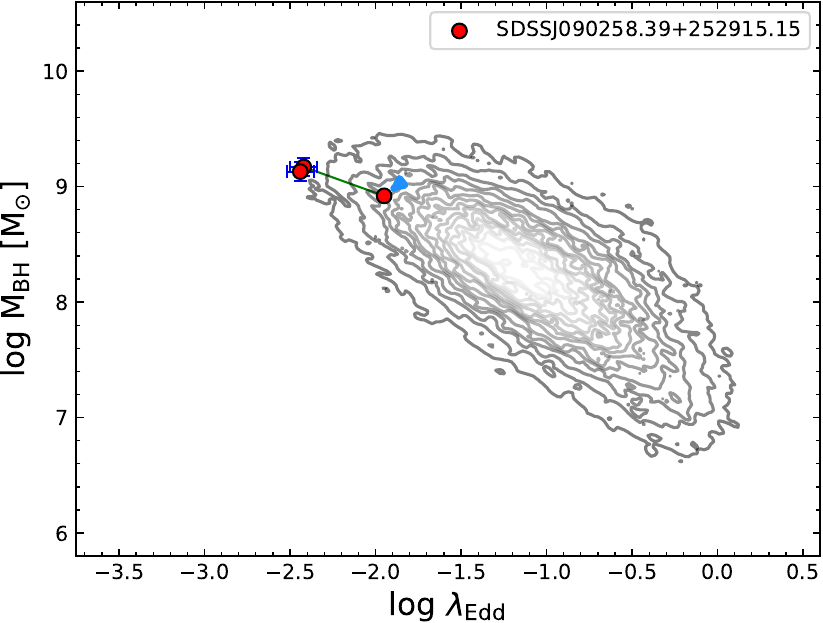}
    \includegraphics[width=0.21\textwidth]{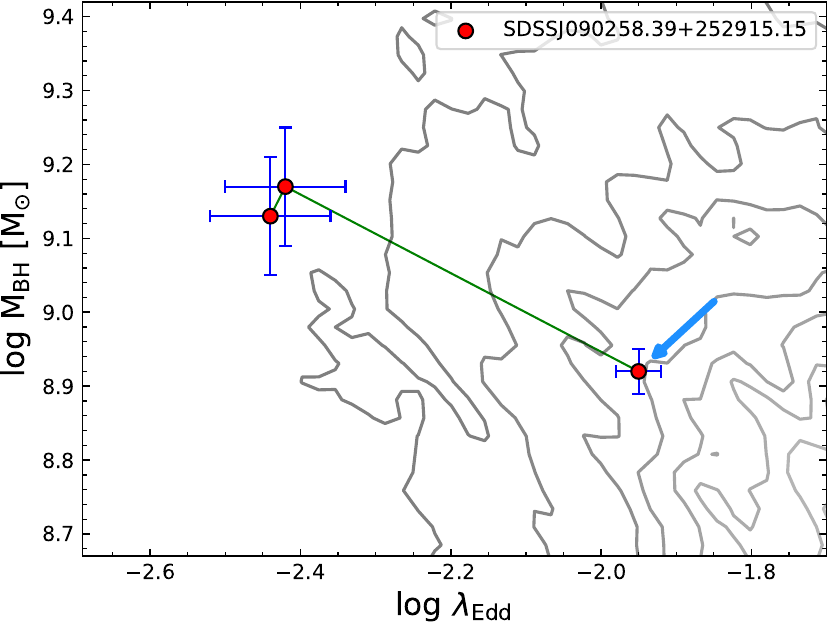}

    \includegraphics[width=0.21\textwidth]{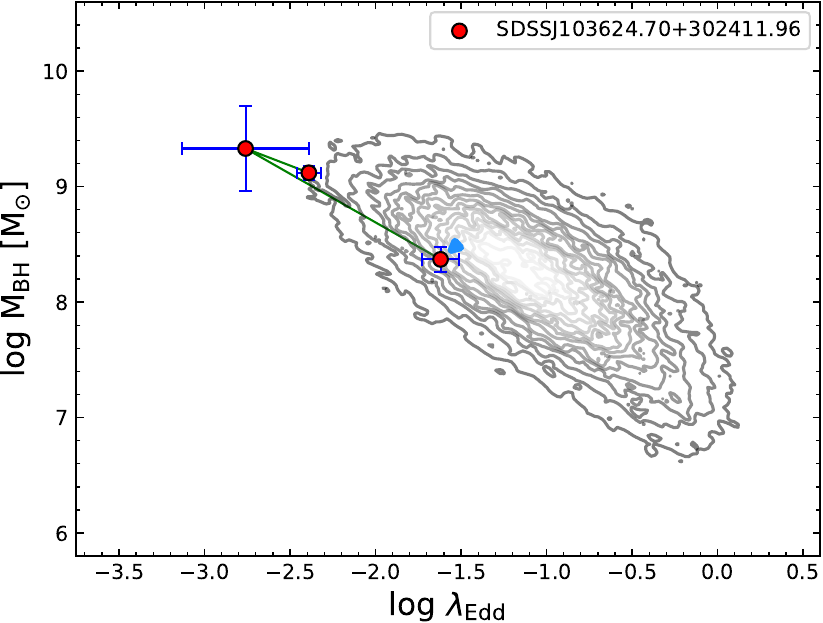}
    \includegraphics[width=0.21\textwidth]{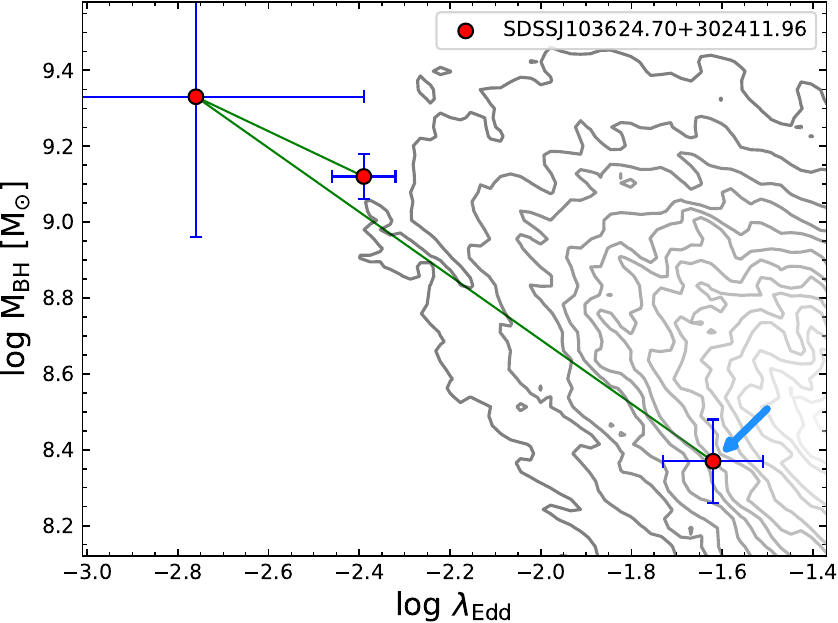}
    \includegraphics[width=0.21\textwidth]{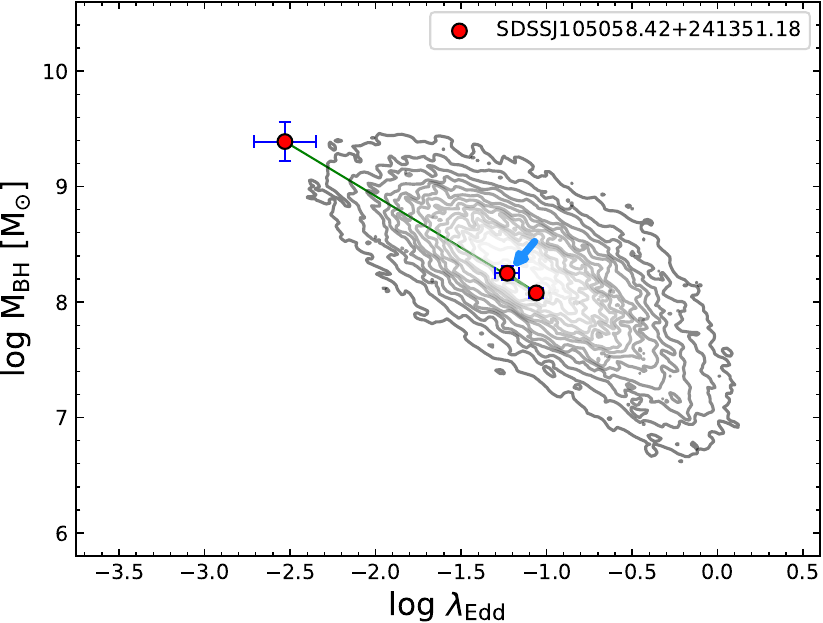}
    \includegraphics[width=0.21\textwidth]{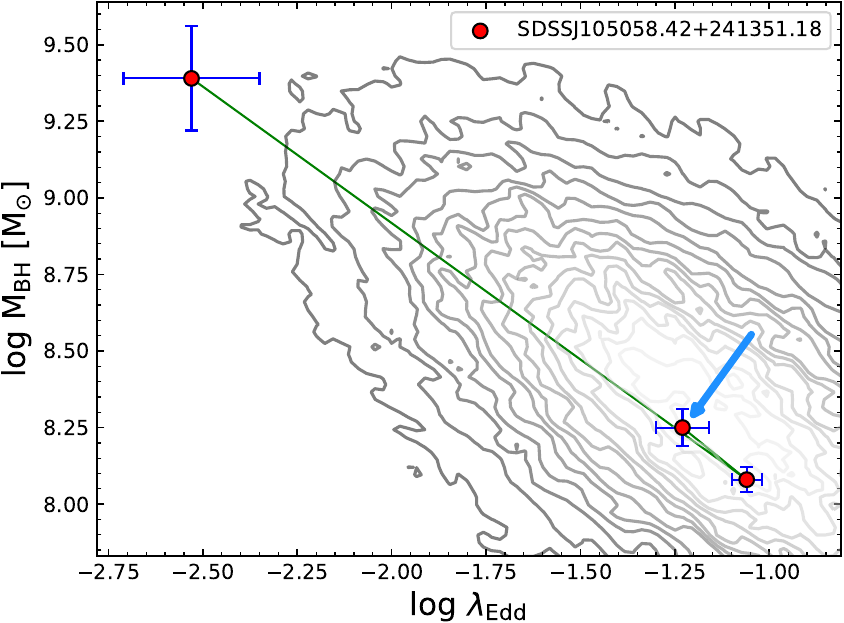}

    \includegraphics[width=0.21\textwidth]{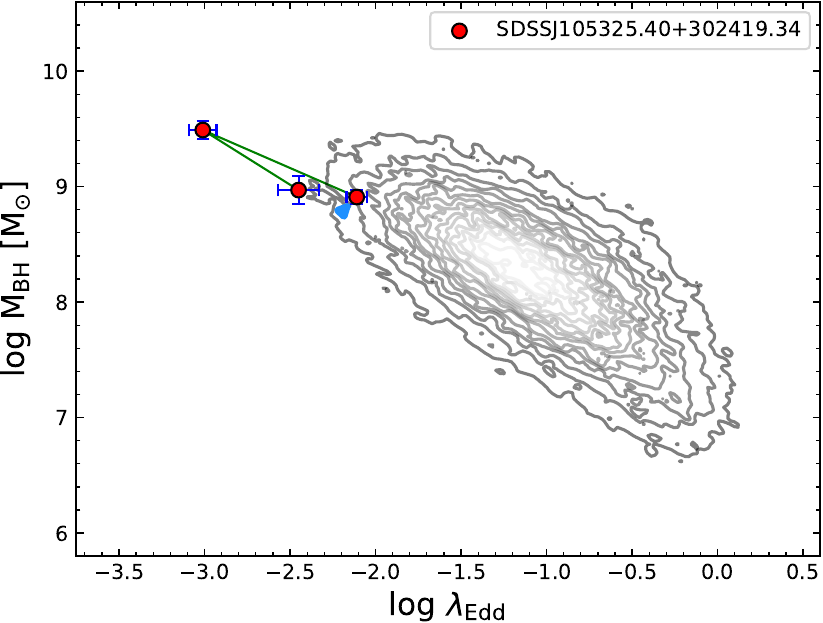}
    \includegraphics[width=0.21\textwidth]{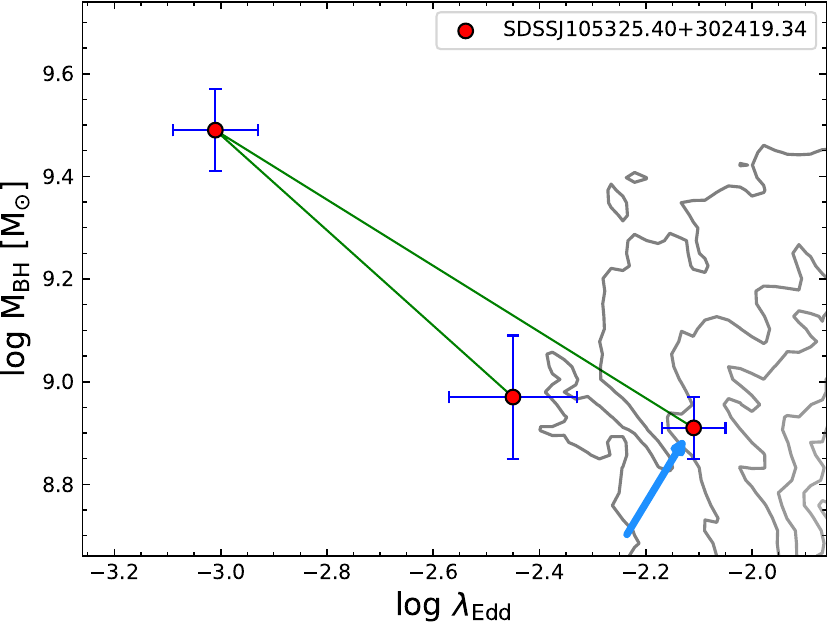}
    \includegraphics[width=0.21\textwidth]{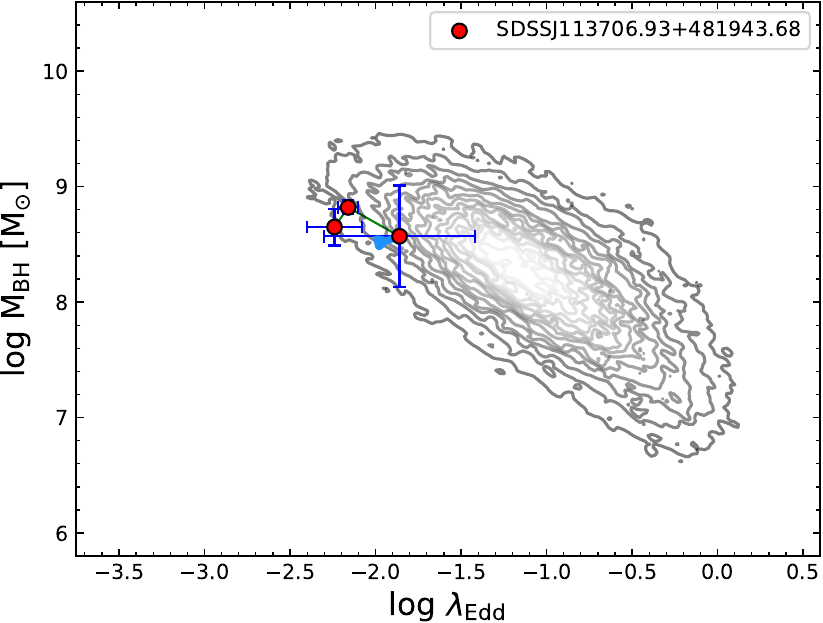}
    \includegraphics[width=0.21\textwidth]{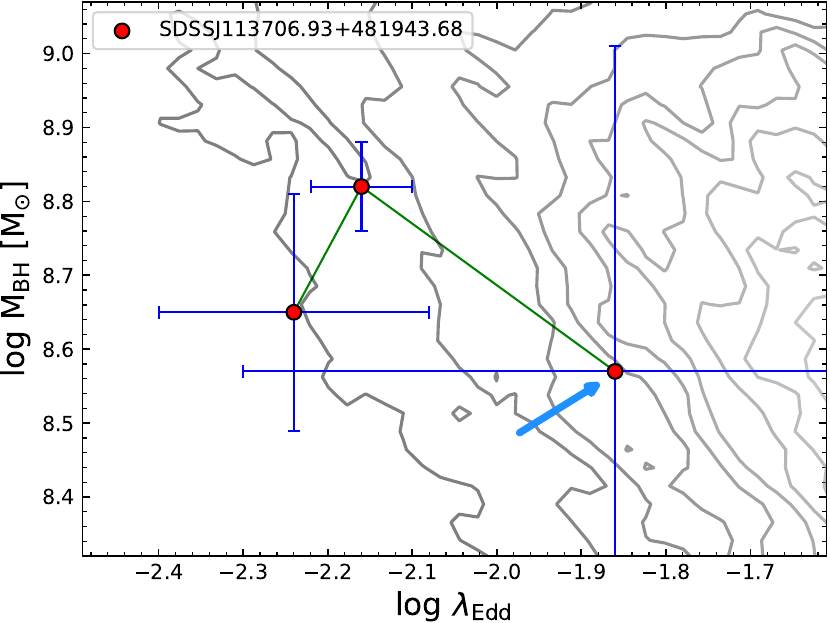}

    \includegraphics[width=0.21\textwidth]{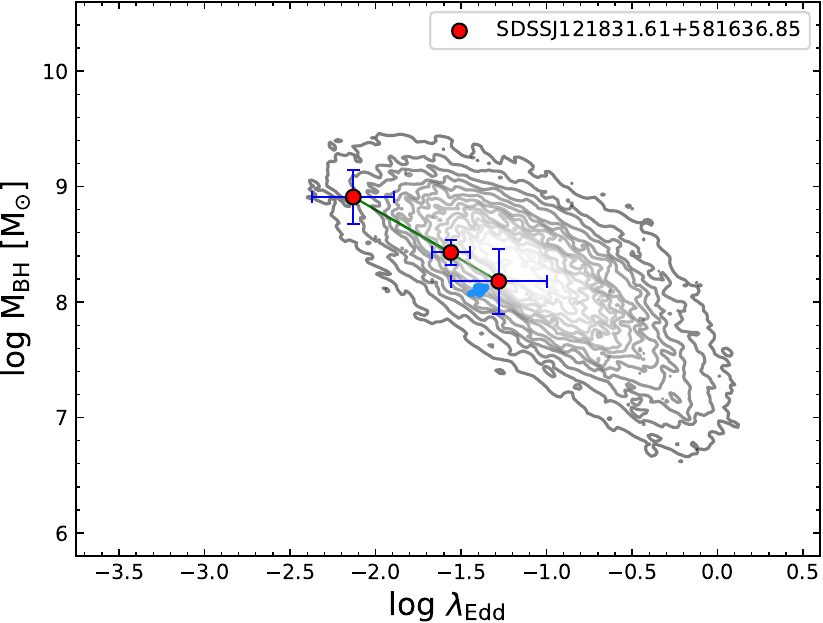}
    \includegraphics[width=0.21\textwidth]{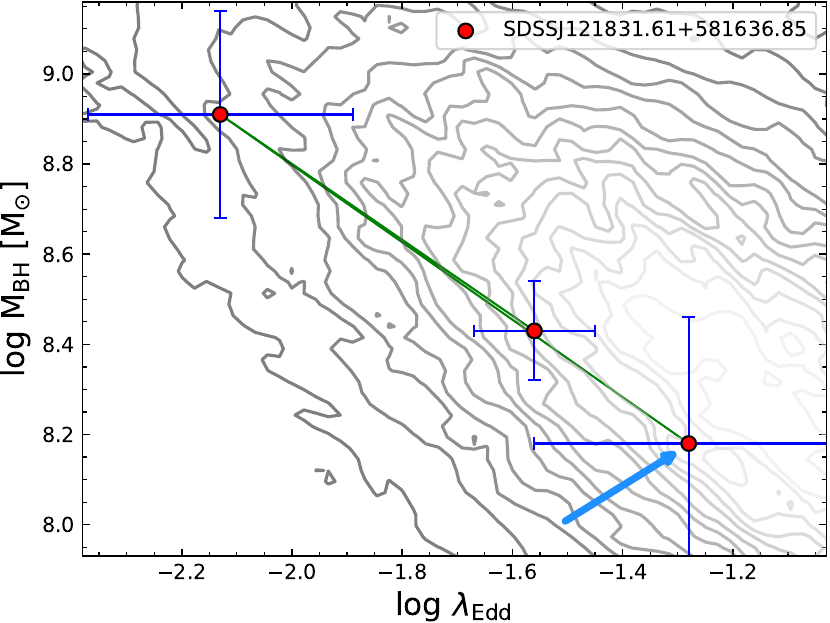}
    \includegraphics[width=0.21\textwidth]{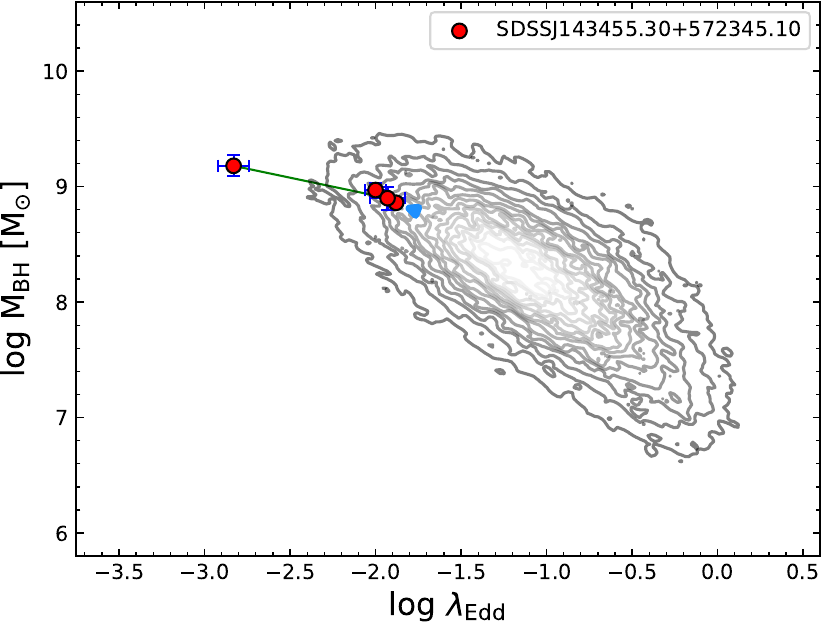}
    \includegraphics[width=0.21\textwidth]{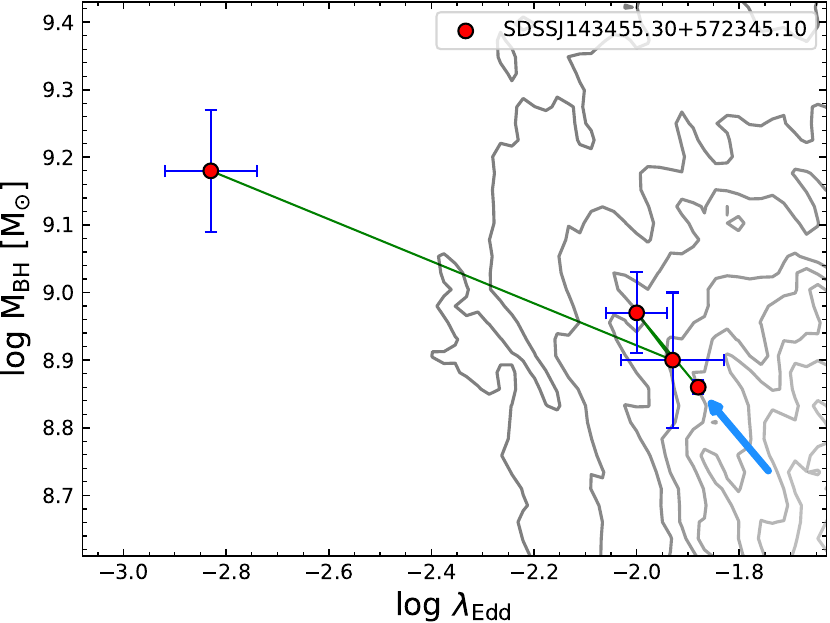}

    \includegraphics[width=0.21\textwidth]{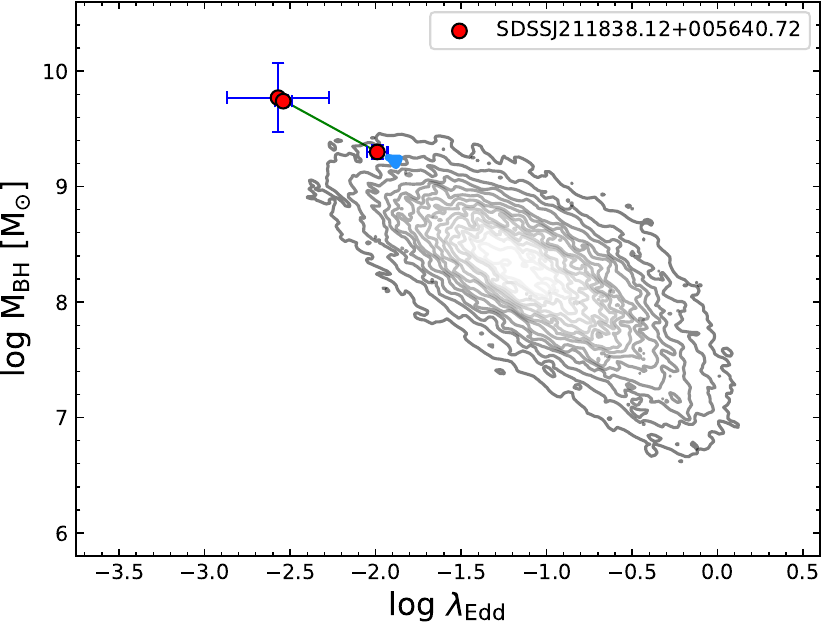}
    \includegraphics[width=0.21\textwidth]{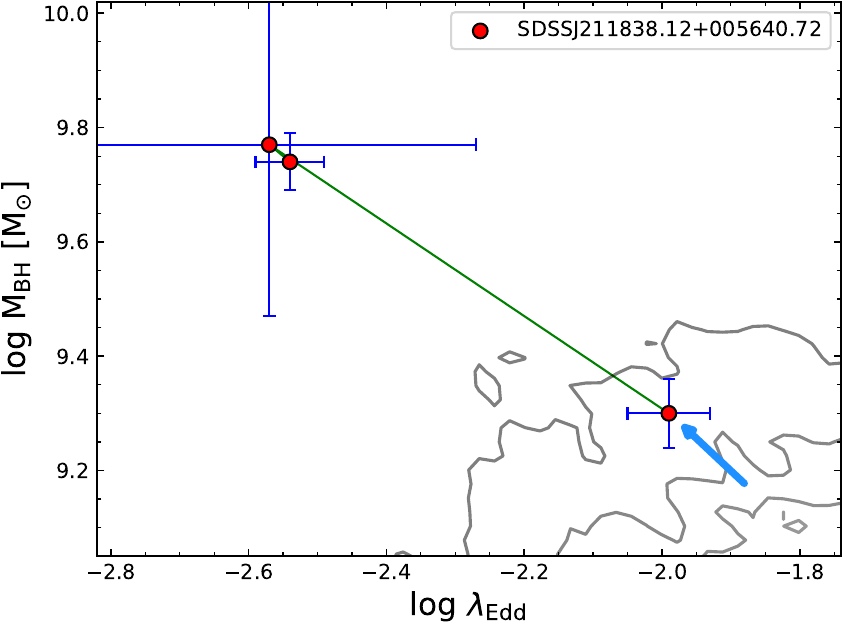}
    \includegraphics[width=0.21\textwidth]{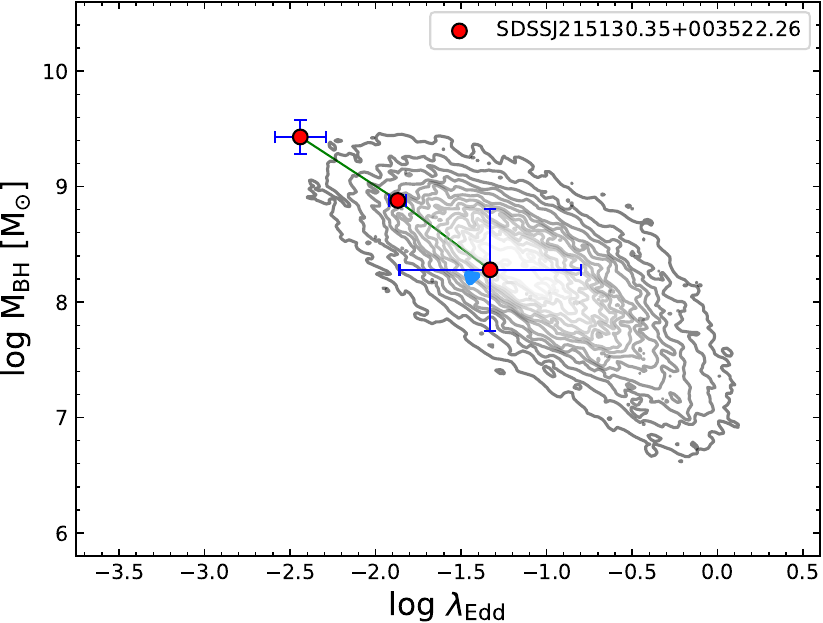}
    \includegraphics[width=0.21\textwidth]{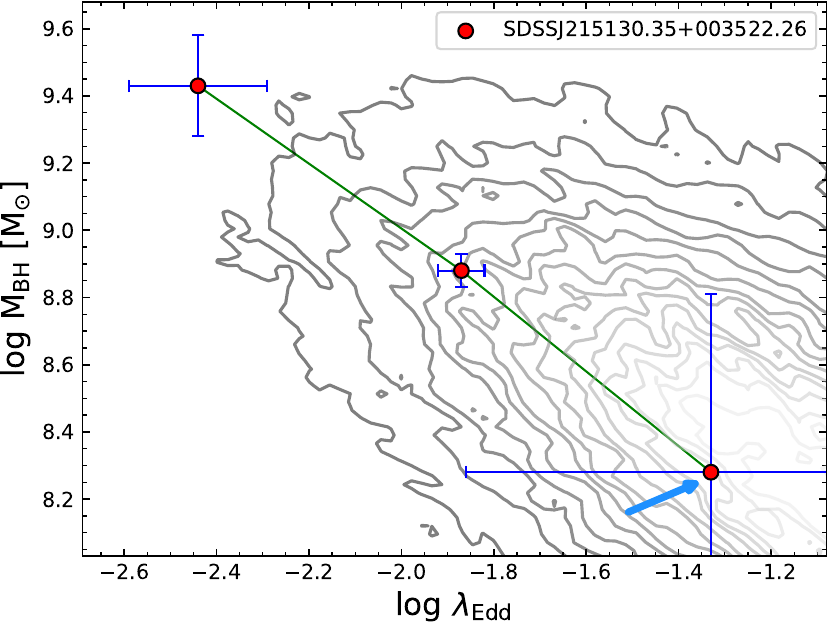}

    \includegraphics[width=0.21\textwidth]{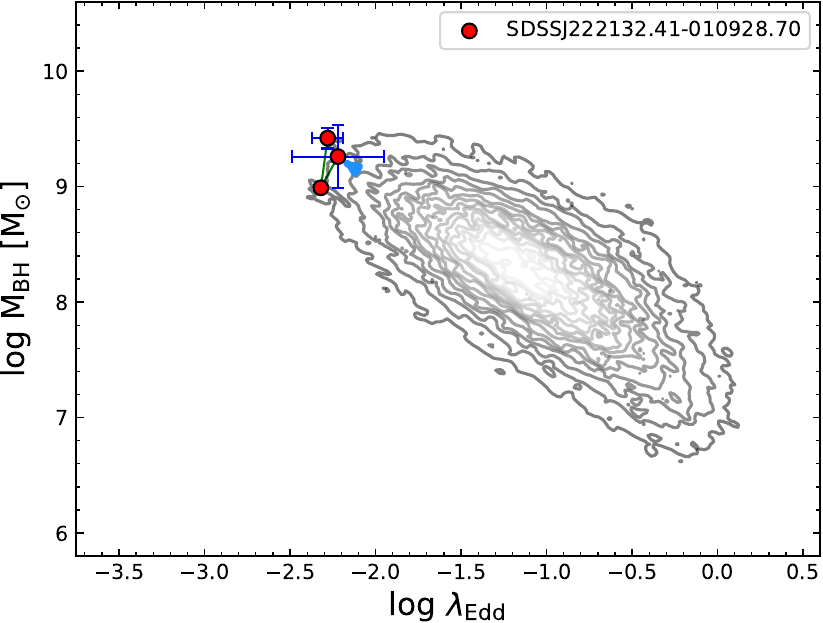}
    \includegraphics[width=0.21\textwidth]{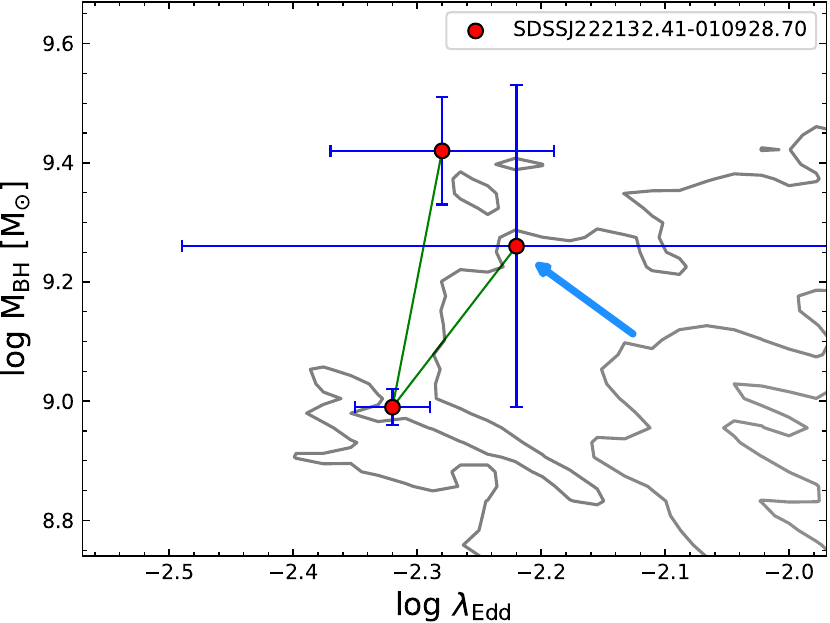}
    \includegraphics[width=0.21\textwidth]{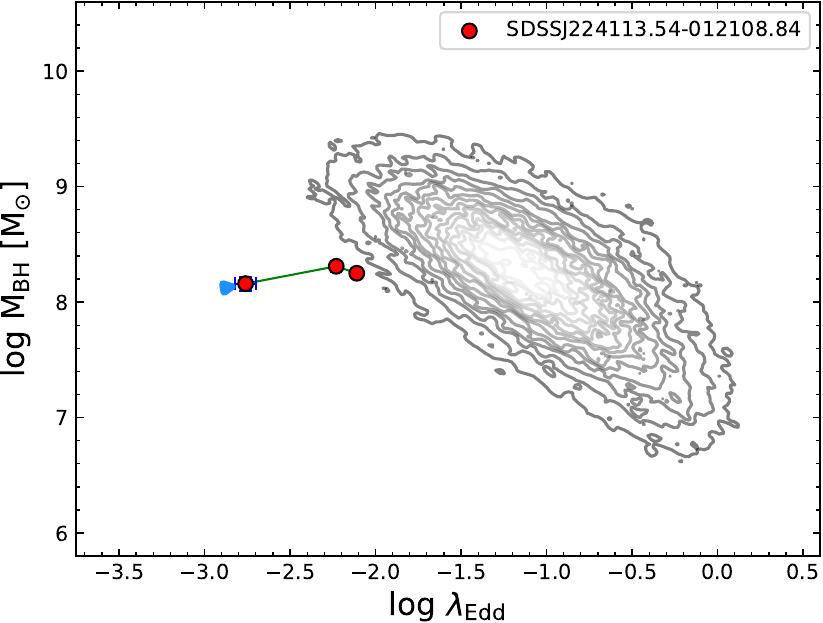}
    \includegraphics[width=0.21\textwidth]{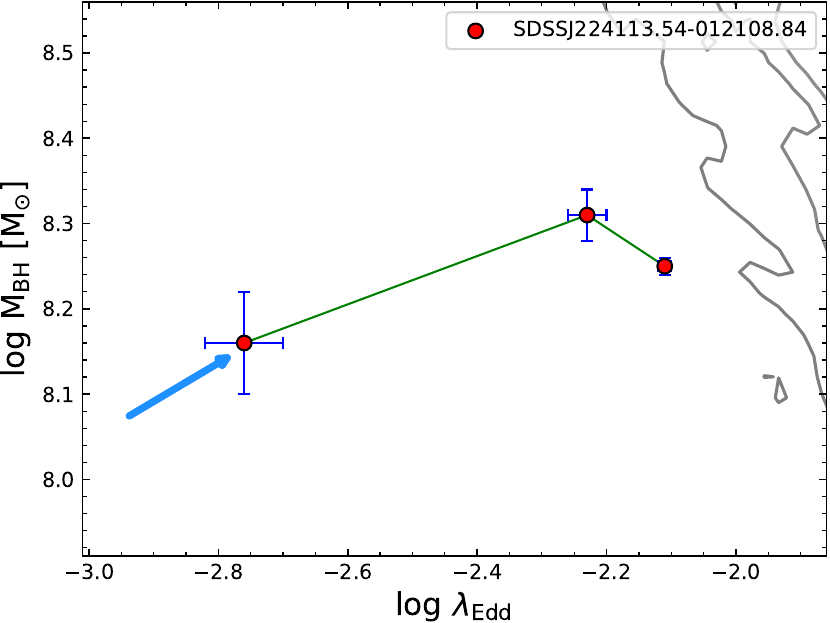}

    \includegraphics[width=0.21\textwidth]{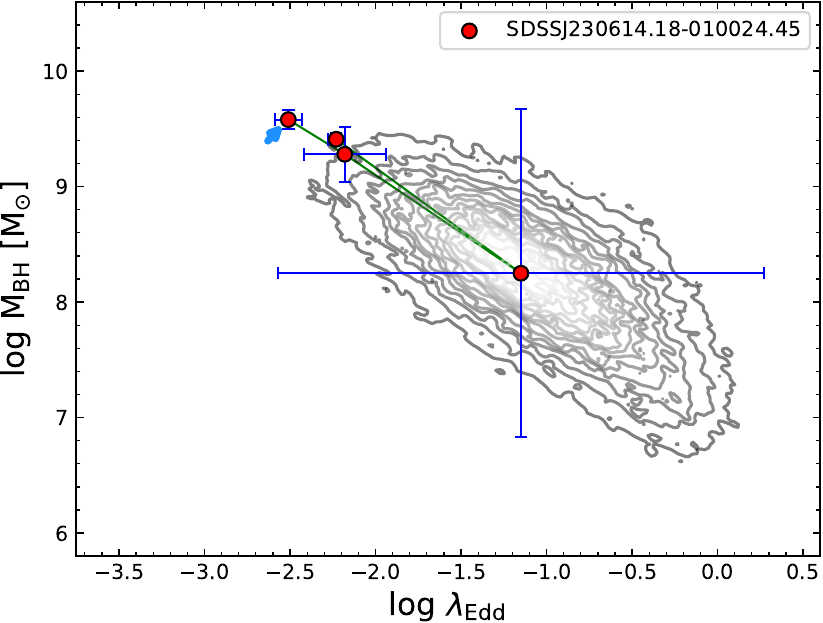}
    \includegraphics[width=0.21\textwidth]{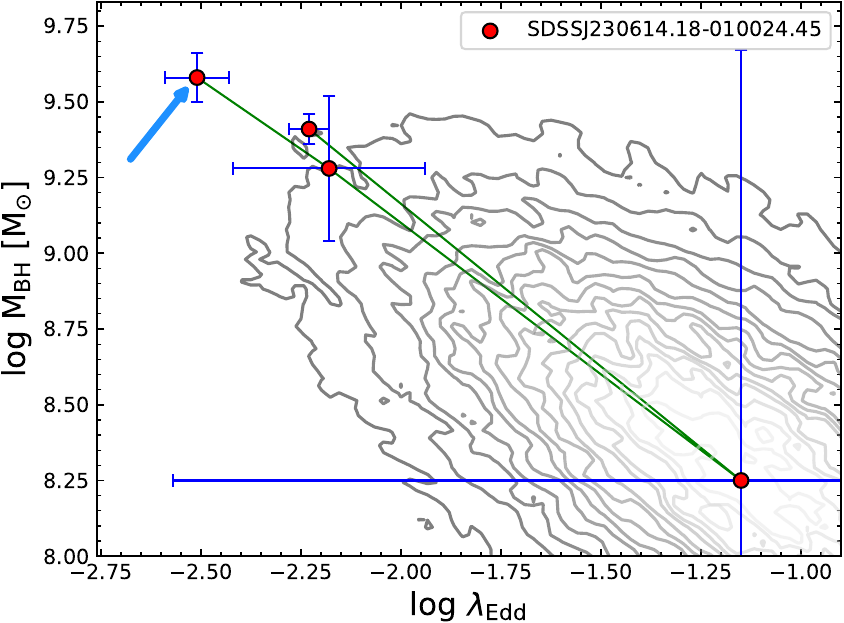}
    \includegraphics[width=0.21\textwidth]{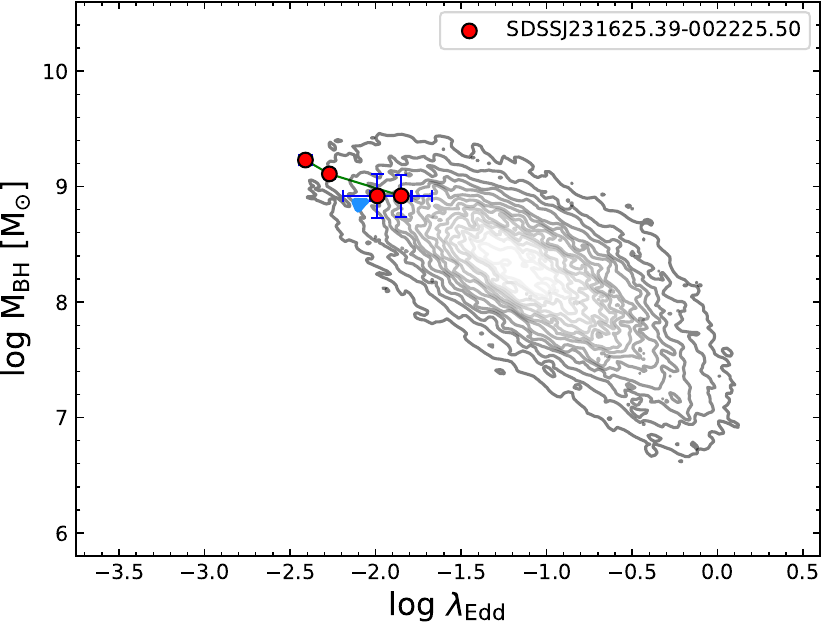}
    \includegraphics[width=0.21\textwidth]{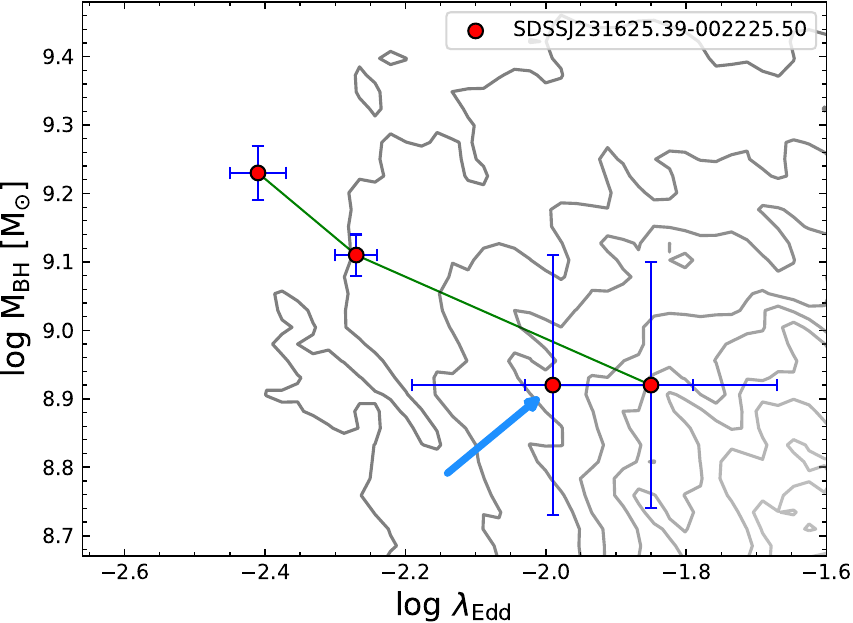}

    \caption{Panels are similar to Figures \ref{fig:M-Edd-others} for the remaining sources.}
    \label{fig:M-Edd-others2}
\end{figure*}


\begin{figure*}[!htb]
    \centering
    \includegraphics[width=0.21\textwidth]{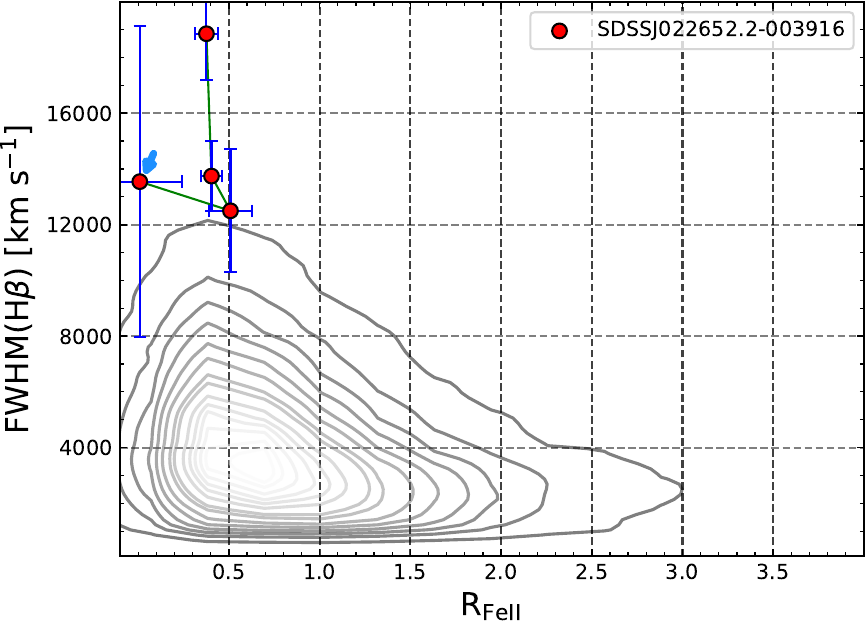}
    \includegraphics[width=0.21\textwidth]{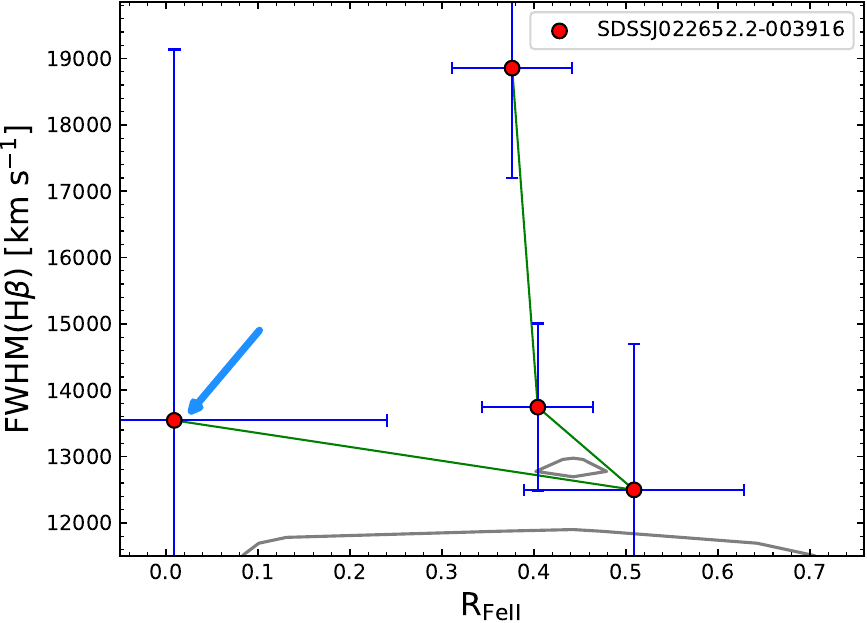}    
    \includegraphics[width=0.21\textwidth]{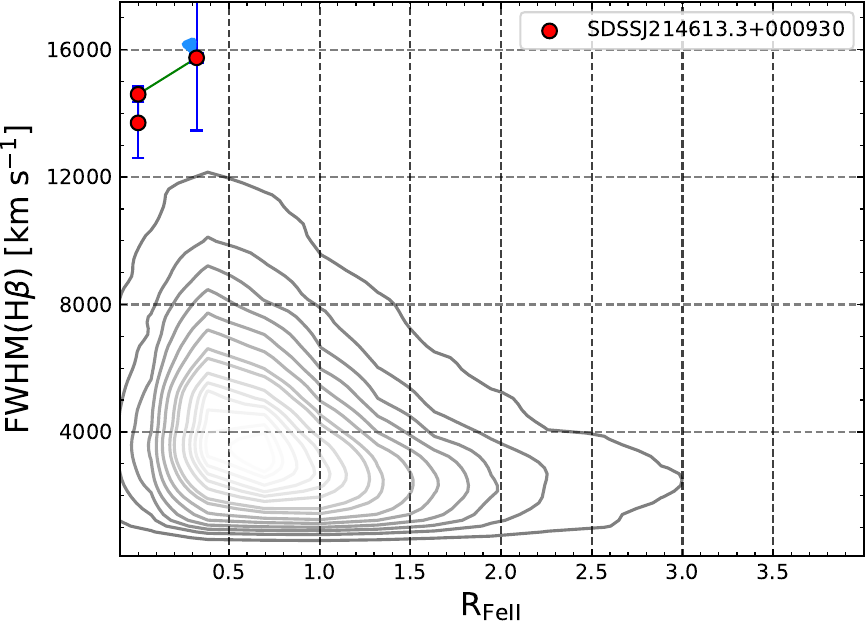}
    \includegraphics[width=0.21\textwidth]{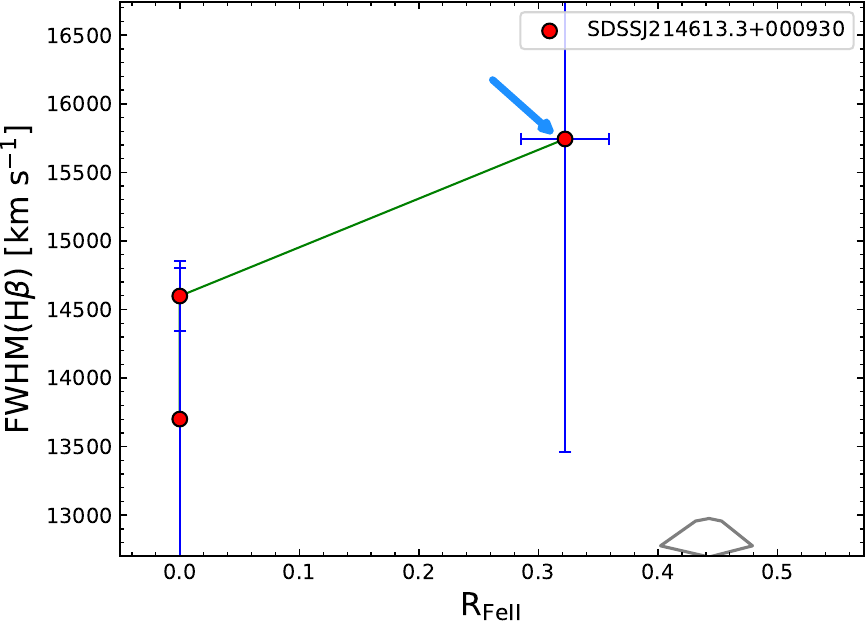}\\

    \includegraphics[width=0.21\textwidth]{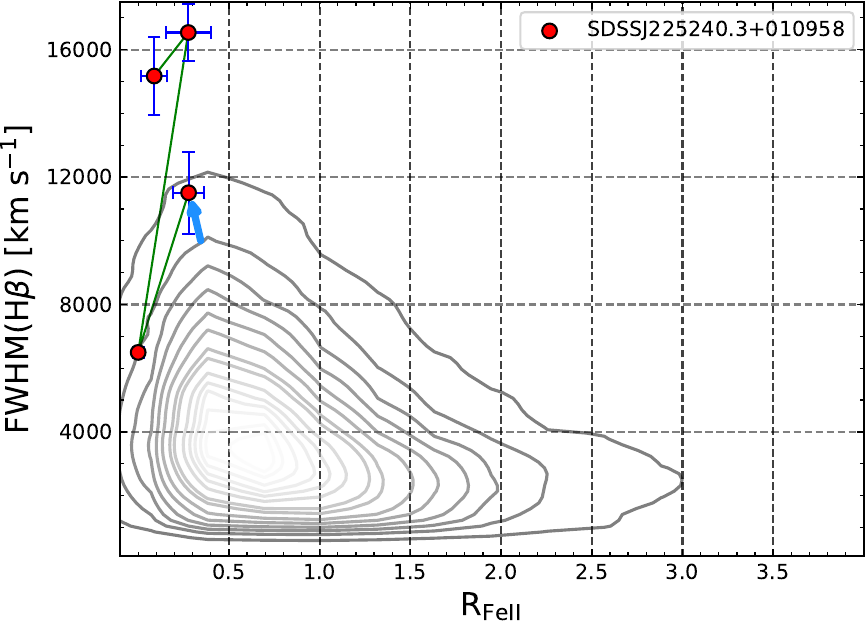}
    \includegraphics[width=0.21\textwidth]{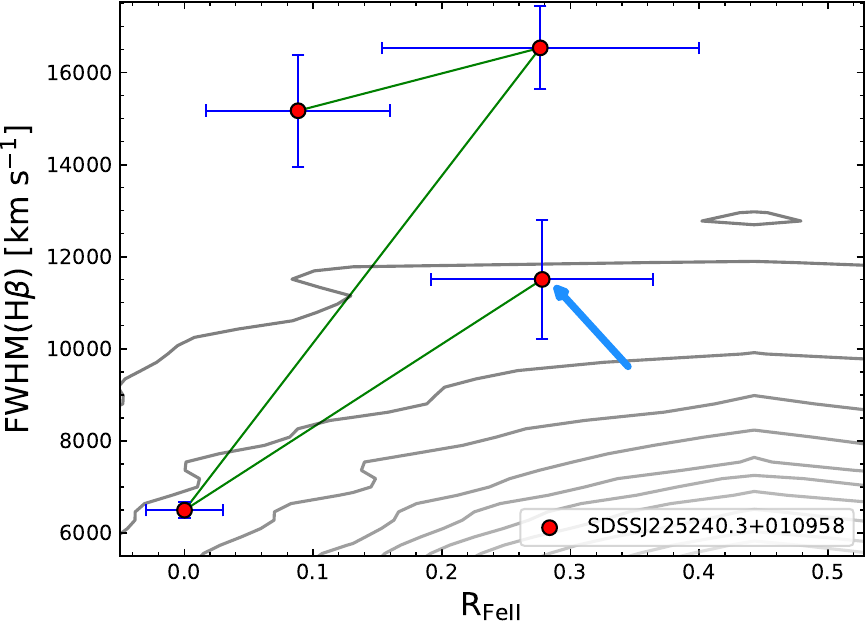}    
    \includegraphics[width=0.21\textwidth]{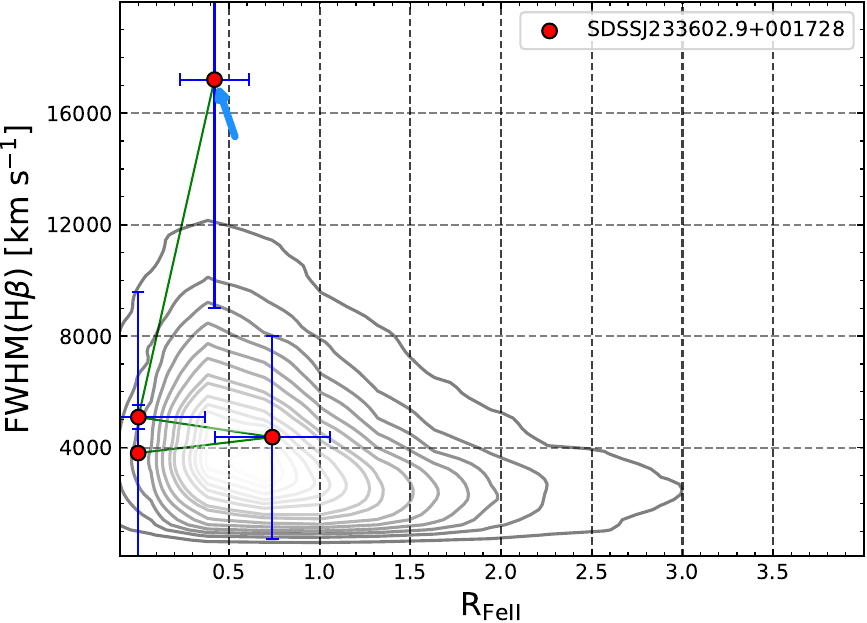}
    \includegraphics[width=0.21\textwidth]{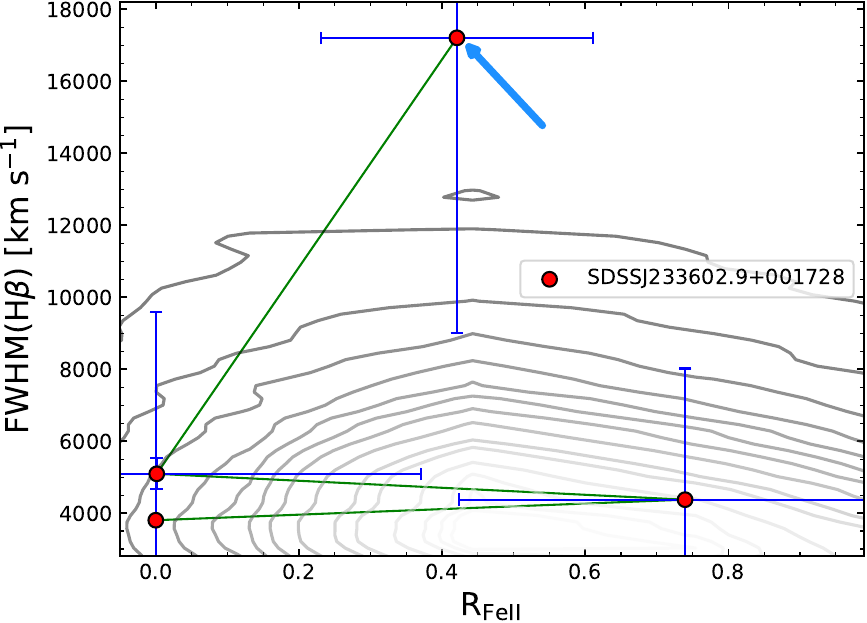}\\
   
    \includegraphics[width=0.21\textwidth]{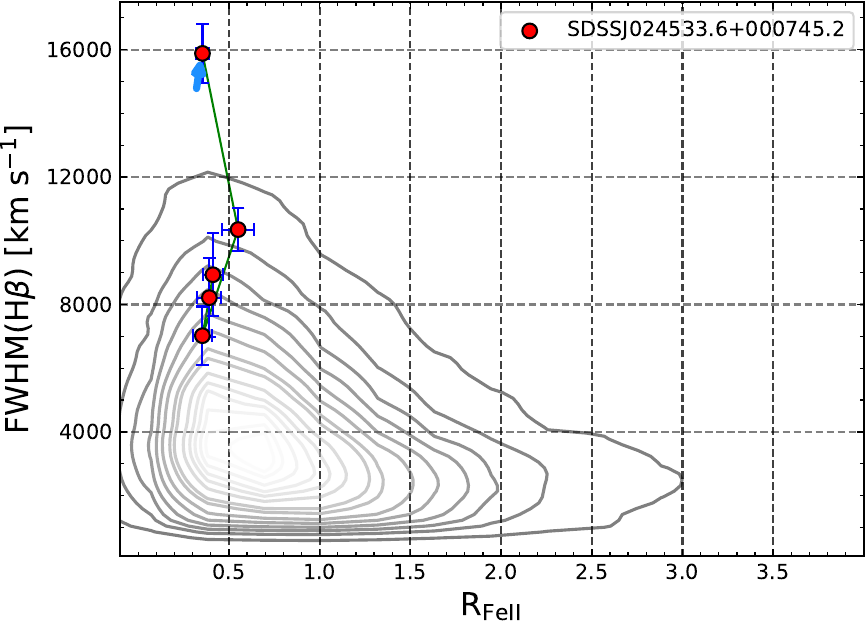}
    \includegraphics[width=0.21\textwidth]{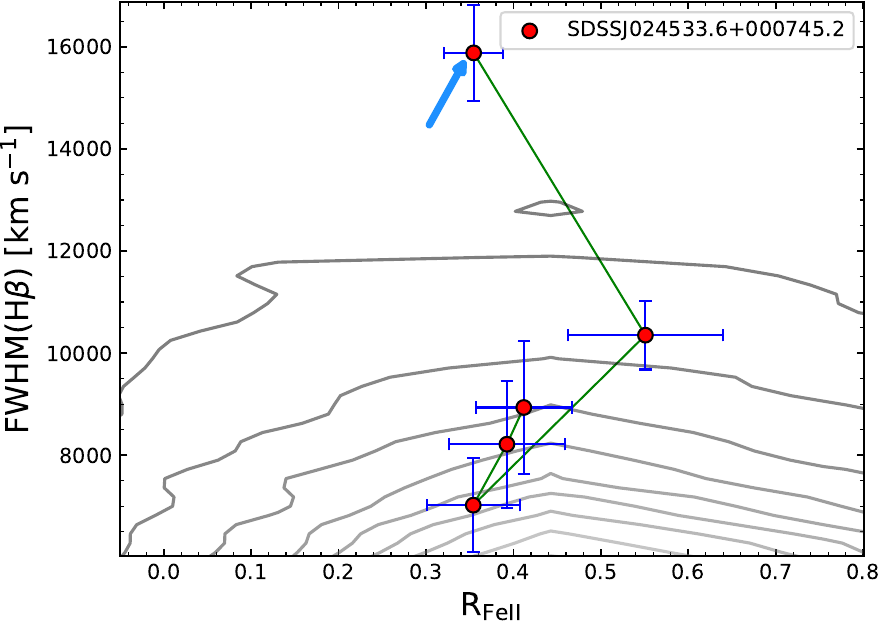}  
    \includegraphics[width=0.21\textwidth]{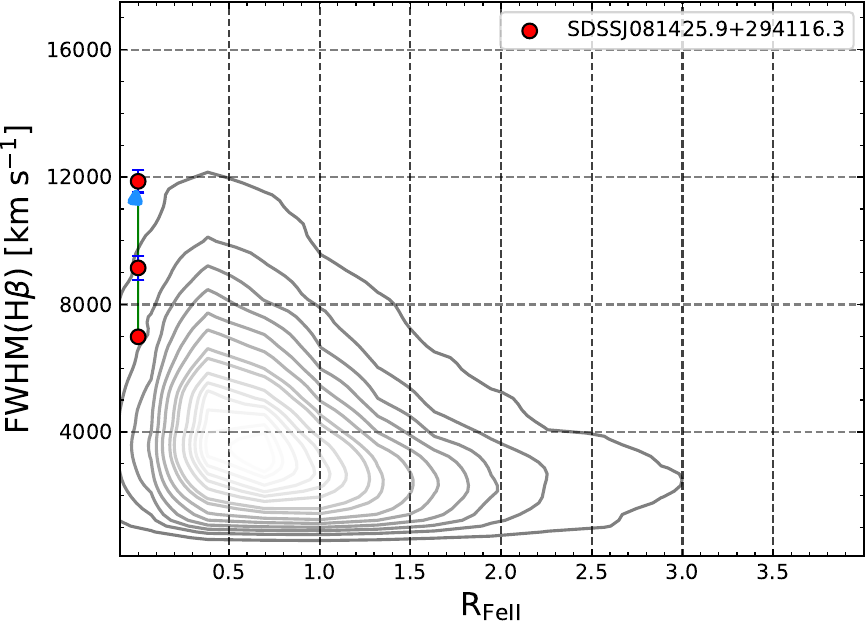}
    \includegraphics[width=0.21\textwidth]{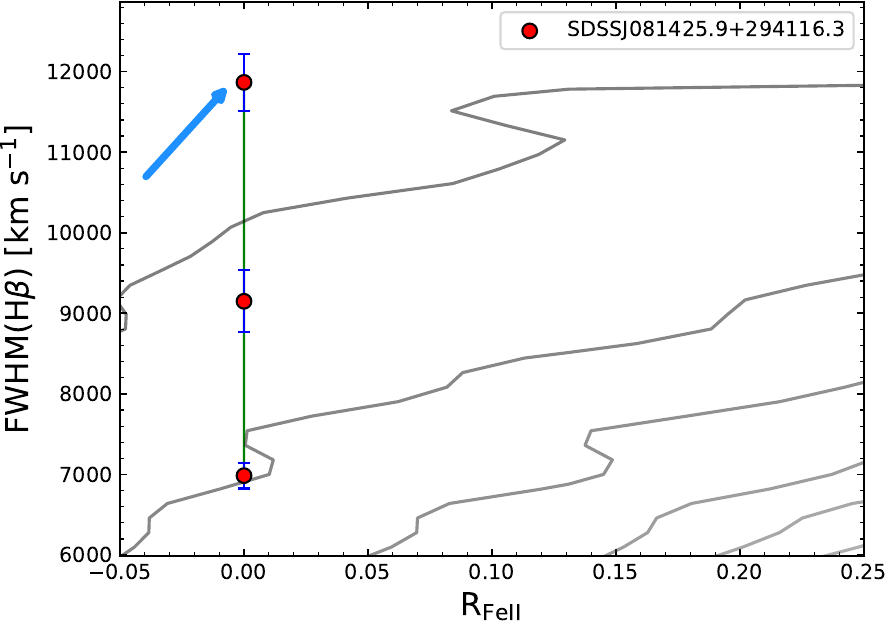}\\

    \includegraphics[width=0.21\textwidth]{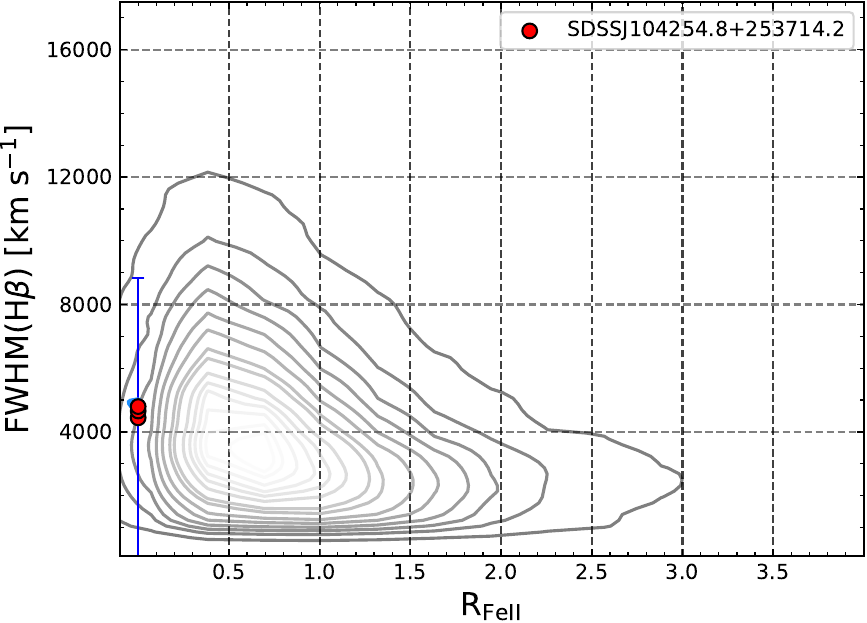}
    \includegraphics[width=0.21\textwidth]{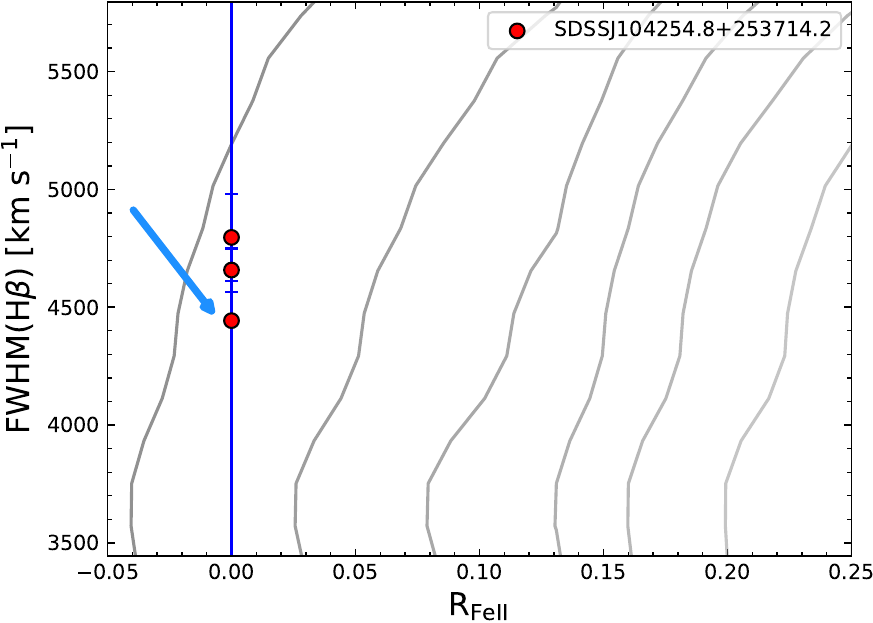}    
    \includegraphics[width=0.21\textwidth]{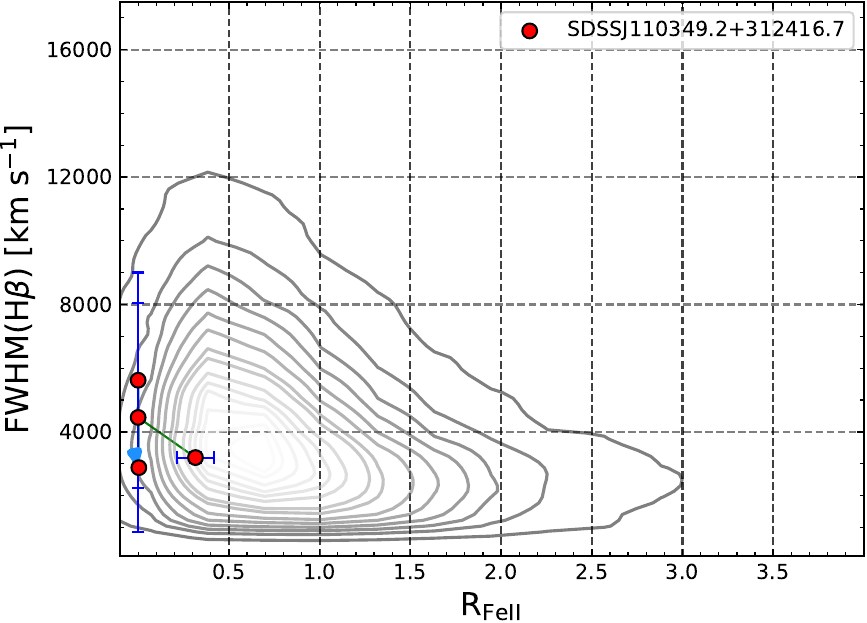}
    \includegraphics[width=0.21\textwidth]{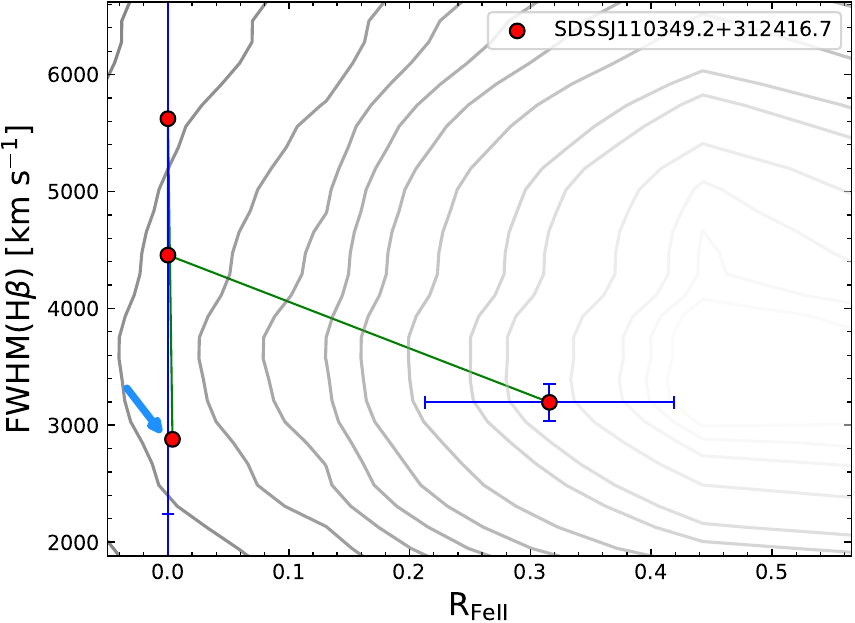}\\

    \includegraphics[width=0.21\textwidth]{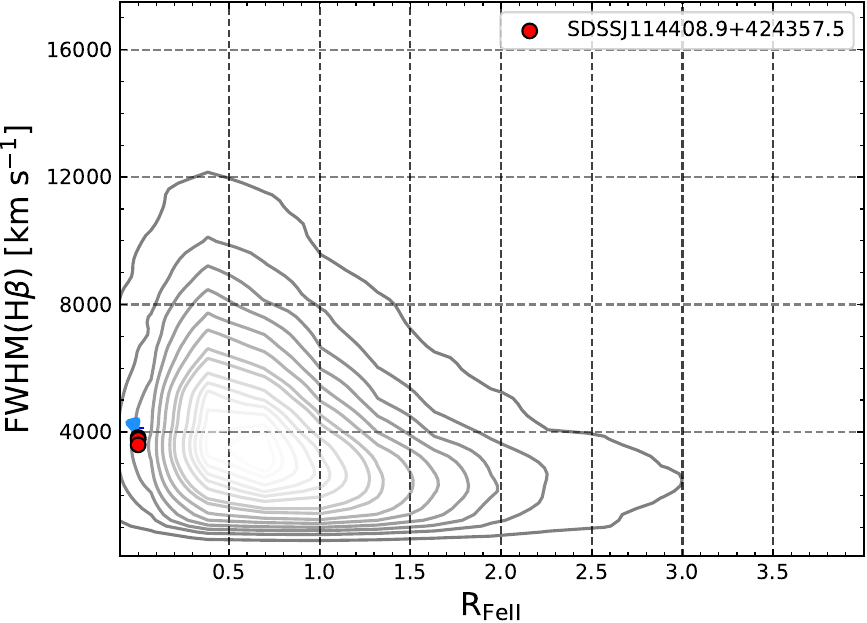}
    \includegraphics[width=0.21\textwidth]{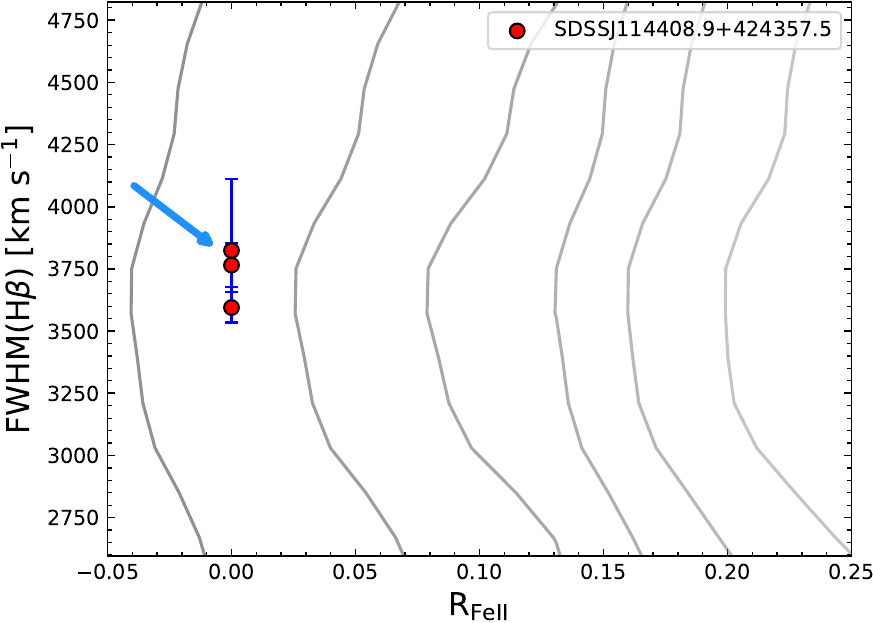}    
    \includegraphics[width=0.21\textwidth]{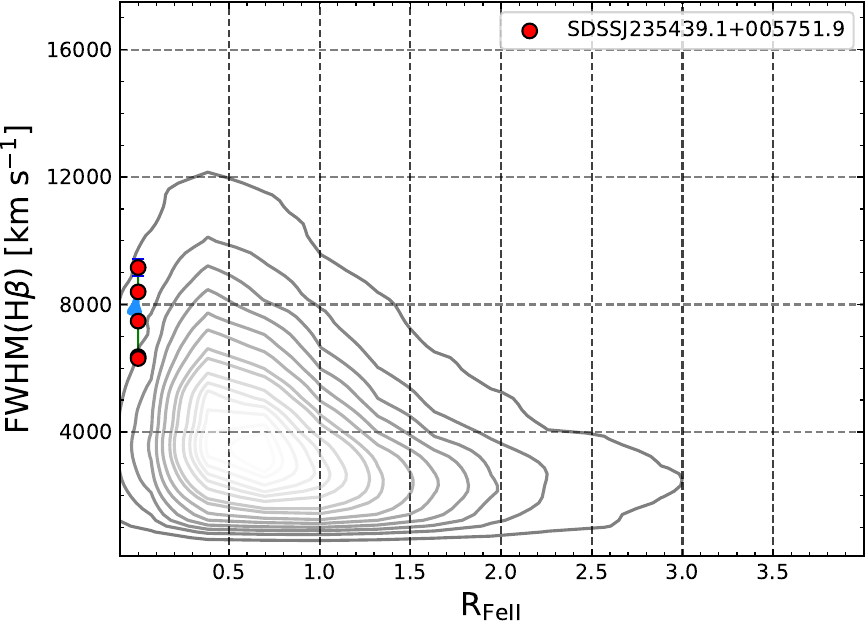}
    \includegraphics[width=0.21\textwidth]{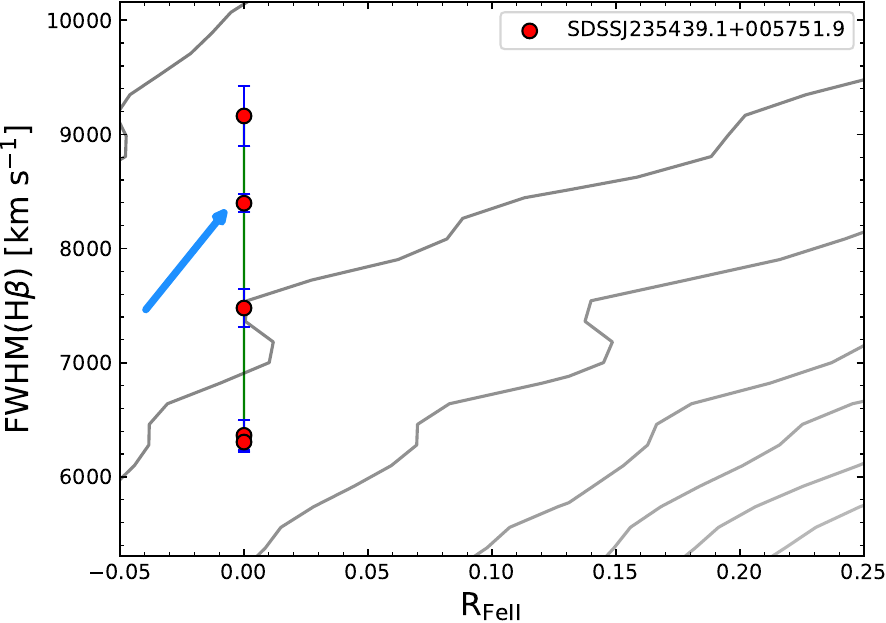}\\
    
    \includegraphics[width=0.21\textwidth]{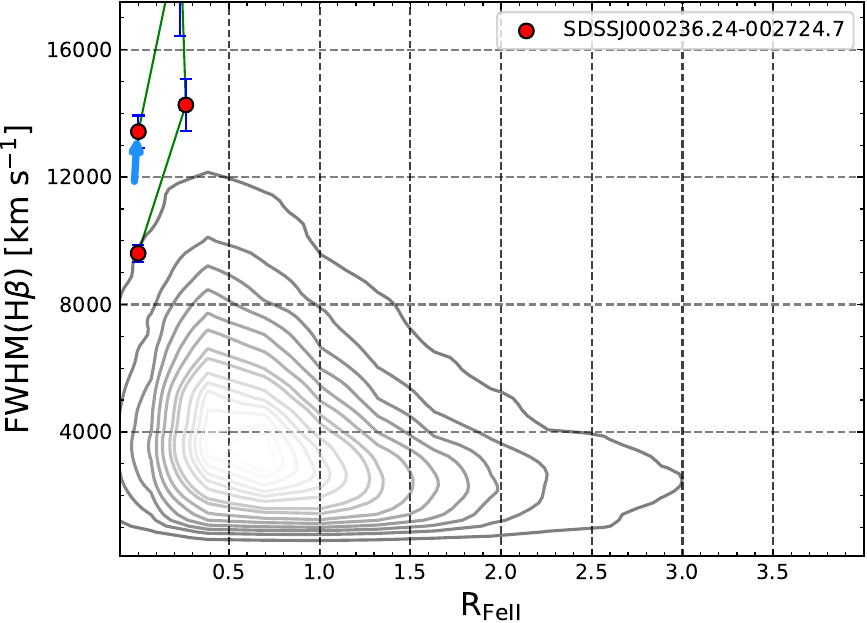}
    \includegraphics[width=0.21\textwidth]{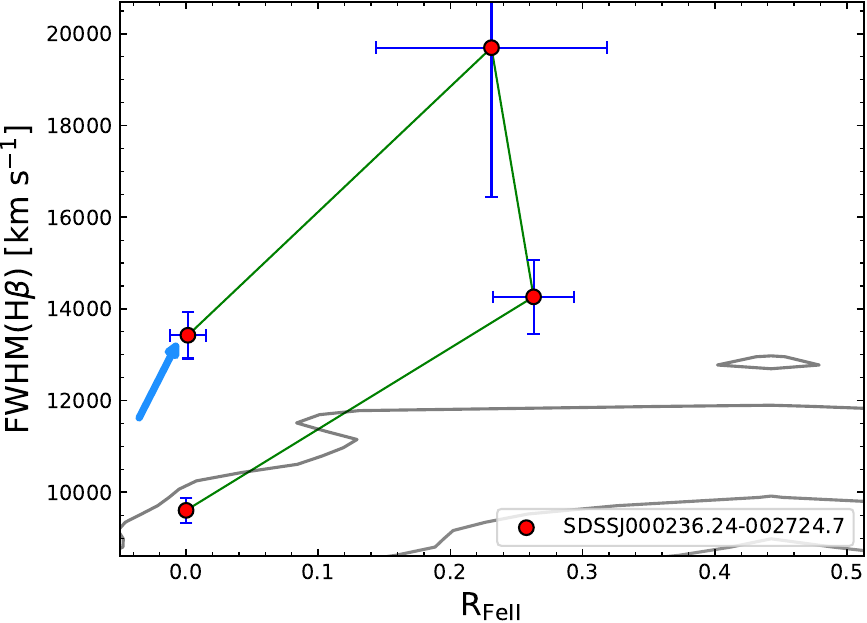}
    \includegraphics[width=0.21\textwidth]{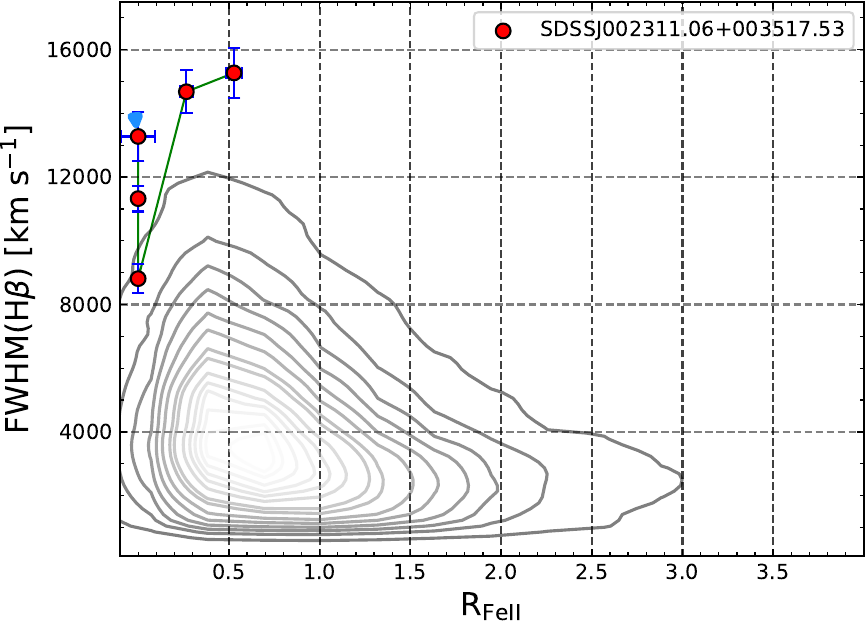}
    \includegraphics[width=0.21\textwidth]{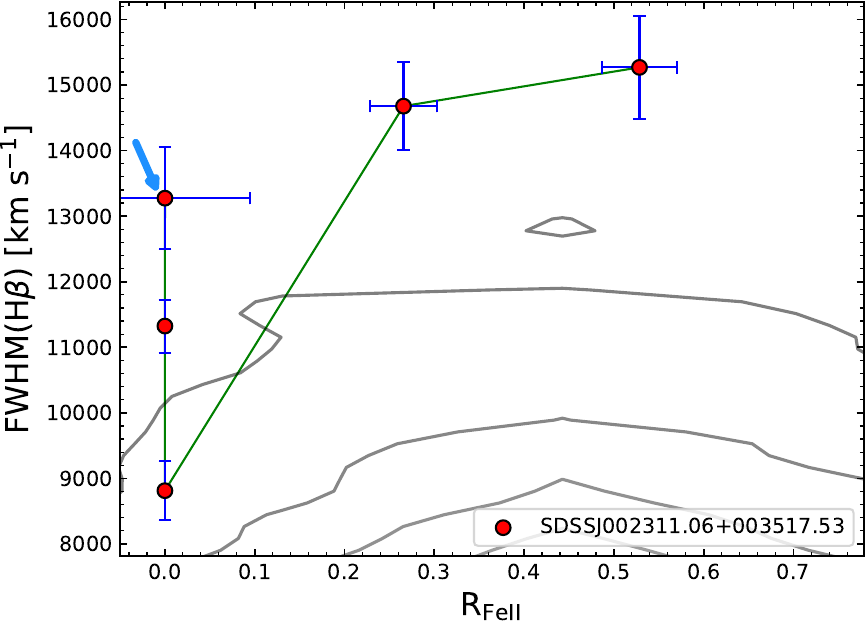}\\

    \includegraphics[width=0.21\textwidth]{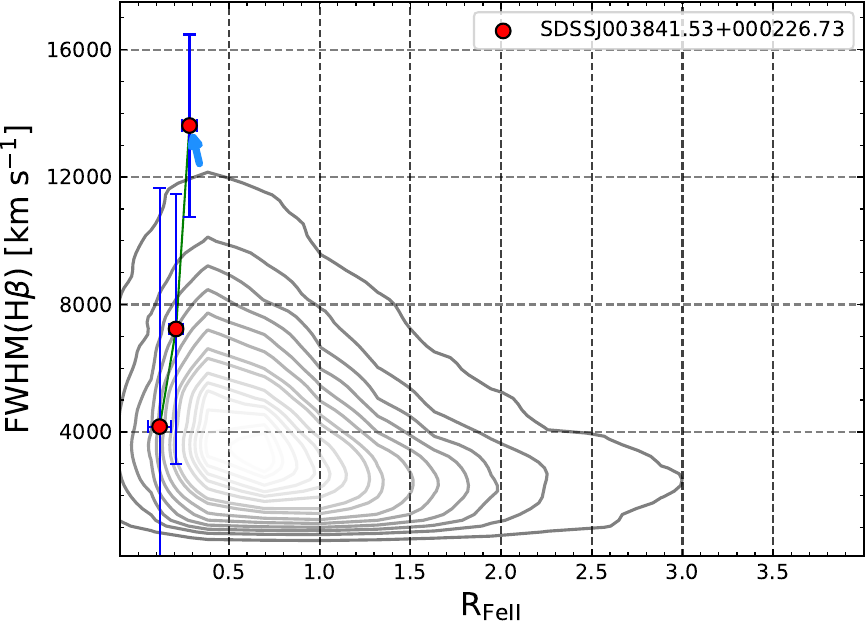}
    \includegraphics[width=0.21\textwidth]{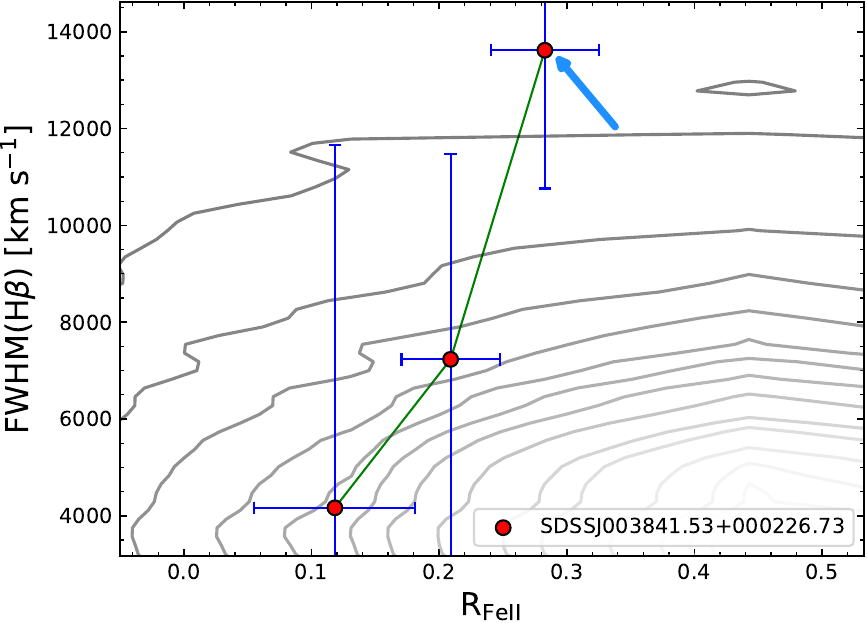}
    \includegraphics[width=0.21\textwidth]{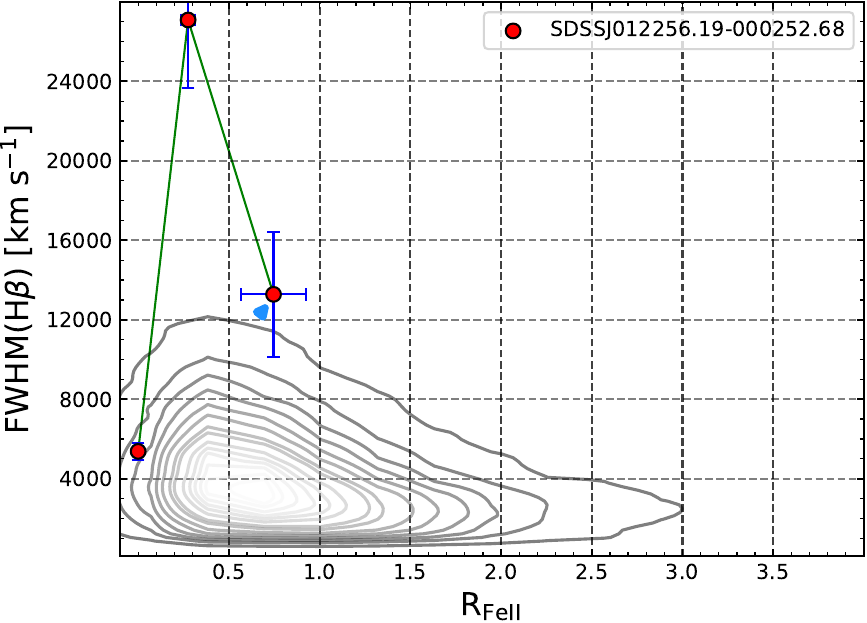}
    \includegraphics[width=0.21\textwidth]{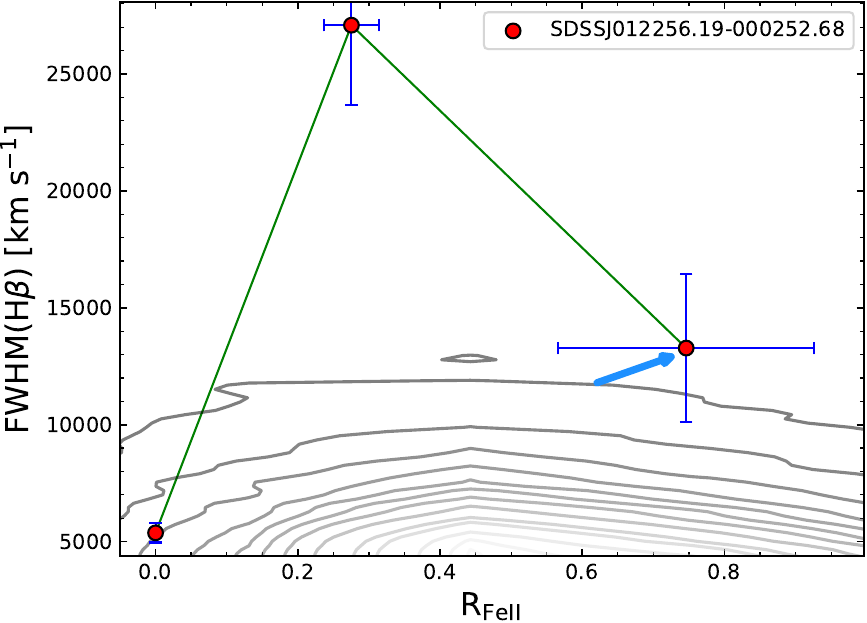}\\

    \includegraphics[width=0.21\textwidth]{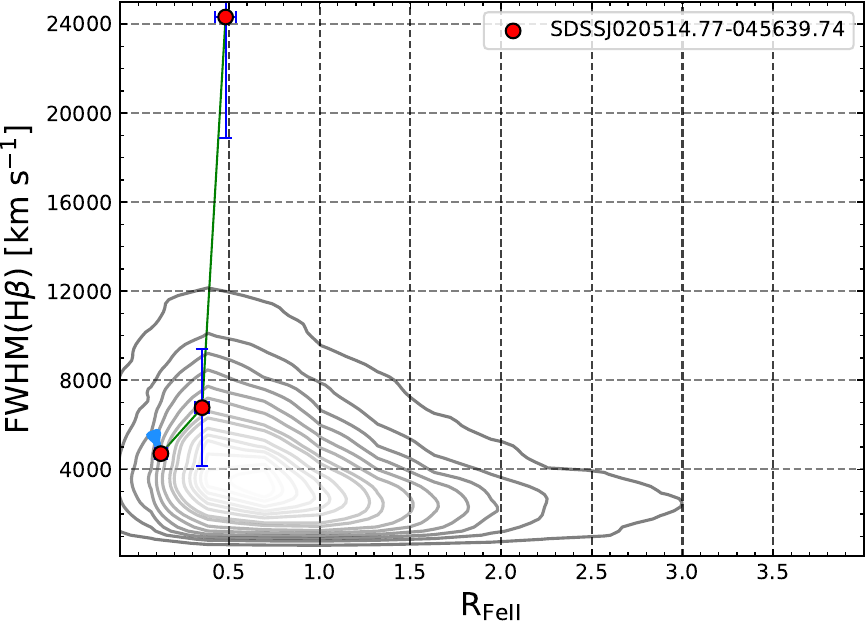}
    \includegraphics[width=0.21\textwidth]{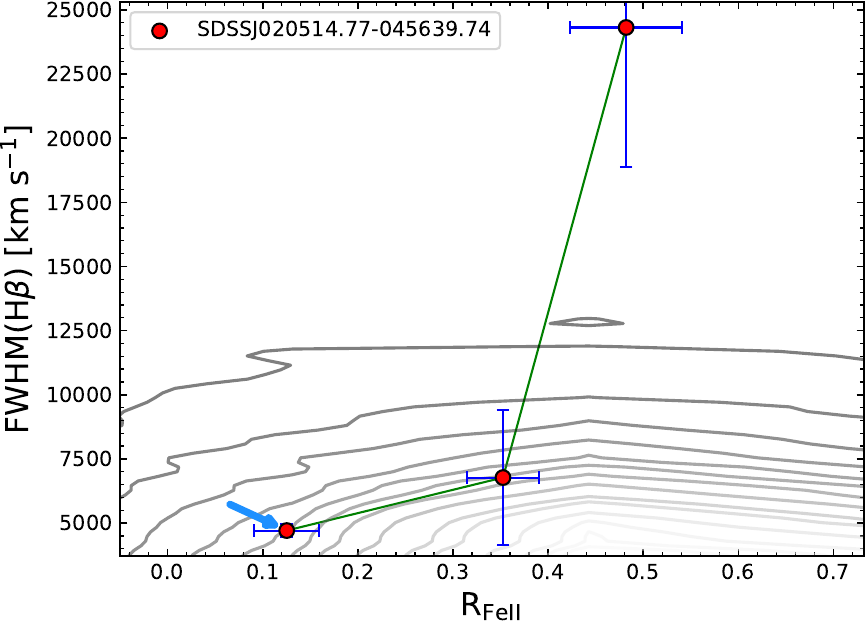}
    \includegraphics[width=0.21\textwidth]{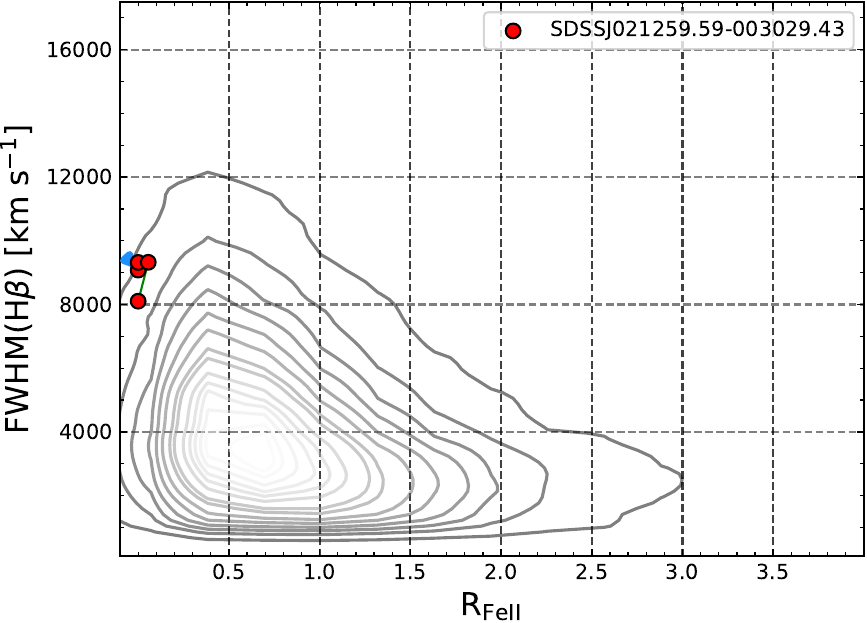}
    \includegraphics[width=0.21\textwidth]{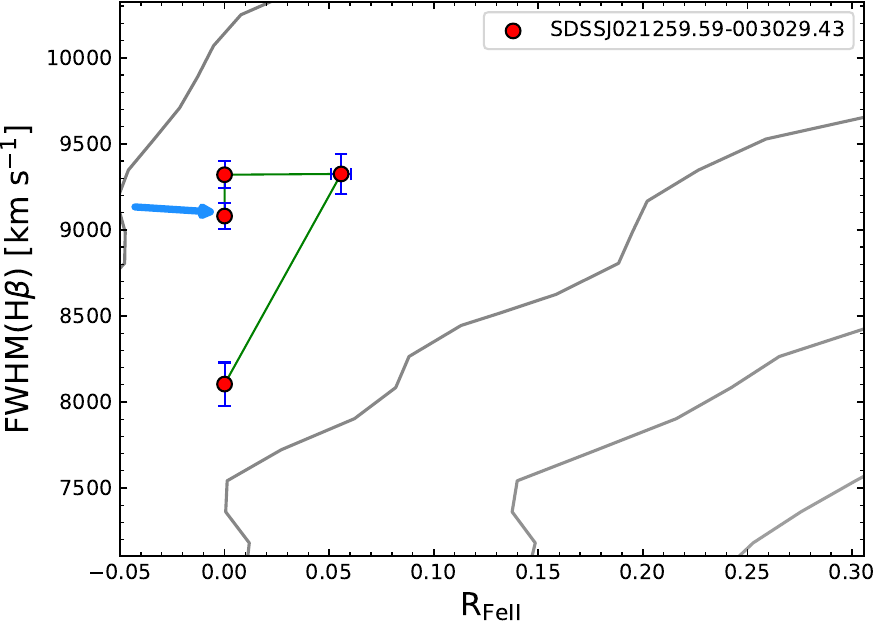}\\
    
    \caption{Optical plane of the main sequence of quasars for the remaining sources with at least 3 spectroscopic epochs in our sample with \rfe{}$>$0 and FWHM(\hb{})$>$0. The remaining parameters and annotations are identical to Figure \ref{fig:EV1}.}
    \label{fig:EV1-others}
\end{figure*}

\begin{figure*}[!htb]
    \centering  
    
    \includegraphics[width=0.21\textwidth]{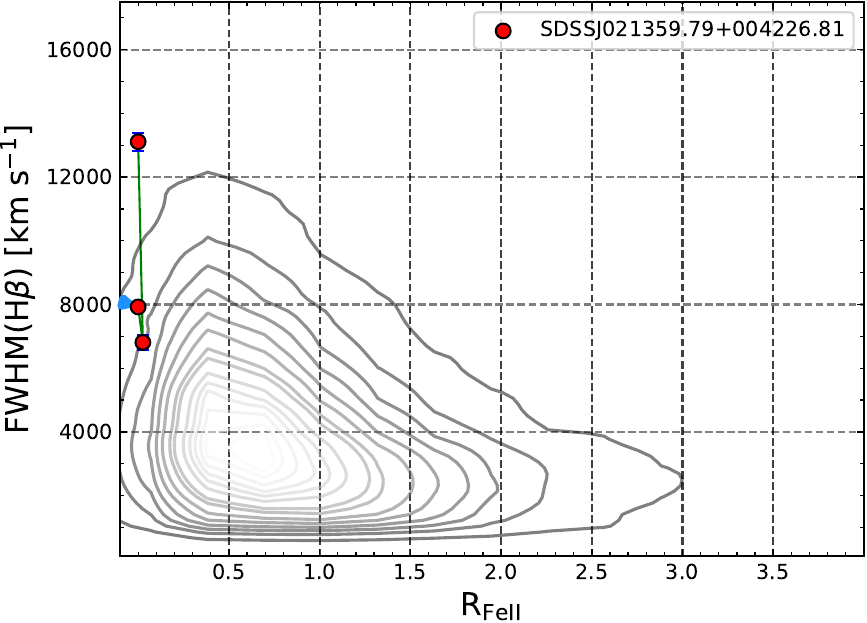}
    \includegraphics[width=0.21\textwidth]{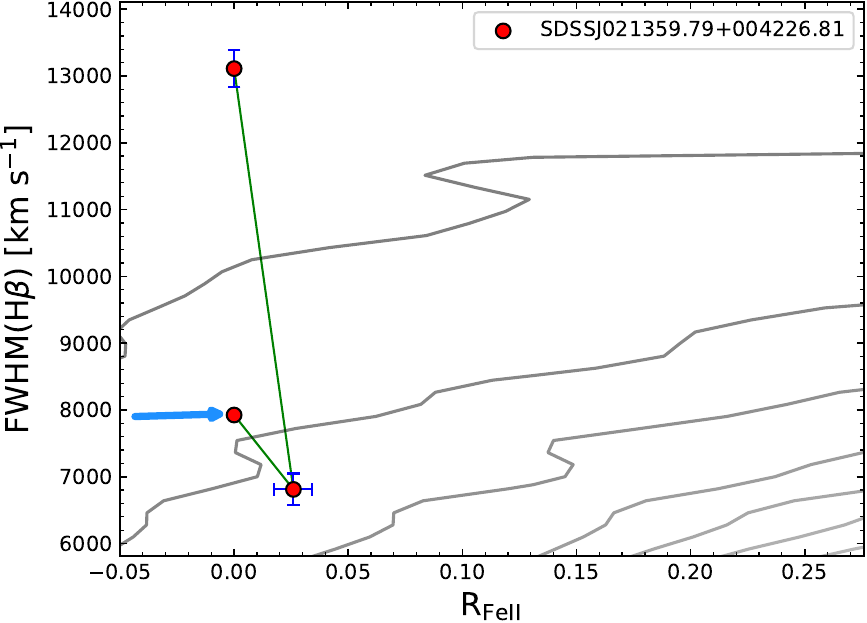}
    \includegraphics[width=0.21\textwidth]{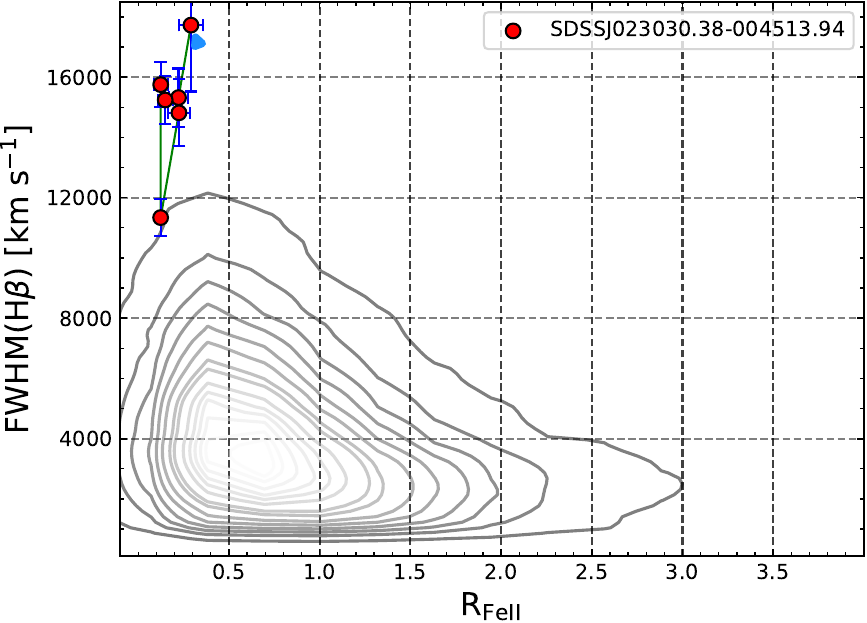}
    \includegraphics[width=0.21\textwidth]{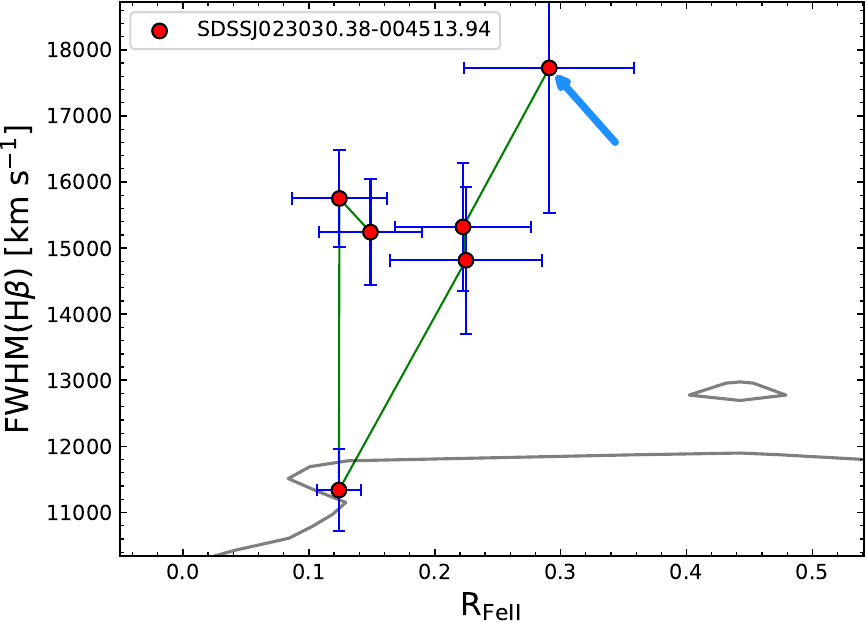}\\    
  
    \includegraphics[width=0.21\textwidth]{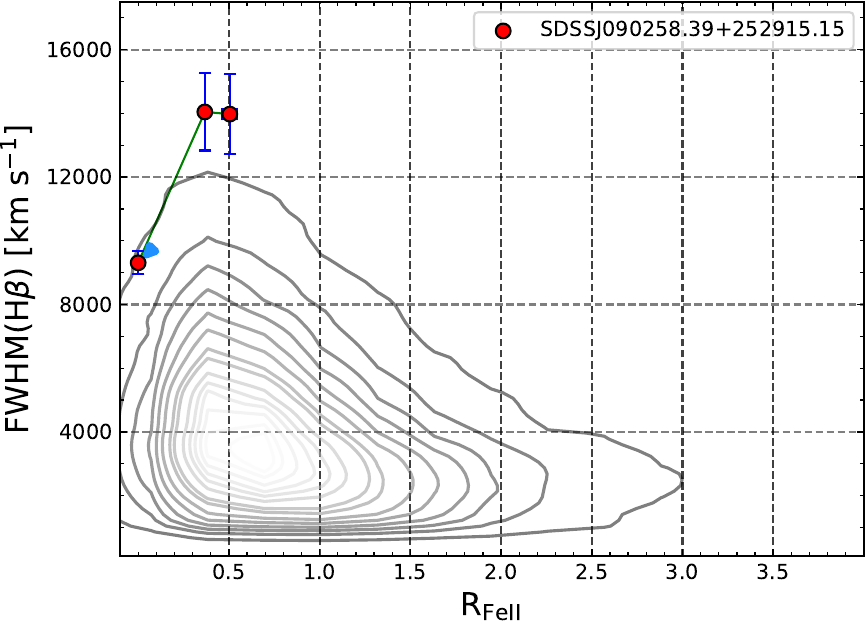}
    \includegraphics[width=0.21\textwidth]{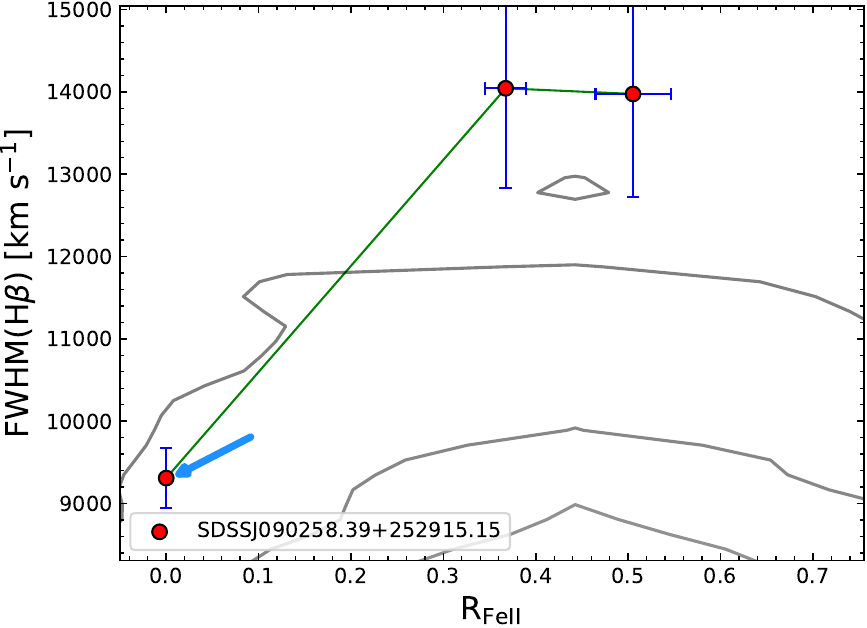}   
    \includegraphics[width=0.21\textwidth]{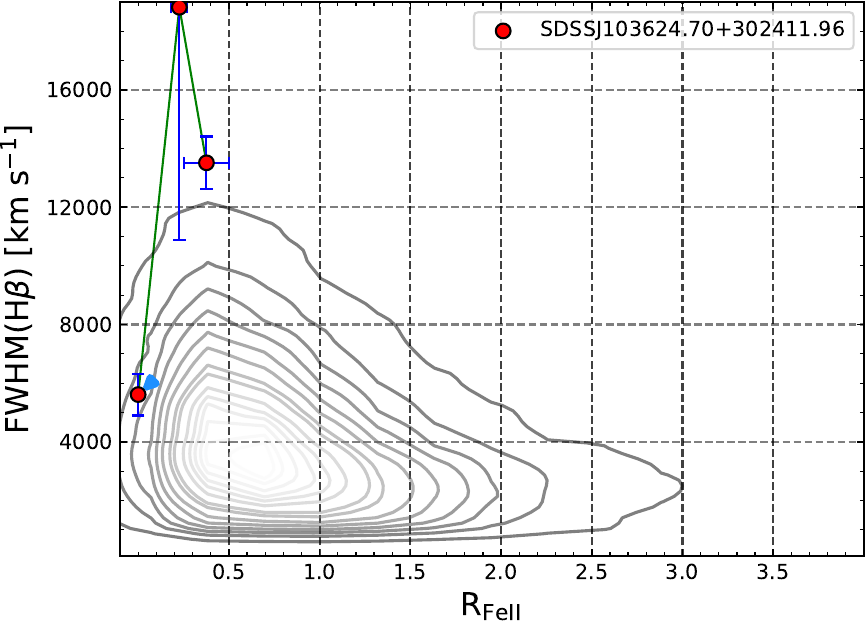}
    \includegraphics[width=0.21\textwidth]{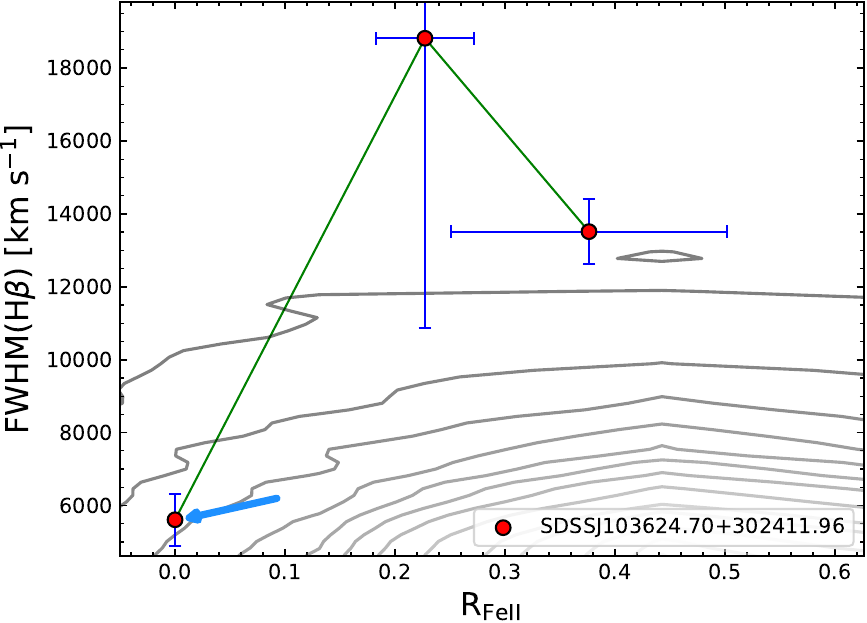}\\
    
    \includegraphics[width=0.21\textwidth]{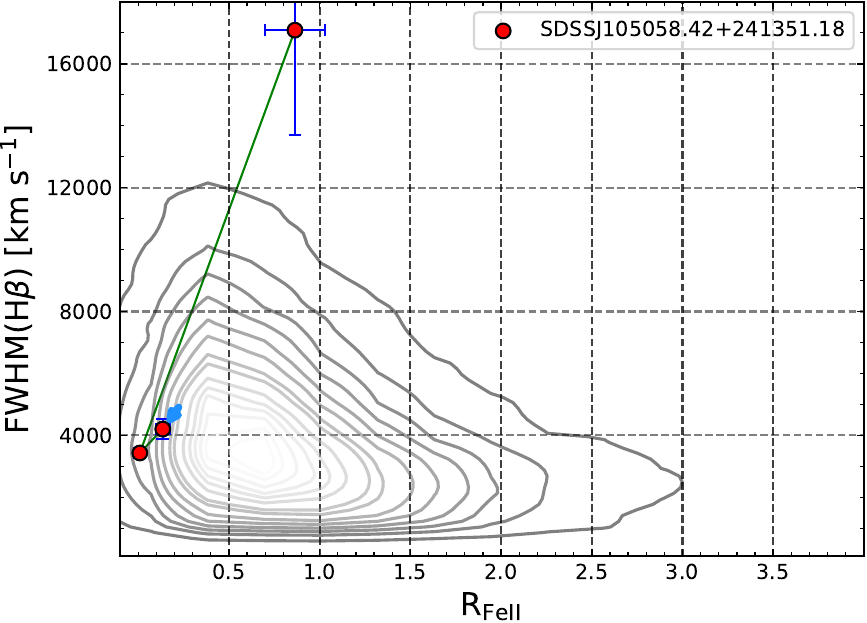}
    \includegraphics[width=0.21\textwidth]{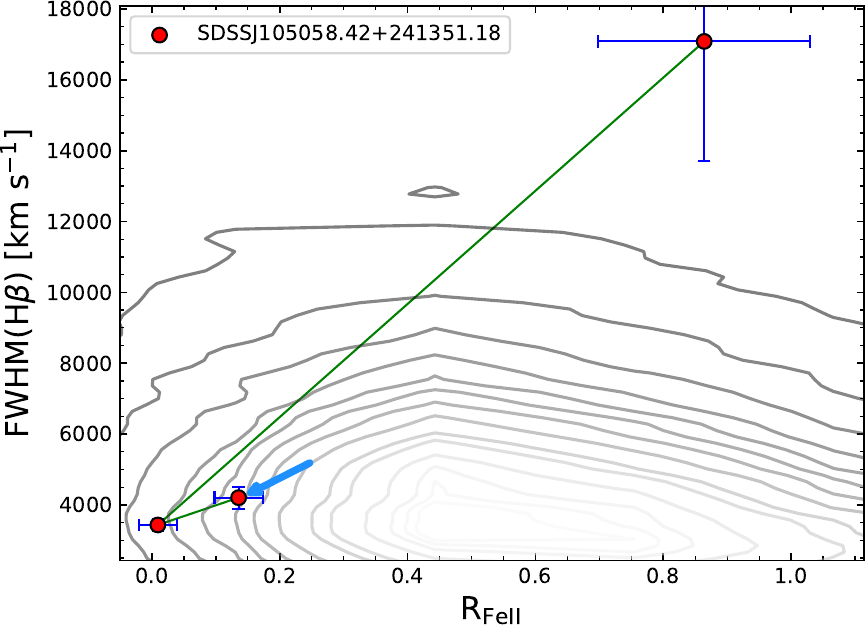}
    \includegraphics[width=0.21\textwidth]{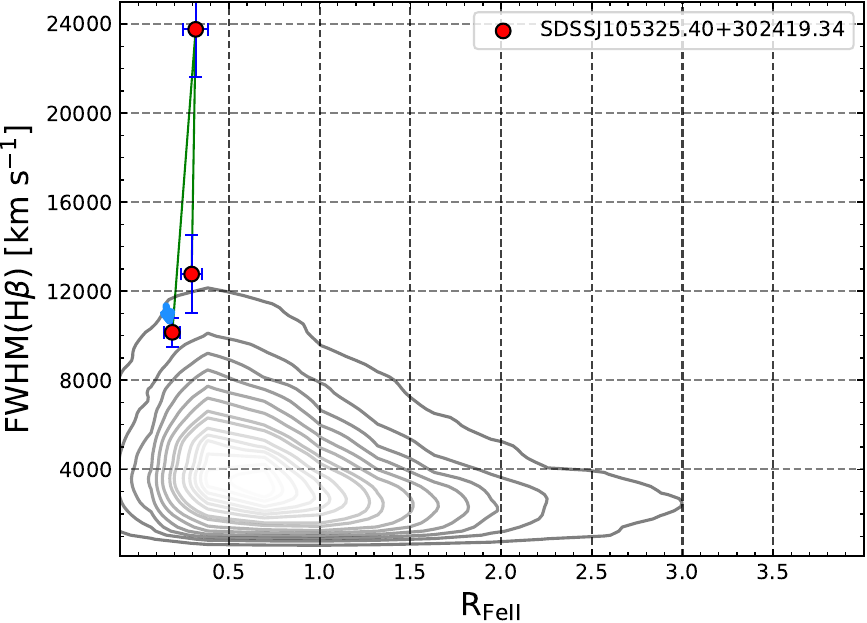}
    \includegraphics[width=0.21\textwidth]{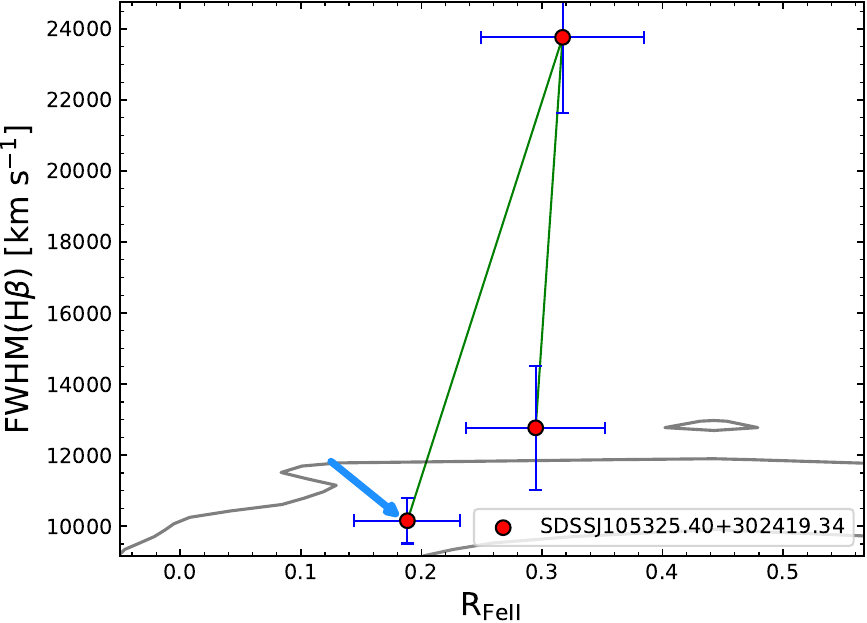}\\

    \includegraphics[width=0.21\textwidth]{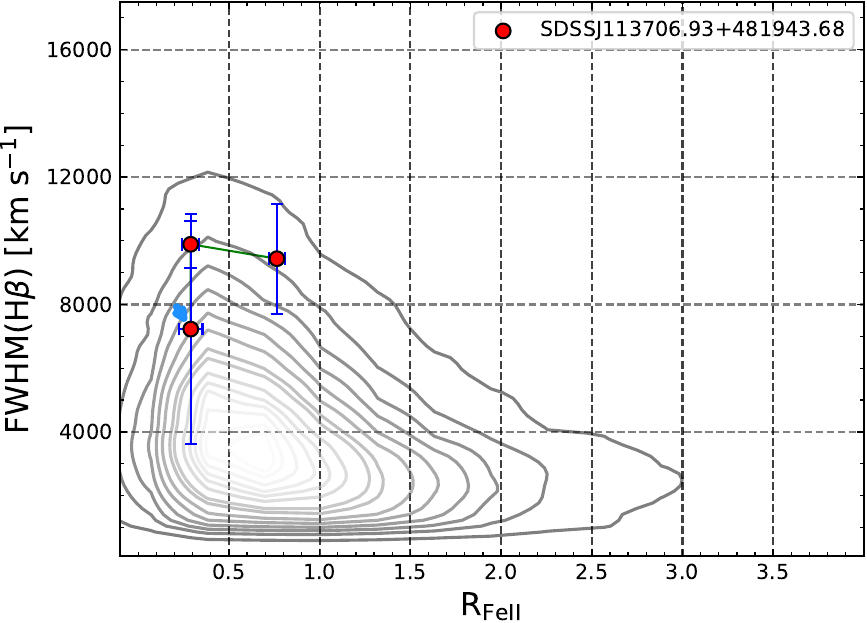}
    \includegraphics[width=0.21\textwidth]{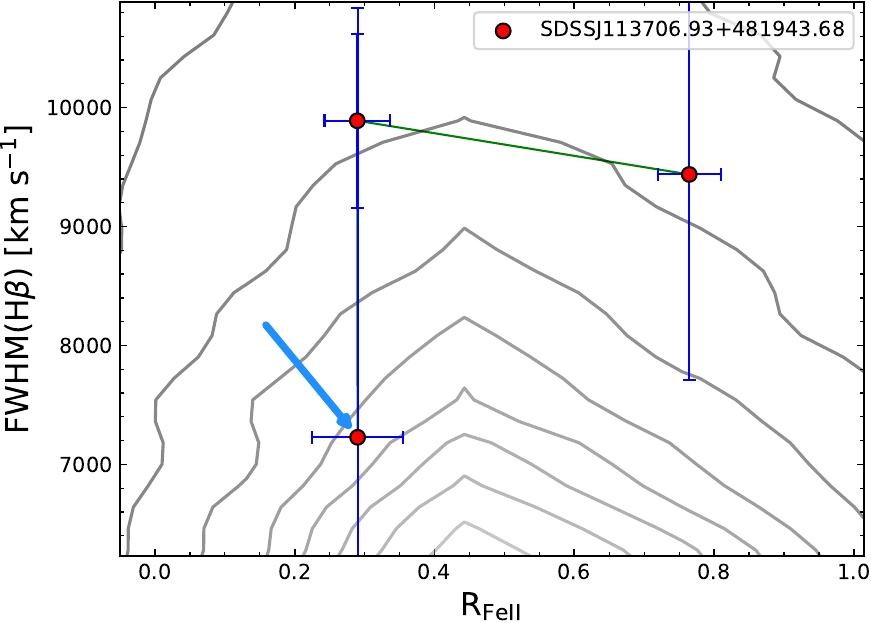}
    \includegraphics[width=0.21\textwidth]{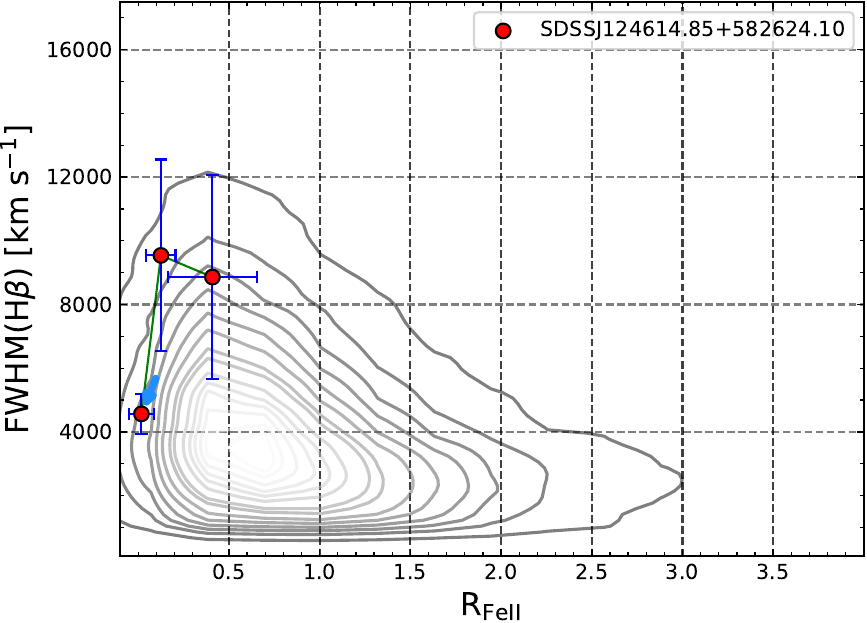}
    \includegraphics[width=0.21\textwidth]{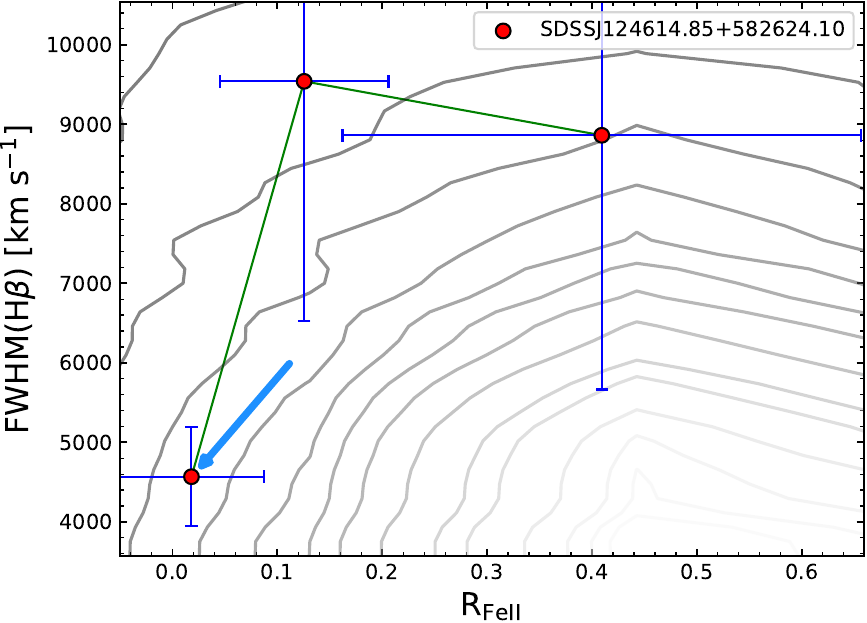}\\
    
    \includegraphics[width=0.21\textwidth]{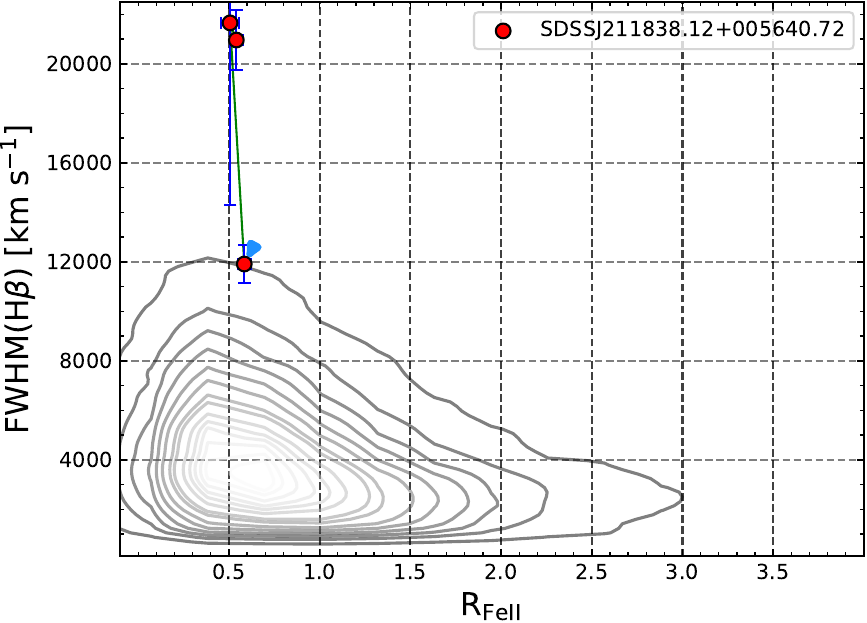}
    \includegraphics[width=0.21\textwidth]{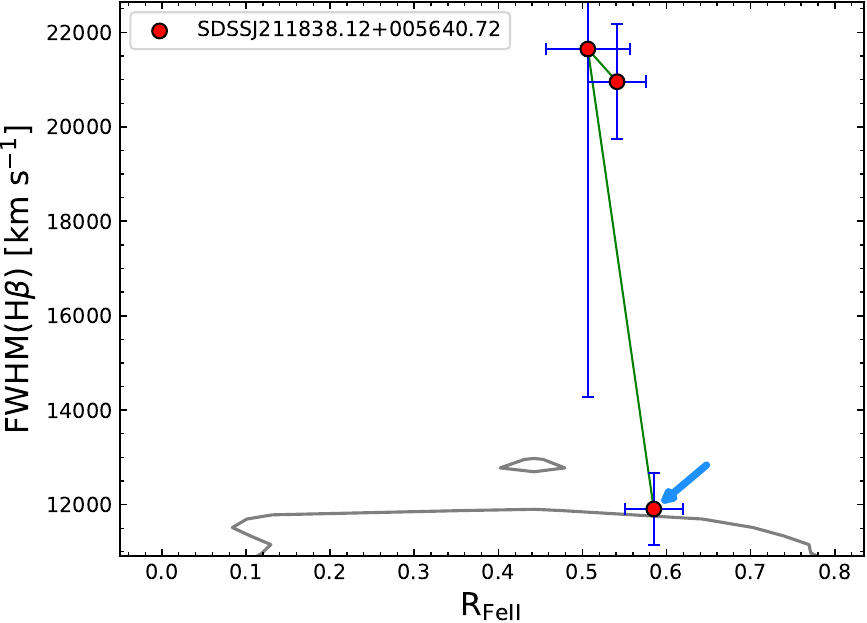}    
    \includegraphics[width=0.21\textwidth]{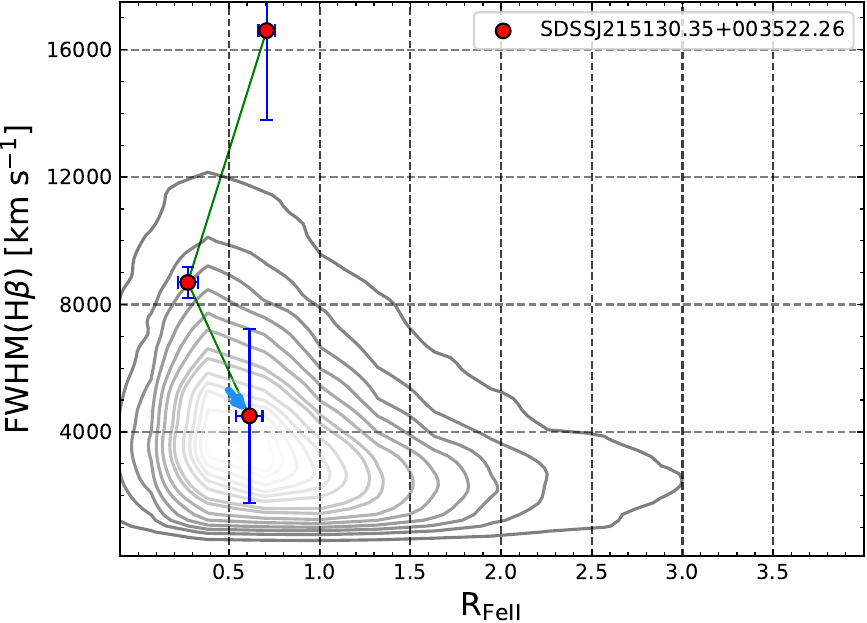}
    \includegraphics[width=0.21\textwidth]{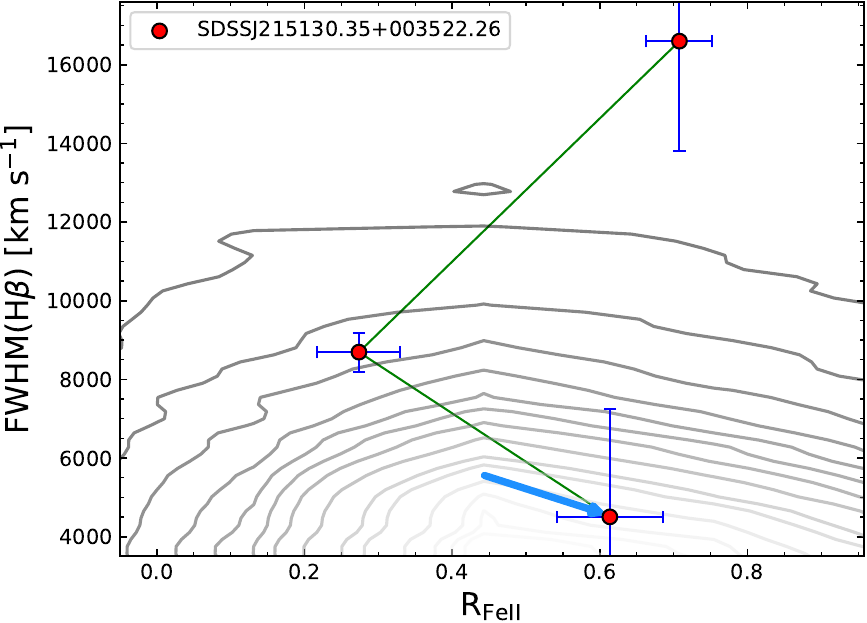}\\
    
    \includegraphics[width=0.21\textwidth]{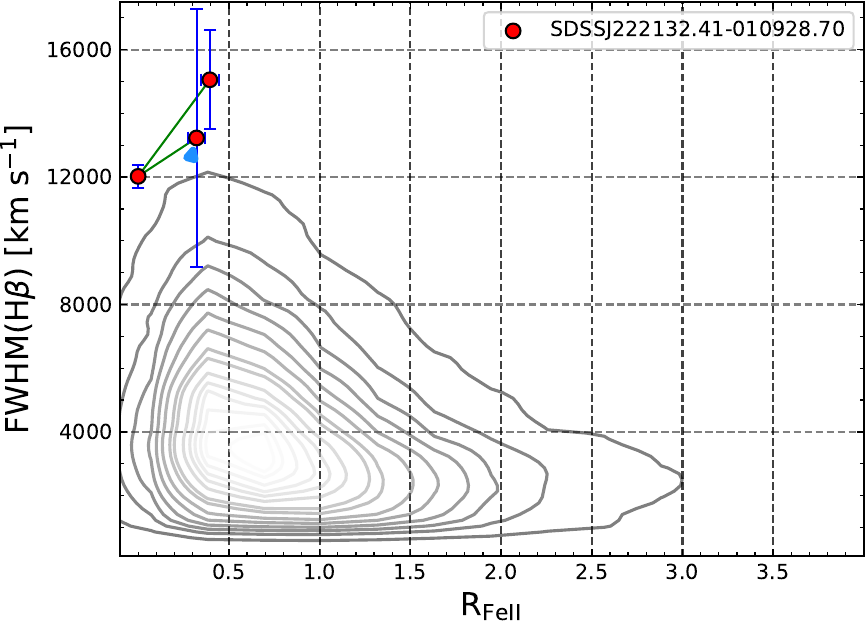}
    \includegraphics[width=0.21\textwidth]{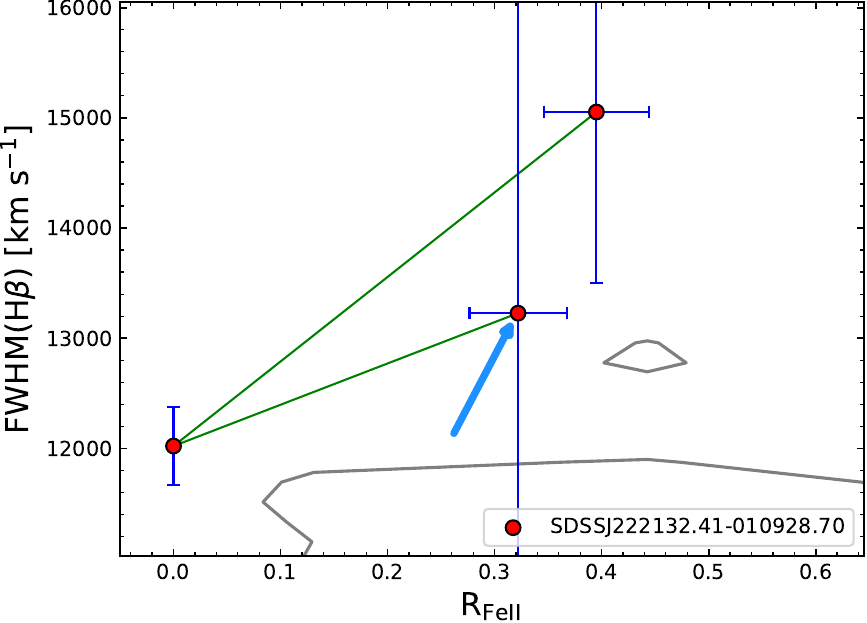}    
    \includegraphics[width=0.21\textwidth]{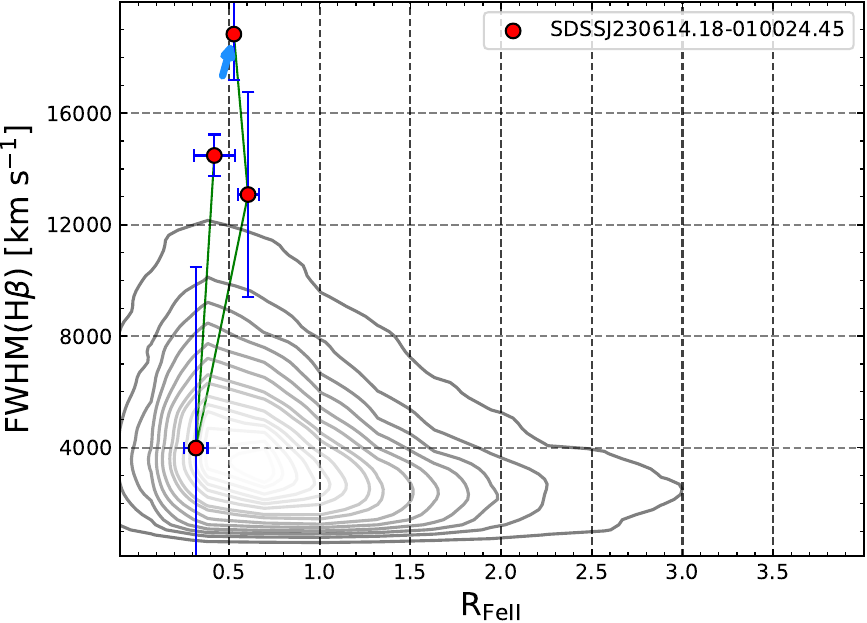}
    \includegraphics[width=0.21\textwidth]{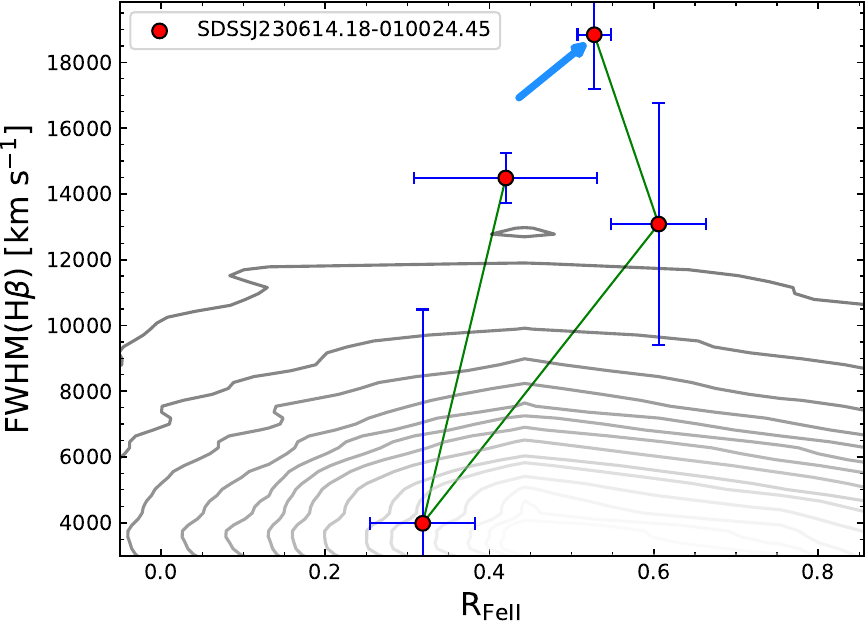}\\    

    \includegraphics[width=0.21\textwidth]{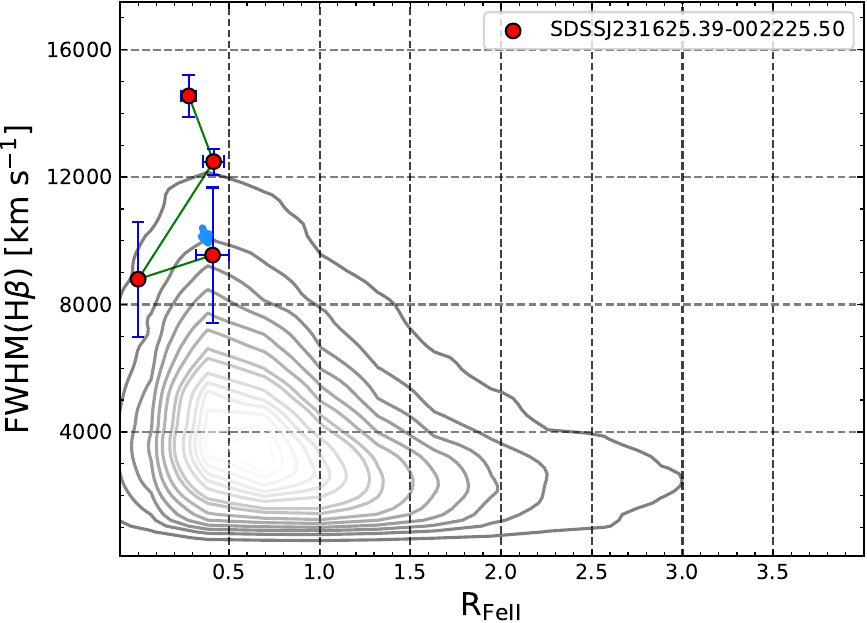}
    \includegraphics[width=0.21\textwidth]{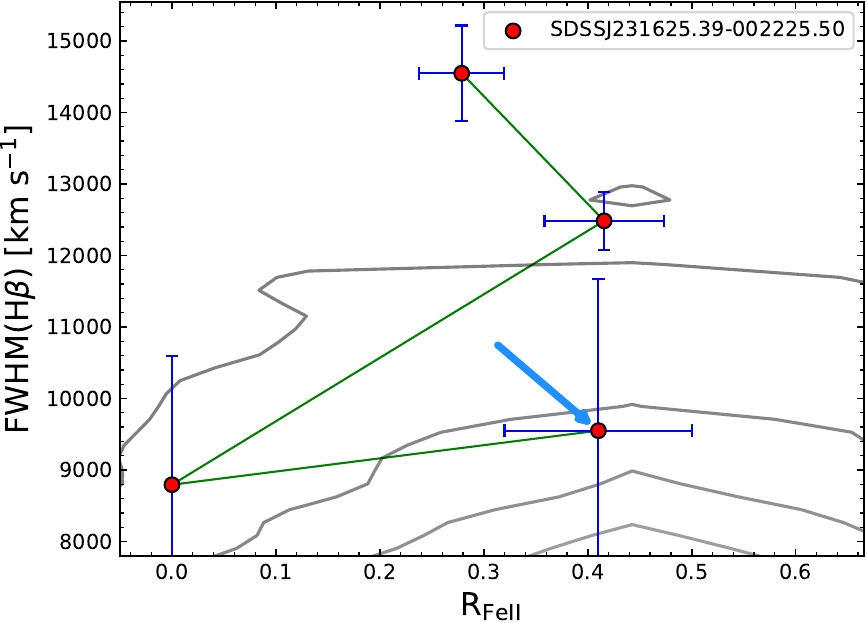}
    \includegraphics[width=0.21\textwidth]{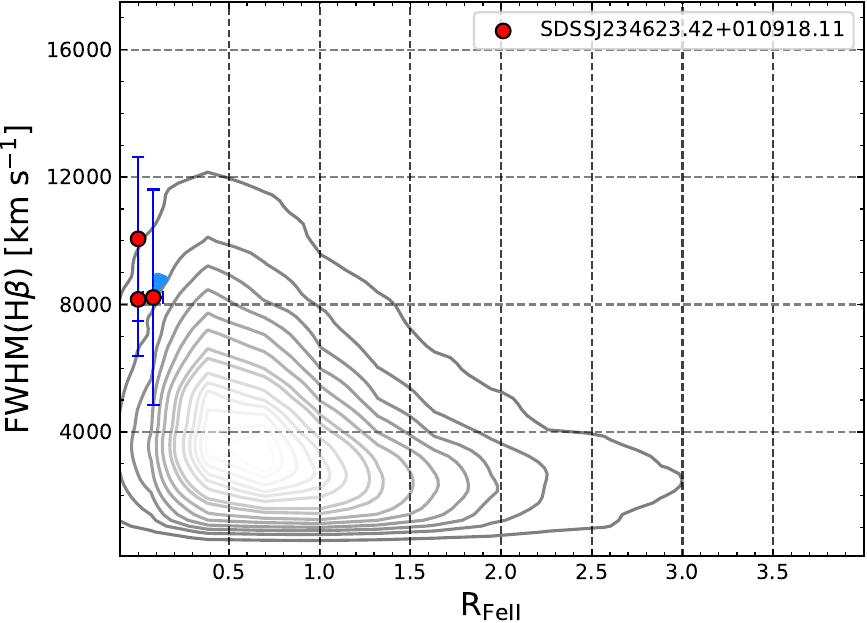}
    \includegraphics[width=0.21\textwidth]{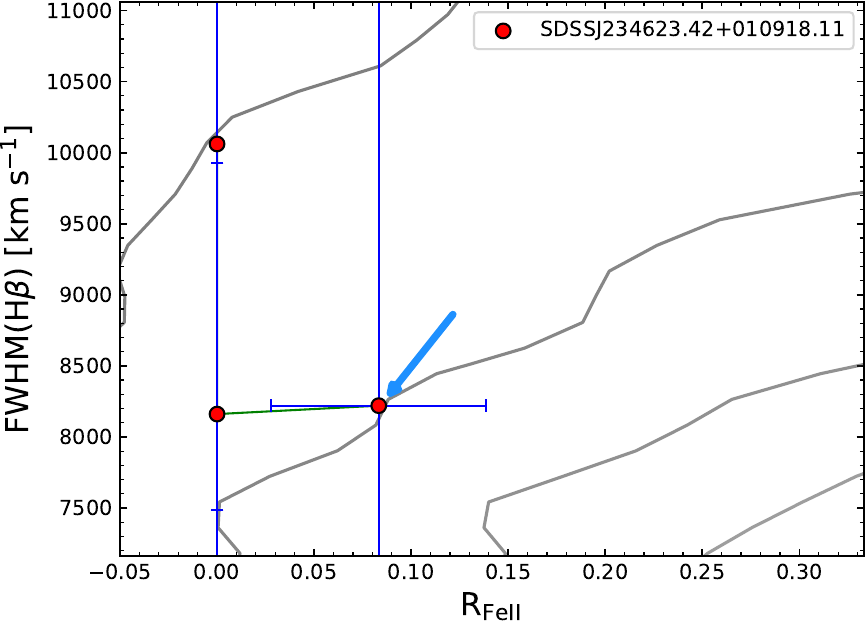}\\

    \caption{Panels are similar to Figures \ref{fig:EV1-others} for the remaining sources.}
    \label{fig:EV1-others2}
\end{figure*}


\begin{figure*}[!htb]
    \centering
    \includegraphics[width=0.21\textwidth]{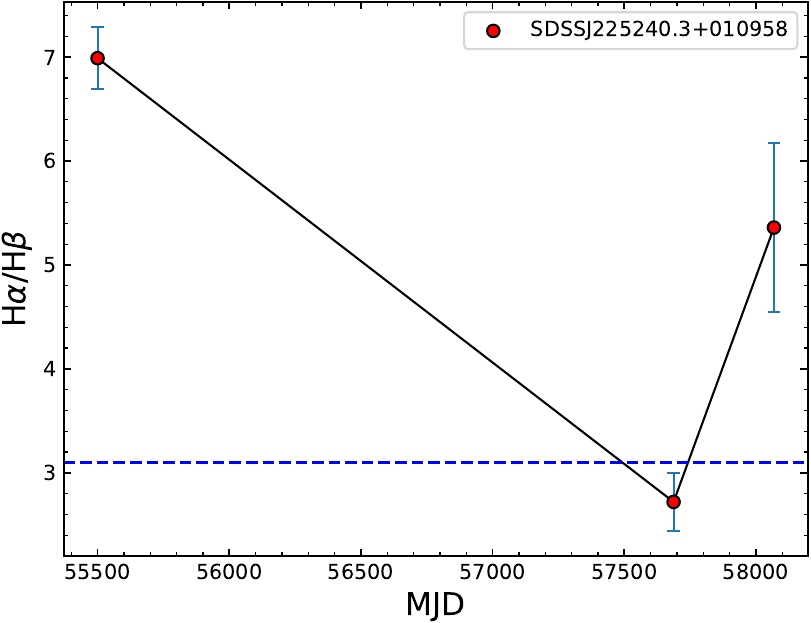}
    \includegraphics[width=0.21\textwidth]{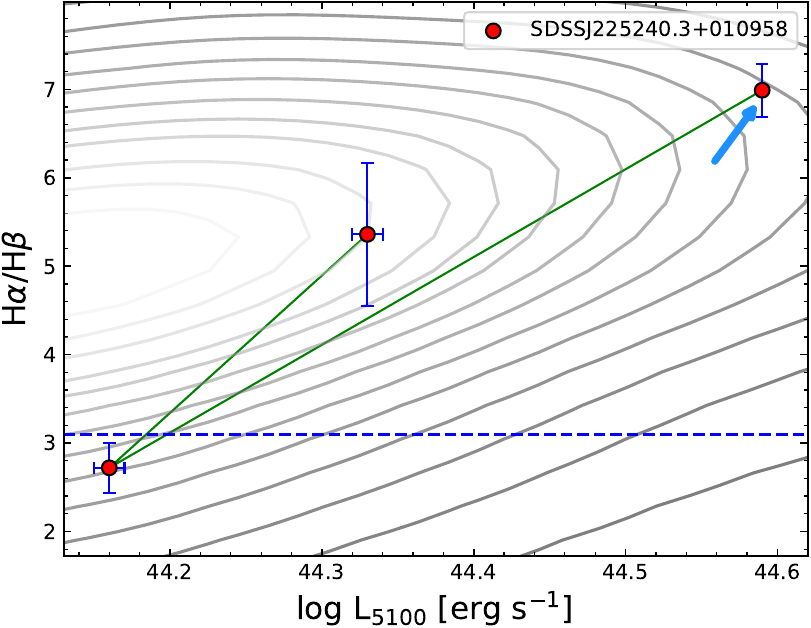}
    \includegraphics[width=0.21\textwidth]{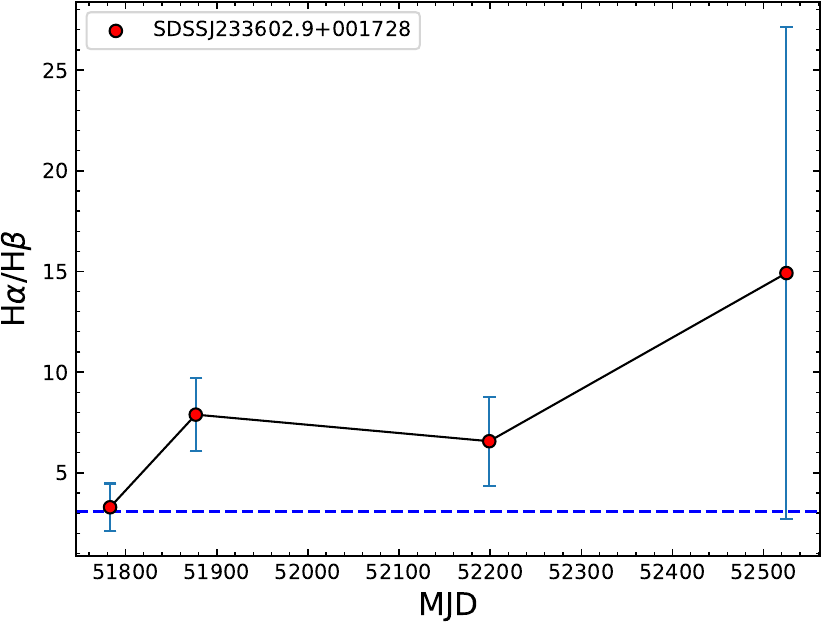}
    \includegraphics[width=0.21\textwidth]{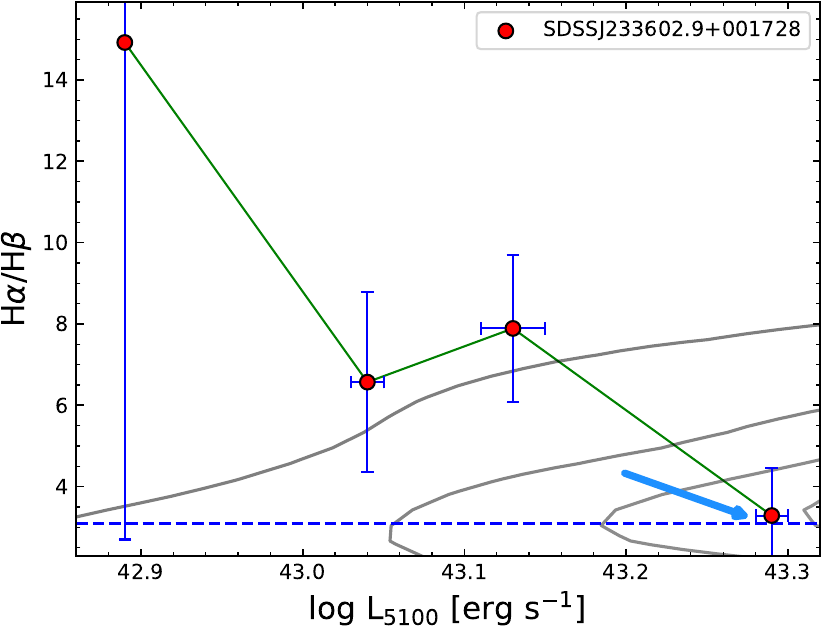}

    \includegraphics[width=0.21\textwidth]{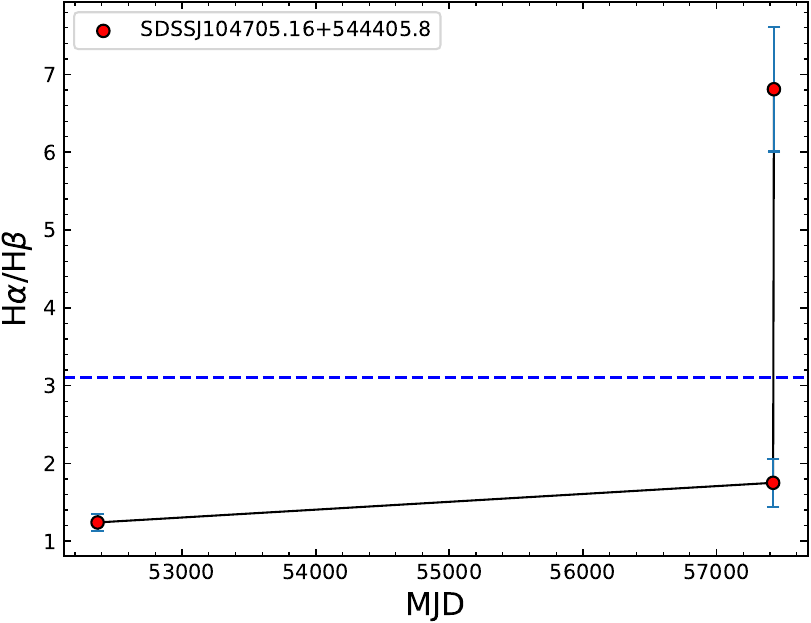}
    \includegraphics[width=0.21\textwidth]{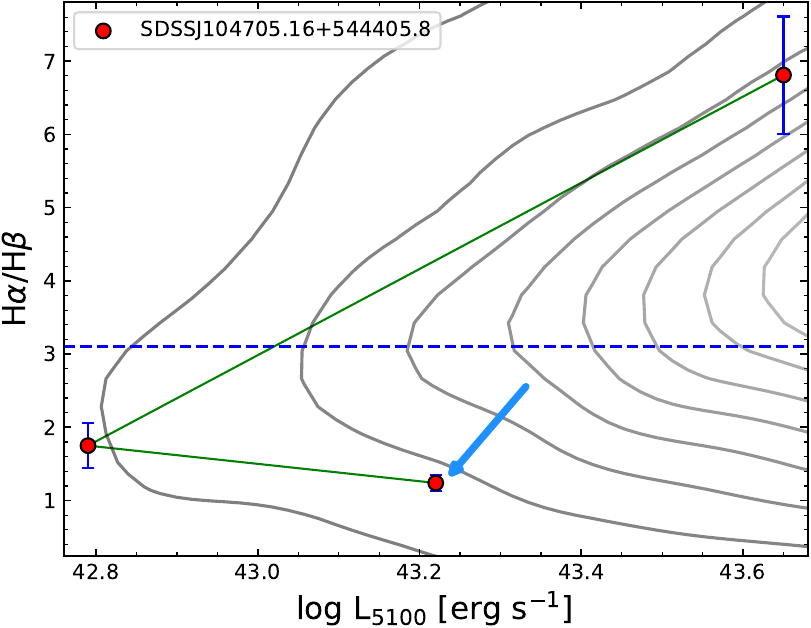}
    \includegraphics[width=0.21\textwidth]{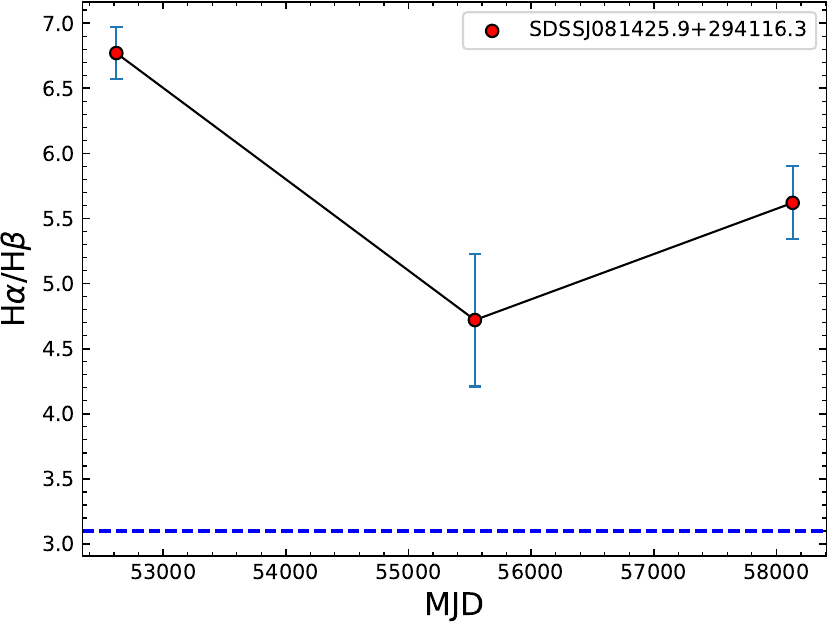}
    \includegraphics[width=0.21\textwidth]{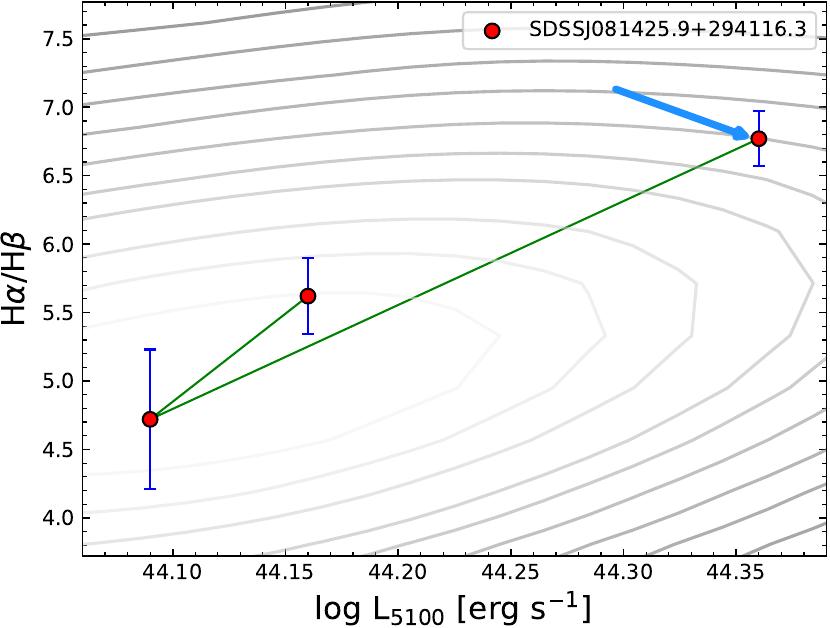}

    \includegraphics[width=0.21\textwidth]{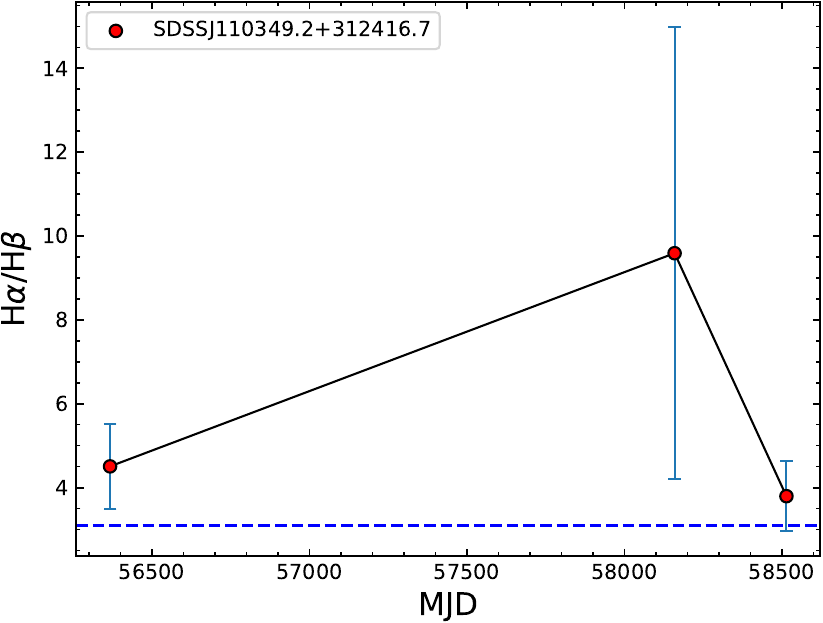}
    \includegraphics[width=0.21\textwidth]{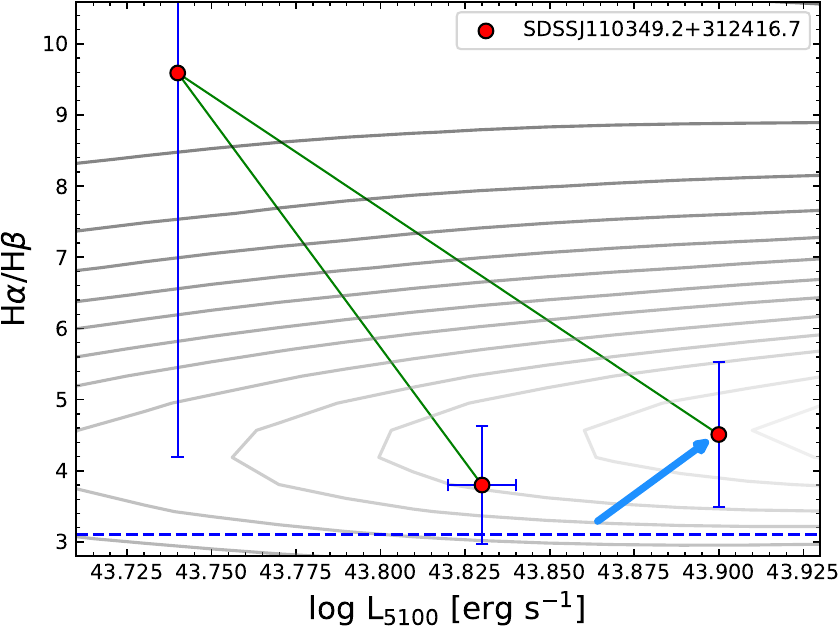}
    \includegraphics[width=0.21\textwidth]{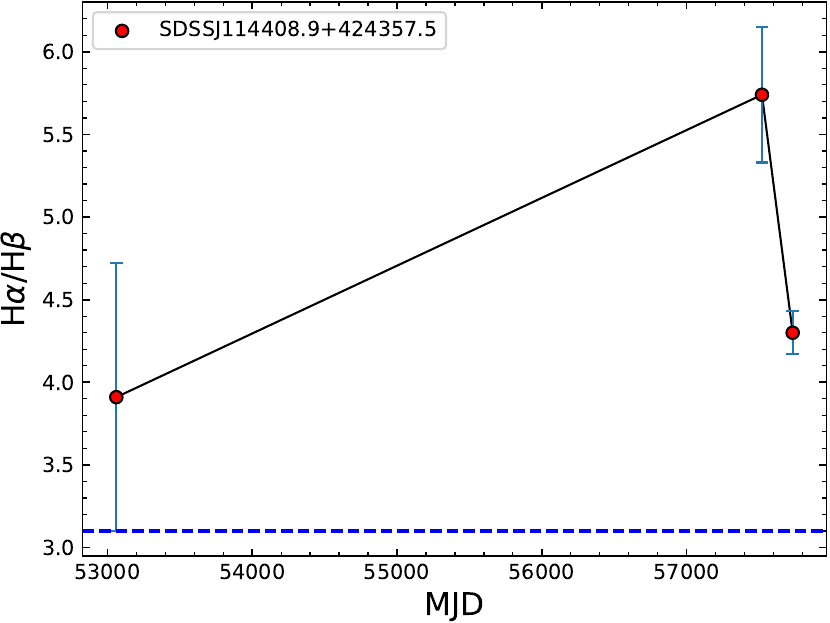}
    \includegraphics[width=0.21\textwidth]{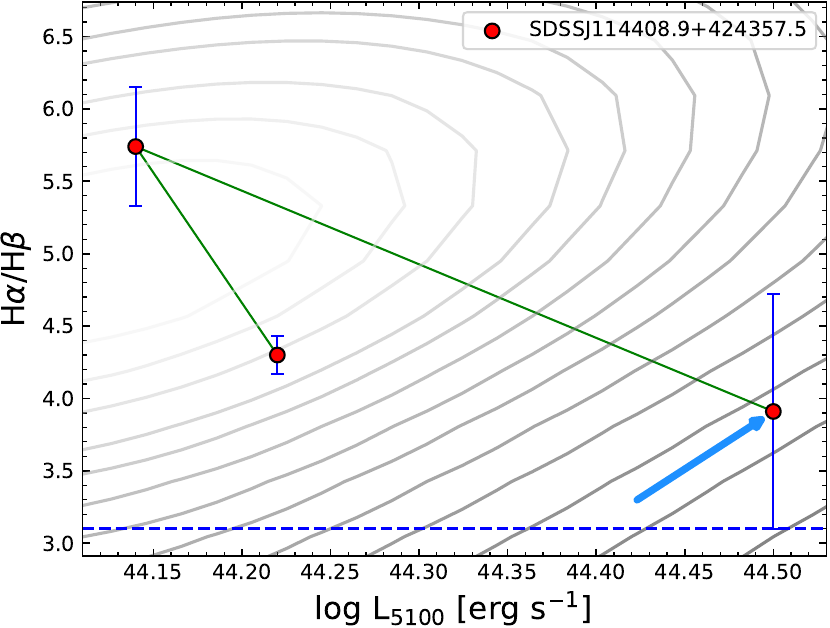}

    \includegraphics[width=0.21\textwidth]{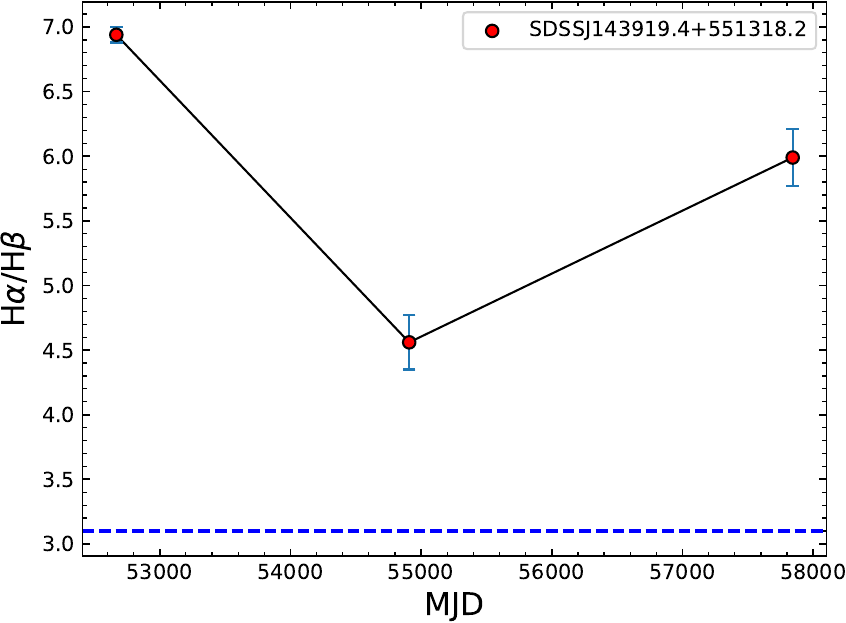}
    \includegraphics[width=0.21\textwidth]{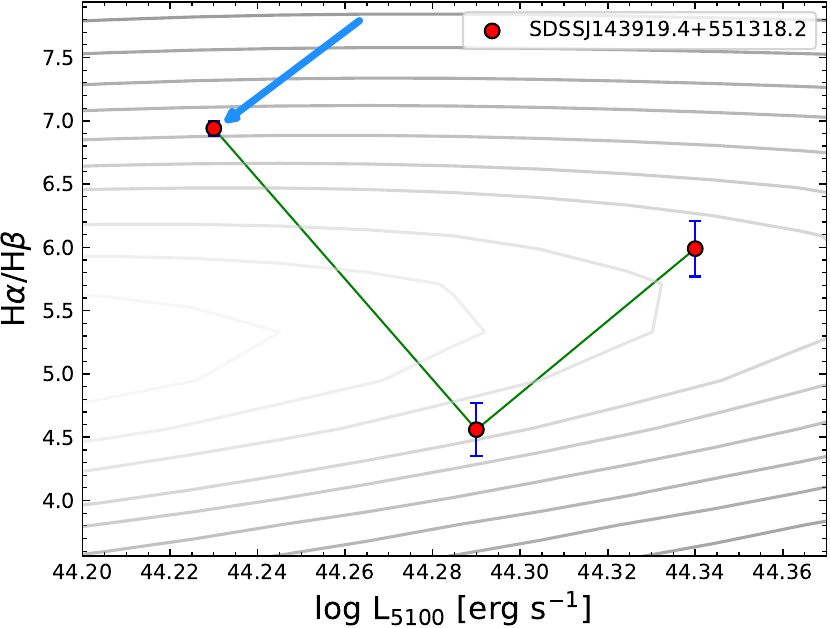}
    \includegraphics[width=0.21\textwidth]{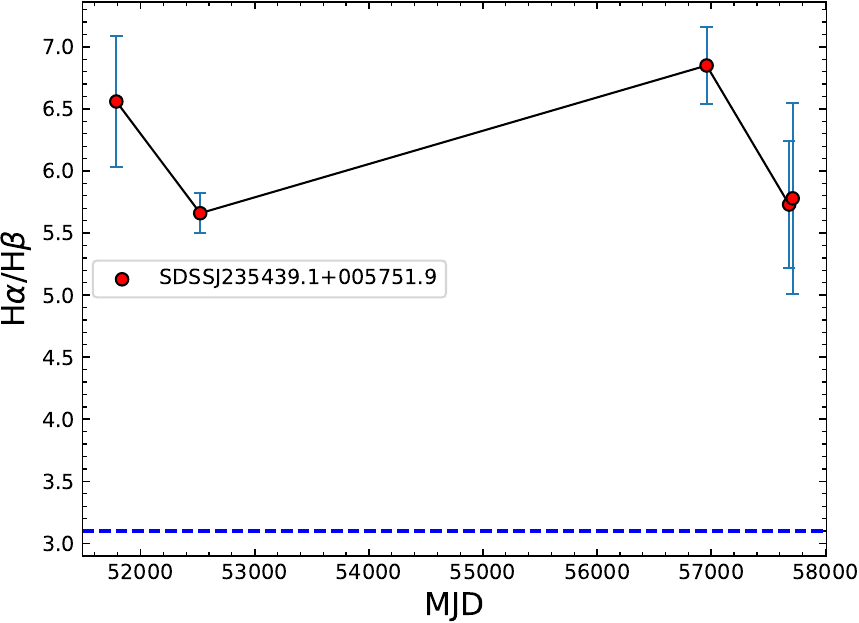}
    \includegraphics[width=0.21\textwidth]{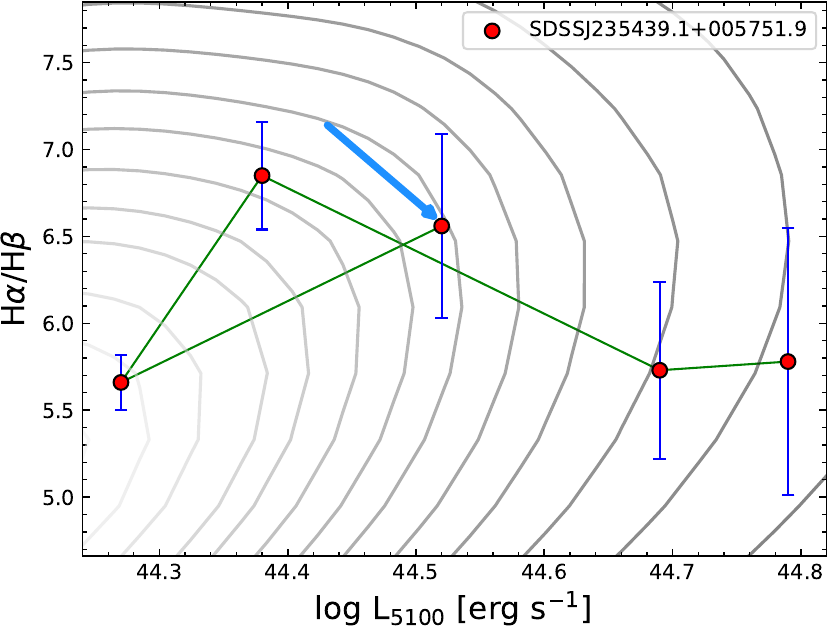}

    \includegraphics[width=0.21\textwidth]{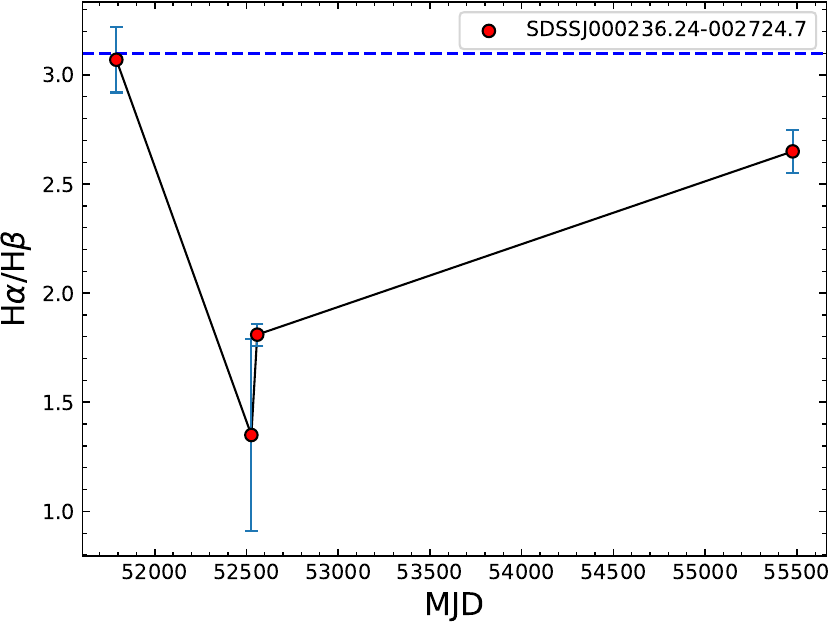}
    \includegraphics[width=0.21\textwidth]{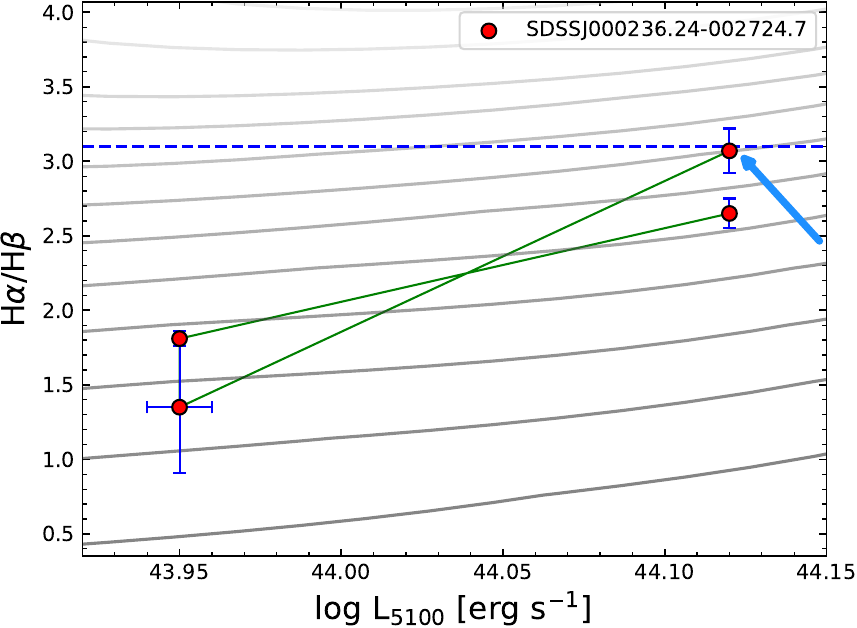}
    \includegraphics[width=0.21\textwidth]{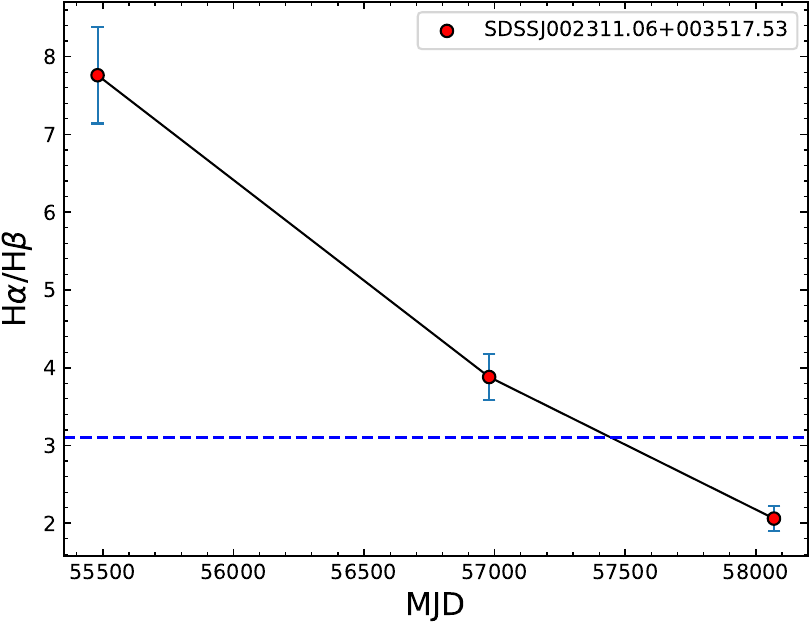}
    \includegraphics[width=0.21\textwidth]{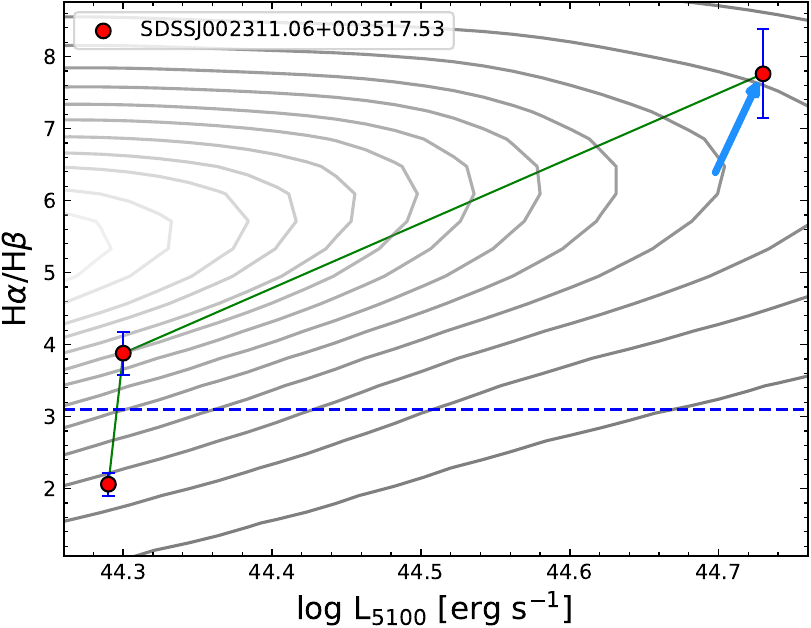}

    \includegraphics[width=0.21\textwidth]{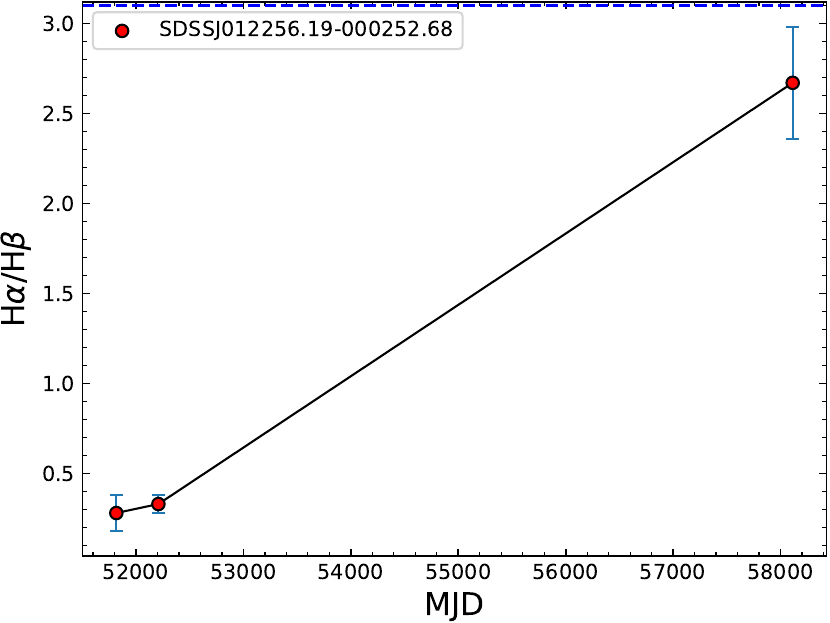}
    \includegraphics[width=0.21\textwidth]{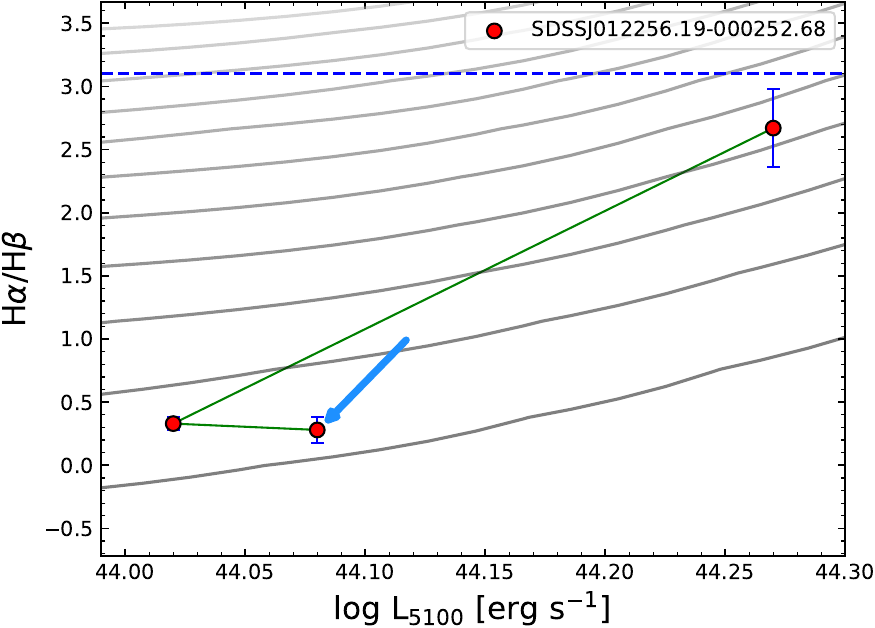}
    \includegraphics[width=0.21\textwidth]{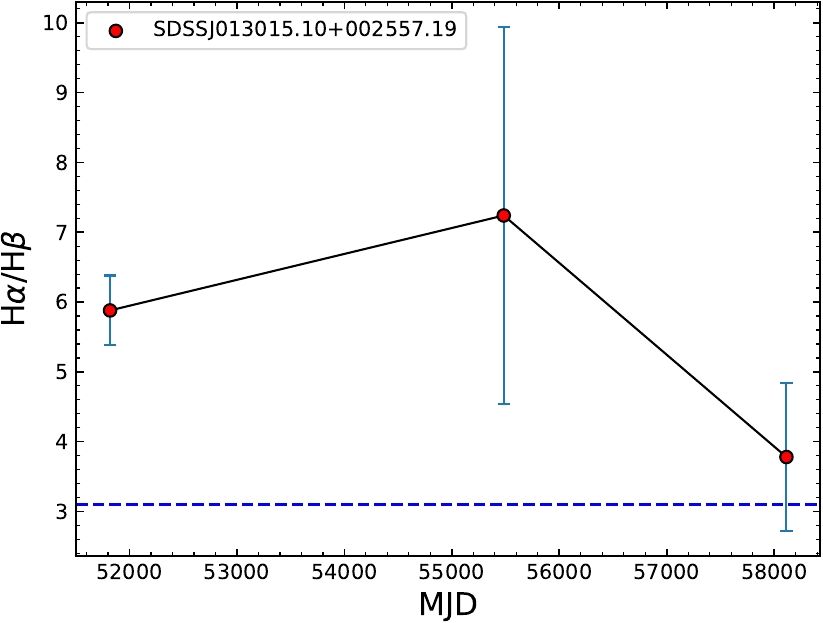}
    \includegraphics[width=0.21\textwidth]{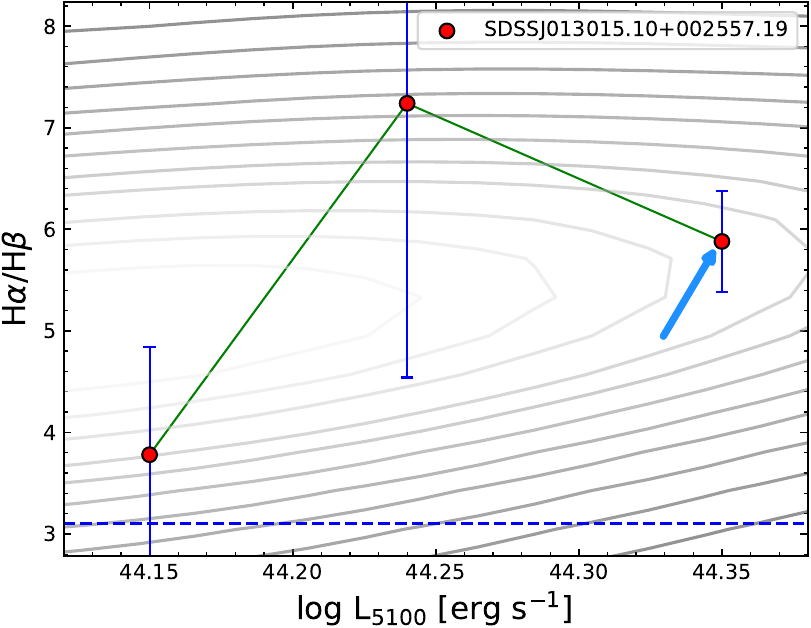}

    \includegraphics[width=0.21\textwidth]{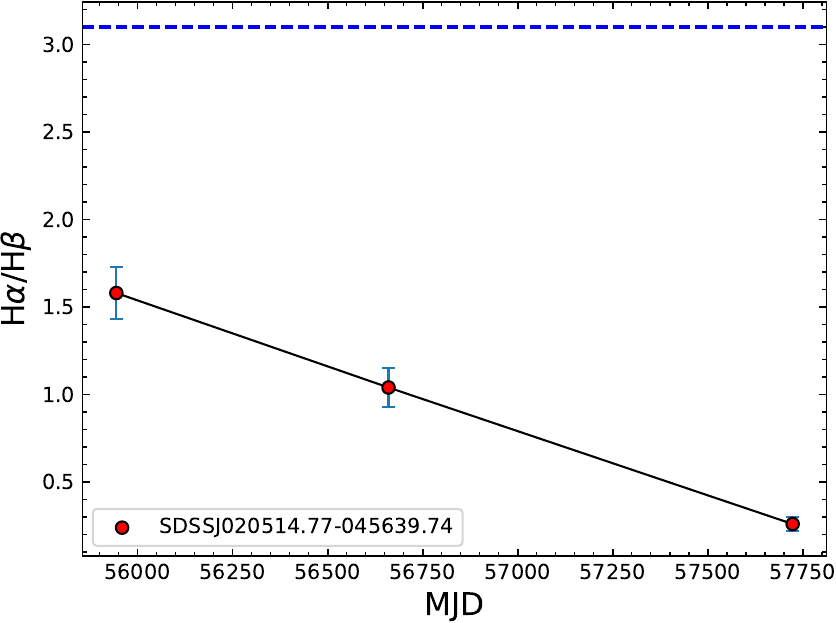}
    \includegraphics[width=0.21\textwidth]{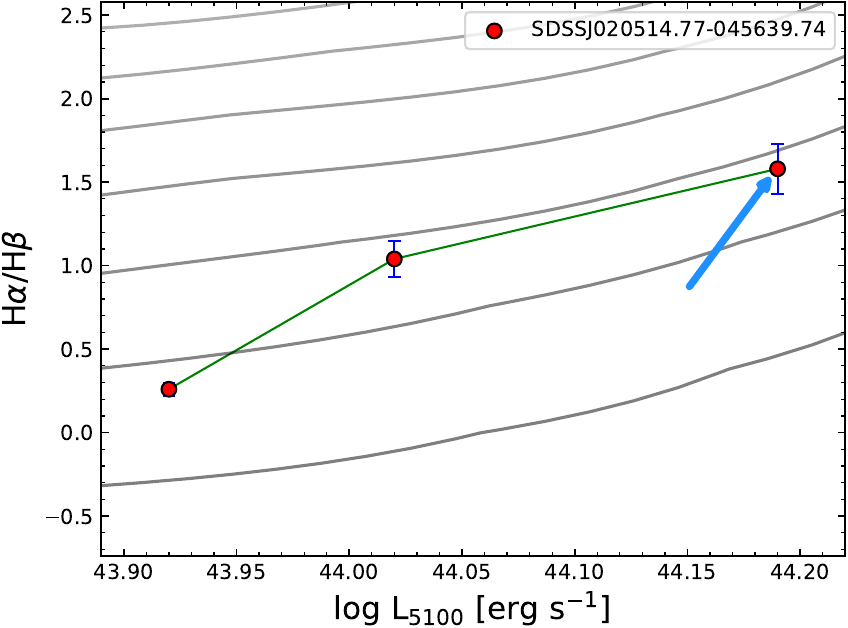}
    \includegraphics[width=0.21\textwidth]{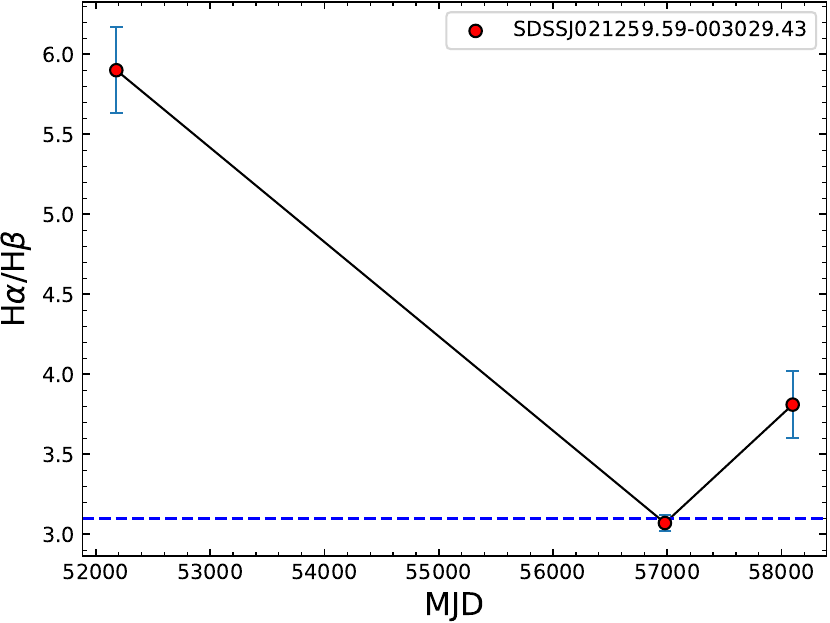}
    \includegraphics[width=0.21\textwidth]{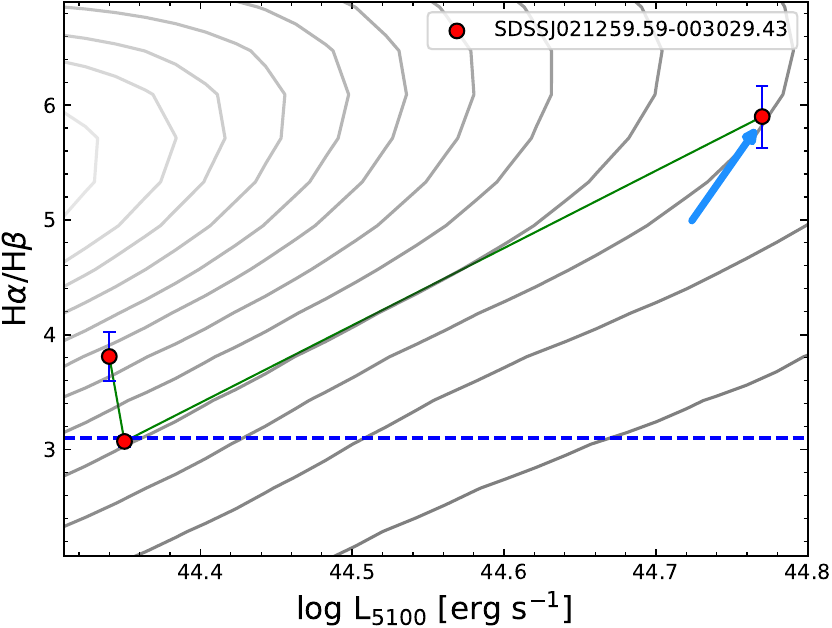}

    \caption{Panels are similar to Figures \ref{fig:balmer_decre} and \ref{fig:balmer_decre-lagn} for the remaining sources. Here, we show only the trends of \ha{}/\hb{} ratio with respect to observed time and \lagn{} for brevity.}
    \label{fig:balmer_decre_others}
\end{figure*}

\begin{figure*}[!htb]
    \centering
    \includegraphics[width=0.21\textwidth]{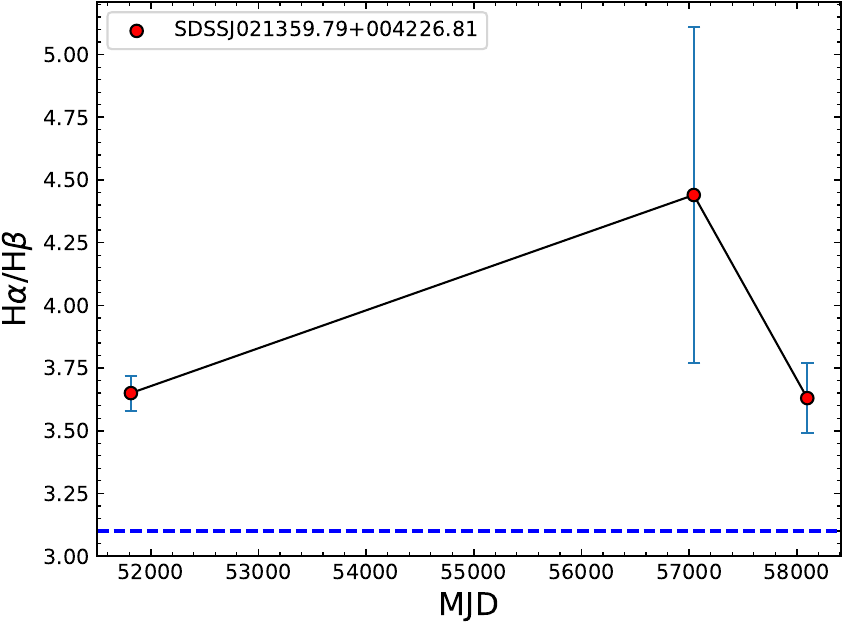}
    \includegraphics[width=0.21\textwidth]{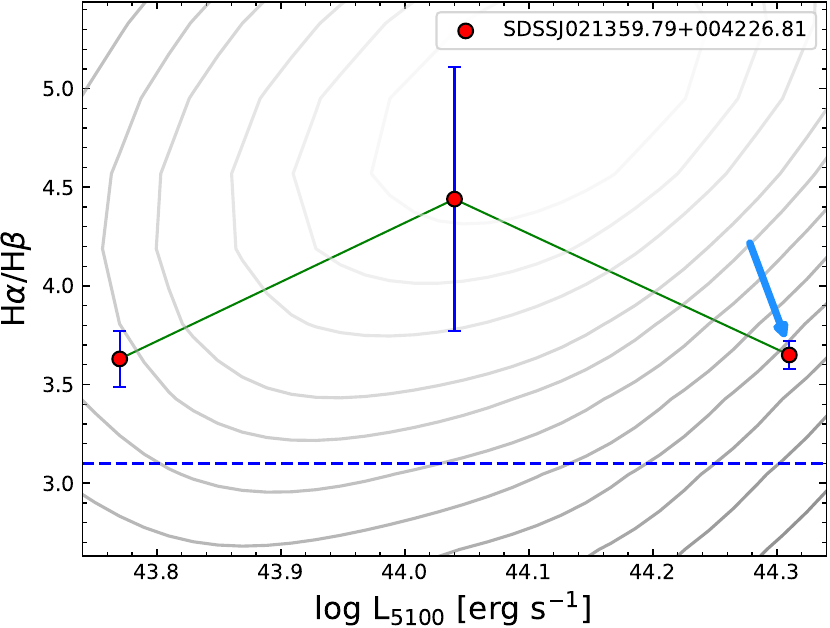}
    \includegraphics[width=0.21\textwidth]{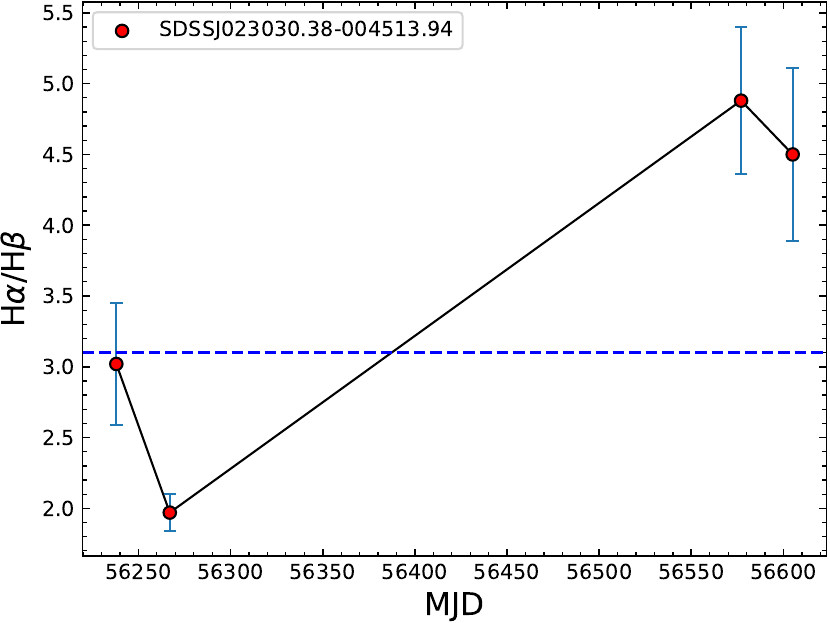}
    \includegraphics[width=0.21\textwidth]{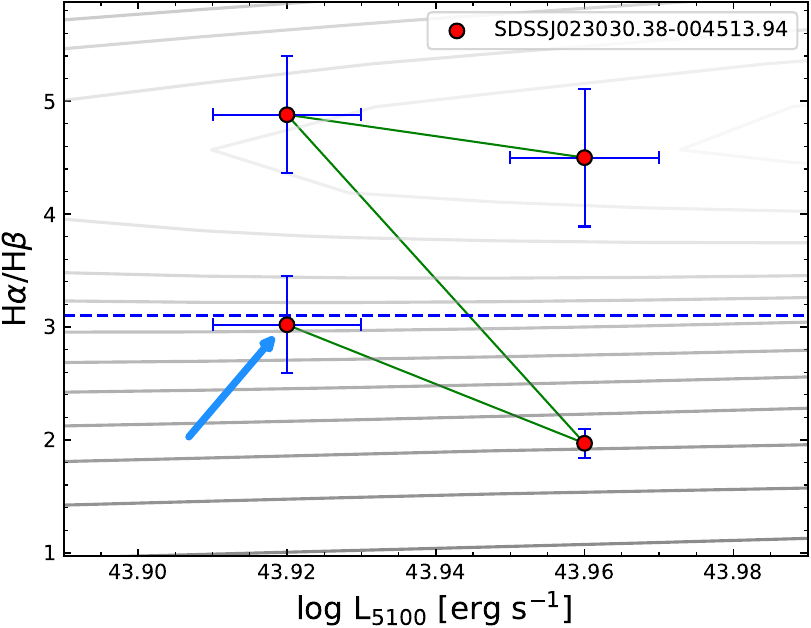}

    \includegraphics[width=0.21\textwidth]{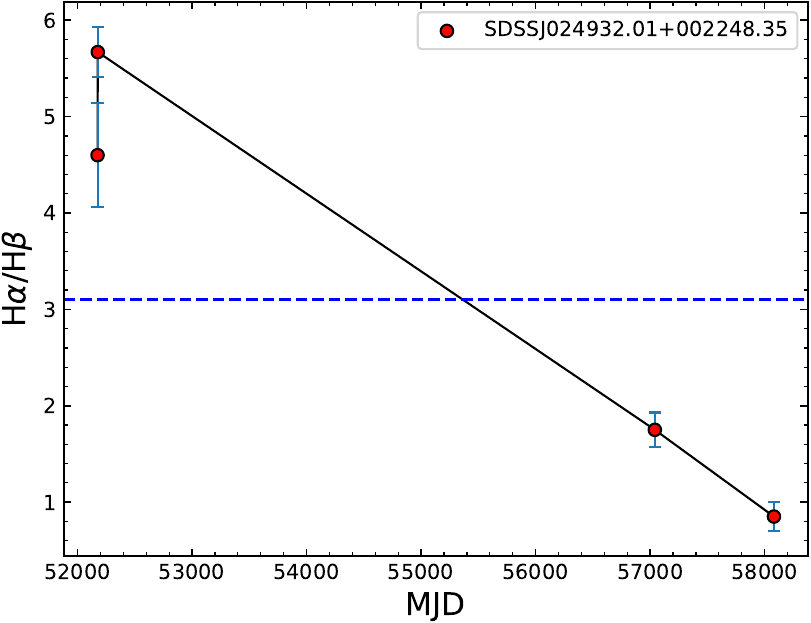}
    \includegraphics[width=0.21\textwidth]{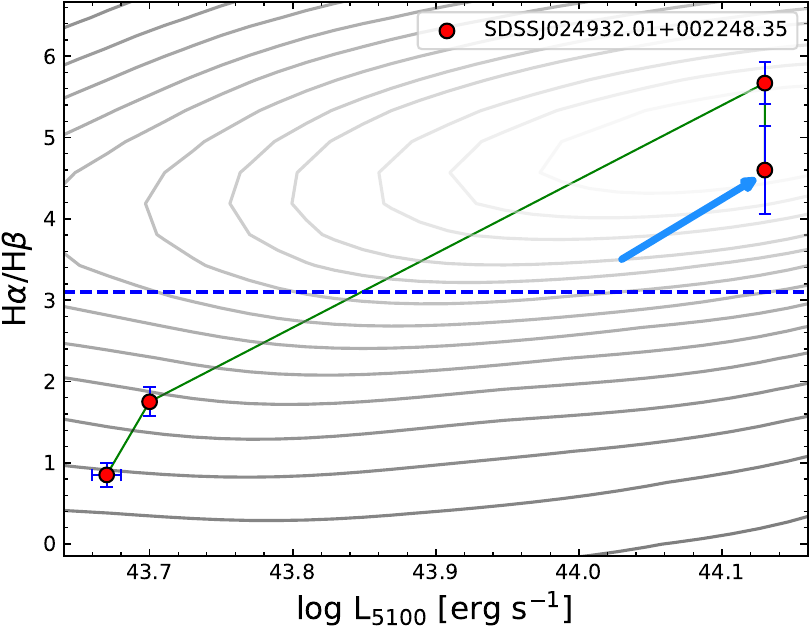}
    \includegraphics[width=0.21\textwidth]{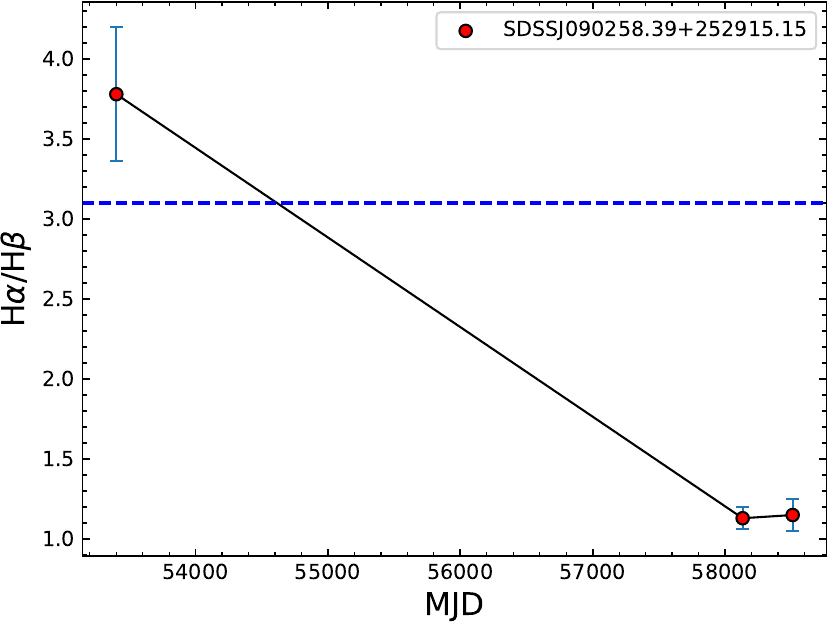}
    \includegraphics[width=0.21\textwidth]{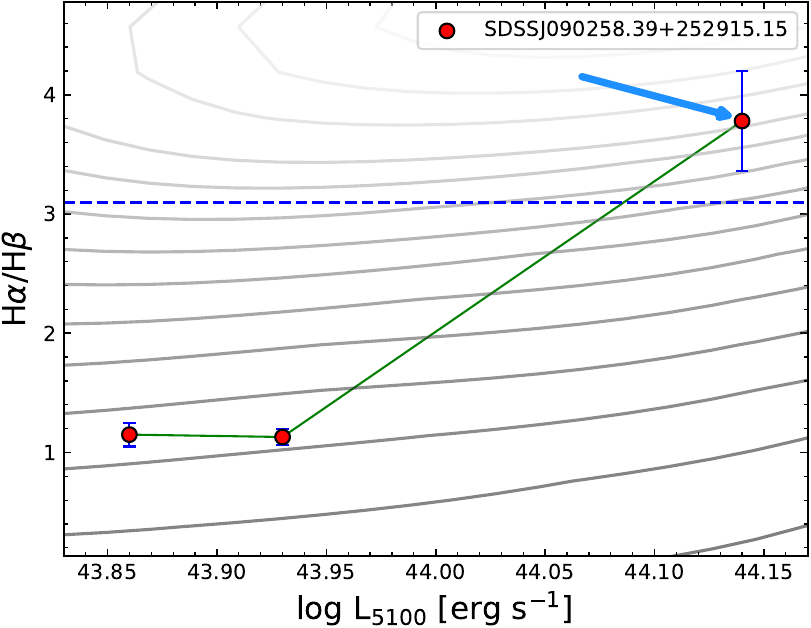}

    \includegraphics[width=0.21\textwidth]{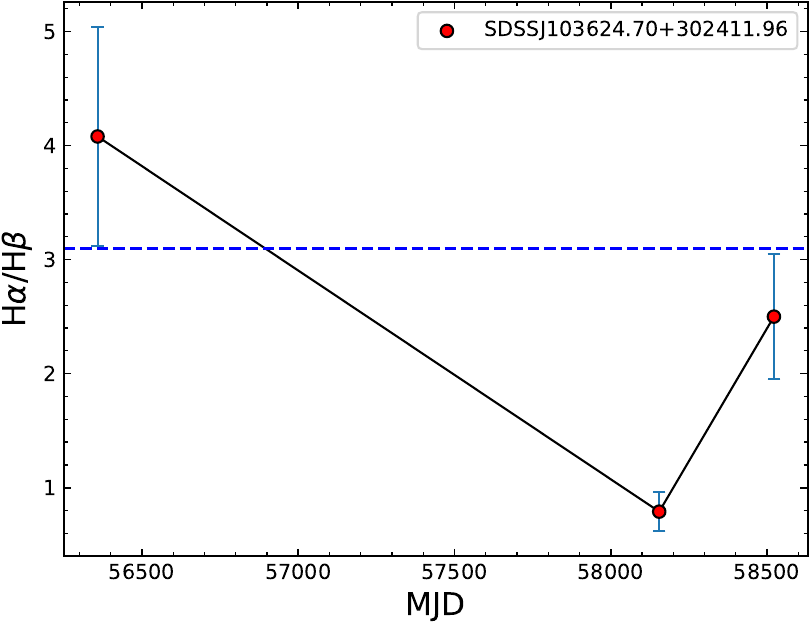}
    \includegraphics[width=0.21\textwidth]{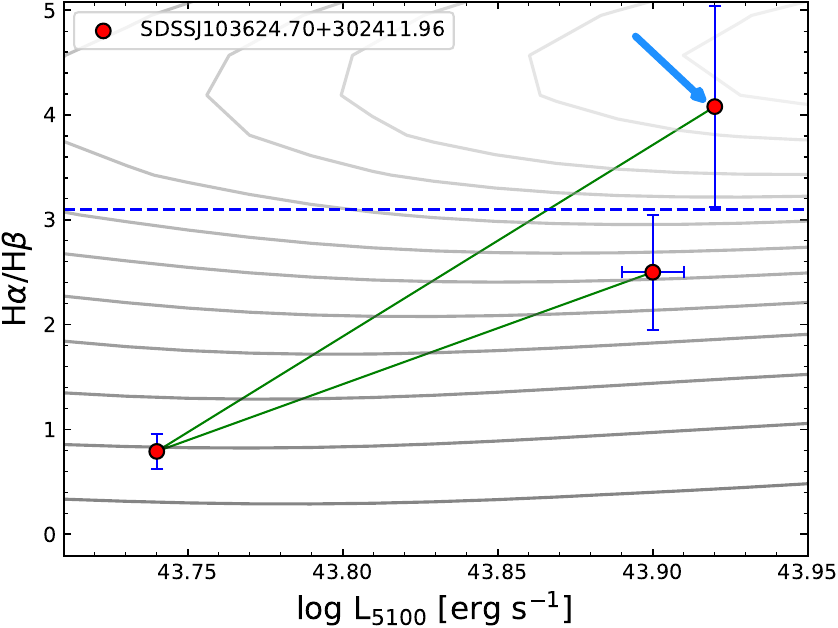}
    \includegraphics[width=0.21\textwidth]{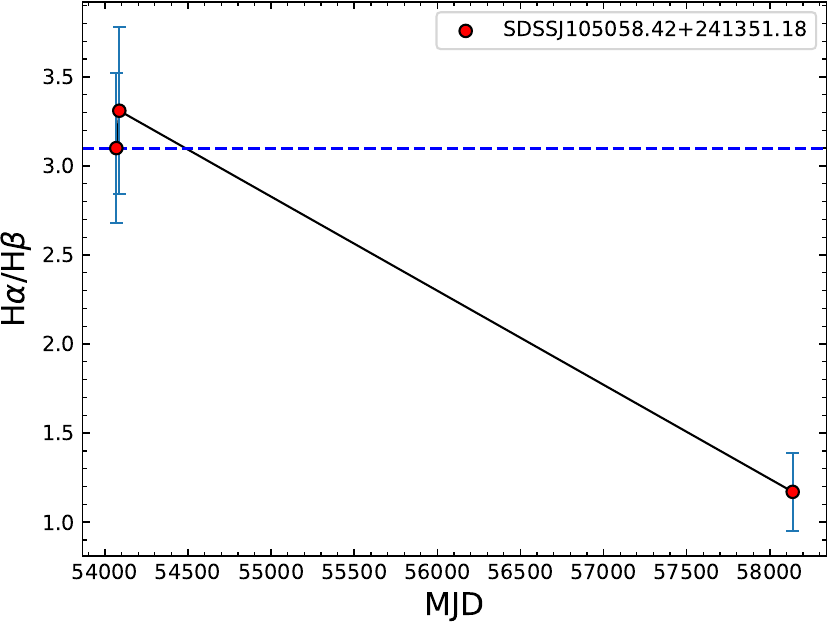}
    \includegraphics[width=0.21\textwidth]{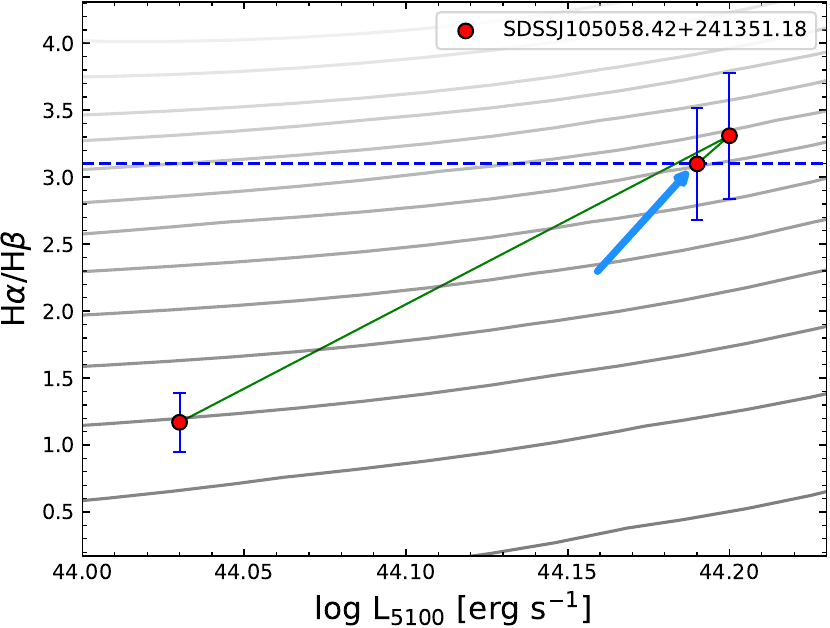}

    \includegraphics[width=0.21\textwidth]{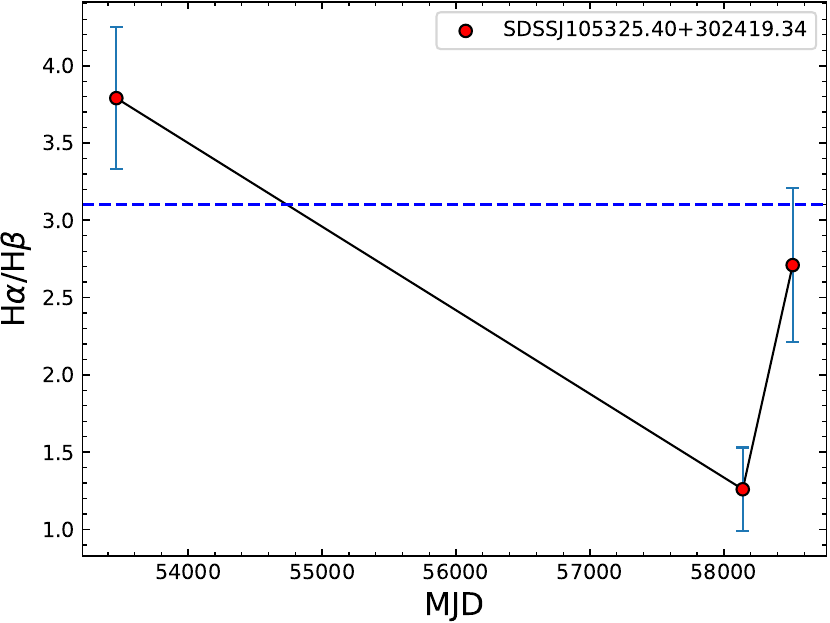}
    \includegraphics[width=0.21\textwidth]{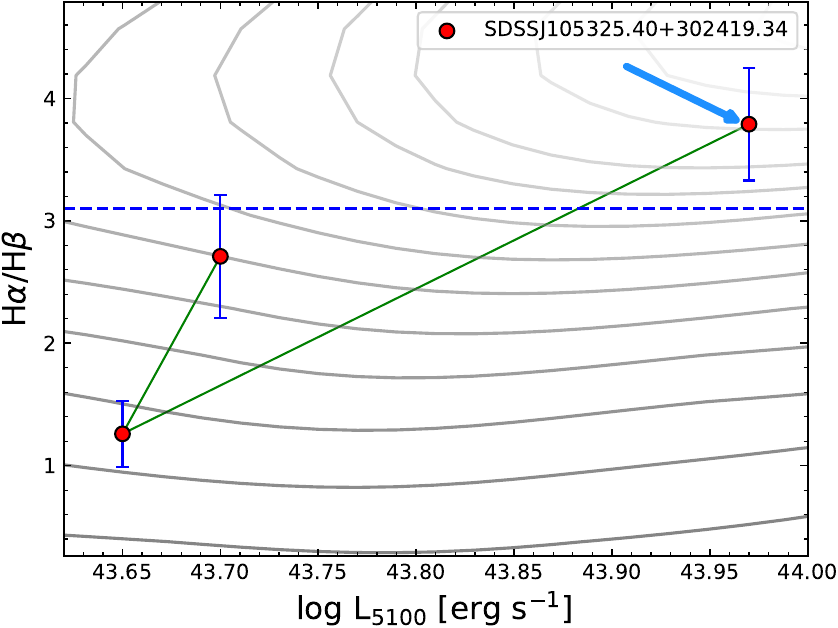}
    \includegraphics[width=0.21\textwidth]{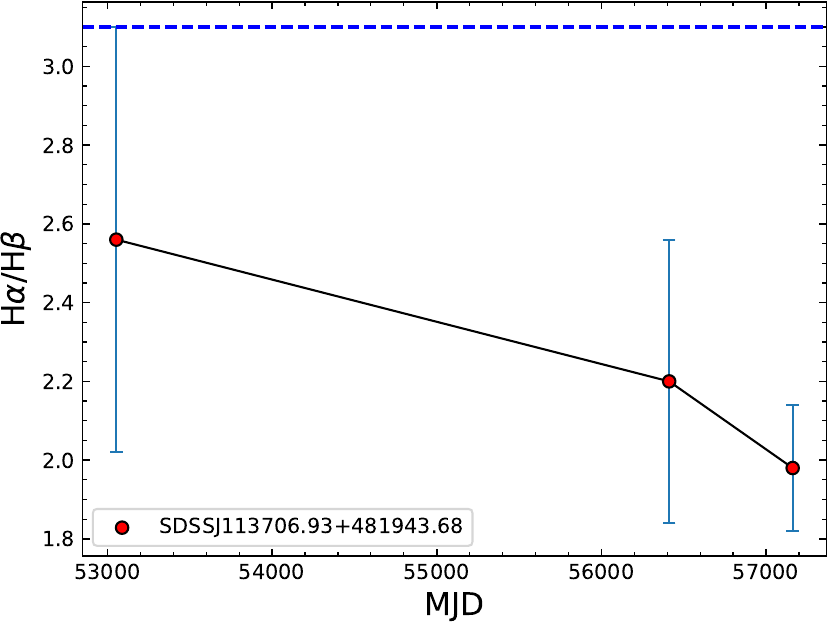}
    \includegraphics[width=0.21\textwidth]{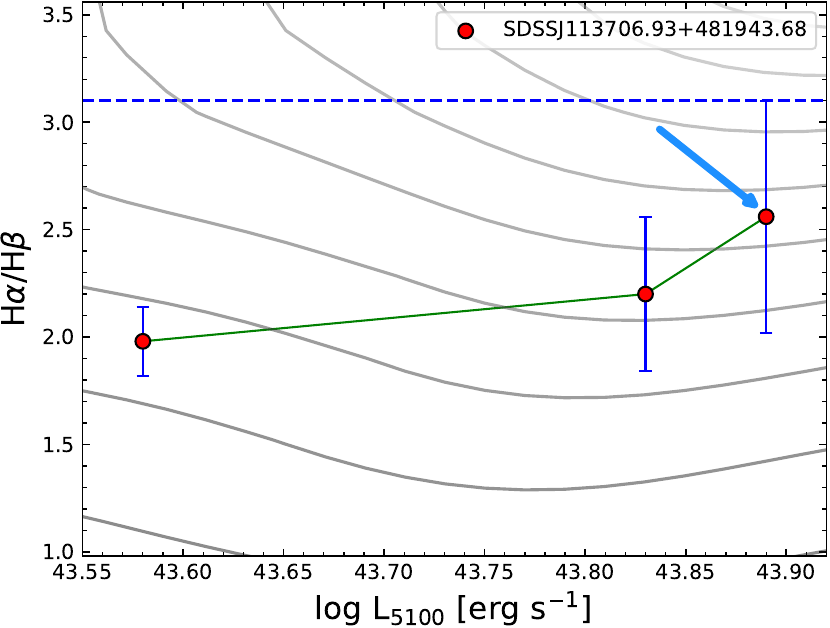}

    \includegraphics[width=0.21\textwidth]{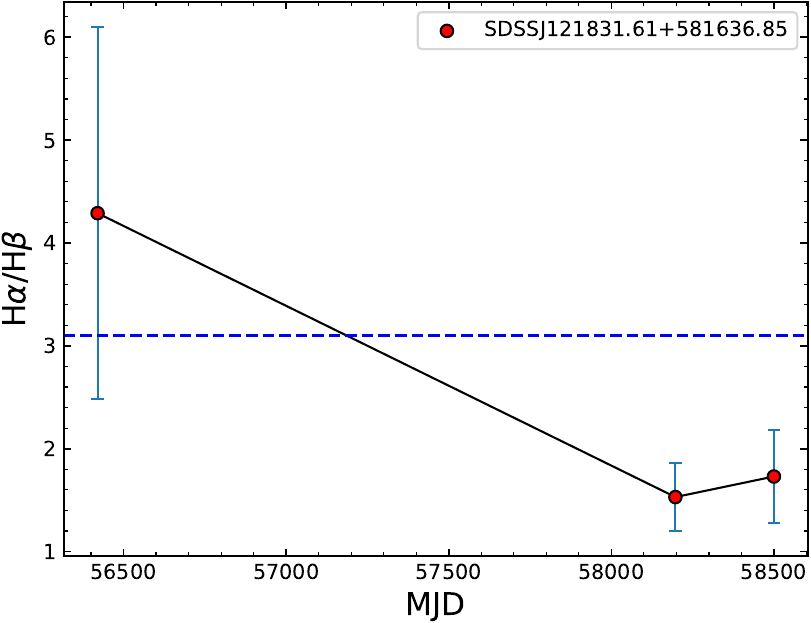}
    \includegraphics[width=0.21\textwidth]{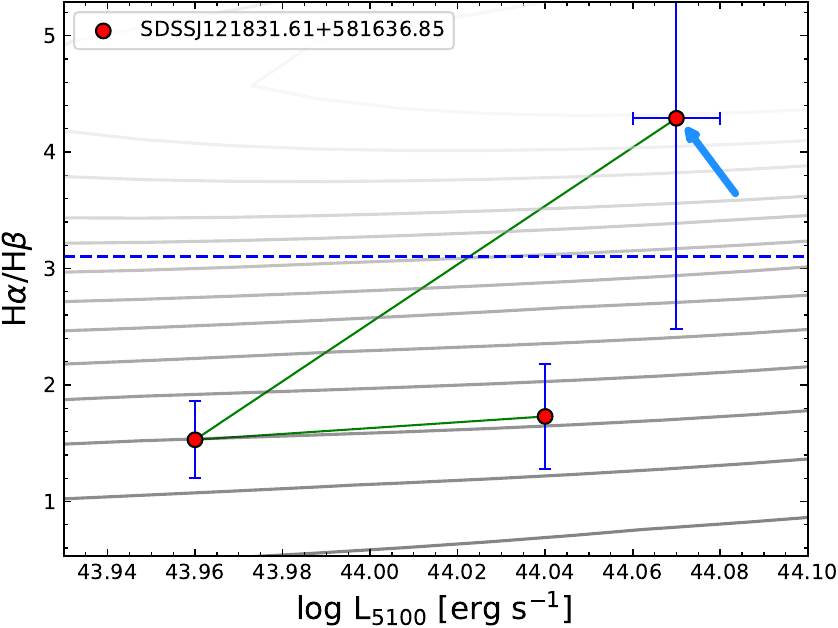}
    \includegraphics[width=0.21\textwidth]{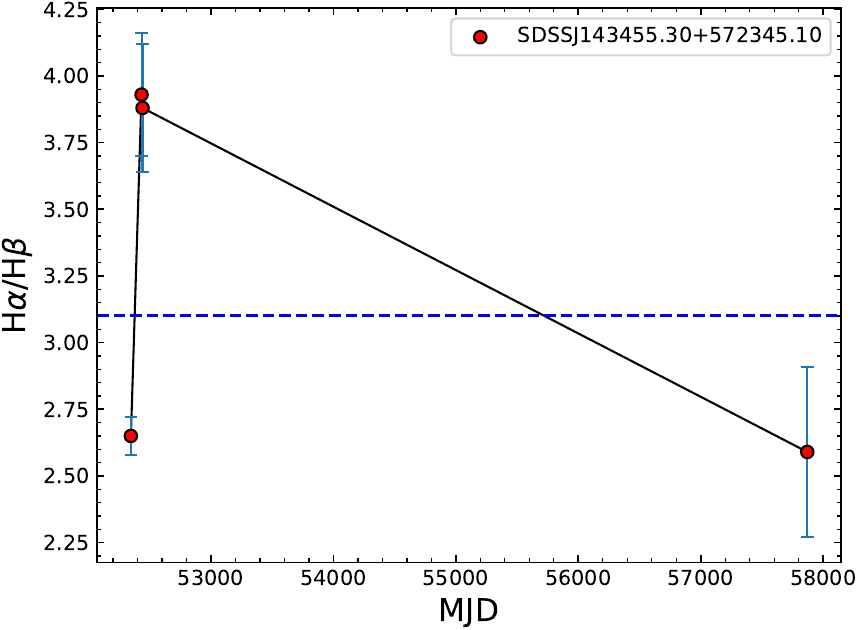}
    \includegraphics[width=0.21\textwidth]{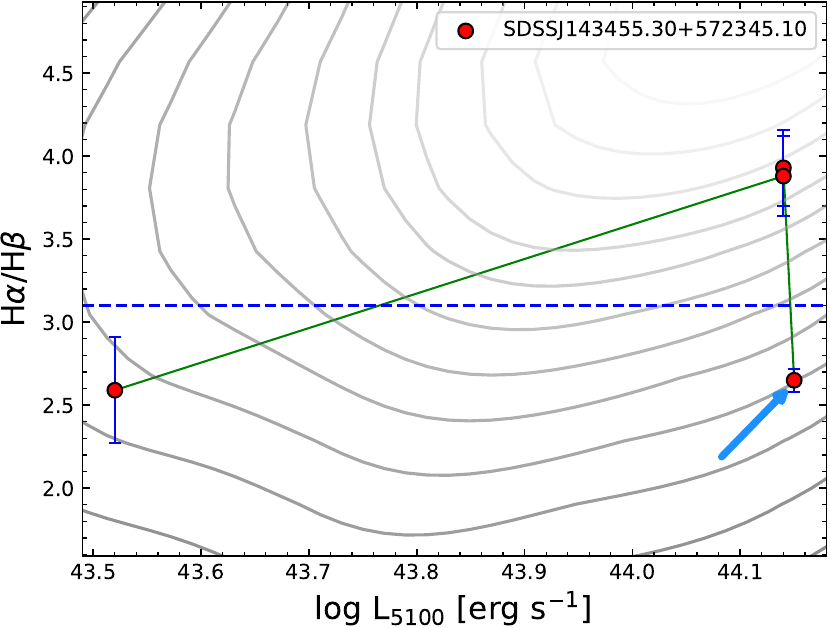}

    \includegraphics[width=0.21\textwidth]{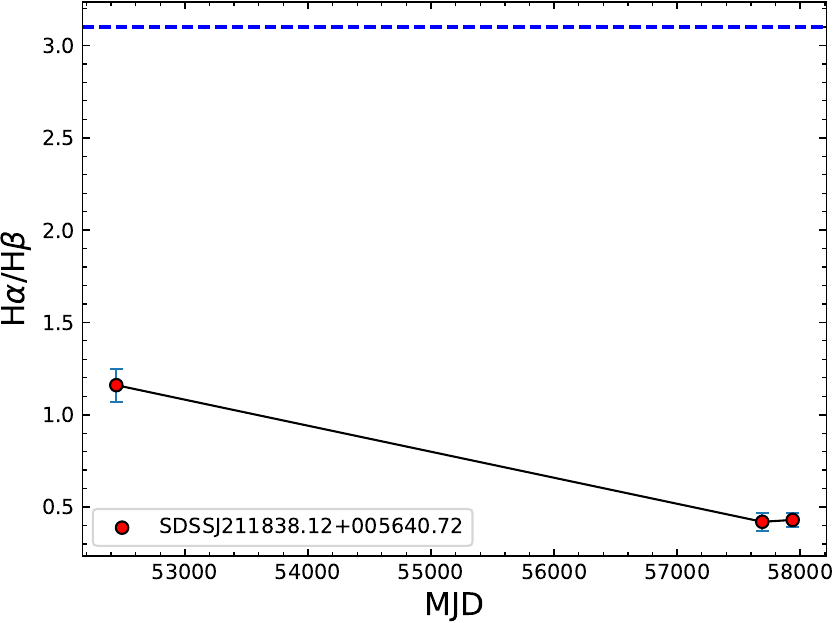}
    \includegraphics[width=0.21\textwidth]{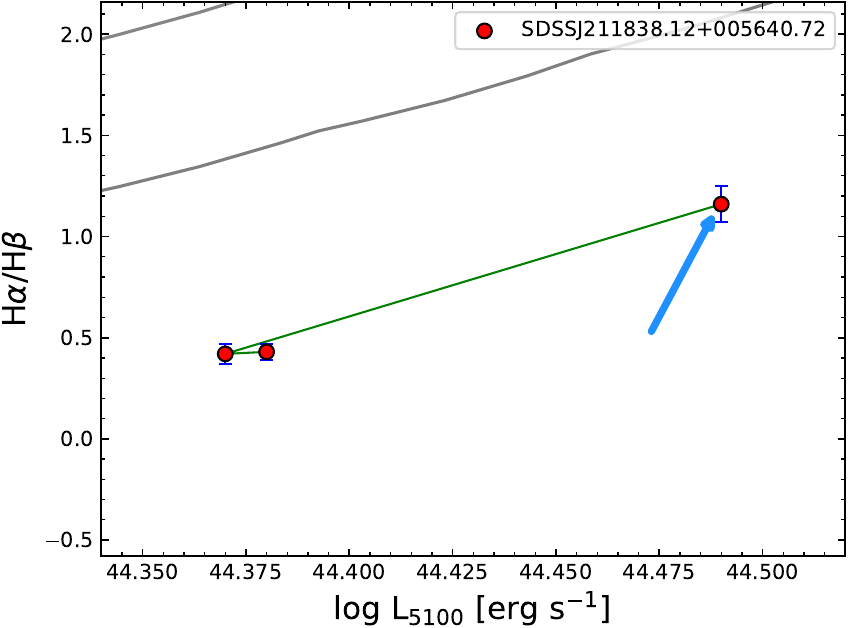}
    \includegraphics[width=0.21\textwidth]{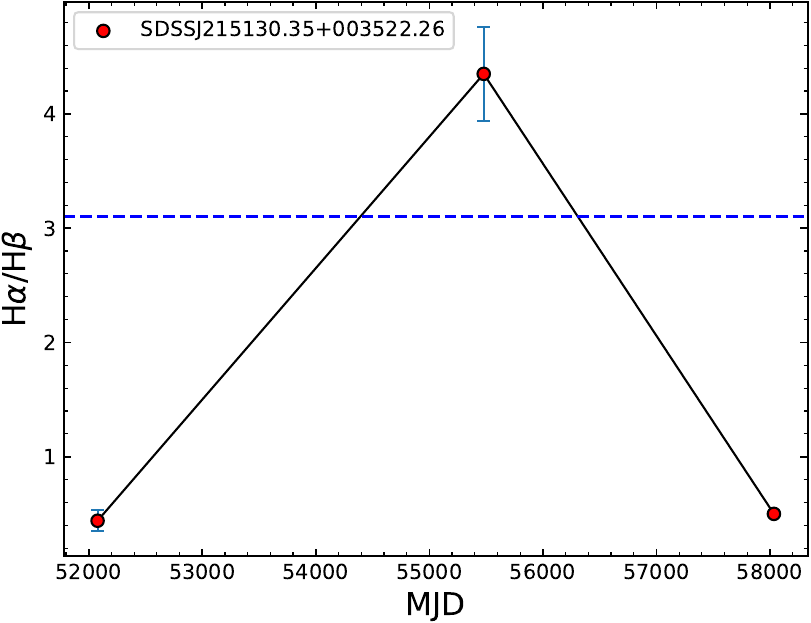}
    \includegraphics[width=0.21\textwidth]{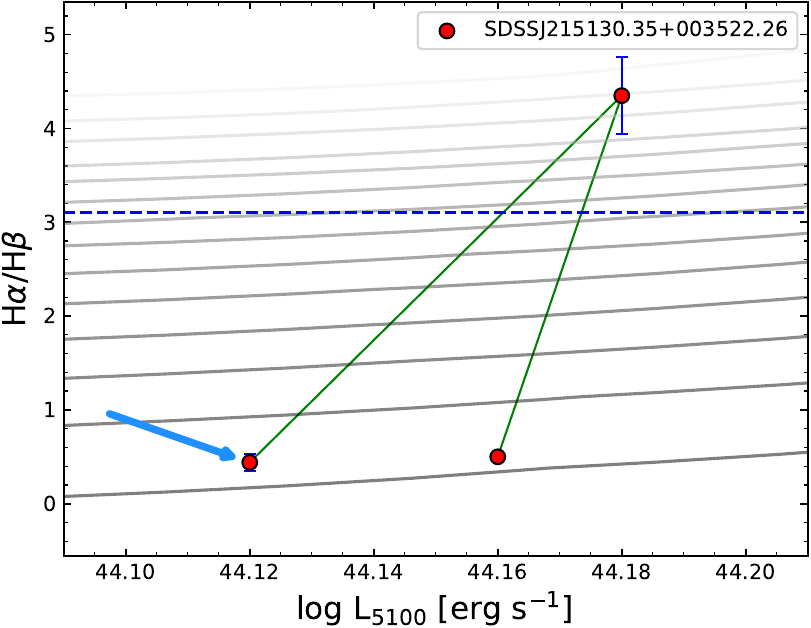}

    \includegraphics[width=0.21\textwidth]{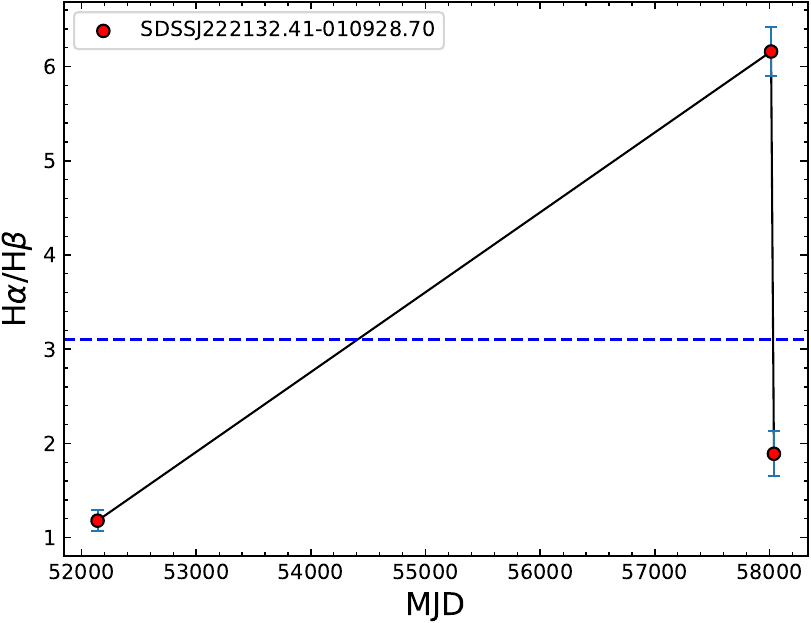}
    \includegraphics[width=0.21\textwidth]{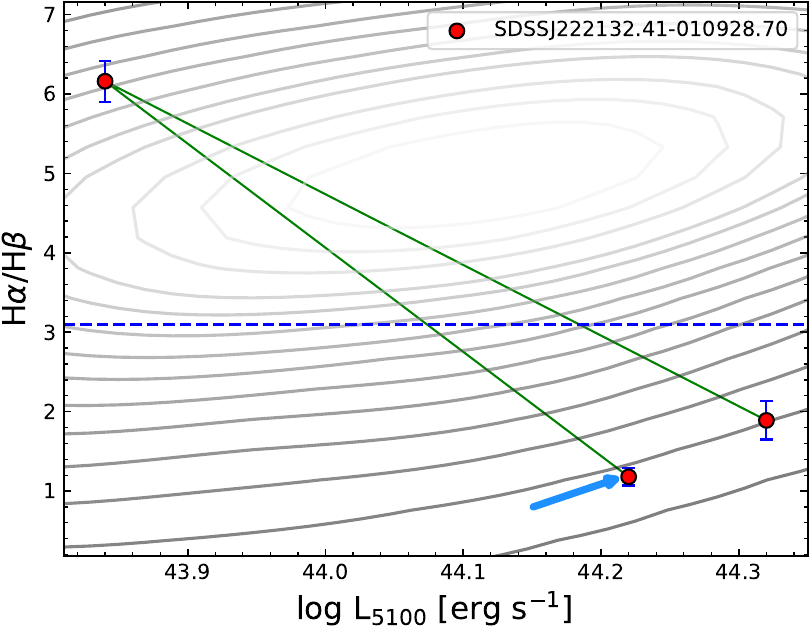}
    \includegraphics[width=0.21\textwidth]{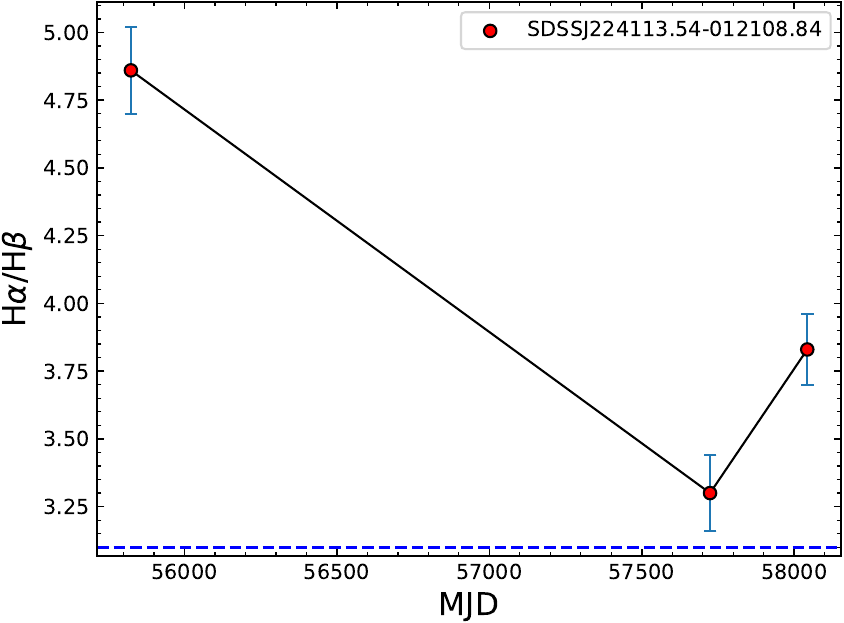}
    \includegraphics[width=0.21\textwidth]{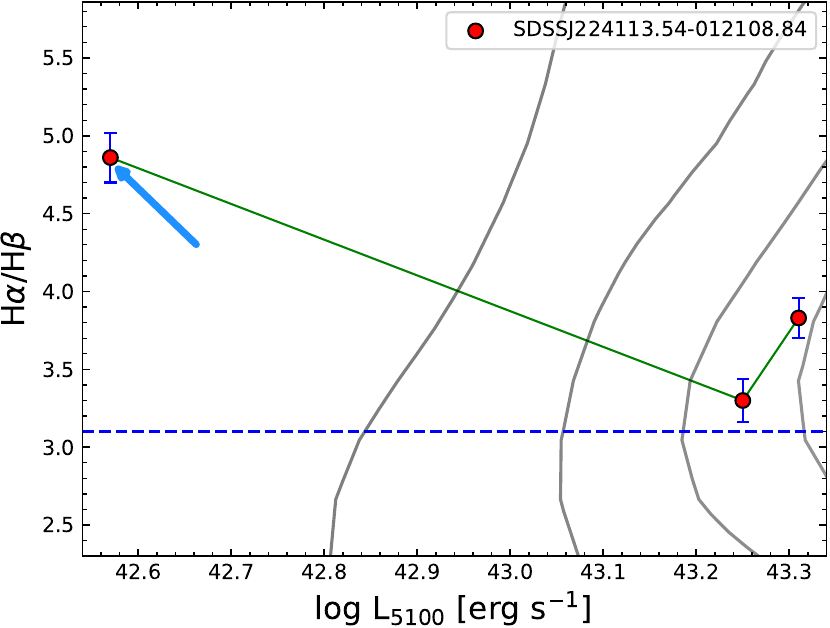}

    \includegraphics[width=0.21\textwidth]{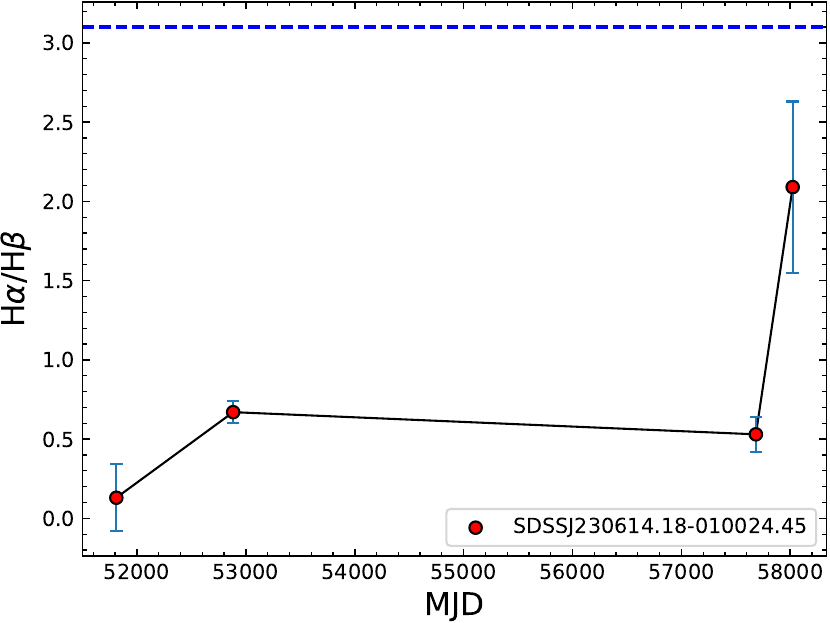}
    \includegraphics[width=0.21\textwidth]{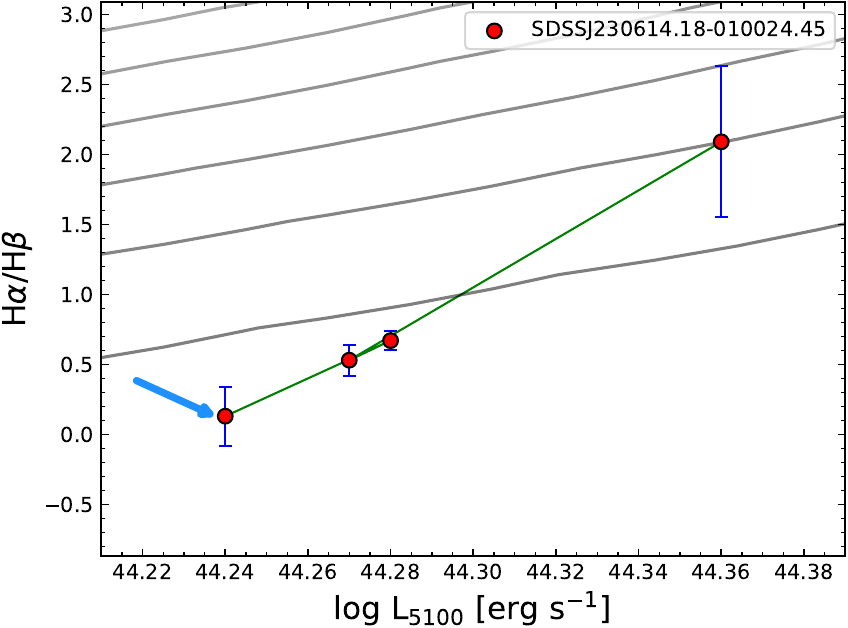}
    \includegraphics[width=0.21\textwidth]{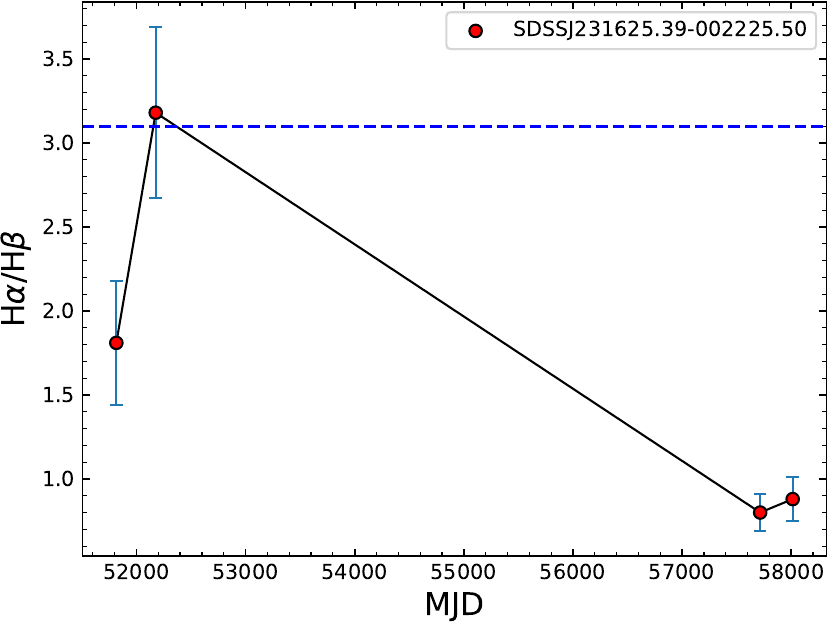}
    \includegraphics[width=0.21\textwidth]{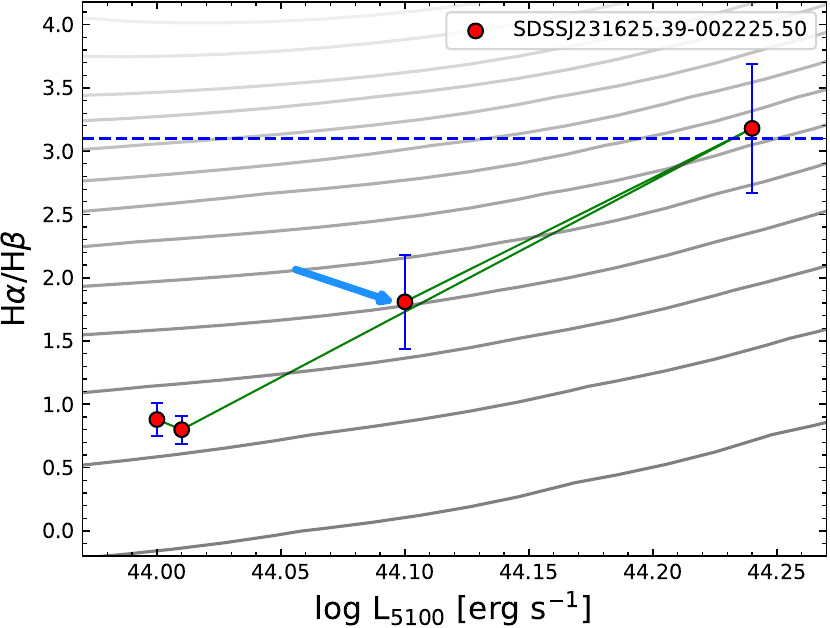}

    \caption{Panels are similar to Figures \ref{fig:balmer_decre_others}, but for the remaining sources.}
    \label{fig:balmer_decre_others2}
\end{figure*}

\begin{longrotatetable}
\begin{longtable}{llccccccr}
\caption{Details of the sources in our sample (full table can be found in the electronic version)}
\label{tab:table-observe}\\
\hline
Sample & Source & \multicolumn{1}{c}{Plate} & \multicolumn{1}{c}{MJD} & \multicolumn{1}{c}{Fiber} & \multicolumn{1}{c}{RA} & \multicolumn{1}{c}{Dec} & \multicolumn{1}{c}{z} & \multicolumn{1}{c}{Ref.} \\ \hline
\endfirsthead
\multicolumn{9}{c}%
{{\bfseries Table \thetable\ continued from previous page}} \\
\hline
Sample & Source & \multicolumn{1}{c}{Plate} & \multicolumn{1}{c}{MJD} & \multicolumn{1}{c}{Fiber} & \multicolumn{1}{c}{RA} & \multicolumn{1}{c}{Dec} & \multicolumn{1}{c}{z} & \multicolumn{1}{c}{Ref.} \\ \hline
\endhead
\hline
\endfoot
\endlastfoot
\citet{lamassa2015} & SDSSJ015957.6+003310 & 403 & 51871 & 549 & 29.990148 & 0.55291729 & 0.31182525 & legacy \\
 &  & 3609 & 55201 & 524 & 29.990161 & 0.55290301 & 0.3120502 & commissioning \\
 &  & 9384$^{\dagger{}}$ & 58080 & 605 & 29.990187 & 0.55289999 & 0.3121399 & ELG\_SGC \\ \hline

\citet{Runnoe2016} & SDSSJ101152.98+544206.4 & 945 & 52652 & 22 & 152.97075 & 54.701776 & 0.2463892 & legacy \\
 &  & 8181 & 57073 & 827 & 152.97078 & 54.701786 & 0.24651496 & eboss \\ \hline

\citet{macleod2016} & SDSSJ022556.0+003026 & 705 & 52200 & 342 & 36.483662 & 0.50746587 & 0.50387114 & southern \\
 &  & 1508 & 52944 & 556 & 36.483654 & 0.507422 & 0.50425416 & southern \\
 &  & 3615 & 55179 & 641 & 36.483664 & 0.50742591 & 0.5040848 & commissioning \\
... & ... & ... & ... & ... & ... & ... & ... & ... \\ 
... & ... & ... & ... & ... & ... & ... & ... & ... \\ \hline
\end{longtable}
\footnotesize{{\sc Notes.} Columns are as follows: (1) Reference paper. (2) SDSS identifier for the source. (3) SDSS Plate ID. (4) SDSS date of observation (in modified Julian day). (5) SDSS Fiber ID. (6) right ascension (RA) in degrees. (7) declination in degrees. (8) redshift. (9) original source used to collect the spectrum. Sources marked with $\dagger{}$ in column (3) have either wrong redshift, noisy spectrum with no broad \hb{} detected, or problems with the FITS file. Such sources are removed from further analysis.}
\end{longrotatetable}

\begin{longrotatetable}
\begin{longtable}{llccccccr}
\caption{Spectral parameters for the sample using {\sc PyQSOFit} - I (full table can be found in the electronic version)}
\label{tab:table-pyqsofit-ew}\\
\hline
Sample & Source & \multicolumn{1}{c}{MJD} & \multicolumn{1}{c}{\begin{tabular}[c]{@{}c@{}}log L$_{3000}$\\ {[}erg/s{]}\end{tabular}} & \multicolumn{1}{c}{\begin{tabular}[c]{@{}c@{}}log L$_{5100}$\\ {[}erg/s{]}\end{tabular}} & \multicolumn{1}{c}{\begin{tabular}[c]{@{}c@{}}log L$_{6000}$\\ {[}erg/s{]}\end{tabular}} & \multicolumn{1}{c}{\begin{tabular}[c]{@{}c@{}}EW(MgII)\\ {[}\AA{]}\end{tabular}} & \multicolumn{1}{c}{\begin{tabular}[c]{@{}c@{}}EW(H$\beta$)\\ {[}\AA{]}\end{tabular}} & \multicolumn{1}{c}{\begin{tabular}[c]{@{}c@{}}EW(H$\alpha$)\\ {[}\AA{]}\end{tabular}} \\ \hline
\endfirsthead
\multicolumn{9}{c}%
{{\bfseries Table \thetable\ continued from previous page}} \\
\hline
Sample & Source & \multicolumn{1}{c}{MJD} & \multicolumn{1}{c}{\begin{tabular}[c]{@{}c@{}}log L$_{3000}$\\ {[}erg/s{]}\end{tabular}} & \multicolumn{1}{c}{\begin{tabular}[c]{@{}c@{}}log L$_{5100}$\\ {[}erg/s{]}\end{tabular}} & \multicolumn{1}{c}{\begin{tabular}[c]{@{}c@{}}log L$_{6000}$\\ {[}erg/s{]}\end{tabular}} & \multicolumn{1}{c}{\begin{tabular}[c]{@{}c@{}}EW(MgII)\\ {[}\AA{]}\end{tabular}} & \multicolumn{1}{c}{\begin{tabular}[c]{@{}c@{}}EW(H$\beta$)\\ {[}\AA{]}\end{tabular}} & \multicolumn{1}{c}{\begin{tabular}[c]{@{}c@{}}EW(H$\alpha$)\\ {[}\AA{]}\end{tabular}} \\ \hline
\endhead
\hline
\endfoot
\endlastfoot
\citet{lamassa2015} & SDSSJ015957.6+003310 & 51871 & 44.09$\pm$0.01 & 44.18$\pm$0.01 & 44.14$\pm$0.02 & -- & 75.65$\pm$4.63 & 109.55$\pm$12.22 \\
 &  & 55201 & 43.59$\pm$0.00 & 43.92$\pm$0.00 & 43.95$\pm$0.01 & 173.30$\pm$48.02 & 71.91$\pm$3.90 & 28.18$\pm$2.34 \\ \hline

\citet{Runnoe2016} & SDSSJ101152.98+544206.4 & 52652 & -- & 43.75$\pm$0.00 & 43.54$\pm$0.01 & -- & 155.16$\pm$8.21 & 825.03$\pm$10.38 \\
 &  & 57073 & -- & 43.50$\pm$0.00 & 43.40$\pm$0.02 & -- & 29.33$\pm$15.54 & 33.35$\pm$3.69 \\ \hline

\citet{macleod2016} & SDSSJ022556.0+003026 & 52200 & 44.09$\pm$0.00 & 44.26$\pm$0.01 & 44.24$\pm$0.02 & 99.62$\pm$25.88 & 18.95$\pm$4.42 & -- \\
 &  & 52944 & 44.35$\pm$0.01 & 44.38$\pm$0.01 & 44.32$\pm$0.02 & 113.59$\pm$13.34 & 35.63$\pm$3.69 & -- \\
 &  & 55179 & 43.55$\pm$0.01 & 44.06$\pm$0.01 & 44.15$\pm$0.02 & 139.68$\pm$23.14 & 66.35$\pm$2.80 & 13.93$\pm$3.35 \\
... & ... & ... & ... & ... & ... & ... & ... & ... \\
... & ... & ... & ... & ... & ... & ... & ... & ... \\ \hline
\end{longtable}
\footnotesize{{\sc Notes.} Columns are as follows: (1) Reference paper. (2) SDSS identifier for the source. (3) Date of observation (in modified Julian day). (4) AGN continuum luminosity at 3000\AA~. (5) AGN continuum luminosity at 5100\AA~. (6) AGN continuum luminosity at 6000\AA~. (7) Equivalent width (EW) of \mg{}. (8) EW of \hb{}. (9) EW of \ha{}.}
\end{longrotatetable}

\begin{longrotatetable}
\begin{longtable}{llcccccccr}
\caption{Spectral parameters for the sample using {\sc PyQSOFit} - II (full table can be found in the electronic version)}
\label{tab:table-pyqsofit}\\
\hline
Sample & Source & MJD & \begin{tabular}[c]{@{}c@{}}FWHM(H$\beta_{\rm broad}$)\\ {[}km s$^{-1}${]}\end{tabular} & R$_{\rm FeII}$ & \begin{tabular}[c]{@{}c@{}}log L$_{5100}$\\ {[}erg s$^{-1}${]}\end{tabular} & \begin{tabular}[c]{@{}c@{}}log L$_{\rm bol}$\\ {[}erg s$^{-1}${]}\end{tabular} & \begin{tabular}[c]{@{}c@{}}log M$_{\rm BH}$\\ {[}M$_{\odot}${]}\end{tabular} & log $\lambda_{\rm Edd}$ & \begin{tabular}[c]{@{}c@{}}$\Delta$L$_{5100}$\\ {[}10$^{44}$ erg s$^{-1}${]}\end{tabular} \\ \hline\\
\endfirsthead
\multicolumn{10}{c}%
{{\bfseries Table \thetable\ continued from previous page}} \\
\hline
Sample & Source & MJD & \begin{tabular}[c]{@{}c@{}}FWHM(H$\beta_{\rm broad}$)\\ {[}km s$^{-1}${]}\end{tabular} & R$_{\rm FeII}$ & \begin{tabular}[c]{@{}c@{}}log L$_{5100}$\\ {[}erg s$^{-1}${]}\end{tabular} & \begin{tabular}[c]{@{}c@{}}log L$_{\rm bol}$\\ {[}erg s$^{-1}${]}\end{tabular} & \begin{tabular}[c]{@{}c@{}}log M$_{\rm BH}$\\ {[}M$_{\odot}${]}\end{tabular} & log $\lambda_{\rm Edd}$ & \begin{tabular}[c]{@{}c@{}}$\Delta$L$_{5100}$\\ {[}10$^{44}$ erg s$^{-1}${]}\end{tabular} \\ \hline\\
\endhead
\hline
\endfoot
\endlastfoot
\citet{lamassa2015} & SDSSJ015957.6+003310 & 51871 & 6220$\pm$435 & 0.22$\pm$0.06 & 44.18$\pm$0.01 & 45.15$\pm$0.01 & 8.59$\pm$0.06 & -1.58$\pm$0.07 & 0 \\
 &  & 55201 & 16866$\pm$1114 & 1.06$\pm$0.06 & 43.92$\pm$0.00 & 44.89$\pm$0.00 & 9.32$\pm$0.06 & -2.58$\pm$0.06 & -0.692 \\ \hline

\citet{Runnoe2016} & SDSSJ101152.98+544206.4 & 52652 & 2826$\pm$122 & 0.60$\pm$0.04 & 43.75$\pm$0.00 & 44.72$\pm$0.00 & 7.69$\pm$0.00 & -1.11$\pm$0.01 & 0 \\
 &  & 57073 & 17862$\pm$9079 & -- & 43.50$\pm$0.00 & 44.47$\pm$0.00 & 9.16$\pm$0.44 & -2.84$\pm$0.44 & -0.245 \\ \hline

\citet{macleod2016} & SDSSJ022556.0+003026 & 52200 & 7181$\pm$3545 & 0.00$\pm$0.06 & 44.26$\pm$0.01 & 45.23$\pm$0.01 & 8.75$\pm$0.43 & -1.67$\pm$0.44 & 0 \\
 &  & 52944 & 7715$\pm$1518 & 0.38$\pm$0.08 & 44.38$\pm$0.01 & 45.35$\pm$0.01 & 8.88$\pm$0.17 & -1.67$\pm$0.18 & 0.605 \\
 &  & 55179 & 16218$\pm$1009 & 1.11$\pm$0.06 & 44.06$\pm$0.01 & 45.03$\pm$0.01 & 9.36$\pm$0.06 & -2.47$\pm$0.06 & -0.667 \\ 
... & ... & ... & ... & ... & ... & ... & ... & ... & ... \\
... & ... & ... & ... & ... & ... & ... & ... & ... & ... \\ \hline
\end{longtable}
\footnotesize{{\sc Notes.} Columns are as follows: (1) Reference paper. (2) SDSS identifier for the source. (3) Date of observation (in modified Julian day). (4) Full-width at half-maximum for the broad-component of \hb{}. (5) the ratio \rfe{}. (6) Optical continuum luminosity at 5100\AA~. (7) Bolometric luminosity. (8) Black hole mass. (9) Eddington ratio. (10) difference between the L$_{5100}$ of the source observed at a given MJD to the first observation of the same source.}
\end{longrotatetable}

\end{document}